\documentclass[truedoublespace,plainchapterheads]{ccw_chithesis} 
\usepackage{astroextras}
\usepackage{natbib}
\usepackage{amsmath}
\usepackage{fancyvrb}
\usepackage{color}
\usepackage{svn-multi}
\usepackage{threeparttablex}
\usepackage{longtable}
\usepackage{titlesec}
\usepackage{epigraph}
\usepackage[letterpaper, includefoot, top=1in, bottom=1in, left=1in, right=1in]{geometry}
\usepackage[bookmarks=true, breaklinks=true, implicit=true, debug=false, plainpages=false, pagebackref=true, colorlinks=true, 
linkcolor=blue, urlcolor=blue, citecolor=blue, filecolor=blue, 
linktocpage=true, hyperindex=true,pdftitle={Buder Thesis}, bookmarksnumbered=true]{hyperref}
\usepackage{breakurl}
\graphicspath{{./figures/}}
\newcommand{\thread}[1]{\texttt{#1}}
\newcommand{\process}[1]{\texttt{#1}}
\newcommand{\fixme}[1]{\typeout{FIXME: #1 on input line \the\inputlineno}}
\newcommand{\rinterval}[0]{\ensuremath{1.48^{+1.06}_{-0.92}}}
\newcommand{\rupper}[0]{3.3}
\DeclareMathOperator{\erf}{erf}

\makeatletter
\let\oldfigure\figure
\def\figure{\@ifnextchar[\figure@i \figure@ii}
\def\figure@i[#1]{\oldfigure[#1]\centering}
\def\figure@ii{\oldfigure\centering}
\makeatother

\svnid{$Id: thesis.tex 148 2012-07-19 22:42:10Z ibuder $}

\titleformat{\subsection}{\normalfont\large\center}{\thesubsection}{1em}{}
\begin{document}

\title{Measurement of the CMB Polarization at 95\,GHz from QUIET}
\author{Immanuel David Buder}
\department{Physics}
\division{Physical Sciences}
\degree{Doctor of Philosophy}
\date{AUGUST 2012}
\maketitle
\makecopyright

\dedication
\begin{center}
to Mr. Wallin, who never doubted\\
and\\
to Bruce, who made it possible
\end{center}

\pagestyle{plain} 
\tableofcontents
\listoffigures
\listoftables

\acknowledgments{
\svnid{$Id: acknowledgments.tex 149 2012-07-23 20:51:50Z ibuder $}
\epigraph{Here friend, have this!\\  Okay, thanks.}{Team Potato}

I could never have written this thesis without the support of many people.
Here I thank as many as memory will permit.

First thanks go to my adviser Bruce Winstein, who was also the PI of QUIET.
Without him I never would have gotten involved in CMB physics.
Bruce taught me how to be a really careful scientist.
He was also incredibly generous with his time and resourceful at getting me whatever I needed to get that science done.
My greatest regret is that he never got to see this work finished.

My adviser Stephan Meyer stepped in at a time when we were all reeling from Bruce's death.
Steve helped me understand all the things I thought I understood---but didn't.
Steve was also really good at keeping me focused on the next steps at every stage in this process.

Akito Kusaka took on responsibility beyond what is normally expected of a post-doc.
He kindly pointed out mistakes in every memo I sent him.
Any ones remaining here are my own.

My thesis committee, Paolo Privitera, Michael Turner, and David Biron, provided many insightful comments and suggestions. This thesis is much better as a result.

Thanks to the rest of the Chicago QUIET group, Colin Bischoff, Alison Brizius, Yuji Chinone, Dan Kapner, and Osamu Tajima, for keeping things fun.
My Fermilab colleagues, Hogan Nguyen and Donna Kubik, kept up the enthusiasm when the Chicago team was burnt out.

Thanks to the Q-band deployment team for putting up with me even when I was too low on oxygen to make any sense.
Special thanks to the engineers and technician who maintained and operated the telescope, Jos\'e Cort\'es, Cristobal Jara, Freddy
Mu\~noz, and Carlos Verdugo.
Without them, we never would have gotten past the first mount stall.
Thanks also to the other Atacama experiments, especially APEX, ACT, and ALMA, for supporting our operations in divers ways.

Thanks to everyone else in the QUIET Collaboration for numerous comments, discussions, and a whole lot of science.

My office mates, Cora Dvorkin, Sam Leitner, Denis Erkal, and Yin Li, provided job-seeking support, laughs, and daily encouragement. 

Thanks to my classmates and friends in Chicago, who were uniformly, point-wise, and L2 supportive.
Thanks to my Tpot friends, whose support was neither closed nor bounded.

My family thought I was doing something worthwhile, even when they didn't understand it.

Some work was performed on the Joint Fermilab-KICP
Supercomputing Cluster, supported by grants from Fermilab,
the Kavli Institute for Cosmological Physics, and
the University of Chicago.
Some work was performed on the
Central
Computing System, owned and operated by the Computing
Research Center at KEK.
This research used resources of the National Energy Research Scientific Computing Center, which is supported by the Office of Science of the U.S. Department of Energy under Contract No. DE-AC02-05CH11231. 
Some of the results in this paper have been derived using the HEALPix \citep{gorski_healpix} package.

Last and least, I thank the mount, which lasted much longer than anyone thought it would.

}

\begin{abstract}
\svnid{$Id: abstract.tex 135 2012-06-19 22:43:19Z ibuder $}

Despite the great success of precision cosmology, cosmologists cannot fully explain the initial conditions of the Universe.
Inflation, an exponential expansion in the first $\approx 10^{-36}$\,s, is a promising potential explanation.
A generic prediction of inflation is odd-parity (B-mode) polarization in the cosmic microwave background (CMB).
The Q/U Imaging ExperimenT (QUIET) aimed to limit or detect this polarization.

We built a coherent pseudo-correlation microwave polarimeter.
An array of mass-produced ``modules'' populated the focal plane of a 1.4-m telescope.
Each module had a sensitivity to polarization of 756\,$\mu$K$\sqrt{\textrm{s}}$ with a bandwidth of $10.7\pm 1.1$\,GHz centered at $94.5\pm0.8$\,GHz; the combined sensitivity was $87\pm7\,\mu$K$\sqrt{\textrm{s}}$.
We incorporated ``deck'' rotation, an absorbing ground screen, a new time-stream ``double-demodulation'' technique, and optimized optics into the design to reduce instrumental polarization.
We observed with this instrument at the Atacama Plateau in Chile between August 2009 and December 2010.
We collected 5336.9\,hours of CMB observation and 1090\,hours of astronomical calibration.

This thesis describes the analysis and results of these data.
We characterized the instrument using the astronomical calibration data as well as purpose-built artificial sources.
We developed noise modeling, filtering, and data selection following a blind-analysis strategy.
Central to this strategy was a suite of 32 null tests, each motivated by a possible instrumental problem or systematic effect.
We also evaluated the systematic errors in the blind stage of the analysis before the result was known.
We then calculated the CMB power spectra using a pseudo-$C_\ell$ cross-correlation technique that suppressed contamination and made the result insensitive to noise bias.
We measured the first three peaks of the E-mode spectrum at high significance and limited B-mode polarization ($r= \rinterval$ at 68\% confidence and $r < \rupper$ at 95\% confidence).
Systematic errors were well below ($r < 0.01$) our B-mode polarization limit.
This systematic-error reduction was a strong demonstration of technology for application in more sensitive, next-generation CMB experiments.

\end{abstract}

\mainmatter
\svnid{$Id: introduction.tex 149 2012-07-23 20:51:50Z ibuder $}

\chapter{Introduction}
\label{sec:intro}
\epigraph{In the beginning the Universe was created. This has made a lot of people very angry and been widely regarded as a bad move.}{Douglas Adams}

Q/U Imaging ExperimenT (QUIET) measured the cosmic microwave background (CMB) polarization with the goal of constraining a signal from inflation.
The standard model of cosmology begins with inflation, an accelerated expansion of the Universe in the first $\approx 10^{-36}$\,s that sets its initial conditions.
Inflation generated a stochastic background of gravitational waves.  These gravitational waves created a small polarization anisotropy in the CMB, the radiation released when the early Universe became un-ionized.
We\footnote{Herein ``we'' means the members of the QUIET Collaboration, \url{http://quiet.uchicago.edu/}.  ``I'' indicates particular contributions of the author.} measured the CMB polarization and 
isolated the signal potentially bearing  evidence of inflation.
This thesis describes the QUIET instrument, observations, data analysis, and results including a limit ($r < \rupper$) on the inflation signal.

\section{Inflationary Cosmology}
Inflation provides the initial conditions for the standard model of cosmology ($\Lambda$CDM, see e.g. \cite{ryden}).
In the standard model, the Universe is assumed to be homogeneous and isotropic on large scales ($\gtrsim100$\,Mpc).
In General Relativity, these conditions require a metric of the form
\begin{equation}
ds^2 = -c^2dt^2 + a(t)^2\left(dx^2+dy^2+dz^2\right).
\end{equation}
This form is only valid for a nearly flat Universe, one of the assumptions of the standard model.
The Universe expanded from a hot, dense initial state; thus $a(t)$\footnote{Often called the ``scale factor.''} increases with time.
In addition to the background metric, the standard model specifies the energy content and how it behaves.
The evolution is governed by
\begin{eqnarray}
\label{eq:friedmann}
\left(\frac{1}{a}\frac{da}{dt}\right)^2 &=& \frac{8\pi G}{3c^2}\rho \\
\frac{1}{a}\frac{d^2a}{dt^2} &=& -\frac{4\pi G}{3c^2}(\rho + 3p),
\end{eqnarray}
where $\rho$ is the energy density and $p$ is the pressure.

Inflation was originally proposed by \cite{guth_inflation} to explain puzzling observations (see also \S\ref{sec:CMB}):
the Universe is homogeneous even across regions that were causally disconnected.
The Universe is (almost) flat, but a nearly flat Universe is an unstable solution.
Guth's final motivation was the absence of magnetic monopoles, which are generally expected in grand unified theories.
The generation of initial density perturbations is an additional motivation for an inflationary origin.

Although there are many models for the details of inflation, a simple, single-field, slow-roll model\footnote{This discussion follows \cite{Linde:2007fr}.} demonstrates the general features important for this thesis.
Suppose there is a scalar field $\phi$ with mass $m$\footnote{In this example I chose units $M_p^{-2}=8\pi G = 1$.} and potential
\begin{equation}
V(\phi) = m^2\phi^2/2.
\end{equation}
The corresponding equations of motion are
\begin{equation}
\ddot{\phi} + 3H\dot{\phi} = -m^2\phi
\end{equation}
and
\begin{equation}
H^2 + \frac{k}{a^2} = \left(\dot{\phi}^2 + m^2\phi^2\right)/6,
\end{equation}
where $k$ is $-1$, 0, or 1 for an open, flat, or closed Universe, respectively.
An overdot denotes a derivative with respect to time coordinate $t$, and $H\equiv\dot{a}/a$.
If the initial value of $\phi$ is $\gg1$ then the evolution is in the ``slow-roll'' regime with
\begin{equation}
\ddot{\phi} \ll 3H\dot{\phi},
\end{equation}
\begin{equation}
\label{eq:inflation_curvature}
H^2 \gg k/a^2,
\end{equation}
and
\begin{equation}
\dot{\phi}^2 \ll m^2\phi^2.
\end{equation}
Eq. \ref{eq:inflation_curvature} implies that spatial curvature becomes negligible.
In this regime the equations of motion simplify to
\begin{equation}
\label{eq:Hslowroll}
H = m\phi/\sqrt{6}
\end{equation}
 and
\begin{equation}
\dot{\phi} = -m\sqrt{2/3}.
\end{equation}
In the slow-roll regime, the change of $\phi$ is negligible so 
\begin{equation}
a(t) = e^{Ht} = e^{m\phi t/\sqrt{6}}.
\end{equation}
This exponential expansion continues until $\phi\approx1$.
If the initial scale factor and field values are $a_0$ and $\phi_0$, respectively, then the total expansion caused by inflation is\footnote{See \S1.7 of \cite{Linde:2005ht} for a detailed derivation.}
\begin{equation}
\label{eq:efold}
a/a_0 = e^{\phi_0^2/4}.
\end{equation}

This model can explain the ``puzzling observations'' above.
For realistic\footnote{As shown below, the perturbation amplitude $\delta\rho/\rho$ depends on $m$.
To be realistic, $m$ must be consistent with the observed perturbation amplitude.}
 $m\approx10^{-6}$, $a/a_0\approx 10^{10^{10}}$.
This expansion occurred in $\approx 10^{-36}$\,s.
Under a ``worst case'' assumption that the Universe before inflation had radius of curvature approximately the Planck length ($10^{-33}$\,cm), after inflation it would grow to $\approx10^{10^{10}}$\,cm.
This is much larger than the present observable Universe, $10^{28}$\,cm.
Thus all observations cosmologists have made correspond to a small patch that was a tiny fraction of the total volume of the initial Universe and easily causally connected.
The initial curvature is now unobservable because the present radius of curvature is much larger than any scale we can observe.
For similar reasons, the expansion erases any evidence of initial inhomogeneity and so reduces the density of monopoles that it is unlikely to find one in our Hubble volume.
At the end of inflation, $\phi$ decays into other elementary particles which interact and eventually reach thermal equilibrium.

Due to quantum fluctuations there were perturbations in the scalar field
\begin{equation}
\label{eq:delta_phi}
\delta\phi\approx \frac{H}{2\pi}.
\end{equation}
Differences in the value of $\phi$ caused inflation to end slightly earlier or later than the average value.
The change in the time of the end of inflation is
\begin{equation}
\delta t = \delta\phi/\dot{\phi} \approx \frac{H}{2\pi\dot{\phi}}.
\end{equation}
From Eq.~\ref{eq:friedmann},
\begin{equation}
\rho \propto H^2.
\end{equation}
After the end of inflation, the Universe was radiation-dominated so
\begin{equation}
\label{eq:Hrad}
H \propto t^{-1}.
\end{equation}
Combining Eqs. \ref{eq:delta_phi}--\ref{eq:Hrad},
\begin{equation}
\delta H/H \approx \delta t / t \approx \delta\rho / \rho \approx \frac{H^2}{2\pi\dot{\phi}}.
\end{equation}
The perturbation amplitude for a given wavelength depends on the fluctuations occurring when that wavelength was comparable to the horizon size.
Since $\frac{H^2}{2\pi\dot{\phi}}$ changed slowly during the slow-roll regime, the spectrum of initial perturbations was nearly scale-invariant.
Thus the combination of inflation and quantum fluctuations naturally produced the small anisotropy we observe today.

In addition to the density perturbations, inflation will in general create tensor perturbations (gravitational waves).
I adopted the convention \citep{wmap7_cosmology}
\begin{eqnarray}
\Delta^2_R(k) &\equiv& \frac{k^3<|R_k|^2>}{2\pi^2} = \Delta^2_R(k_0)\left(k/k_0\right)^{n_s-1} \\
\Delta_h^2(k) &\equiv& \frac{4k^3<|h_k|^2>}{2\pi^2} = \Delta_h^2(k_0)\left(k/k_0\right)^{n_t}
\end{eqnarray}
for nearly scale-invariant\footnote{$n_s\approx1$ and $n_t\ll1$.} spectra, where $R_k$ is a perturbation of the scalar curvature with comoving wavenumber $k$, $k_0$ is a pivot scale\footnote{In this thesis, $k_0=0.002$\,Mpc$^{-1}$.}, $n_s$ is the scalar spectral index, $h_k$ is a tensor perturbation, and $n_t$ is the tensor spectral index.

There is no observational evidence for tensor perturbations; however, many inflationary models predict them.
The tensor-to-scalar ratio
\begin{equation}
r\equiv \frac{\Delta^2_h(k_0)}{\Delta^2_R(k_0)}
\end{equation}
is a figure of merit for comparing tensor-perturbation searches
and comparing experimental results to different inflationary models.
In the slow-roll regime,
\begin{equation}
r = 16\epsilon \equiv 8 \left(\frac{1}{V}\frac{dV}{d\phi}\right)^2.
\end{equation}
In the $V=m^2\phi^2/2$ model, $r = 32/\phi^2$.
Using Eq.~\ref{eq:efold}, $r = 8/N$, where $N$ is the number of e-folds 
of inflation for scales corresponding to the size of the observable Universe.
For typical $N\approx60$, $r\approx0.13$.
Regardless of the model, detection of non-zero $r$ would be new, strong evidence for inflation.
Moreover, a measurement of $r$ would give information about the energy scale of inflation \citep{Liddle19931, baumann_cmbpol}
\begin{equation}
V^{1/4} = 1.06\times10^{16} \textrm{~GeV~} (r/0.01)^{1/4}.
\end{equation}
The best existing limit is $r < 0.2$ \citep{wmap7_cosmology}.
For reasons that will be explained below (\S\ref{sec:cmb:pol}), only $r\gtrsim 0.001$ is detectable in CMB polarization.
Therefore, a measurement of $r$ would indicate that inflation occurred at an energy comparable to the grand unification scale of particle physics.
Although models with arbitrarily small $r$ exist, 
current evidence and naturalness considerations suggest $r\geq 0.01$ \citep{Boyle06}.
Thus we, along with many other cosmologists, are interested in detecting tensor perturbations from inflation and measuring $r$ from CMB observations.

\section{The Cosmic Microwave Background}
\label{sec:CMB}

The CMB is relic radiation from when the Universe was younger and hotter.
The CMB was created when the formation of Hydrogen made the Universe transparent to photons.
By observing this radiation, cosmologists learn about the conditions of the early Universe.
The small anisotropy in the radiation intensity has constrained cosmological models.
The CMB polarization is the current frontier for probes of inflation because it has the smallest contaminating signals.

The CMB was formed at redshift $z\approx1100$ 
when the temperature became low enough (3000\,K) for the ionization fraction to decrease rapidly\footnote{Eqs. 9.9, 9.29, and 9.37 of \cite{ryden}}.
Once the Universe was un-ionized, the photon mean free path became much greater than the Hubble volume.
Most CMB photons last scattered at the time of recombination\footnote{Reionization at $z\approx10$ caused 10\% \citep{wmap7_cosmology} of photons to scatter again.}.
Thus the measured properties of the CMB directly correspond to early Universe physics.
The CMB has a blackbody spectrum \citep{fixsen_firas} with temperature $T_\textrm{cmb} = 2.72548 \pm 0.00057$\,K \citep{Fixsen_cmb_09}.
Spatially, the CMB is nearly uniform with $\Delta T/T\approx10^{-5}$ \citep{smoot_dmr}.
These small anisotropies are the result of the initial perturbations from inflation modulated by plasma physics at recombination \citep{Hu:2001bc, scott_review}.

\subsection{CMB Measurements and Their Implications}
CMB measurements, particularly of the temperature anisotropy, are one of the observational foundations of the standard cosmological model.
The basic result of a temperature anisotropy experiment is a map of the CMB temperature $T(\theta, \phi)$.
To extract the cosmologically relevant information, cosmologists typically decompose the map in spherical harmonics
\begin{equation}
\sum_{\ell,m}a_{\ell m}Y_{\ell m}(\theta, \phi) = T(\theta,\phi).
\end{equation}
If the fluctuations are Gaussian and isotropic then all cosmological information is contained in the power spectrum
\begin{equation}
C_\ell\equiv ~<|a_{\ell m}^2|>.
\end{equation}
At low $\ell$ ($\lesssim 100$) the Sachs--Wolfe effect \citep{sachs_wolfe} dominates, and $C_\ell\ell(\ell+1)$ is nearly constant.
At $\ell\gtrsim100$, the oscillations of the photon--baryon fluid create acoustic peaks.
At $\ell\gtrsim1000$, the non-zero width of the last-scattering surface damps the anisotropies.

Cosmologists can only observe a single realization of the sky.
At each $\ell$ there are only $2\ell+1$ modes.
Thus any measurement of the ensemble average $C_\ell$ has an irreducible sample variance\footnote{Also called ``cosmic variance.''} 
$\frac{2}{2\ell+1}C_\ell^2$.
Moreover, practical\footnote{Ground-based experiments rarely cover the full sky.  
Even space-based experimenters reduce the effective fraction of the sky to avoid strong Galactic emission.} experiments only cover a fraction of the sky, $f_\textrm{sky}<1$.
Limited sky fraction reduces the effective number of modes measured so the sample variance increases to $\frac{1}{f_\textrm{sky}}\frac{2}{2\ell+1}C_\ell^2$ \citep{scott94}.
Sample variance fundamentally limits the cosmological information that can be extracted from the CMB.

Existing CMB observations \citep{quad_2009, bicep_2009,wmap7_spectra, Reichardt:2011yv, Keisler:2011aw,Dunkley:2010ge}
combined with other observations (especially, galaxy clusters, weak gravitational lensing, supernova luminosity distances, and baryon acoustic oscillations (BAO). See e.g. \cite{wmap7_cosmology} for combined analyses)
have shown that
\begin{enumerate}
\item The Universe is nearly flat.  \cite{wmap7_cosmology} measured the curvature fraction $-0.0133 < \Omega_k < 0.0084$.
\item The initial perturbations were Gaussian and adiabatic (within measurement uncertainty).
\item The spectral index $n_s$ is almost, but not exactly 1.
\item The initial perturbations at a given wavelength were in phase \citep{Dodelson:2003ip}.
\item The Universe became un-ionized at $z\approx1100$ and reionized at $z\approx10$.
\item The current energy of the Universe is in dark energy ($72.8^{+1.5}_{-1.6}$\%) and dark matter ($22.7\pm1.4$\%).
\item Structure formed from small initial perturbations which grew because of gravitational collapse.
\end{enumerate}
These observations form the basis for the standard model of cosmology.
Particularly, inflation predicts 1--4.
To provide even more compelling evidence, we searched for the signal inflation generates in the CMB polarization.

\subsection{CMB Polarization Measurements}
\label{sec:cmb:pol}
\begin{figure}
\includegraphics[width=0.75\textwidth]{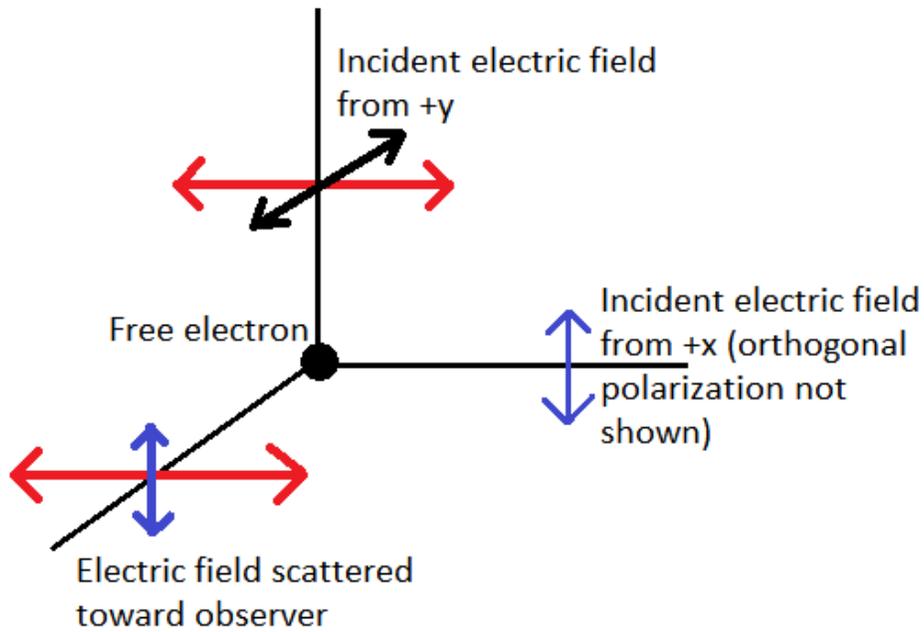}
\caption[Creation of Linear Polarization in
the CMB]{\label{fig:quad_linear_polarization_gen}
Quadrupole temperature anisotropies at decoupling created linear polarization in the CMB.
Because electromagnetic waves are transverse, only one polarization state from the $y$ direction can scatter toward us ($z$ direction).
Therefore if there was a difference in the electric field amplitudes in the $x$ and $y$ directions (equivalent to a radiation temperature difference), the scattered radiation will be polarized even though the incident radiation was not.
}
\end{figure}

The CMB is partially polarized
\citep{hu_white_review, Hu:2001bc}.
During decoupling, the quadrupole anisotropy in the local temperature led to a net linear polarization of the scattered radiation (Figure \ref{fig:quad_linear_polarization_gen}).
In Thomson scattering, incoming radiation polarized parallel to the outgoing direction does not contribute to scattering.
If the incoming radiation were isotropic, there would be no net polarization.
Because the incoming radiation had a quadrupole anisotropy, the intensity was different in the two outgoing polarization states.

The Stokes parameters
\begin{eqnarray}
\label{eq:I}
I &=& LL^* + RR^*\\
\label{eq:Q}
Q &=& 2\Re [L^*R]\\
\label{eq:U}
U &=& -2\Im [L^*R]\\
V &=& LL^* - RR^*
\end{eqnarray}
quantify the polarization properties of an electric field with components $E_x$ and $E_y$.  For later convenience I have introduced circular polarization components  
\begin{equation}
L = \frac{E_x + iE_y}{\sqrt{2}}
\end{equation}
and
\begin{equation}
R = \frac{E_x - iE_y}{\sqrt{2}};
\end{equation}
although the net circular polarization ($V$) is negligible for the CMB in theory \citep{cooray_circular_polarization}.
The Stokes parameters depend on the choice of coordinates\footnote{This discussion follows \cite{zaldariagga_spin2_spherical_harmonics}.}.
If the coordinate system is rotated by an angle $\psi$ so that $x\rightarrow x'$ and $y\rightarrow y'$,
\begin{eqnarray}
Q' &=& Q\cos{2\psi} + U\sin{2\psi} \\
U' &=& -Q\sin{2\psi} + U\cos{2\psi}.
\end{eqnarray}
The quantity
\begin{equation}
(Q\pm iU)' = e^{\mp2i\psi} (Q\pm iU)
\end{equation}
has spin $\pm2$, and I decomposed it into
\begin{equation}
Q\pm iU = \sum_{\ell,m} a^{\pm}_{\ell m}Y^{\pm}_{\ell m}
\end{equation}
with spin-2 spherical harmonics $Y^{\pm}$.
(The superscripts in $a^{\pm}$ are part of the names of the expansion coefficients, not exponents.)
The linear combinations
\begin{eqnarray}
a^{E}_{\ell m} &=& -\frac{a^{+}_{\ell m} + a^{-}_{\ell m}}{2}\\
a^{B}_{\ell m} &=& i\frac{a^+_{\ell m} - a^{-}_{\ell m}}{2}
\end{eqnarray}
have definite parity: $a^{E}$ (E mode) has even parity and $a^{B}$ (B mode) has odd parity.
The corresponding power spectra are
\begin{eqnarray}
C^{EE}_\ell &=& ~<a^{E*}_{\ell m}a^E_{\ell m}>\\
C^{BB}_\ell &=& ~<a^{B*}_{\ell m}a^B_{\ell m}>\\
C^{EB}_\ell &=& ~<a^{E*}_{\ell m}a^B_{\ell m}>.
\end{eqnarray}
Since E and B modes have different parity, $C^{EB}_\ell=0$ unless there is parity violation.
For the same reason, primordial scalar perturbations do not produce B modes.
However, tensor perturbations produce both E and B modes.
Since $r$ is small, scalar perturbations gave the dominant contribution to $C^{EE}_\ell$.
Therefore $C^{BB}_\ell$ is the most promising for a search for inflationary tensor perturbations.

Many experiments have measured non-zero $C^{EE}_\ell$ \citep{dasi_first_detection, cbi_2005, capmap_season3, quad_2009, bicep_2009, quiet_qband_result},
and the results agreed with the standard cosmological model. 
No experiment has detected B modes.
The best upper limit on inflationary B modes is $r < 0.72$ \citep{bicep_2009}.
Our main science goal in QUIET was to improve this limit.
Astrophysical foregrounds\footnote{The most important are synchrotron and dust emission from the Galaxy and gravitational lensing of the CMB.} make it extremely difficult to detect $r\lesssim0.001$ \citep{cmbpol_foreground_removal, kendrick_cmbpol_lensing}.

\section{The QUIET Experiment}
QUIET \citep{buder:77411D, quiet_instrument} measured the CMB polarization to limit the B-mode signal from inflation.
The QUIET instrument was a correlation polarimeter designed for high sensor density and low systematic errors.
We completed two seasons of observations with this instrument; this thesis describes results from the second season.

\begin{figure}
\includegraphics[width=1.0\textwidth]{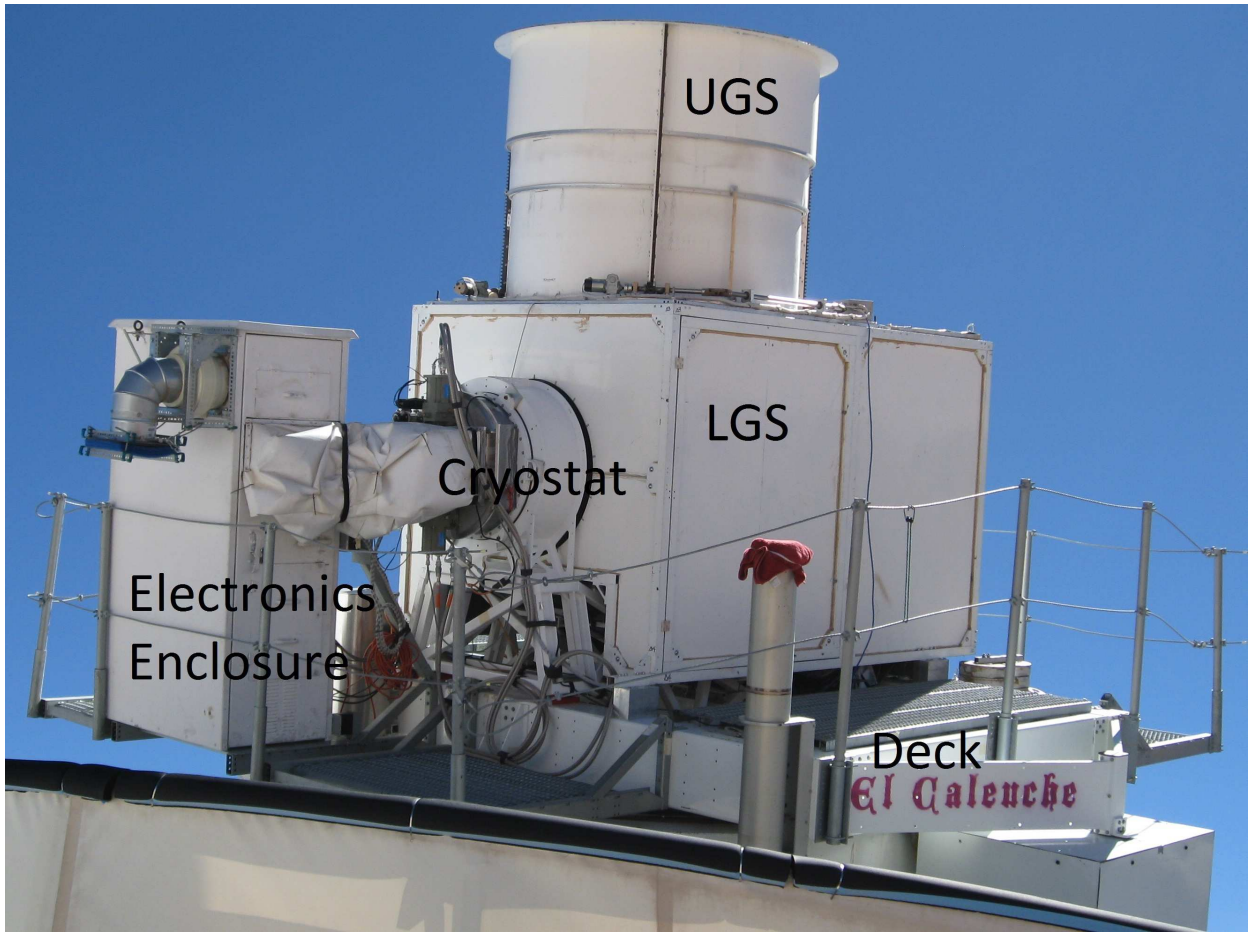}
\caption[QUIET Instrument Deployed]{\label{fig:quiet}
The QUIET instrument deployed to the Chajnantor Plateau in Chile's Atacama Desert to reduce atmospheric emission.
}
\end{figure}

The QUIET instrument (Figure \ref{fig:quiet}) was a correlation polarimeter based on coherent-amplification technology.
A 1.4-m side-fed Dragonian telescope collected the CMB radiation.
An array of feed horns at the telescope focus formed nearly Gaussian beams for each of 90 pixels\footnote{``Module'' means the polarization detectors associated with each pixel and, in a more general sense, the corresponding feed horn, septum polarizer, data, etc.
We attached six modules to orthomode transducers that made them sensitive to temperature anisotropies; I excluded these modules (``differential-temperature'') from analysis in this thesis.
Six of the remaining 84 modules were not functional or had no data passing cuts (\S\ref{sec:cuts}).
An additional four modules had a non-functional diode.
Thus the total diode yield for polarization was $308/336=92$\%.}.
At the end of each feed horn, the radiation entered a septum polarizer
that separated it into two circular polarization components ($L$ and $R$).
We used high--electron-mobility transistor (HEMT) based polarimeter modules to convert the two circular
polarization components into a measurement of the Stokes linear polarization parameters $Q$ and $U$. 
We cooled the feed horns, septum polarizers, and detectors to 20\,K.
We built a custom electronics system to bias the module active components and record the data.
(``Receiver'' means the cryostat, the components inside it, and the associated electronics hardware.)

We observed with this instrument from the Chajnantor Plateau, Chile from October 2008 until December 2010.
In the first season (October 2008--June 2009) we measured the radiation in the Q band (centered at 43\,GHz).
\cite{colin_thesis, quiet_qband_result,ali_thesis, yuji_thesis} described this measurement including the resulting limit $r < 2.2$.
In the second season (August 2009--December 2010) we measured the radiation in the W band (centered at $94.5\pm0.8$\,GHz with $10.7\pm 1.1$\,GHz bandwidth).
Each receiver (Q or W) worked at only one frequency band, and we could only observe with one receiver at a time.
\cite{quiet_wband_result} summarizes the CMB polarization results from the second season; this thesis describes them in more detail.
Details about the instrument, observations, analysis, and results apply to W band unless otherwise noted.

This thesis contains the following elements.
Chapter \ref{sec:intro} introduces QUIET.
Chapter \ref{sec:electronics} describes the electronics system.
Chapter \ref{sec:DAQ} describes data-management software and practices.
Chapter \ref{sec:observations} describes the W-band observation from Chile.
Chapter \ref{sec:analysis} describes the data analysis.
Chapter \ref{sec:results} describes the results and my conclusions.

\subsection{Module Operation Principles}
\label{sec:module_principles}
\begin{figure}
\includegraphics[width=1.0\textwidth]{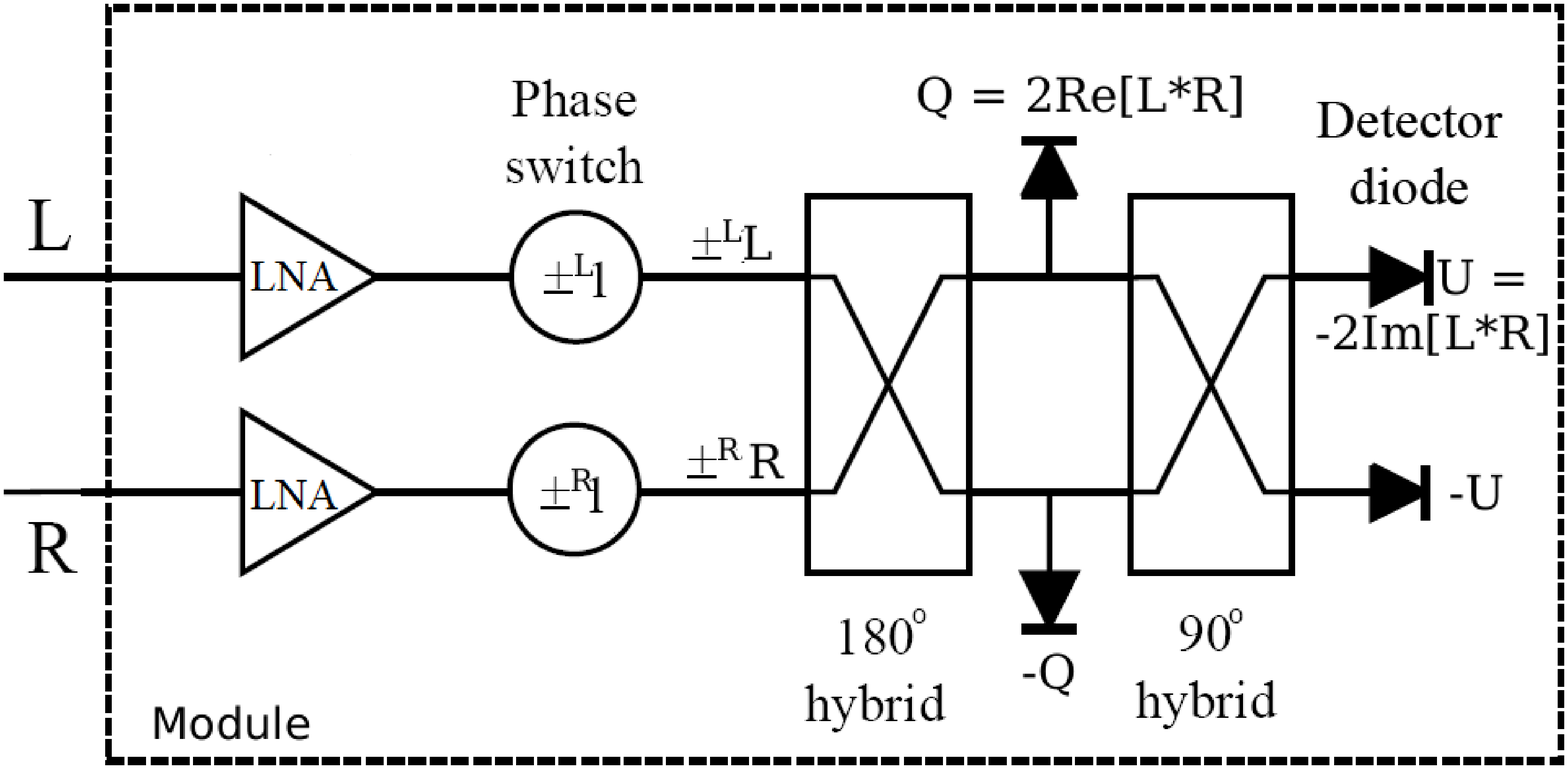}
\caption[QUIET Module Diagram]{\label{fig:module_processing_schematic}
The QUIET module converted $L$ and $R$ into measurements of $Q$ and $U$.
We modulated each measurement twice: once with each phase switch.
}
\end{figure}

The QUIET module implemented our polarization-detection scheme.
Here I outline the module operation; \S\ref{sec:modules} gives implementation details.
The following discussion applies to an ``ideal'' module---one with no systematic errors.
\cite{colin_thesis, quiet_instrument}; and \S\ref{sec:modules} of this thesis describe module non-idealities.
The module was a pseudo-correlation polarimeter \citep{gaier:296}: the polarization signal appeared as the product of two independently-processed polarization states (Figure \ref{fig:module_processing_schematic}).
The module operated entirely at radio frequency (RF).
There was no intermediate frequency or local oscillator.
The final stage of the module rectified and integrated the signal.

The module had two input ports.  Normally, these received the left and right circular-polarization components ($L$ and $R$) of the CMB.
Each input fed into a ``leg'' of the module.
Each leg contained a phase switch that applied $0\degr$ or $180\degr$ phase shift to the incoming radiation.
I denote the phase switch outputs by $\pm^L L$ and $\pm^R R$\footnote{The two phase switches can operate independently so there are four possible choices for the $\pm$ signs, denoted by ``$\pm^L$'' and ``$\pm^R$.''}.
The two legs then merged in a $180\degr$ hybrid coupler.
The coupler output $\frac{\pm^L L \pm^R R}{\sqrt{2}}$ and $\frac{\pm^L L \mp^R R}{\sqrt{2}}$.

Detector diodes\footnote{I use ``diode'' to refer to any of the detector diodes or their associated data.}  (Q1 and Q2) rectified half of each signal with resulting voltages
\begin{eqnarray}
V_\textrm{Q1} &\propto& \frac{\pm^L L \pm^R R}{2}\frac{\pm^L L^* \pm^R R^*}{2} =  
\frac{LL^{*} \pm^L\pm^R (L^*R + LR^{*} ) + RR^*}{4}  \nonumber\\ 
&=& I/4 \pm^L\pm^R \Re[L^*R]/2 = \frac{I \pm^L\pm^R Q}{4}\\
V_\textrm{Q2} &\propto& \frac{\pm^L L \mp^R R}{2}\frac{\pm^L L^* \mp^R R^*}{2} = \frac{LL^{*} \pm^L\mp^R (L^*R + LR^{*} ) + RR^*}{4} \nonumber\\
&=& \frac{I - \pm^L\pm^R Q}{4}
\end{eqnarray}
after using Eqs. \ref{eq:I} and \ref{eq:Q}.
The other halves of the signal entered a second, $90\degr$ hybrid coupler which output $\frac{\pm^L L\mp^R R-i(\pm^L L  \pm^R R)}{\sqrt{8}}$ and  $\frac{\pm^L L\mp^R R+i(\pm^L L  \pm^R R)}{\sqrt{8}}$.
Two more detector diodes (U1 and U2) rectified these outputs with resulting voltages
\begin{eqnarray}
V_\textrm{U1} &\propto& \frac{\pm^L L\mp^R R-i(\pm^L L  \pm^R R)}{\sqrt{8}} \frac{\pm^L L^*\mp^R R^*+i(\pm^L L^*  \pm^R R^*)}{\sqrt{8}} \nonumber\\
&=& \frac{1}{8} \left( LL^* - \pm^L\pm^R LR^* + iLL^* + \pm^L\pm^R iLR^* \right. \nonumber\\
&&- \pm^L\pm^R RL^* + RR^* - \pm^L\pm^R iRL^* - iRR^* \nonumber\\
&&- iLL^* + \pm^L\pm^R iLR^* + LL^* + \pm^L\pm^R  LR^* \nonumber\\
&&- \left.  \pm^L\pm^R iRL^* +iRR^* + \pm^L\pm^R RL^* + RR^* \right) \nonumber\\
&=& \frac{1}{8} \left( 2LL^* + \pm^L\pm^R  2iLR^*  + 2RR^* - \pm^L\pm^R 2iRL^* \right) \nonumber\\
&=& \frac{I -\pm^L\pm^R 2\Re[iL^*R]}{4}
= \frac{I -\pm^L\pm^R 2\Im[L^*R]}{4} \nonumber\\
&=& \frac{I -\pm^L\pm^R U}{4}
\end{eqnarray}

\begin{eqnarray}
V_\textrm{U2} &\propto& \frac{\pm^L L\mp^R R+i(\pm^L L  \pm^R R)}{\sqrt{8}} \frac{\pm^L L^*\mp^R R^*-i(\pm^L L^*  \pm^R R^*)}{\sqrt{8}} \nonumber\\
&=& \frac{1}{8} \left( LL^* - \pm^L\pm^R LR^* - iLL^* - \pm^L\pm^R iLR^* \right. \nonumber\\
&&- \pm^L\pm^R RL^* + RR^* + \pm^L\pm^R iRL^* + iRR^* \nonumber\\
&&+  iLL^* - \pm^L\pm^R iLR^* + LL^* + \pm^L\pm^R  LR^* \nonumber\\
&&+\left.  \pm^L\pm^R iRL^* -iRR^* + \pm^L\pm^R RL^* + RR^* \right) \nonumber\\
&=& \frac{1}{8} \left( 2LL^* - \pm^L\pm^R  2iLR^*  + 2RR^* + \pm^L\pm^R 2iRL^* \right) \nonumber\\
&=& \frac{I \pm^L\pm^R 2\Re[iL^*R]}{4}
= \frac{I -\pm^L\pm^R 2\Im[L^*R]}{4} \nonumber\\
&=& \frac{I \pm^L\pm^R U}{4}
\end{eqnarray}
after using Eqs. \ref{eq:I} and \ref{eq:U}.

The electronics (\S\ref{sec:electronics}) modulated the $L$ phase switch at 4\,kHz and the $R$ phase switch at 50\,Hz.
After digitization we recorded the average (``TP'') and demodulation (``DE'') of the data in both 4-kHz phase switch states.
For the Q1 diode these are
\begin{eqnarray}
\textrm{TP} &\propto& (I\pm^L\pm^RQ) + (I\mp^L\pm^RQ) \propto I
\\
\textrm{DE} &\propto& (I\pm^L\pm^RQ) - (I\mp^L\pm^RQ) \propto \pm^RQ.
\end{eqnarray}
We stored the data at 100\,Hz so adjacent data samples had opposite signs of $\pm^R$.
By subsequently differencing two adjacent samples we obtained ``double-demodulated'' (DD) data $\propto Q$.
(The Q2 diode yielded the same except with $Q\rightarrow-Q$.
The U diodes yielded the same except $Q\rightarrow U$.)
Double demodulation reduced the effect of amplifier gain variations (1/f noise) and eliminated instrumental polarization otherwise caused by gain mismatch between the two legs \citep{quiet_instrument}.
Furthermore each module (and therefore each pixel) fully characterized the linear polarization.

\svnid{$Id: instrument.tex 149 2012-07-23 20:51:50Z ibuder $}

\subsection{Optics}
Our optical system consisted of the telescope, ground screen, feed horns, and septum polarizers.
We chose a 1.4-m side-fed Dragonian antenna for the telescope.
A comoving, absorbing ground screen limited the sidelobes.
We used corrugated feed horns to form the beams for each module.
Each feed horn sent its radiation to a septum polarizer, which separated $L$ and $R$ into different waveguides for the module. 

\begin{figure}
\includegraphics[width=1.0\textwidth]{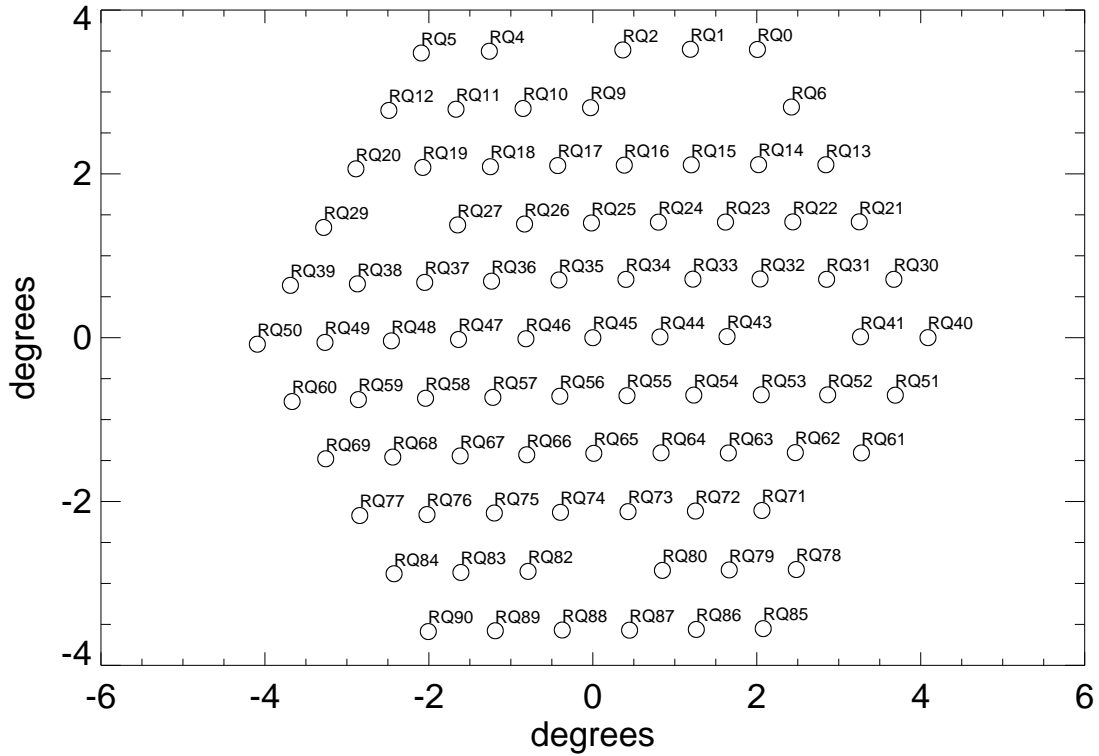}
\caption[Field of View]{\label{fig:array_layout}
The module array had a large ($8\degr$) field of view on the sky.
Each circle represents one module beam with diameter equal to the FWHM of $11\farcm66$.
The modules and their corresponding positions are labeled RQ0--RQ90.
Modules RQ7, 8, 28, and 42 did not function.
Modules RQ3 and RQ70 did not have any data passing selection.
There was no module at position RQ81.
}
\end{figure}

The telescope used a two-reflector design \citep{quiet_telescope} that satisfied the Mizuguchi condition \citep{mizugutch}.
The result had a wide field of view (Figure \ref{fig:array_layout}), low instrumental polarization, and low sidelobes.
We made the two mirrors out of aluminum, the backs
of which we light-weighted and connected by hexapods with adjustable turnbuckles to a steel support structure.
This support structure attached the telescope (and the remainder of the instrument) to the existing Cosmic Background Imager (CBI) mount \citep{cbi_instrument}.
The mount had three rotation axes (azimuth, elevation, and ``deck'').
The azimuth axis was capable of fast scanning ($\approx5\degr/$s) so that the typical scan frequency was well above the typical 1/f knee frequency.
By regularly (\S\ref{sec:obs:cmb}) rotating the third axis (deck), we modulated the instrumental-polarization axis, thereby suppressing instrumental polarization.

We enclosed the optics with a comoving, ambient-temperature, absorbing ground screen.
The ground screen had two main parts: lower (LGS) and upper (UGS).
The LGS was an aluminum box surrounding the mirrors with a hole for the cryostat.
The UGS was a telescoping cylinder that surrounded the path of the main beam.
Several smaller pieces formed the bottom ground screen (BGS) that covered the floor of the LGS and the gap between the LGS and the cryostat.
We coated the entire inner surface of the ground screen with microwave absorber Emmerson Cummings HR-10 Eccosorb \citep{ali_thesis}.
We sealed the Eccosorb with Volara (a microwave-transparent polyethylene foam) for weatherproofing.
We coated the outer surface with white paint to minimize radiative loading.

\begin{figure}
\includegraphics[width=0.75\textwidth]{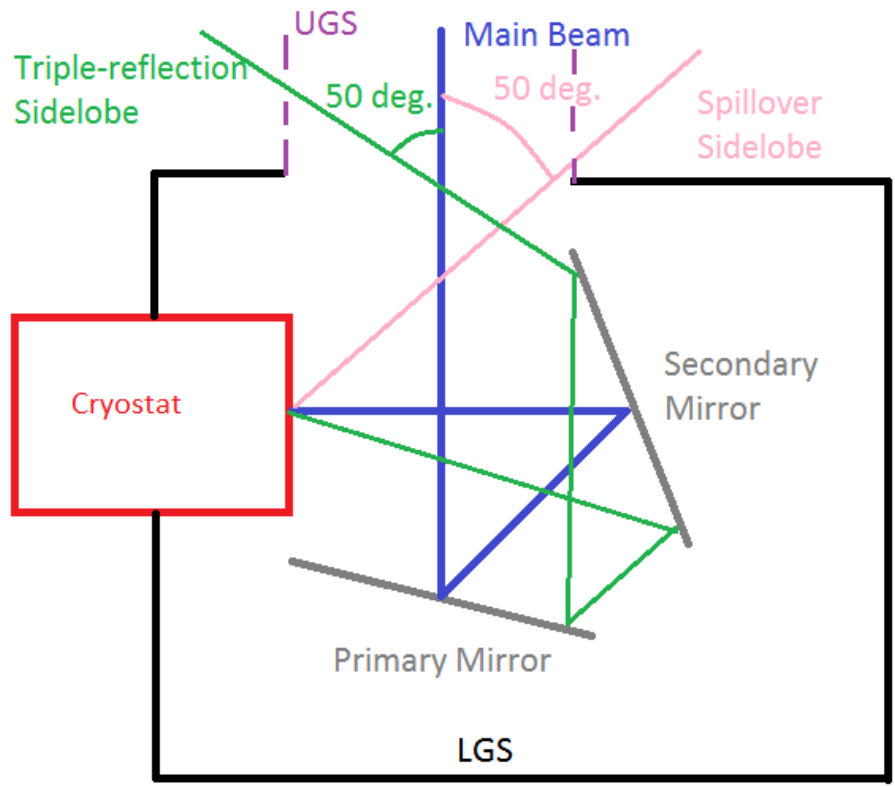}
\caption[Sidelobe Diagram]{\label{fig:sidelobe_cartoon}
Before installation of the upper ground screen (in the middle of the observing season) two significant far sidelobes escaped the ground screen.
The upper ground screen captured both sidelobes on a microwave absorber.
}
\end{figure}

Due to manufacturing delays, we did not install the UGS until January 2010.
As a result, two significant sidelobes were present in the 2009 data.
The ``triple-reflection'' sidelobe was due to a portion of the beam making an additional, unintended reflection off the secondary mirror.
The result was a narrow ($1\degr$) sidelobe $50\degr$ away from the main beam \citep{yuji_far_sidelobe}. 
The ``spillover'' sidelobe resulted from part of the main beam escaping above the secondary mirror and LGS (Figure \ref{fig:sidelobe_cartoon}).
The spillover appeared at $\approx50\degr$ from the main beam in the opposite direction from the triple-reflection sidelobe; however, the spillover sidelobe was much more extended ($\pm20\degr$ in the direction towards the main beam).
Holes in the BGS caused additional spillover \citep{sidelobe_measurement} (see also \S\ref{sec:noteworthy_events}).
We mitigated these effects with data selection (\S\ref{sec:cuts}).

A corrugated, linear-flare feed horn formed the beam of each module.
We constructed these feed horns as a hexagonally-close-packed ``platelet array'' \citep{platelet}.” We machined many large aluminum plates so that each plate formed a slice of all the feed horns,
transverse to the optical axis.  We then stacked the plates along the optical axis and diffusion bonded them together
to form a single component containing all the feed horns.
 The platelet design incorporated light-weighting holes
that also functioned as access holes for attachment screws. The fabrication costs of these arrays were at least an
order of magnitude less than for the same number of electroformed feed horns. Measurements of the return loss 
and beam characteristics of the platelet arrays showed performance comparable to an equivalent electroformed
design.
The feed-horn full-width half-maximum (FWHM) was $7\fdg0$--$8\fdg6$ across the array, and instrumental polarization was $-29$\,dB or better.

Radiation exited each feed horn into a septum polarizer \citep{bornemann1995, omt_memo}, which separated $L$ and $R$ for input to the QUIET detector modules.
The polarizer itself was a square waveguide with a stepped-thickness septum in the center.
The presence of the septum added a $90\degr$ phase shift to the TE$_{01}$ mode in the waveguide\footnote{\S2.2.3 of \cite{ali_thesis}}.
A waveguide splitter followed the polarizer and physically separated the two outputs into rectangular WR-10 waveguides with spacing compatible with the module waveguide inputs.
Although these waveguides carried components corresponding to circular polarization on the sky, the radiation in each waveguide was in a single, linearly-polarized (TE) mode.
Polarizer imperfections caused temperature to polarization ($I\rightarrow Q/U$) leakage\footnote{Appendix~10.3 of \cite{quiet_instrument}}.
Our lab measurements \citep{omt_measurements} implied leakages of $\approx0.3$\%.

\subsection{Modules}
\label{sec:modules}
\begin{figure}
\includegraphics[width=1.0\textwidth]{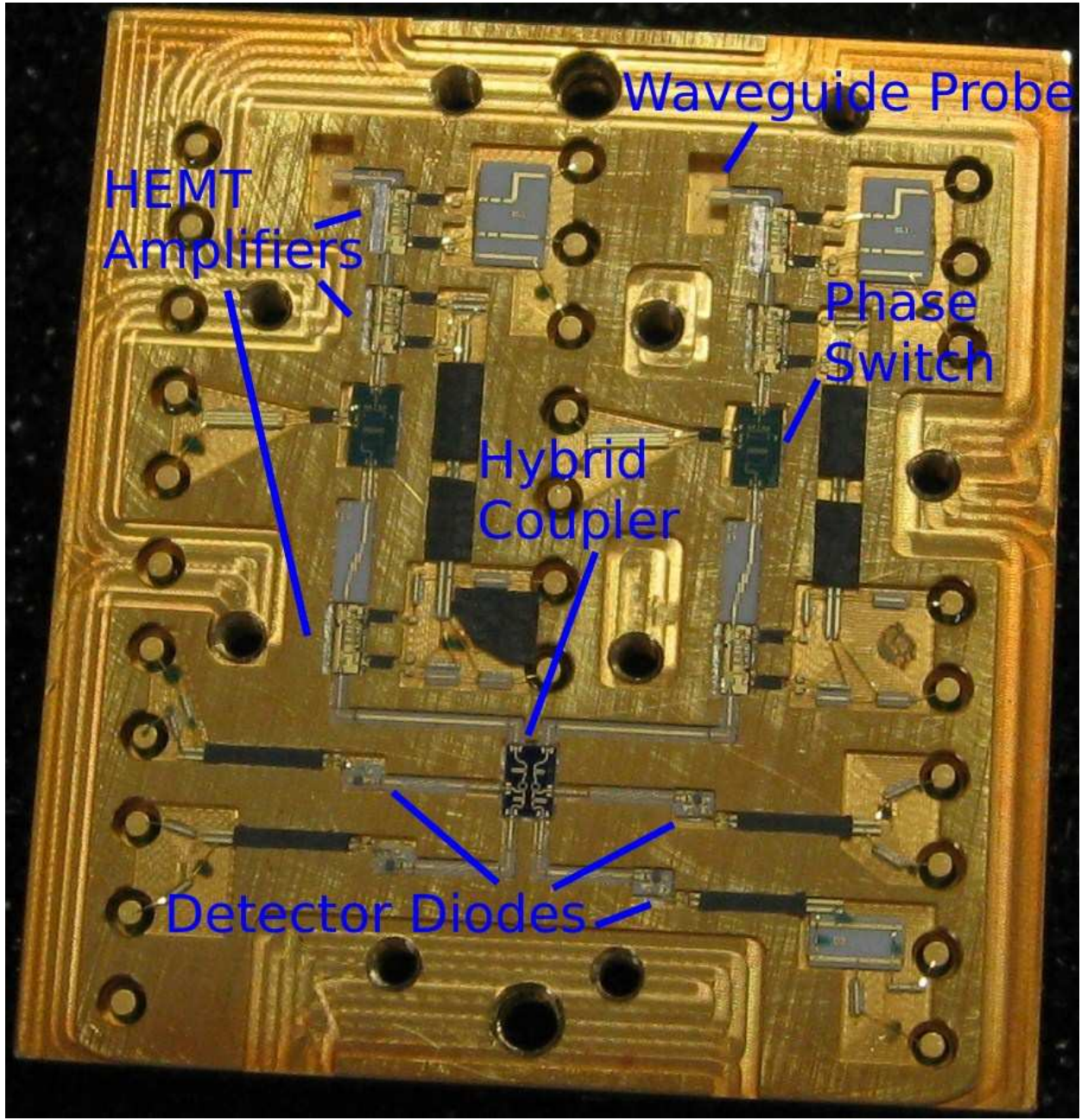}
\caption[Module Photograph]{\label{fig:module_picture}
The QUIET module was a miniaturized correlation polarimeter.
We used microstrip transmission lines to couple different components.
This reduced the size by a factor of 10 compared to a traditional waveguide-based design.
}
\end{figure}

The QUIET detectors were miniaturized polarimeter modules \citep{cleary:77412H} that converted $L$ and $R$ into $Q$ and $U$ as described in \S\ref{sec:module_principles}.” Each module was based on
HEMT amplifiers. 
Previous experiments including 
PIQUE \citep{pique2001},
DASI \citep{Leitch02},
CBI \citep{cbi_instrument},
\textit{WMAP} \citep{jarosik2003}, 
and CAPMAP \citep{capmap_instrument} also
used HEMT-based polarimeters, but we used advances
in millimeter-wave circuit technology and packaging \citep{Gaier20031167} to replace waveguide-block components and connections
with strip-line--coupled devices. The resulting modules were 3.2\,cm $\times$ 2.9\,cm (Figure \ref{fig:module_picture}), almost an order of magnitude smaller than a comparable waveguide design. 
Each module had two waveguide inputs matched to the two septum-polarizer outputs.
A probe in each port coupled the radiation into (50-$\Omega$) microstrip transmission lines.
The first active component in each leg was a HEMT-based monolithic-microwave-integrated-circuit (MMIC) amplifier\footnote{Herein, ``MMIC'' appearing alone means one of the QUIET MMIC amplifiers.}.
A second MMIC provided additional amplification before the phase switch.
After the phase switch, the signal passed through a bandpass filter before the third (final) MMIC.
After the final MMIC, the two legs combined in the hybrid coupler.
(For conceptual simplicity, \S\ref{sec:module_principles} separates the hybrid coupler into a first $180\degr$ coupler and second $90\degr$ coupler.  We manufactured both couplers as a single six-port device.)
Each coupler output (Q1, Q2, U1, and U2) passed through a second bandpass filter before rectification at its detector diode.

The MMICs were low-noise amplifiers with four HEMT stages manufactured in an Indium-Phosphide (InP) process \citep{module_doc}.
Each MMIC provided $\approx25$\,dB gain with noise temperature 50--80\,K
when cooled to a physical temperature of 20\,K\footnote{The improvement in noise temperature below a physical temperature of 40\,K was percent-level \citep{lab_noiseT_vs_physicalT}.
Including all the active and passive components, the typical module noise temperature was 100\,K.
See Appendix~\ref{app:array_sensitivity} for the total array sensitivity.
}.
At 20\,K the MMICs required $\approx$ 10\,mA of drain current bias, 0.5\,V drain voltage, and 0.1\,V gate voltage for optimum performance.
Because the number of module electrical connections was limited, we connected MMIC drains together\footnote{The second and third MMIC drains were common.  Some modules had two independent gate controls for the first MMIC and a single common gate for the second and third MMICs.  Other modules had one gate control per MMIC.  }
so that each leg had three gate connections and two drain connections.  Thus each module had 10 MMIC bias connections in total.
We optimized the sensitivity over these 10 parameters with a sparse--wire-grid calibration source \citep{ltd_wiregrid} and downhill-simplex algorithm\footnote{\S2.4.4 of \cite{ali_thesis} and \cite{becker_opt}}.

The phase switches were two-path circuits also fabricated in InP.  
One of the paths had an additional half-wavelength length so that radiation passing through it acquired a $180\degr$ phase shift. 
A PiN (p-type, intrinsic, n-type) diode controlled whether radiation could pass through each path.
Reverse biasing the diode put it into a high-impedance state, blocking that path.
When forward biased to 400\,$\mu$A, the transmission was $\approx$ unity (Figure \ref{fig:phase_switch_transmission_curve}).
Typically we forward biased one diode and reverse biased the other so the phase shift was either $0\degr$ or $180\degr$ as described in \S\ref{sec:module_principles}.
However, we periodically (\S\ref{sec:regular_observations}) made a diagnostic ``offset'' measurement by reverse biasing both diodes to block all RF power.
Since each module had two phase switches and each switch had two diodes, there were four phase switch bias currents per module. 
Differences in transmission between the two paths would cause $I\rightarrow Q/U$ leakage if we did not use double demodulation.
Because we used double demodulation, we did not attempt to balance the transmission by adjusting the diode biasing.

The hybrid coupler was a passive, planar InP device that summed its two inputs with either $0\degr$, $90\degr$, or $180\degr$ phase shifts as described in \S\ref{sec:module_principles}.
Our design used Lange couplers and Schiffman phase delay lines to perform these functions \citep{module_doc}.
Because the phase shifts were not exactly $90\degr$ or $180\degr$, the detector angles were different from their ideal values\footnote{e.g. the Q diodes acquired some $U$ sensitivity} \citep{sugarbaker_supp}, and 
the noise correlations among the diodes also changed from their ideal values \citep{white_noise_correlation}.

The detector diodes were Schottky diodes (Agilent HSCH-9161) which we operated in the square-law regime.
At 20\,K the diodes required bias ($\approx30$\,$\mu$A) to make the impedance reasonable ($200\,\Omega$)\footnote{See \S3.1.7 of \cite{colin_thesis}; \cite{kapner_diode, colin_diode} for more details on diode biasing.}.
The diode bias and readout was differential so each diode required two connections for a total of eight diode connections per module.

A brass housing contained the components of each module.
Inside the housing, microstrip transmission lines connected the RF components.
Ribbon bonds connected DC bias and readout points to pins leading out of the module.
Each module had 23 such pins: 10 MMIC, 4 phase switch, 8 diode, and 1 shared module ground.
Since all external module connections were DC, the modules were easy to test and optimize in a repeatable way.


\subsection{Cryostat}
\begin{figure}
\includegraphics[width=0.75\textwidth]{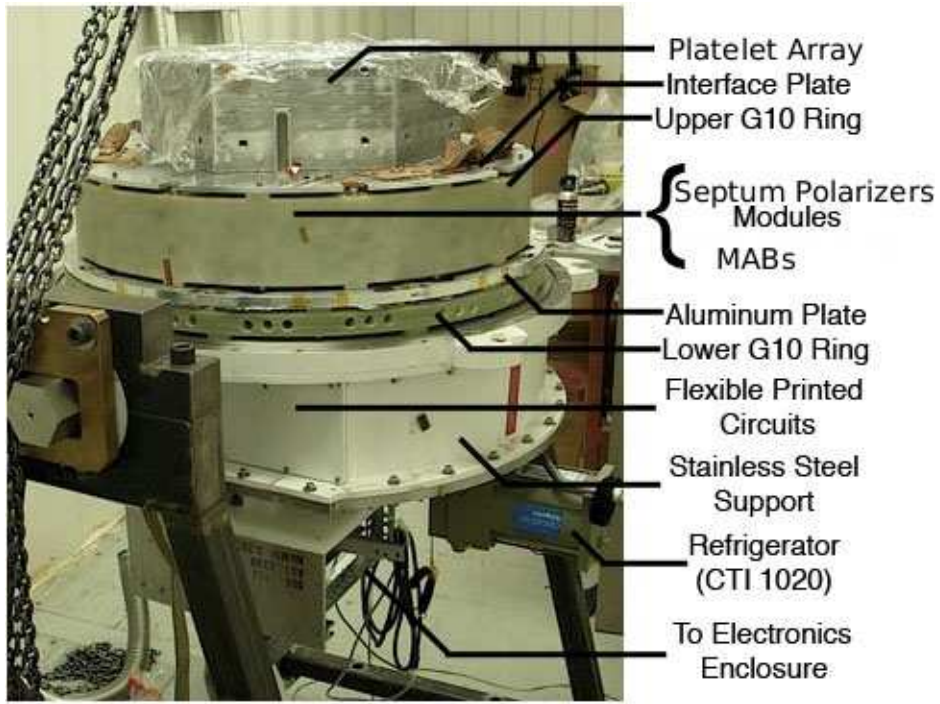}
\caption[Cryostat Internals]{\label{fig:cryostat_photo}
The cryostat contained the platelet array, septum polarizers, and modules.
We cooled them to $\approx25$\,K to reduce the system noise.
}
\end{figure}

\begin{figure}
\includegraphics[width=0.75\textwidth]{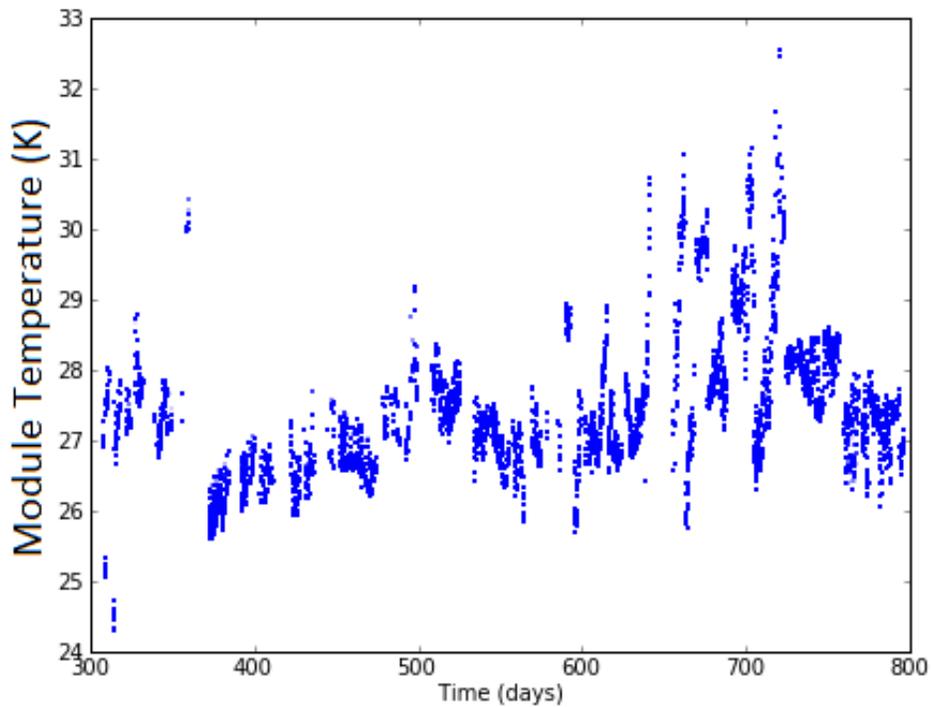}
\caption[Cryostat Temperature Throughout the Season]{\label{fig:cryo_temp_vs_time}
The cryostat temperature varied throughout the season due to problems with the cold heads and vacuum.
The main effect was that the responsivity changed; we took it into account in the responsivity model and associated systematic error.
(I show the temperature of the central module, RQ45.)
}
\end{figure}

The cryostat mounted and supported the platelet array, septum polarizers, and modules (Figure \ref{fig:cryostat_photo}).
Radiation entered the cryostat through an ultra-high--molecular-weight---polyethylene
window with an expanded-Teflon anti-reflection coating\footnote{\S2.6.5 of \cite{laura_thesis}}. With this coating the loss through the window was at
the percent level (The loss increased the system noise temperature by 4\,K.) for microwave frequencies. 
A polystyrene filter blocked infrared radiation, reducing thermal
loading on the cold stage. 
We made an interface plate\footnote{\cite{cryostat_document} and \S2.3.3 of \cite{ali_thesis}} in the cold stage between the platelet array and septum polarizers; the interface plate included waveguide transitions between the circular waveguide at the end of each feed horn and the square waveguide at the input of each septum polarizer.
We cooled the cold stage containing the modules and optical components to $\approx$ 25\,K\footnote{We had difficulty keeping the temperature stable.  The standard deviation of the module temperature during the season was 1\,K.  See \S\ref{sec:noteworthy_events} and Figure \ref{fig:cryo_temp_vs_time} for more details.} using two CTI Cryogenics Cryodyne 1020 Gifford-McMahon refrigerators. 
We attached a heater to each cold head, and a temperature controller (Cryo-con model 32B, hereafter ``CPID'') regulated the cold-head temperatures using a feedback loop.
We put silicon diode
thermometers on the detectors, platelet array, refrigerator heads, and secondary 80-K stage (The CTI 1020 has two cold stages.  The 80-K stage cooled a radiation shield.) to continuously monitor
the cooling and temperature regulation performance. 
The refrigerators had a mechanical cycle frequency of 1.2\,Hz.
I\footnote{``I'' indicates particular contributions of the author.} found a response at this frequency in the DD data (\S\ref{sec:cuts:ps_spikes}).
Stycast-epoxy--sealed hermetic pass-throughs allowed the bias
and readout connections (\S\ref{sec:electronics:passive}) to be brought out of the cryostat. 

\svnid{$Id: electronics.tex 149 2012-07-23 20:51:50Z ibuder $}

\chapter{Electronics}
\label{sec:electronics}

\epigraph{They said it couldn't be done, but sometimes it doesn't work out that way.}{Casey Stengel}

\begin{figure}
\includegraphics[width=0.75\textwidth]{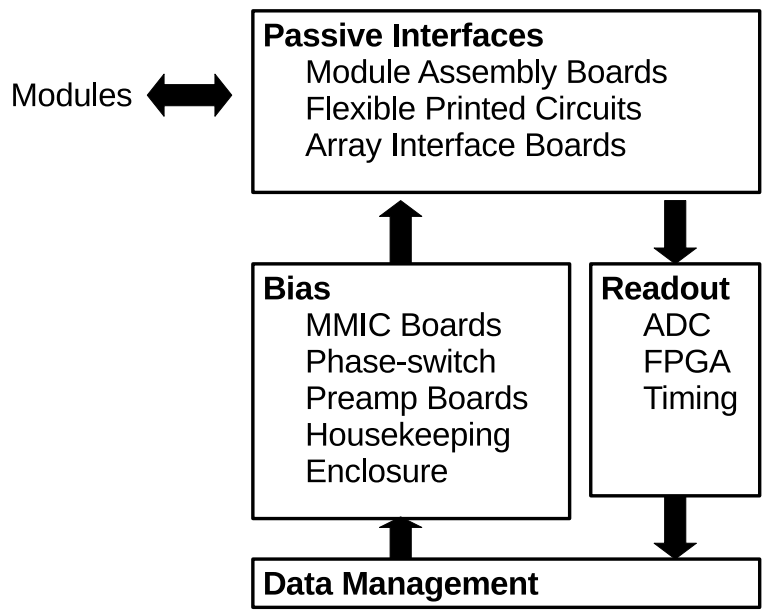}
\caption[Electronics Systems]{\label{fig:electronics_block_diagram}
We divided the electronics into four systems: Passive Interfaces, Bias, Readout, and Data Management.
Commands flowed from the observers, through the Bias and Passive Interfaces systems, and to the modules.
The module outputs flowed through the Passive Interfaces system to the Readout system, which digitized them, and the Data Management system, which recorded them.
}
\end{figure}

The electronics system provided module biasing, timing synchronization, and data acquisition (DAQ) for the receiver.
We divided and assigned these tasks to four systems (Figure \ref{fig:electronics_block_diagram}): 
(1) Passive Interfaces, (2) Bias, (3) Readout, and (4) Data Management. 
The Passive Interfaces system (\S\ref{sec:electronics:passive}) created
an interface between the modules, the biasing electronics, and the DAQ.
The Bias system (\S\ref{sec:electronics:bias}) generated the voltage or current  necessary to operate each module active component.
The Readout system (\S\ref{sec:electronics:readout}) amplified
and digitized the module outputs.
It included the timing synchronization hardware.
The Data Management system (\S\ref{sec:DAQ}) commanded the other systems
and recorded the data.

\section{Passive Interfaces}
\label{sec:electronics:passive}
The main function of the Passive Interfaces system was to bring the module electrical connections out of the cryostat.
We accomplished this function in three stages: first, Module Assembly Boards (MAB) connected directly to the modules.
Second, Flexible Printed Circuits (FPC) connected to the MABs and passed through the cryostat vacuum seal.
Third, Array Interface Boards (AIB) interfaced between the FPCs and standard ribbon cables.

\begin{figure}
\includegraphics[width=0.75\textwidth]{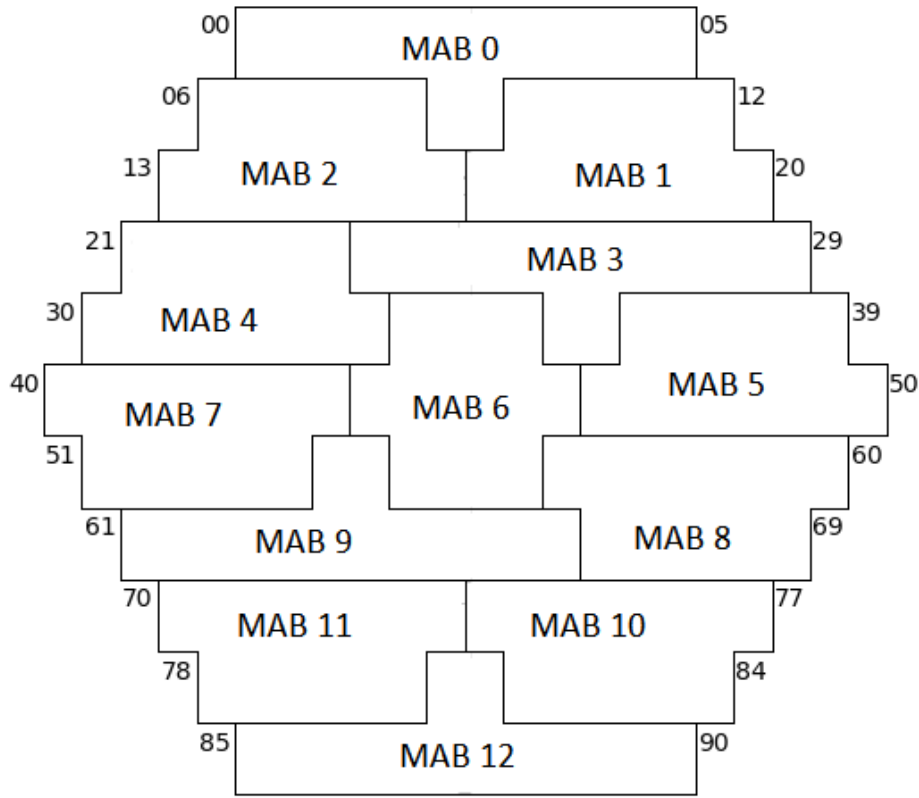}
\caption[Module Assembly Boards Tiling the Focal Plane]{\label{fig:focal_plane_MAB}
We tiled 13 MABs to fill the focal plane.
Numbers to the left and right indicate module position numbers.
MABs 0 and 3 had the same shape and were physically identical circuit boards.
Similarly, MABs 1, 2, 4, and 5; 7, 8, 10, and 11; and 9 and 12 were physically identical.
Thus we needed only five MAB circuit layouts.
The modules could not rotate $180\degr$ in the focal plane so a duplicate MAB 4 cannot serve in place of MAB 7, for example.
}
\end{figure}

Each MAB was a printed circuit board with pin
sockets for seven modules. 
(Sometimes ``MAB'' refers to the associated modules e.g. ``MAB3 has high noise.'')
We made five electrically identical MAB designs with different physical shapes as needed to fill the focal plane (Figure \ref{fig:focal_plane_MAB}); there were 13 MABs in total.
Each module had 27 pins (see \S\ref{sec:modules} for their functions; four pins were unused).
Voltage clamps and RC low-pass filters on the MAB protected the sensitive components inside the modules from damage caused by transients or accidental overvoltage\footnote{See \S2.5.1 of \cite{ali_thesis}; \cite{mab_spec_2, mab_measurements} for details about the MAB protection circuitry for each module component.}.
After protection, the MAB circuitry connected the (23 used) module pins to 40-contact Hirose connectors.
Each MAB had 5 such connectors: 2 for MMIC bias pins, 2 for detector diodes, and 1 for phase switches.
Each Hirose connector accepted one FPC.

The FPCs were high-density (40 conductors each) 32''-long flexible circuit boards\footnote{Figure 2.25 in \S2.5.1 of \cite{ali_thesis}}.
One end of each FPC connected to an MAB. 
Each FPC carried either MMIC, diode, or phase-switch signals for that MAB\footnote{Table 3 of \cite{fpc_redesign}; \cite{preamp_FPC_pinout, ps_FPC_pinout} list which signal is on which FPC conductor.}.
We grouped the five FPCs for each MAB together for cable routing (there were $5\times 13=65$ FPCs in total).
We clamped each FPC group to the cryostat 80-K stage to reduce thermal loading.
Each FPC group passed through the cryostat wall in a Stycast-epoxy--filled custom hermetic seal\footnote{\S4.3.2 of \cite{colin_thesis}}.
The other end of each FPC connected to an AIB.

\begin{figure}
\includegraphics[width=1.0\textwidth]{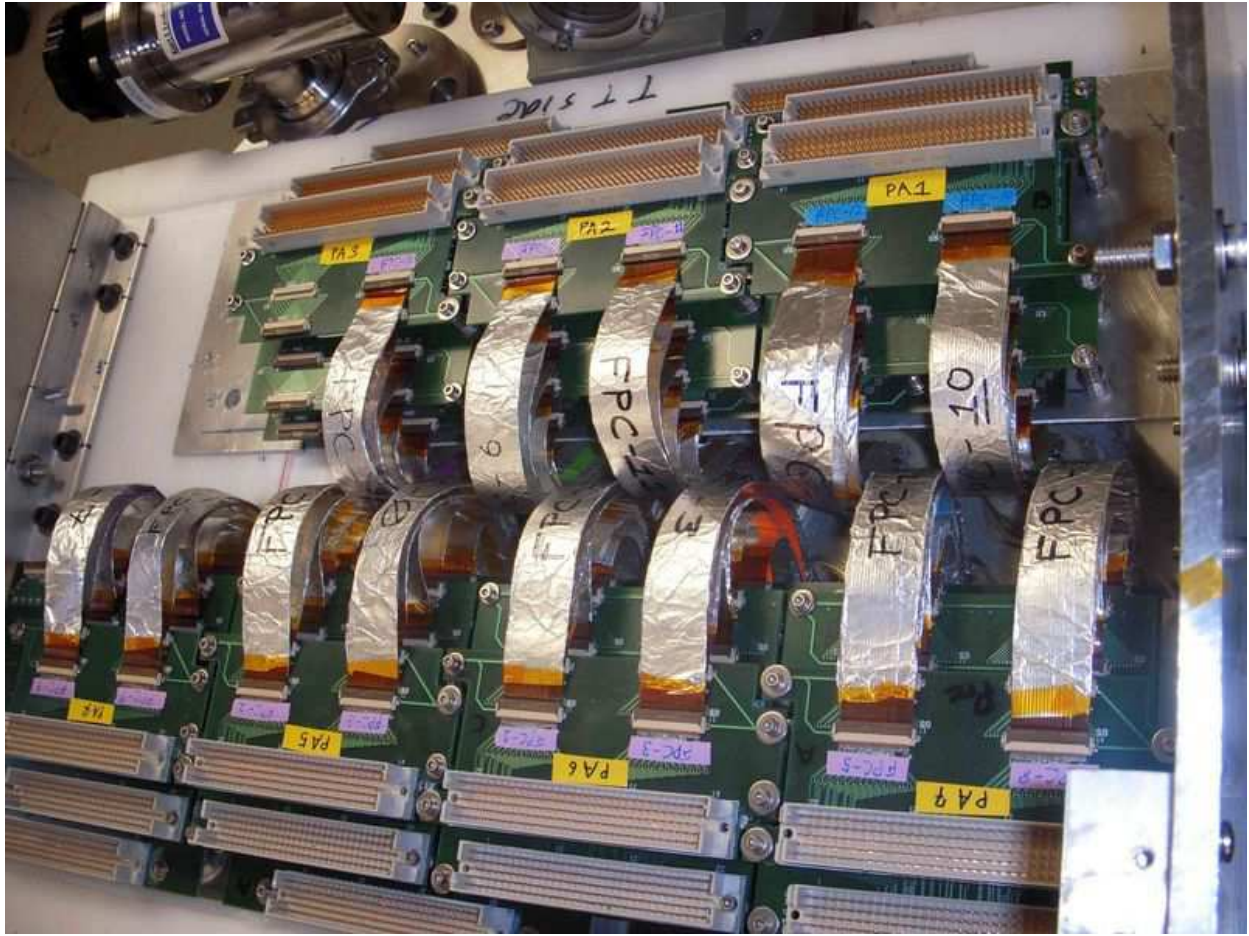}
\caption[Array Interface Boards]{\label{fig:AIB_photo}
We attached the Array Interface Boards (AIB) to the back of the cryostat.
Each AIB connected one bias board to two or three MABs.
The AIBs also provided electrical protection for the modules.
SOURCE: Alison Brizius
}
\end{figure}

\begin{figure}
\includegraphics[width=1.0\textwidth]{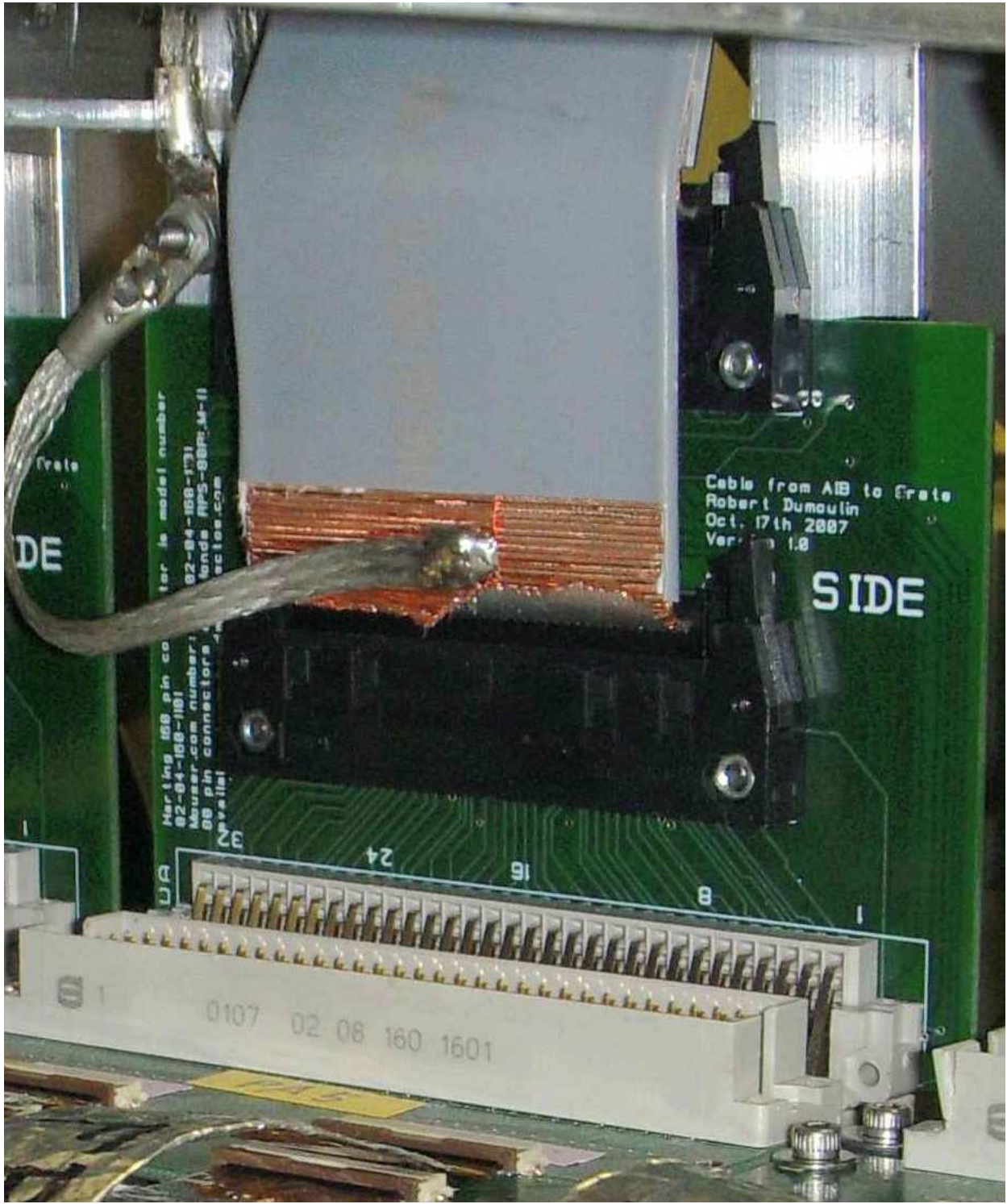}
\caption[AIB Cable Boards]{\label{fig:AIB_cable_board_photo}
AIB Cable Boards interfaced between the Array Interface Board (AIB) Harting connectors (bottom) and the two ribbon cables making up each AIB Cable.
SOURCE: Alison Brizius
}
\end{figure}

\begin{figure}
\includegraphics[width=0.5\textwidth]{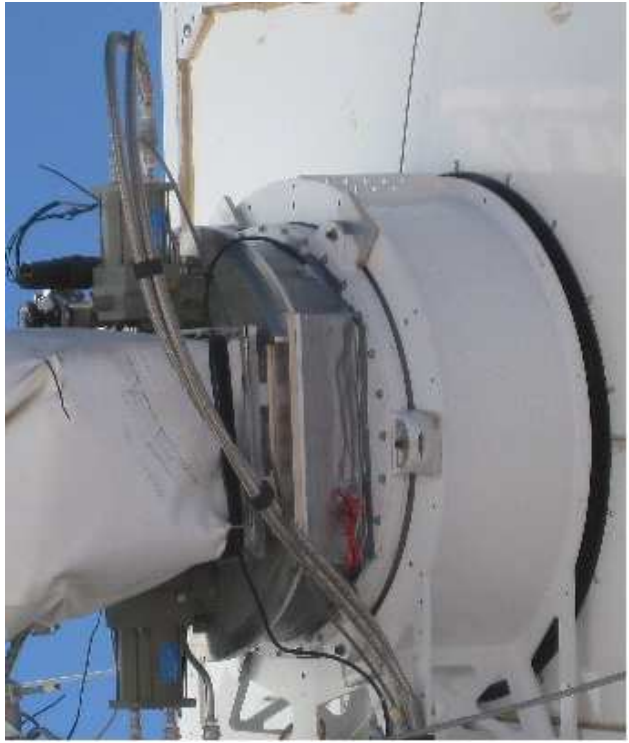}
\caption[Array Interface Board Weatherproofing]{\label{fig:AIB_weatherproofing_photo}
Weatherproofing protected the Array Interface Boards (AIB) and AIB Cable connections.
}
\end{figure}

The AIBs provided a second layer of electrical protection for the modules.
We attached them to the back of the cryostat (Figure \ref{fig:AIB_photo}).
When cold, the diodes comprising the voltage clamps on the MABs had drastically different I-V curves than when warm and were no longer effective at protecting the modules.
(The filter resistances and capacitances could change as well.)
Therefore, we duplicated the module protection circuitry on the AIBs. 
Each AIB accepted only MMIC, diode, or phase-switch FPCs.
A MMIC AIB accepted four MMIC FPCs (from two MABs) and combined and routed the signals (after protection) to one AIB Cable (Harting) connector.
The AIB Cables connected the AIBs to the bias boards (\S\ref{sec:electronics:bias}).
We made each AIB Cable from two 80-pin shielded ribbon cables; AIB Cable Boards (Figure \ref{fig:AIB_cable_board_photo}, see also \cite{AIB_cable_proposal})  interfaced between the Harting connectors and the ribbon cables.
A ``Preamp'' AIB (corresponding to ``Preamp Boards,'' \S\ref{sec:electronics:bias}) accepted four diode FPCs (from two MABs), and a Phase-switch AIB accepted three phase-switch FPCs (from three MABs).
Each AIB connected one AIB Cable, which connected to one bias board (see Appendix~\ref{app:receiver_db_tables} for the mapping between modules, MABs, and bias boards).
We surrounded the AIBs and AIB Cable connections with a weatherproofing structure (Figure \ref{fig:AIB_weatherproofing_photo}) to protect them from damage due to moisture and dust. 

\section{Bias}
\label{sec:electronics:bias}
We made custom integrated-circuit boards to provide the bias signals needed to operate the modules.
We apportioned the module components that require biasing to three types of bias boards: MMIC, Phase-switch, and Preamp.
We monitored the biasing with a single Housekeeping Board.
The boards communicated with other electronics systems through the Bias-Board Backplane.
A weatherproof Electronics Enclosure housed and protected the Bias system (the Readout system shared this enclosure).

MMIC Boards provided voltage and current to power the MMIC amplifiers in the modules.
Each MMIC Board could bias two MABs (14 modules); there were seven MMIC Boards in total.
The boards provided voltage sources for the MMIC gates and drains.
We stabilized each drain-voltage source with an op-amp feedback loop \citep{mmic3_prototype}.
A 10-bit ($2^{10}$ settings) digital-to-analog converter (DAC) controlled each voltage.
(The DACs were Linear Technologies LTC1660 chips.  Each LTC1660 has eight independent DACs.  We used an ``address'' 1--8 to distinguish different DACs on the same chip.
The Preamp and Phase-switch Boards used the same chips.  Unless otherwise noted, ``DAC'' means one of the DACs controlling a bias-board output.)
Using a DAC to control each bias setting allowed us to reliably save and reapply the biasing state of the entire receiver.
Moreover, with digital control we could optimize the performance upon adjusting the settings (\S\ref{sec:modules}).

\begin{figure}
\includegraphics[width=0.5\textwidth]{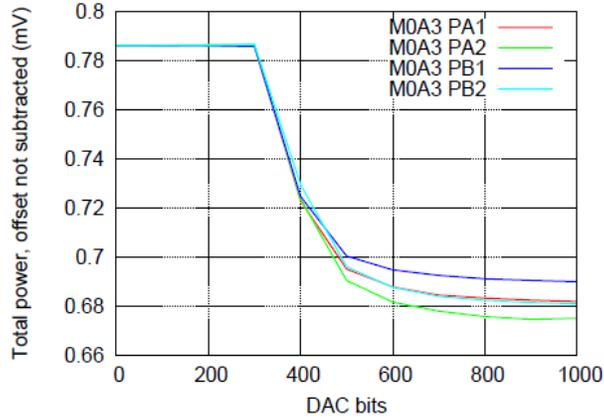}
\caption[Phase-switch Transmission Curve]{\label{fig:phase_switch_transmission_curve}
The phase-switch diodes required forward bias (positive DAC setting) to transmit power.
Lower voltage corresponds to higher transmission power.
The four phase-switch diodes (different colors) in each module had different transmission plateaus.
We biased all phase-switch diodes in the plateau region and used double demodulation to remove the systematic effects of the transmission differences.
SOURCE: \cite{ps_curve-20090117}
}
\end{figure}

\begin{figure}
\includegraphics[width=1.0\textwidth]{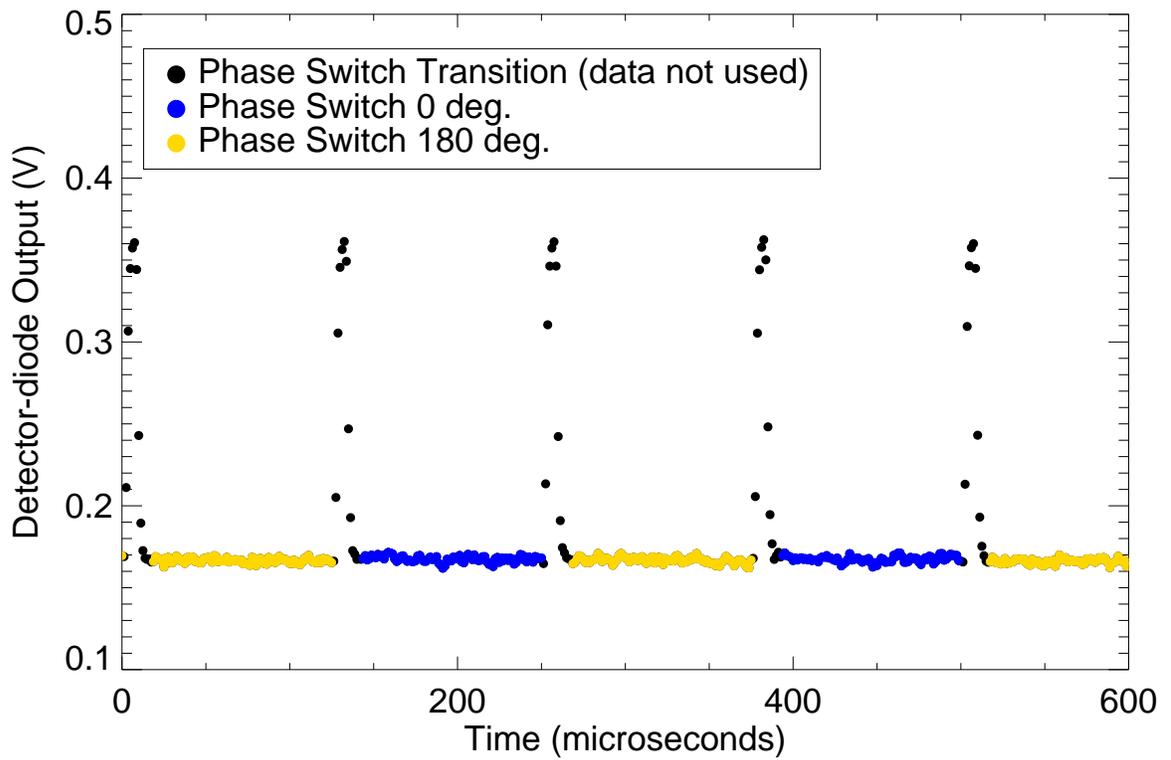}
\caption[Phase-switch Modulated Data]{\label{fig:snapshot_example}
A phase switch modulated the sign of polarization at 4\,kHz.
We discarded the data during the phase-switch transition.
}
\end{figure}

Phase-switch Boards provided control to the PiN diodes in the module phase switches.
Each Phase-switch Board could bias three MABs (21 modules); there were five Phase-switch Boards in total.
For each phase switch, the Board forward biased one PiN diode and reverse biased the other diode.
The reverse bias was always $-1.8$\,V\footnote{\S2.5.3 of \cite{ali_thesis}}.
We adjusted the forward bias (controlled by a DAC) to make the phase-switch transmission nearly independent of the bias setting (Figure \ref{fig:phase_switch_transmission_curve} and   \S\ref{sec:modules}).
Because different diodes had different transmission in this plateau region, we could not balance the transmission between the two phase-switch states\footnote{We did balance the transmission in Q band.}.
Instead, we used double demodulation to remove the $I\rightarrow Q/U$ leakage that would otherwise be caused by the imbalance \citep{quiet_instrument}.
At 4\,kHz (50\,Hz) the Phase-switch Board switched the bias direction of both diodes in the $L$ ($R$) phase switch.
The Readout system provided synchronized clocks to synchronize the switching of the two switches.
Because the AIB filtered the phase-switch bias, the switching transition was not instantaneous.
We made the Readout system reject 17.5\,$\mu$s of data during each transition when the signal was unstable (Figure \ref{fig:snapshot_example}).

Preamp Boards biased and read out the detector diodes.
Each Preamp Board could bias two MABs (14 modules); there were seven Preamp Boards in total.
The bias circuit for each diode combined a floating\footnote{i.e. neither side of the diode was at the module ground} current source with a differential voltage amplifier\footnote{\S4.4.3 of \cite{colin_thesis} and \cite{colin_diode}}.
A DAC controlled the bias current.
The amplifier circuit subtracted a (DAC-adjustable) voltage offset so that the diode voltage, output by the Preamp Board, was centered in the dynamic range of the Readout system\footnote{Because of Type-B glitching (\S\ref{sec:electronics:readout}) the DD data acquired sensitivity to the voltage offset.}.
\S\ref{sec:electronics:readout} provides details about the amplification and filtering applied by the Preamp Board.

One Housekeeping Board monitored all the MMIC and phase-switch biases.
(We monitored the diode bias by recording the DAC settings and diode voltages.)
We added several current sense resistors and voltage-probe points to each bias board.
Multiplexers on the bias boards selected one of these measurements to send to the Housekeeping Board.
The Housekeeping Board selected the measurement from one bias board or a thermometer (recording the temperature in the cryostat or Electronics Enclosure) to digitize with a 16-bit analog-digital converter (ADC).
We monitored 1774 biases and temperatures (Appendix~\ref{app:receiver_db_tables}), and the multiplexing rate was 500\,Hz.
Thus we measured each every 3.5\,s.
The multiplexing cycled only during phase-switch transitions (during which we already discarded the data) so the multiplexing transition did not affect the module signal.
Analog opto-isolators separated the bias circuitry from the monitoring circuitry.
The Readout and Data Management systems controlled this multiplexing.

\begin{figure}
\includegraphics[width=0.5\textwidth]{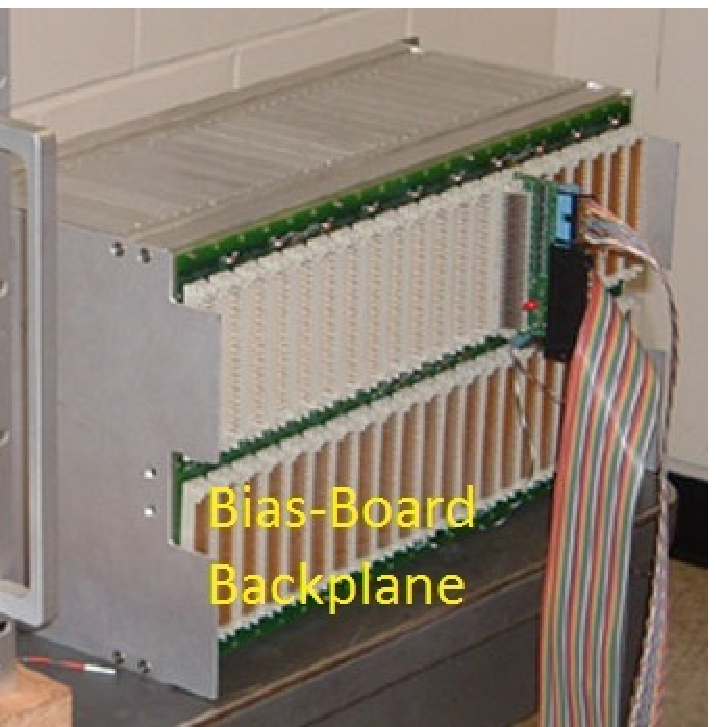}
\caption[Bias-Board Backplane]{\label{fig:electronics_box_photo}
The bias boards entered through the face hidden in this photo.
The AIB Cables connected to the lower set of connectors shown.
}
\end{figure}

The Bias-Board Backplane housed the bias boards and provided an interface to them.
It was a modified VME (VERSAmodule Eurocard bus) 6U backplane with a partial metal enclosure that supported the bias boards (Figure \ref{fig:electronics_box_photo}).
The backplane P2 (bottom) pins were straight-through and connected the AIB Cables to the bias boards \citep{electronics_box}.
We connected the P1 (top) pins connect together so the bias boards could communicate.
These pins provided a path for the bias boards to send a monitored voltage to the Housekeeping Board, a path for the housekeeping digital output to the Readout system, distribution of the housekeeping multiplex address, distribution of the phase-switch clocks, and digital control of the DACs.
Table \ref{tab:bias_board_locations} lists the location of each board in the backplane.

\begin{deluxetable}{lp{5in}}
\tablecolumns{2}
\tablecaption{Bias-board Locations}
\label{tab:bias_board_locations}
\tablehead{\colhead{Slot} & \colhead{Board}}
\startdata
1 & Preamp 1\\
2 & Preamp 2\\
3 & Preamp 3\\
4 & Preamp 4\\
5 & Preamp 5\\
6 & Preamp 6\\
7 & Preamp 7\\
8 & thermometer for Electronics Enclosure regulation\\
9 & Housekeeping\\
10 & MMIC 1\\
11 & MMIC 2\\
12 & MMIC 3\\
13 & MMIC 4\\
14 & MMIC 5\\
15 & MMIC 6\\
16 & MMIC 7\\
17 & Phase-switch 1\\
18 & Phase-switch 2\\
19 & Phase-switch 3\\
20 & Phase-switch 4\\*
21 & Phase-switch 5\\*
\enddata
\end{deluxetable}

\begin{figure}
\includegraphics[width=0.5\textwidth]{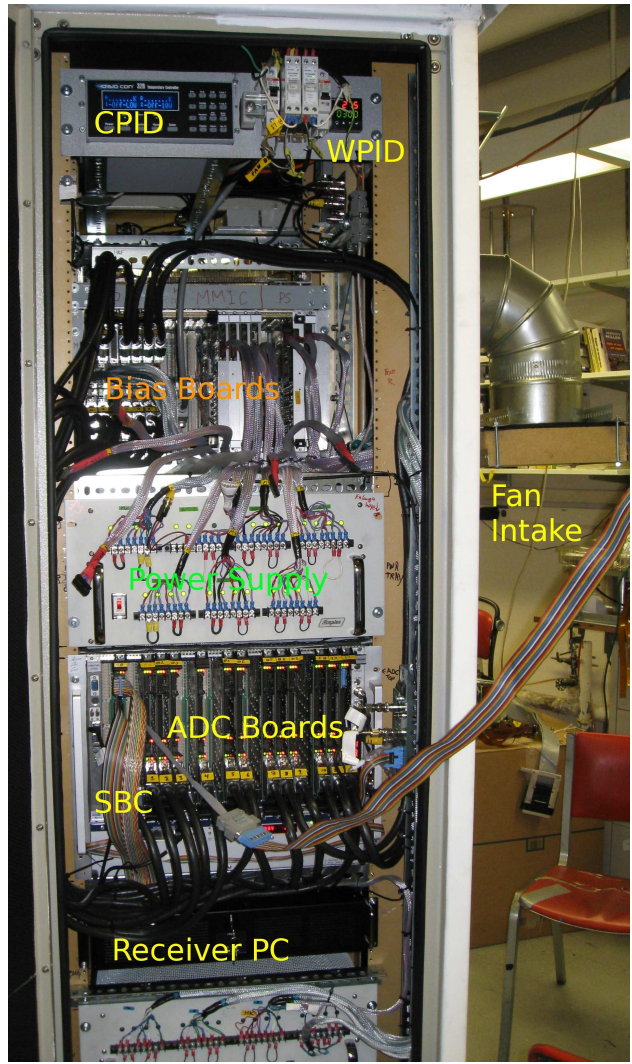}
\caption[Electronics Enclosure]{\label{fig:enclosure_photo}
The Electronics Enclosure housed and protected the Bias and Readout systems.
}
\end{figure}

The Electronics Enclosure housed and protected the Bias and Readout systems.
It was an insulated 54''$\times$25''$\times$25'' box we mounted next to the cryostat (DDB Unlimited OD-78DDXC, Figure \ref{fig:enclosure_photo} and \S2.5.2 of \cite{ali_thesis}).
It contained the Bias-Board Backplane, power supplies, CPID, and Receiver PC\footnote{a rackmount computer that ran software for the Data Management system} (Personal Computer).
Because the responsivity could vary with the electronics temperature\footnote{We limited this effect to 0.3\%/K \citep{gain_systematics}.
In Q band, the effect was larger due to a different MMIC Board design \citep{colin_thesis, ali_thesis}.},
we regulated the temperature of the Electronics Enclosure.

\subsection{Temperature Regulation}
\begin{figure}
\includegraphics[width=1.0\textwidth]{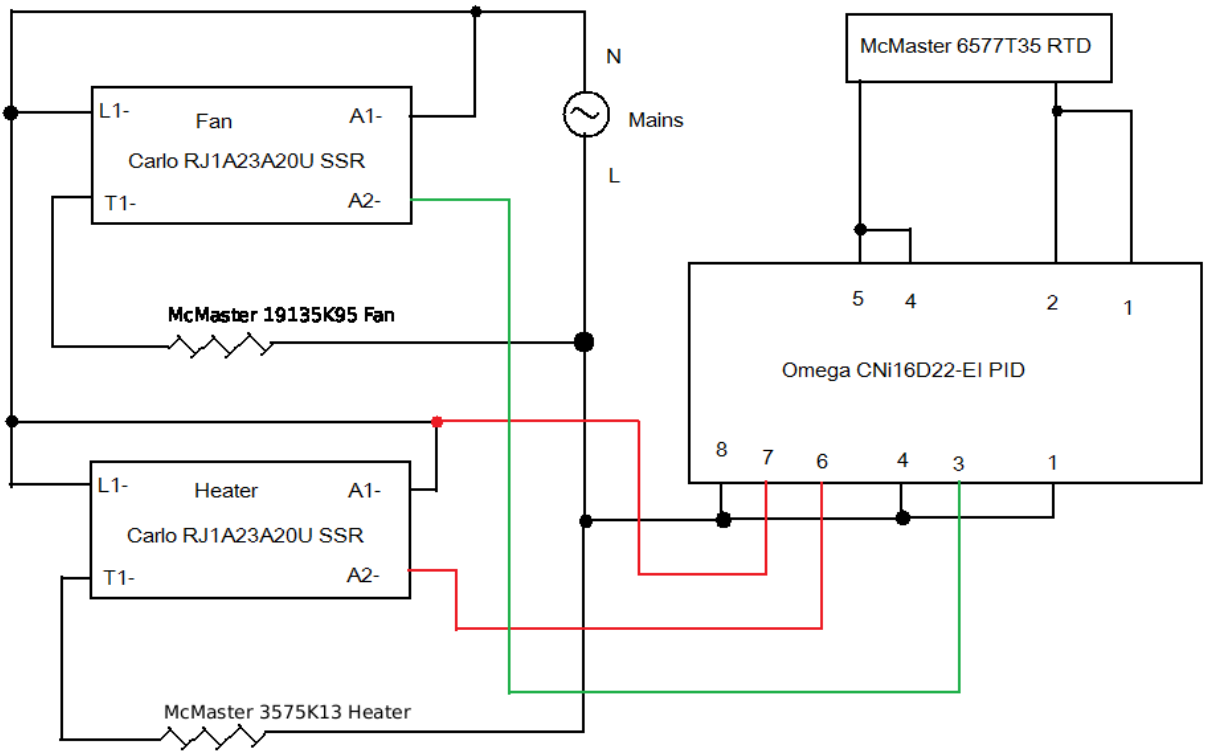}
\caption[Electronics Enclosure Temperature Regulation System]{\label{fig:PID_wiring}
I made a proportional-integral-derivative (PID) system to regulate the Electronics Enclosure temperature.
The numbers on the PID controller correspond to the physical numbered
terminals. Wires of different colors do not connect except at filled circles. ``L''
represents the live electrical phase. When the PID turns on output 1, terminals 4 and 6
become connected. This connects the live phase to A2 on the Carlo SSR, activating the
relay. The SSR connects L1 to T1, providing power to the heater. The same sequence
applies to the fan (output 2) with terminals 1 and 3.
}
\end{figure}

I designed a proportional-integral-derivative (PID) feedback-loop temperature-regulation system for the Electronics Enclosure (Figure \ref{fig:PID_wiring} and \cite{enclosure_regulation_system}).
The system consisted of a PID controller (Omega CNi16D22-EI, denoted ``WPID''), thermometer (McMaster-Carr 6577T35 resistance temperature detector), solid state relays (SSR, Carlo RJ1A23A20U), heater (McMaster-Carr 3575K13), and fan (McMaster-Carr 19135K95).
I achieved good regulation performance ($<1\degr$C) on the time scale of a single observation (1\,hour); however, the temperature changed significantly ($20\degr$C) throughout the season (\S\ref{sec:routine_checks:enclosure}).
I made a  monitoring circuit to let the Data Management system record the heater and fan activity.

The WPID controlled the Enclosure temperature by alternately activating the heater or using the fan to bring cold air into the Enclosure.
The thermometer measured the temperature in the Bias-Board Backplane.
Based on this temperature, the WPID activated one of the outputs to bring the temperature to the setpoint ($35\degr$C or $40\degr$C depending on the time of year, see \S\ref{sec:routine_checks:enclosure}).
Each output activated an SSR that supplied power to the heater or fan.
I mounted the heater directly above the Bias-Board Backplane; the heater supplied up to 1.1\,kW.
The fan brought cold ($-10$--$+15\degr$C) air in above the  Bias-Board Backplane. 
Air circulation fans (McMaster-Carr 1985K11) kept the air inside the Enclosure well-mixed.
The WPID had an ethernet interface, and I wrote a custom control program\footnote{Source code at \url{https://cmb.uchicago.edu/svn/ibuder/PIDControl}} for remote control and automated operation by the Data Management system.

\subsection{WPID Output Monitoring Board}
\label{sec:elec:WOMB}
\begin{figure}
\includegraphics[width=1.0\textwidth]{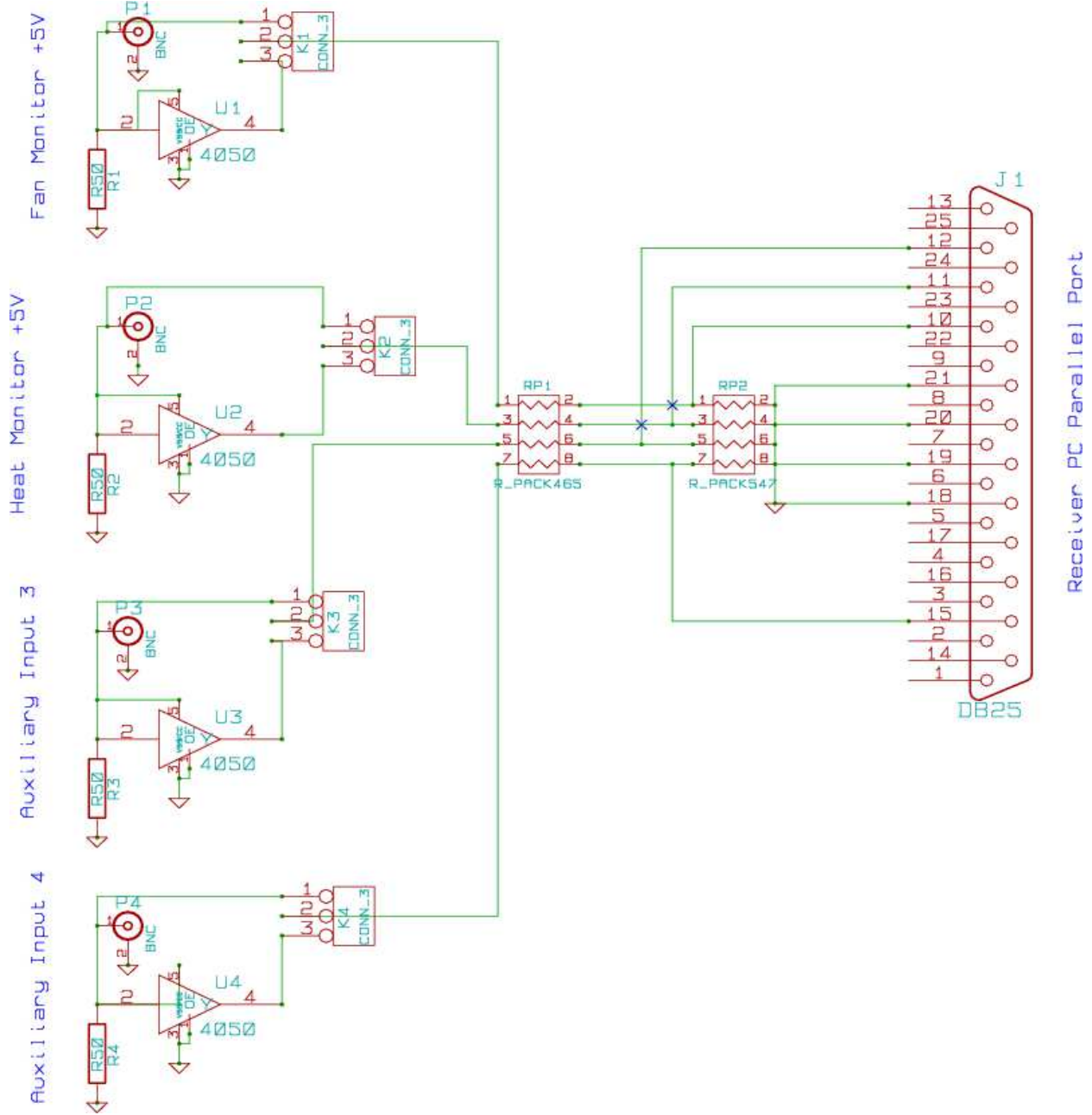}
\caption[WPID Output Monitoring Board Schematic]{\label{fig:WOMB_schematic}
The WPID Output Monitoring Board sent the WPID heater and fan activity to the Data Management system for recording.
}
\end{figure}

The WPID did not report when it was turning on the fan and heater; therefore, I built an auxiliary monitoring circuit (``WPID Output Monitoring Board,'' Figure \ref{fig:WOMB_schematic}).
Linear power supplies converted convert the AC fan and heater voltages to 5\,V DC.
The WPID Output Monitoring Board supplied this voltage to the Receiver PC parallel port. 
A buffer (ON Semiconductor MC74VHC1GT125DT1G) and voltage divider (resistances 470\,$\Omega$ and 560\,$\Omega$) protected the port from overvoltage damage.
I made the Data Management system sample the parallel-port input to record the heater and fan activity (\S\ref{sec:datacompilation}).

\section{Readout}
\label{sec:electronics:readout}
The Readout system amplified, filtered, and demodulated the signal.
Preamp Boards provided analog amplification and low-pass filtering.
ADC Boards digitized the module signals and demodulated them synchronously with the phase switching.
A Time-code Reader provided the time and reference clocks for all electronics systems.

We used the Preamp Boards to measure the detector-diode voltage.
Three analog amplifiers provided a gain of $\approx130$\footnote{\S4.4.3 of \cite{colin_thesis}. 
Late in the design stage we moved a factor of two in gain from the ADC Board to the Preamp Board \citep{preamp_gain_change}.  As a result, some QUIET documents incorrectly give the Preamp Board gain of 65.}.
This amplifier chain contained low-pass filtering with three poles at 3-dB frequencies of 160\,kHz, 340\,kHz, and 32\,MHz.
This filtering prevented aliasing with the 800-kHz digitization sampling.
As described in \S\ref{sec:electronics:bias}, the board subtracted a voltage offset to compensate for the diode-bias voltage.
The Preamp Board output stage sent the amplified, filtered voltage to ADC Boards using one pair of wires (differential signaling) per diode.

ADC Boards digitized the diode voltages.
Each ADC Board had 32 channels; however, we left four channels unused so each board digitized 28 diodes (seven modules, one MAB).
Each channel had a low-pass filter (3\,dB at 600\,kHz) before digitization by 18-bit Analog Devices AD7674 Successive Approximation Register ADCs.
The ADCs sampled at 800\,kHz.
Each ADC Board had a field-programmable gate array
(FPGA), which accumulated the samples from the 32
ADCs on that Board. 
The FPGA on one ADC Board,
designated the ``Master ADC Board\footnote{The Master ADC Board did not accept signals from the modules.  Therefore, the total number of ADC Boards was 14.},''”generated the 4-kHz and 50-Hz signals used by the Bias system to control the phase-switch modulation. 
We distributed these signals to all ADC Boards, and the FPGA on each ADC
Board used them to demodulate the detector-diode signals synchronously with the phase-switch modulation.

\begin{figure}
\includegraphics[width=1.0\textwidth]{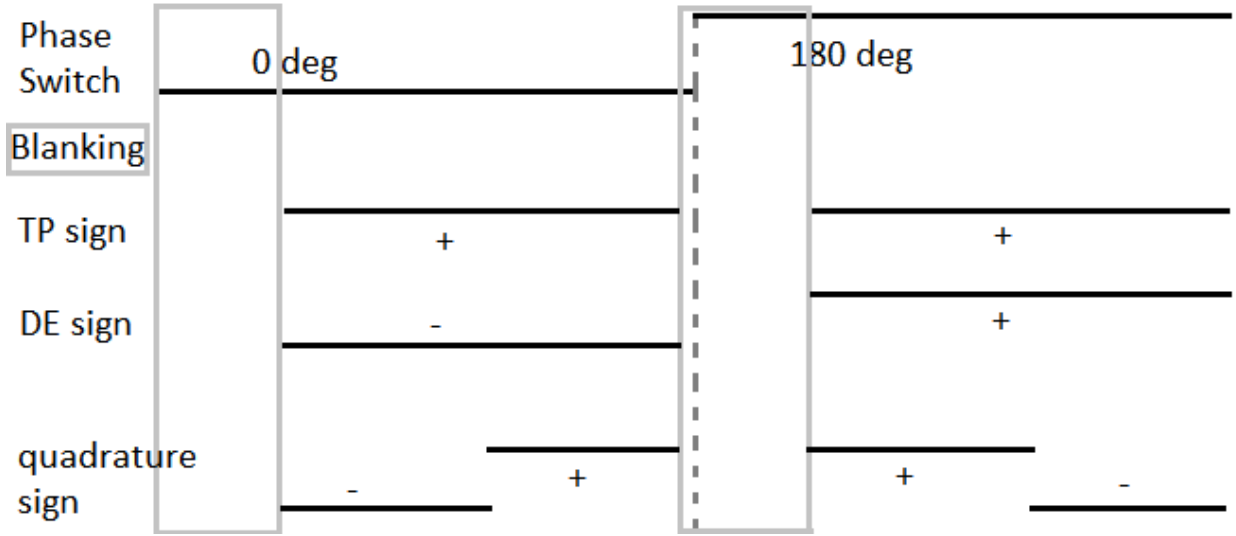}
\caption[Demodulation Cartoon]{\label{fig:demodulation_cartoon}
We implemented digital demodulation on the ADC Boards.
We discarded data during phase-switch transitions (``Blanking'').
The TP stream summed all data regardless of the phase-switch state.
The DE stream differenced data between the (two 4-kHz) phase-switch states.
Quadrature data was $90\degr$ out of phase so that we canceled the signal.
(I exaggerated the fraction of data blanked, 14\%, for ease of viewing.)
}
\end{figure}

Normally we recorded three 100-Hz data streams for each diode.
We recorded short (1024-sample) ``snapshots'' of 800-kHz data for diagnostic and monitoring purposes; however, there was insufficient storage to record at that rate continuously.
To create the TP data stream (\S\ref{sec:module_principles}), the FPGA summed all 800-kHz samples, regardless of phase-switch modulation.
The
TP stream was sensitive to Stokes $I$, and we used it for calibration and monitoring.
To create the DE data stream, the FPGA differenced samples, assigning different signs to the two 4-kHz phase-switch states.
We differenced two adjacent DE samples in post-processing to demodulate the 50-Hz phase state switching and create the DD stream.
In the ``quadrature'' stream, we differenced 4-kHz samples  $90\degr$
out of phase with the 4-kHz switching (Figure \ref{fig:demodulation_cartoon}). The quadrature
data had the same noise as DE data and no
signal. We used quadrature data  to monitor potential
contamination and understand the detector noise properties. 
The three streams (TP, DE, and quadrature) together constituted the ``radiometer data.''
As mentioned in \S\ref{sec:electronics:bias}, the FPGA did not accumulate data in any stream when the phase-switch state was changing.
This blanking rejected 14 out of every 100 samples, reducing the effective integration time.

\begin{figure}
\includegraphics[width=0.5\textwidth]{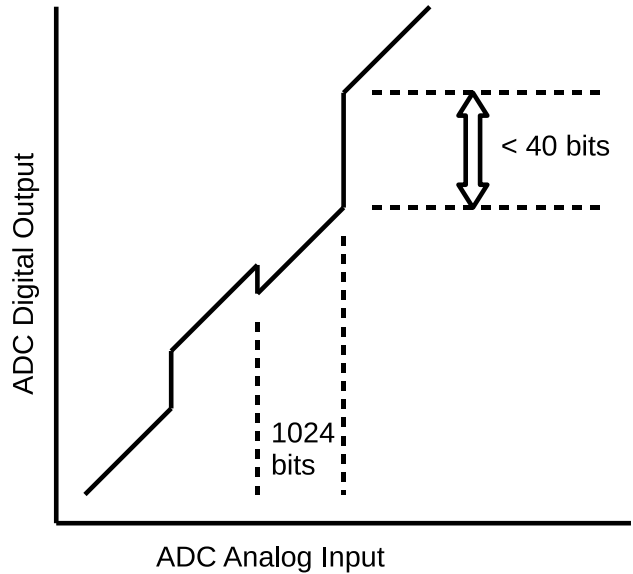}
\caption[Type-B--glitch Cartoon]{\label{fig:typeb_cartoon}
Type-B glitches were jump discontinuities (differential non-linearity) in the ADC response.
The jumps were evenly spaced every 1024\,bits; however, the jump size and direction were different for different jumps.
When the input was near a jump, the difference between the two phase-switch states caused the jump size to add to the DE data.
}
\end{figure}

The ADC Boards had a small differential non-linearity in their
response. 
At intervals of 1024 bits\footnote{There was one exceptional channel that had a different interval.}, the ADC output
had a jump discontinuity of size $<40$ bits (Figure \ref{fig:typeb_cartoon}). 
The size was negligible compared to the large 1/f noise in the TP data.
However, the effect in DE data was not negligible.
 When the average 800-kHz voltage level was near a discontinuity, 
the voltage difference between the two phase-switch states caused the level to move across the discontinuity.
This caused the discontinuity to be added to the DE data.
The 800-kHz noise (much higher per sample than per 100-Hz sample) smeared out the discontinuity, reducing its amplitude but also increasing its width in bit space (the jump can occur whenever the signal level plus noise encounters the discontinuity).
We called this effect ``Type-B glitching\footnote{We believe Type-B was caused by a combination of bad ADC clock signals and bad ADC chips.
Type-B disappeared when we operated the ADCs at 400\,kHz.
Some ADC channels did not have this glitching, and the jump sizes for the channels that did glitch were not identical.
However, the exact mechanism in the ADC causing Type-B is not known.}.''
We mitigated the effects by using the Preamp Board offsets to move the typical voltage levels away from the discontinuities.
However, some effect remained, and
I corrected for Type-B in analysis (\S\ref{sec:calibration:typeb}). 

The ADC Backplane contained the ADC Boards, Time-code Reader, and a single-board computer (SBC, part of the Data Management system).
The ADC Backplane was a Wiener Series 6000 VME crate; we put it in the Electronics Enclosure. 
The Master ADC Board used the VME interface to distribute the phase-switch control signals (and all other clock signals for synchronization) to the other ADC Boards.

The timing hardware consisted of a Time-code Reader and Auxiliary Timing Board (ATB).
The Time-code Reader synchronized the electronics to an external Global Positioning System (GPS) derived time.
The ATB distributed timing signals from the Time-code Reader to the Master ADC Board, which synchronized the other ADC Boards.

The Time-code Reader was a Symmetricom TTM635VME-OCXO \citep{tcr_manual}.
We used an IRIG-B (Inter-range Instrumentation Group Mod B)
 amplitude-modulated time code from the GPS receiver in the
 control room (\S\ref{sec:observations:site}) as the reference.
From the time code we created synchronized 1-Hz and 10-MHz clock signals.
The ATB distributed these clocks to the Master ADC Board, which used them to synchronize all the ADC Boards\footnote{\S2.5.4 of \cite{ali_thesis}}.

\begin{figure}
\includegraphics[width=1.0\textwidth]{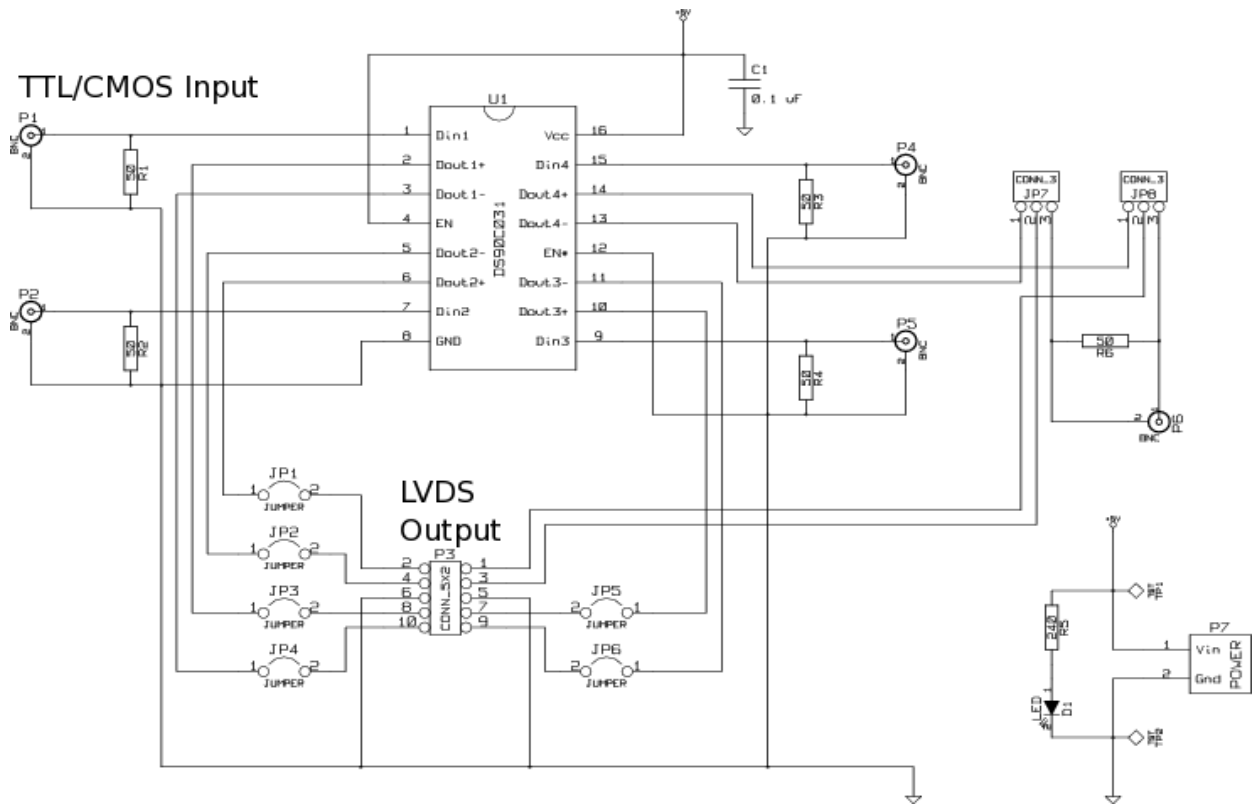}
\caption[Auxiliary Timing Board Schematic]{\label{fig:ATB_schematic}
The Auxiliary Timing Board (ATB) converted logic levels between the Time-code Reader and Master ADC Board.
}
\end{figure}

\label{sec:ATB}
We made the ATB to convert logic levels between the Time-code Reader and Master ADC Board.
The Time-code Reader output levels were TTL/CMOS, but the Master ADC Board could only receive low-voltage differential signals (LVDS).
We designed a small circuit board to convert between them (Figure \ref{fig:ATB_schematic}).
The board used National Semiconductor DS90C031 LVDS Differential Line Drivers. 

\svnid{$Id: DAQ.tex 150 2012-07-24 22:04:36Z ibuder $}

\chapter{Data Management}
\label{sec:DAQ}
\epigraph{The first 90 percent of the code accounts for the first 90 percent of the development time. The remaining 10 percent of the code accounts for the other 90 percent of the development time.}{Tom Cargill}

The Data Management system was responsible for sending commands to the Bias system and mount pointing control system, receiving data from the Readout system, synchronizing data from different readout subsystems, 
and recording the data.
We created several software processes that ran on several computers to accomplish these main functions.
We organized the processes  in a hierarchy.
For example, the lowest level bias control process (\process{adc\_server}) understands and implements only very simple commands e.g. ``change the bias output of DAC \#N to m bits.''
Whereas the highest level control process (\process{Receiver Control Panel}) has a command ``change the bias so the receiver is on.''
Each process translated high level commands to low level commands (or low level data to high level data) appropriate for the other processes it communicated with.
The subsections of this chapter detail these interactions.

\begin{figure}
\includegraphics[width=1.0\textwidth]{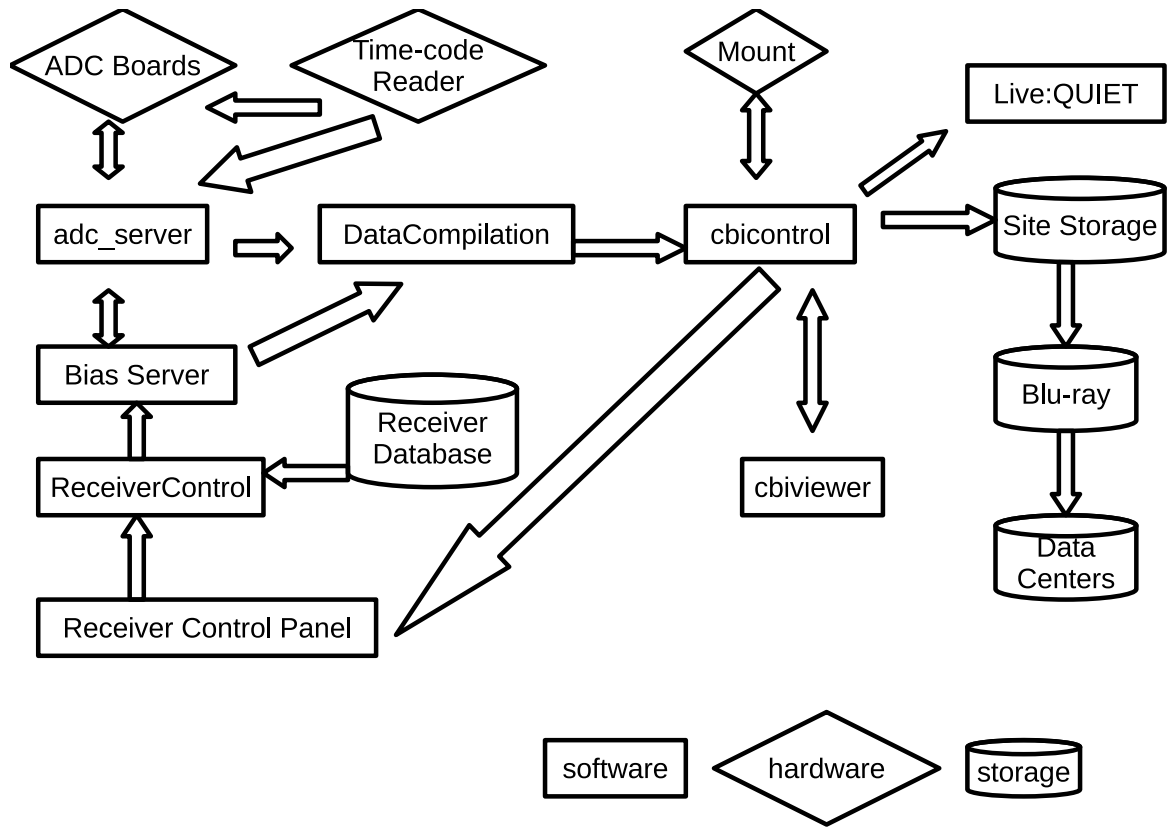}
\caption[Data Management System Overview]{\label{fig:DAQ:overview}
The Data Management system recorded, distributed, and stored the data.
I wrote software processes that were its major components.
}
\end{figure}

We divided software processes  into two broad categories: Online Software and Control Software.
Online Software processes were always running and rarely used by the observers directly.
Control Software processes had user-friendly graphical interfaces.
The observers used them to schedule and start observations or monitor the system status.
Two additional major Data Management components were the data files, whose format is described in \S\ref{sec:DAQ:files}, and Receiver Database.
The Receiver Database contained configuration information, e.g. which electronics board was connected to which module, which was used when translating command levels.
Figure \ref{fig:DAQ:overview} is a simplified diagram of the Data Management system.

\section{Online Software}

The Online Software was responsible for the continuous readout and recording of the data.
In addition, we periodically (or sporadically) generated bias commands.  The Online Software translated these commands to the lowest level and sent them to electronics hardware.
The Online Software helped us monitor the DAQ and command implementation status and automatically detected and reported several types of problems (Appendix~\ref{sec:DataCompilationStatusFlags}) via the Control Software.

We created five processes to accomplish these functions.  The \process{adc\_server} ran on the single-board computer (SBC, a GE Fanuc VME-7700) in the ADC Backplane and was responsible for direct communication with electronics hardware or firmware.
The \process{DataCompilation} process ran on the Receiver PC in the Electronics Enclosure and was responsible for combining data from different readout systems and distributing them to recording and monitoring processes.
The \process{ReceiverControl} process ran on the Receiver PC and allowed us to enter high-level bias commands.
The \process{Bias Server} process, which translated electronics bias commands, and \process{PeripheralServer}, which interfaced with auxiliary pressure sensors, temperature controllers, etc. are described in \cite{colin_thesis}.

\subsection{\process{adc\_server}}
\label{sec:adc_server}

The process \process{adc\_server}\footnote{Source code is available at \url{https://cmb.uchicago.edu/svn/kapner/quiet/crate_fs/branches/quadrature/root/vme_adc_threaded/src}} was the lowest level of the Online Software.  It communicated directly with the ADC Boards and Time-code Reader using the VME64 interface on the ADC Backplane.
The process \process{adc\_server} received commands from and sent data to higher-level Online Software processes via 100BaseTX ethernet \citep{VMEPC_manual}.
The process \process{adc\_server} was the only user process running on the SBC, and it ran continuously as long as the ADC Backplane was powered.
I divided the functions  among software threads.
Since the SBC had only one Central Processing Unit (CPU) core, only one thread was running at any given time.
The operating system \citep{QUIET_DAQ_Crate} could switch rapidly between them, but some synchronization was necessary.
Table \ref{tab:adc_server:threads} provides an overview of the threads and their functions.

\begin{deluxetable}{lp{3.5in}p{1.5in}}
\tablecolumns{3}
\tablecaption{Summary of \process{adc\_server} Threads
\label{tab:adc_server:threads}}
\tablehead{\colhead{Thread Name} & \colhead{Function} & \colhead{Priority}}
\startdata
\thread{msg\_serve} & Send warning and error messages to the network for recording & 8\\
\thread{fourhz\_write} & Read data from ADC Boards, read time from Time-code Reader & 10\\
\thread{data\_serve} & Send radiometer data to \process{DataCompilation} & 8\\
\thread{data} & Send snapshots to \process{DataCompilation} & 8 (9 while sending a snapshot)\\
\thread{bias} & Implement commands from \process{BiasServer}, send biasing status to \process{BiasServer} & 4\\
\thread{hk\_serve} & Send housekeeping data to \process{DataCompilation} & 8\\
\enddata
\end{deluxetable}

The \thread{msg\_serve} thread waited for a network socket connection on port 5005.  After a connection was established, it sent the \process{adc\_server} software version (4 bytes) and ADC Board firmware version (4 bytes) over the network.  (Unless otherwise specified, all network communication began with sending these version numbers.)
The thread then slept until there was a message to send.  Other threads put error, warning, and informational messages into a buffer which can hold 2000 such messages for sending.
I incremented a semaphore  when a new message was put into the buffer.
The thread checked for an acknowledgment that the message was received before deleting it from the buffer.
If there was no acknowledgment, the thread closed the socket connection and waited for a new one.

The \thread{fourhz\_write} thread waited for the Master ADC Board to signal that a new data frame was ready to be read.
It did this by setting the 6th bit\footnote{i.e. the $2^5$ place} of the word at VME memory address 0x0\footnote{All VME memory addresses have an offset that depends on the slot the ADC Board occupies.  I suppressed these offsets here.  See \S5 of \cite{QUIET_DAQ_Crate} for more details.} to 1.
When a frame was ready to be read, the thread first latched the time from the Time-code Reader.
The thread then read the clock status from the Master ADC Board, radiometer data from each ADC Board, housekeeping data from the Master ADC Board, and the frame counter from each ADC Board.
The thread checked that the frame counter was the same for all ADC Boards.  This ensured that data from different ADC Boards were acquired at the same time.

\begin{figure}
\includegraphics[width=1.0\textwidth]{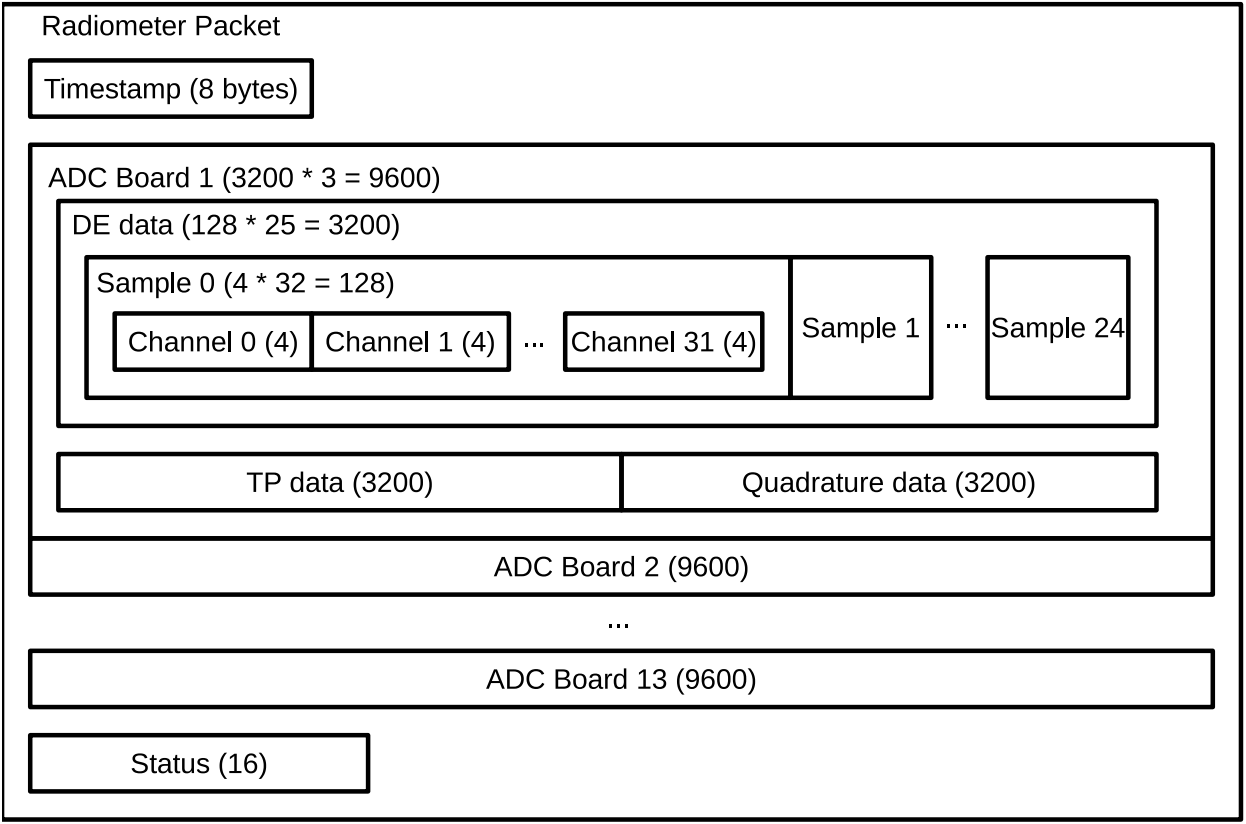}
\caption[\process{adc\_server} Radiometer Data Packets]{\label{fig:fourhz_packet}
I made \process{adc\_server} prepare packets containing 0.25\,s of TP, DE, and quadrature data (collectively, ``radiometer data'').  Numbers in parentheses represent bytes.
}
\end{figure}

I combined the data and status information  in packets shown in Figure \ref{fig:fourhz_packet}.
The first 8 bytes of the packet were the timestamp from the Time-code Reader.
The timestamp was a Unix time; the first 4 bytes were the integer number of seconds since 00:00:00 Coordinated Universal Time (UTC), January 1, 1970; the last 4 bytes were the integer number of 
microseconds\footnote{The use of Unix time here prevented proper representation of the time in the presence of leap seconds.  No QUIET observations contained a leap second.}.
Following the timestamp in each packet were the data.
If the packet contained housekeeping data, these were 125 housekeeping multiplex addresses (16-bit integers) and values (16-bit integers).
If the packet contained radiometer data, these were 25 samples (32-bit integers) each for 32 ADC Board channels for TP, DE, and quadrature per ADC Board.
Finally, each packet ended with 16 bytes of status information (Appendix~\ref{sec:adc_server_status_bytes}).
I temporarily stored the packets in a ring buffer capable of holding 60\,s of data.
After storing each packet, the \thread{fourhz\_write} thread incremented a semaphore to notify the \thread{data\_serve} or \thread{hk\_serve} threads that the packet was ready to be sent.
If network problems or a crash in \process{DataCompilation} caused the ring buffers to become full, data was lost.
In use, a problem that lasted for 60\,s typically required human intervention to solve so the buffer size was not a limiting factor causing data loss.

Every 100\,s the thread used the Time-code Reader time to update the SBC system time.
This procedure maintained the synchronization between them to within 5\,ms.
I only used the SBC time to timestamp monitoring data e.g. snapshots, logs, and 1-Hz housekeeping data.
Thus a possible 5-ms error in this time was acceptable.

The \thread{data\_serve} thread sent radiometer data packets over the network.
It waited for a connection on port 5000.
Upon establishing a connection, it sent the number of ADC Boards recording data\footnote{In W band the Master ADC Board did not acquire radiometer data; thus the number of ADC Boards recording data was the number of Slave ADC Boards.  In Q band, all ADC Boards (Master or Slave) recorded data.}.
Then it waited for a radiometer data packet to become ready to send.
When a packet was ready, it sent it over the network socket.
Then it waited for an acknowledgment from the other end of the connection.
If the acknowledgment byte was `c', then the thread erased the  sent packet from the buffer and returned to waiting for a packet to be ready.
If the acknowledgment byte was `q', then the thread erased the packet, closed the network connection, and returned to waiting for a connection.
If no acknowledgment was received, then the thread closed the connection, but did not erase the packet from the buffer.
This action ensured that data was not removed from the buffer before the data was properly stored in \process{DataCompilation}.

The \thread{hk\_serve} thread sent housekeeping data packets over the network.
It waited for a connection on port 5003 and then waited for a housekeeping data packet to become ready to send.
I used the same packet sending and acknowledgment system as in the \thread{data\_serve} thread.

The \thread{data} thread acquired and sent snapshots.
It waited for a connection on port 5001.
Upon establishing a connection, it sent the number of ADC Boards recording data.
Then it wrote the value 0 to the 4-byte word at VME memory address 0x30 on the Master ADC.
This instructed the ADC Boards to take a snapshot.
The thread waited 10\,ms for the instruction to be processed and then wrote the value 1 to the same VME memory address.
This instructed the ADC Boards to save the snapshot.
The thread waited 1\,ms for the instruction to be processed.
The thread then stored the SBC time to be used as the timestamp for the snapshot.
The thread then sent a packet over the network consisting of:
\begin{itemize}
\item The 8-byte timestamp (see above for format).
\item The $2^{17}$-byte snapshot data for each ADC Board.
\end{itemize}
The thread then waited 1\,ms before writing 0 to the VME memory address, instructing the ADC Boards to release the saved snapshot.
The thread finally closed the connection and returned to waiting for a connection.

The \thread{bias} thread implemented commands from \process{BiasServer} and reported the status of module biasing.
It waited for a connection on port 5002.
After establishing a connection, the action depended on the request received (Appendix~\ref{sec:adc_server_bias_thread_actions}).

\label{sec:DAQ:adc_server:threads}

Since only one \thread{adc\_server} thread can run simultaneously, I assigned priority and implemented synchronization among them.
My goal was to ensure that the \thread{fourhz\_write} thread always read data from the ADC Boards without delay and assigned the proper timestamp while minimizing delay in the tasks of other threads.
The \thread{adc\_server} threads had static scheduling priority; I broke ties within a priority level by round-robin scheduling.
Since the \thread{fourhz\_write} thread had the most critical operation, I assigned it the highest priority.
The \thread{bias} thread had the lowest priority because several bias commands take a long time (10\,s or more), and the other threads should not be blocked while such long commands occur.
The other threads had equal priority except that the \thread{data} thread had higher priority while responding to snapshot requests.
Because the snapshot timestamp was assigned by the thread, it should not be interrupted to provide the most accurate timing.
However, the \thread{fourhz\_write} thread will still interrupt snapshot requests if necessary.
When idle or not in a time-critical operation, I made each thread sleep or yield to allow other threads to run.

The threads used synchronization techniques to share resources.
I protected communication with the ADC Boards by a mutex\footnote{A mutex ensures that multiple threads do not access a shared resource (memory, VME communication, etc.) simultaneously \citep{pthread_mutex}.} so that they need only keep track of one request at a time.
The \thread{fourhz\_write} thread used a semaphore to indicate when new data were ready to be sent.
I used a producer-consumer model to ensure that ring buffers were not accessed unless they were ready to be read or written.
I protected buffers that had multiple producers (e.g. the message buffer)  by a mutex.


I conducted tests to verify that timestamps assigned by \process{adc\_server} were accurate.
I checked the difference between the Time-code Reader timestamp available over VME and the time implied by its 1-Hz output was $<35\,\mu$s. %
This difference included the communication delay between the Time-code Reader and \process{adc\_server}.
To check for delays between the Time-code Reader and ADC Boards, we injected the Time-code Reader 1-Hz clock pulse into an ADC channel and found the pulse in the recorded radiometer data.
This test confirmed a 20-ms delay (understood as ADC firmware buffering) also found from Moon observation \citep{25ms}.
To check that timestamps would be regular, I ran the DAQ system uninterrupted for 20\,hours.  All timestamps were within 1\,ms of the periodic expectation.
The Moon observations limited any overall offset in the timing system to $<1$\,ms \citep{timing_offset_monitor}.


I confirmed that data were not  lost between acquisition and long-term storage.
The frame counter (see above) was a unique identifier for each data packet.
I ran the system for 20\,hours, and there were no missing packets.
This test also checked possible loss in \process{DataCompilation}.
To check that the firmware was assigning the frame counter correctly, I confirmed the timestamp incremented 250\,ms for each frame counter increment, as expected.
I also confirmed that packets were not duplicated\footnote{Duplication was a problem during software development.  Although duplicated packets could be removed in post-processing, to keep the post-processing software simple I eliminated duplicates in the Online Software.}.
In shorter tests, I introduced simulated network problems (e.g. temporarily disconnecting an ethernet cable) and confirmed that the system recovered as long as the problem duration was shorter than the data buffer length.

\subsection{\process{DataCompilation}}
\label{sec:datacompilation}

The process \process{DataCompilation}\footnote{Source code is available at \url{https://cmb.uchicago.edu/svn/ibuder/RCS}} combined multiple data sources into a single stream for later storage.
It ran on the Receiver PC continuously during data taking.
I wrote it as a multi-threaded C++ program.
The threads and their functions are summarized in Table \ref{tab:DataCompilation:threads}.

\begin{deluxetable}{lp{4.5in}}
\tablecolumns{2}
\tablecaption{Summary of \process{DataCompilation} Threads
\label{tab:DataCompilation:threads}}
\tablehead{\colhead{Thread Name} & \colhead{Function} }
\startdata
\thread{demod} & Receive radiometer data from \process{adc\_server}\\
\thread{hk} & Receive housekeeping data from \process{adc\_server}\\
\thread{data\_server} & Send system state information and recent data to other Online Software processes\\
\thread{bias} & Receive bias state information from \process{BiasServer}\\
\thread{archive\_aggregator} & Combine different data sources into a single packet\\
\thread{timestamp} & Timestamp the \process{DataCompilation} log file every second\\
\thread{snap} & Request a snapshot from \process{adc\_server} every hour and write it to disk\\
\thread{per} & Receive auxiliary housekeeping data from \process{PeripheralServer}\\
\thread{wpid\_output} & Record WPID heater and fan output information\\
\thread{pdu\_read} & Record auxiliary housekeeping data from Electronics Enclosure smart power strip\\
\thread{data\_arc\_service} & Send combined packets to \process{DataArc}\\
\thread{cbi\_service} & Send combined packets to \process{cbicontrol}\\ 
\enddata
\end{deluxetable}

The \thread{demod} thread received radiometer data from \process{adc\_server} and performed preliminary processing on it.
It created and maintained a socket connection to the \process{adc\_server} \thread{data\_serve} thread.
Upon receiving a packet, it first checked if the frame counter was the same as the most recently received frame counter; if they were the same, the newly received frame was a duplicate and discarded.
Otherwise, the thread unpacked state information the firmware has stored in the sign bit of the DE data  \citep{QUIET_DAQ_Crate}.
The sign bits of channels 0--11 on each ADC Board comprised the number of samples accumulated when the 4-kHz clock was in the up state.
(Channel 0 contained the 1s place, channel 1 the 2s place, channel 2 the 4s place, etc. in the base-2 representation.)
The sign bits of the quadrature data of channels 0--11 comprised the number of samples accumulated when the quadrature sign was positive.
Channels 16--27 contained the same information for the down state.
Channels 30 and 31 contained the 1-Hz clock and 50-Hz clock, respectively.
After extracting these data, the thread replaced the sign bit with the next most significant bit so that the value was in the standard two's complement representation.
After this processing, the thread stored the resulting frame (250\,ms of data and metadata) in a global buffer that can be accessed by other threads.
It also put a copy in a queue (60\,s long) for inclusion in the archive stream.

The \thread{hk} thread received housekeeping data from \process{adc\_server} and performed preliminary processing on it.
I used the same algorithms  as the \thread{demod} thread except for the sign bit processing, which was not applicable to housekeeping data.

The \thread{data\_server} thread allowed other Online Software processes to receive the latest data over the network.
It listened for a network connection on port 4001\footnote{Online Software port configuration is in RCSports.h in the RCS repository.}.
When it established a connection, it created a new thread (\thread{data\_serve}) to handle it.
In this way, there can be up to five simultaneous connections to this interface.
Each remote process may send a 4-byte request.
The thread's action depended on the request (Appendix~\ref{sec:data_serve_requests}).
Table \ref{tab:DataCompilation:DataInterfaceServer:requests} summarizes the possible requests. 

\begin{deluxetable}{lp{3.5in}c}
\tablecolumns{3}
\tablecaption{Summary of \process{DataCompilation} \thread{data\_server} Requests
\label{tab:DataCompilation:DataInterfaceServer:requests}}
\tablehead{\colhead{Name} & \colhead{Description} & \colhead{Value} }
\startdata
DATA\_STATUS & Send \process{DataCompilation} status flags & 0\\
DATA\_NADC & Send number of ADC Boards with data & 1\\
DATA\_DEMOD & Send latest radiometer data frame& 3\\
DATA\_SNAP & Send latest snapshot & 4\\
DATA\_HK & Send latest housekeeping data frame& 5\\
DATA\_BIAS & Send latest biasing state& 6\\
DATA\_PER & Send latest auxiliary housekeeping data acquired from \process{PeripheralServer}& 8\\
DATA\_NEWSNAP & Cause \process{DataCompilation} to request and acquire a new snapshot from \process{adc\_server}, then send the identification number of the new snapshot& 9\\
DATA\_ENCLOSURE & Send monitoring data from Electronics Enclosure smart power strip& 10\\
DATA\_PID\_OUTPUT1 & Send output status of first WPID output& 11\\
DATA\_PID\_OUTPUT2\tablenotemark{a} & Send output status of second WPID output& 12\\

\enddata

\tablenotetext{a}{There is support in \process{DataCompilation} for additional PID outputs, but they do not exist in hardware.}

\end{deluxetable}

The \thread{bias} thread connected to \process{BiasServer} on port 4002 and received the latest bias state (see DATA\_BIAS in Appendix~\ref{sec:data_serve_requests}) twice per second.

The \thread{archive\_aggregator} thread combined data collected by other threads into a single stream.
This stream contained one frame per second.
In each frame I combined radiometer, housekeeping (from all sources listed above), and biasing data; Online Software status flags; and the most recent snapshot number.
The thread waited for four (the number needed to fill a single archive frame) radiometer and housekeeping frames to be collected by the \thread{demod} and \thread{hk} threads.
I then aligned them according to a timing algorithm (see \S\ref{sec:DAQ:timing_alignment}).
Then I copied the latest biasing data, \process{PeripheralServer} data, WPID outputs, Electronics Enclosure monitoring, and snapshot number to the archive frame being constructed.
I recorded the current version of the Receiver Database in the frame.
I created a new frame number for the archive frame\footnote{The archive frame number was four bytes so there were enough possibilities to avoid duplication during the entire lifetime of QUIET.  However, the frame number reset to zero whenever \process{DataCompilation} restarted.  
Therefore, the archive frame numbers repeated.  However, valid contiguous data cannot include a period when \process{DataCompilation} restarted.  Thus, frame numbers were unique within a single scan.}.
I also included the current Receiver PC time in the frame.
This timestamp was not accurate enough\footnote{I synchronized the Receiver PC time to the SBC time by Network Time Protocol,  
so it was indirectly derived from the same source as other timestamps.  However, the synchronization was only guaranteed within a few seconds.  
Moreover, \process{DataCompilation} was not a real-time process so there can be unexpected delays in assigning the timestamp.} for data alignment or pointing reconstruction but was used for monitoring the Online Software status and debugging.
When the frame was complete, the thread pushed it into two queues (60\,s long) for transmission to \process{DataArc} and \process{cbicontrol}\footnote{As a compile-time option, \process{DataCompilation} could be configured to send to either, neither, or both.}.

The \thread{timestamp} thread wrote the current time (from the Receiver PC clock) to the log every second.

The \thread{snap} thread requested and received snapshots from \process{adc\_server}.
It requested a new snapshot every hour or when signaled by a \thread{data\_serve} thread.
The thread assigned each snapshot a number.
I saved this number to disk on the Receiver PC so that snapshot numbers did not reset even if \process{DataCompilation} restarted.
The thread wrote each snapshot to disk on the Receiver PC.

The \thread{per} thread received the additional housekeeping data from \process{PeripheralServer}.
The thread connected to \process{PeripheralServer} on port 9011 and requested a new data frame every second.

The \thread{wpid\_output} thread acquired the WPID output states.
It sampled the Receiver PC parallel port at 100\,Hz.
The WPID Output Monitoring Board (\S\ref{sec:elec:WOMB}) made the WPID output control signal available at the parallel port.
The thread read the parallel port input (I/O port 0x0379 on the Receiver PC) with inb \citep{inb}\footnote{To access the parallel port in this manner, \process{DataCompilation} required root privileges.  
I typically compiled it to be executable by all users and setuid root.  
After granting access to the parallel port with ioperm \citep{ioperm}, \process{DataCompilation} dropped root privileges.  
Alternatively, \process{DataCompilation} may be run without root privileges; in this case, it did not read the WPID output information.}.
The combination of the WPID Output Monitoring Board wiring and mapping of parallel port pins caused the WPID output 1 to be on bit 7 ($2^7$s place) of the word read; output 2 was on bit 6.
Each sample indicated whether the output was active at the time of sampling.
The thread accumulated 100 samples into a frame and assigned each frame a timestamp based on the Receiver PC time.
The WPID output timestamps should not be used for sub-second timing alignment.

The \thread{pdu\_read} thread collected the monitoring information from the Electronics Enclosure smart power strip.
A separate process (\process{read\_pdu.py}) connected to the power strip by ethernet \citep{PDU_manual} and wrote the resulting data to a file on the Receiver PC.
The \process{DataCompilation} process read this file every second.

The \thread{data\_arc\_service} thread sent archive frames to \process{DataArc}\footnote{During normal W-band operation, we did not use \process{DataArc}.  We only recorded the data with \process{DataArc} for lab tests and initial and special observations that bypassed \process{cbicontrol}.}. 
It waited for an archive frame to be put into the queue by \thread{archive\_aggregator} thread.
Then it listened for a connection on port 4009.
The next action depended on the message received from \process{DataArc} (Table \ref{tab:DAQ:AggInterface} and Appendix \ref{sec:DataCompilationDataArc_communication}).

\begin{deluxetable}{lp{3.5in}c}
\tablecolumns{3}
\tablecaption{Summary of \process{DataCompilation} and \process{DataArc} Communication
\label{tab:DAQ:AggInterface}}
\tablehead{\colhead{Message Name} & \colhead{Description} & \colhead{Value} }
\startdata
AGG\_NOMSG & Invalid message or communication error & 0\\
AGG\_FRAME & Message containing an archive frame (only sent by \process{DataCompilation}) & 1\\
AGG\_ACK & Message acknowledging receipt of frame (only send by \process{DataArc}) & 2\\
AGG\_ACK\_REQ & Message requesting acknowledgment of receipt of frame (only sent by \process{DataCompilation}) & 3\\
AGG\_QUIT & The connection will be closed & 4\\
AGG\_NACK & The most recent frame was not received correctly (only sent by \process{DataArc}) & 5\\
AGG\_FLUSH\_BUFFER & Instruct \process{DataCompilation} to flush the frame buffer& 6\\
\enddata
\end{deluxetable}

The \thread{cbi\_service} thread sent archive frames to \process{cbicontrol}.
It waited for a frame to be put in the queue by the \thread{archive\_aggregator} thread.
Then it waited for \process{cbicontrol} to be ready to receive a frame.
Then it converted the frame into the format acceptable by \process{cbicontrol} \citep{cbi_registers}.
I packed all data types  as unsigned 4-byte integers.
I reorganized the bias data to eliminate the unphysical addresses\footnote{See BiasBitRepacker.h in the RCS repository.}.
I reorganized the radiometer data so that the 100 samples from each channel were in consecutive order.
Finally the thread sent the frame and waited for a new frame.


\subsection{\process{ReceiverControl}}
The process \process{ReceiverControl}\footnote{Source code is available at \url{https://cmb.uchicago.edu/svn/ibuder/RCS}} allowed the Control Software and observers to manage the bias electronics.
It ran on the Receiver PC whenever we changed the bias.
It was a single-threaded C++ program and did not check for other instances.
We  took care to ensure that at most one copy was ever running.
Otherwise, the bias state could be changed in an inconsistent and unexpected way.
The behavior of \process{ReceiverControl} depended on command-line arguments (Table \ref{tab:ReceiverControl:commands}.  Further command details can be found by running \verb|RC_functions_new ?|).
The \process{ReceiverControl} process constructed lower-level commands which it sent to \process{BiasServer} and monitored the command results from \process{DataCompilation}.

The Receiver Database (Appendix~\ref{app:receiver_db_tables}) allowed \process{ReceiverControl} to translate between bias identifiers and addresses.
The database included the mappings between ADC channels, preamp channels, and module diodes; bias identifiers, cards, addresses, and DACs; housekeeping identifiers, addresses, and the corresponding bias boards; card locations, versions, and serial numbers; module serial numbers, types, and locations; and housekeeping thermometer names, locations, types, and calibration constants.
It also stored the opto-isolator conversion tables needed to convert housekeeping voltages to physical units.
It also stored bias settings for rapid reapplication and reuse.
The database stored its own version number for consistency.

\subsection{\process{DataArc}}
The \process{DataArc}\footnote{Source code is at \url{https://cmb.uchicago.edu/svn/ibuder/RCS}}
 process recorded the data to disk on the Receiver PC.  We used it in the lab and for special observations at the site; normal site data were recorded by \process{cbicontrol}. 
I wrote it as a single-threaded C++ program.
It connected to the \process{DataCompilation} \thread{data\_arc\_service} thread and communicated as follows.
First \process{DataArc} sent AGG\_FLUSH\_\\BUFFER to force \process{DataCompilation} to discard any old data.
Then it waited for \\\process{DataCompilation} to send a message.
\begin{enumerate}
\item If the message was AGG\_FRAME, \process{DataArc} wrote the frame to disk and sent AGG\_ACK.

\item If the message was AGG\_ACK\_REQ, \process{DataArc} sent AGG\_ACK.

\item In all other cases or if an error occurred, \process{DataArc} closed the connection and reestablished it.
\end{enumerate}

\section{Timing Alignment}
\label{sec:DAQ:timing_alignment}
I ensured that data were assigned correct timestamps and that data acquired from different sources at the same time were combined in the same frame.
On the hardware side, the Time-code Reader provided absolute time from GPS and clocks that regulated the ADC.
The ADC firmware ensured that frames were aligned with the Time-code Reader time (meaning that a frame began at 0.000\,s) and that the interval between radiometer (housekeeping) samples was 10\,ms (2\,ms).
On the software side, \process{adc\_server} combined the ADC data with timestamps from the Time-code Reader.
The \process{DataCompilation} process aligned the radiometer and housekeeping frames (Appendix~\ref{app:timing_alignment}). 
I verified the alignment between receiver and mount data as part of the Routine Checks (\S\ref{sec:routine_checks:DAQ}).

\label{sec:DAQ:tcr}

The Time-code Reader provided the absolute time reference for the receiver\footnote{The mount electronics time-code reader received the same time code.
Thus we synchronized the receiver and mount systems to each other and to the absolute GPS time.}
by providing the decoded time (year, day, hours, minutes, seconds) to \process{adc\_server}\footnote{Possibly useful information on configuring the Time-code Reader and its VME communication is in \url{https://cmb.uchicago.edu/svn/kapner/quiet/crate_fs/branches/quadrature/root/vme_adc_threaded/src/TFP.c}}.
In addition, it created 1-Hz and 10-MHz clocks, phase-locked to the start of the GPS second, that synchronized the Master ADC.
The 1-Hz clock had a $1/5$ duty cycle, and the 10-MHz clock, a $1/2$ duty cycle (square wave).

\section{Control Software}

The Control Software allowed the observers to implement simple commands and monitor the system status.
The \process{Receiver Control Panel} provided high-level control of the receiver biasing.
The \process{cbiviewer} process provided control and monitoring of the mount and commands for starting observations.
Several stand-alone programs assisted in the creation of observing schedules.
The \process{Live:QUIET} web page provided a graphical summary of the experiment status.

\subsection{\process{Receiver Control Panel}}
The \process{Receiver Control Panel} was a graphical interface for controlling and monitoring the receiver.
It ran on a computer in the control room (control PC).
It displayed the current state of the receiver, which could be
\begin{itemize}
\item \textbf{On} All biases were set for normal operation.
\item \textbf{StandBy} The DAQ system was on, but the modules were off.
\item \textbf{Off} The bias electronics and ADC Boards were off.
\item \textbf{Down} All electronics, including the Receiver PC, were off.
\end{itemize}
We changed the state by clicking a button on \process{Receiver Control Panel}.
It also provided a status display indicating whether the Receiver PC, ADC Backplane, SBC, WPID, and enclosure smart power strip were online; whether the bias electronics were on; and whether data were being properly recorded. 
\cite{receiver_on_off} provides further documentation.

\subsection{Schedule Making}
Several utilities assisted the observers in making the observing script each day.
\cite{observing_priorities} listed the astronomical sources to be observed.
The \process{sched\_maker} utility \citep{scheduling_manual} provided a graphical display of when these sources would be within the elevation limits.
The observers used it to generate a schedule file.
The \process{QUIET Shift Utility} \citep{quiet_shift_utility} checked that observations avoid the Sun\footnote{We maintained at least $30\degr$  separation between the Sun and any source.
The main worry was that radiation focused by the telescope could melt the cryostat vacuum window.}. 
The \process{cbiviewer} process checked that the schedule file could be parsed.
Ideally this check would find invalid schedules before they were submitted.
However, logic errors (e.g. attempting to observe a source below the elevation limit) were not always detected in advance.
Finally \process{cbiviewer} submitted the schedule to \process{cbicontrol}, which moved the mount to follow it.

\subsection{Real-time Monitoring}
\label{sec:LIVE:QUIET}
\begin{figure}
\includegraphics[width=1.0\textwidth]{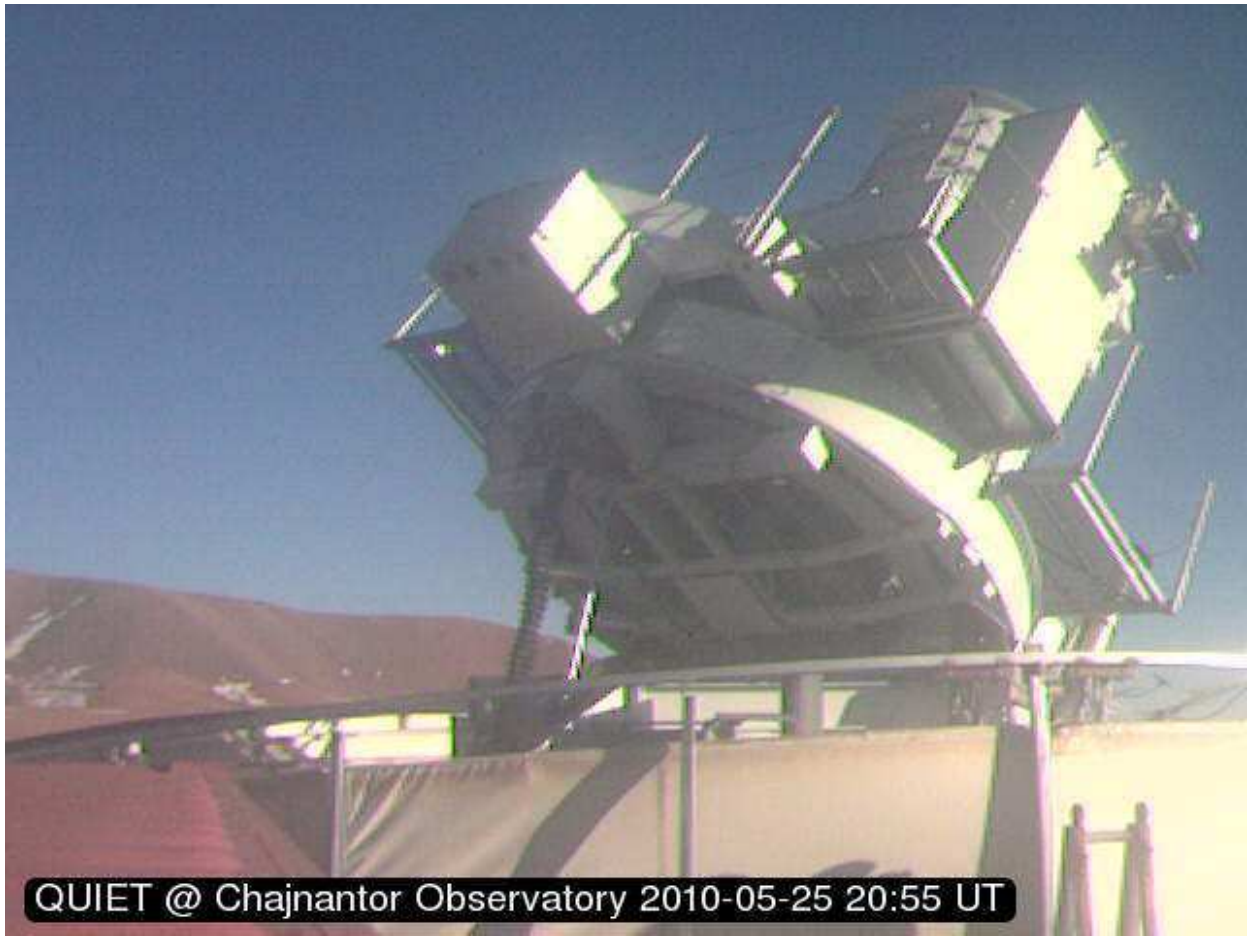}
\caption[\process{Live:QUIET} Automatic Photograph]{\label{fig:webcam_example}
We took an image of the instrument every 5 minutes.
The \process{Live:QUIET} web page displayed the latest image.
We used the images for basic monitoring.
}
\end{figure}

The web page \process{Live:QUIET}\footnote{\url{https://qufs.uchicago.edu/DQM/live_Wband/trunk/html/live2.html}}
summarized the experiment status.
It showed the last day of observation (mount motion and astronomical source), weather statistics, plots of the radiometer data from each module, receiver housekeeping, errors detected by the Online Software, and a picture from a digital camera pointing at the instrument (Figure \ref{fig:webcam_example}).
We performed a detailed check of this information three times per day \citep{checklist_howto}.
We logged the findings in a Google spreadsheet\footnote{\burlalt{http://spreadsheets.google.com/ccc?key=0AgeJEGqcUOLEdE1FVmpoaTIyTkVkeTZDd1ppeFhSdWc&hl=en}{{Online}}}
in case we discovered a problem later.

\section{Data Transfer and Storage}
We transferred data regularly from the site to storage at the analysis centers.
The process \process{cbicontrol} wrote data files\footnote{Typical files stored 15\,minutes of data in 400\,MB.} on the control PC in a proprietary format (``Level-0'').
These files were automatically copied to the Storage PC in the control room, which had an attached RAID with space for 10\,weeks of data.
We copied newly created data files to Blu-ray disks (25\,GB).
Including downtime and compression, typical W-band observation generated 1--2 disks per day.
We sent these disks to Chicago in weekly shipments.
Upon arrival, we copied the data files to a file server with 29\,TB of attached RAID storage.
Furthermore, we copied the files by network to mirror servers in Oslo and the High Energy Accelerator Research Organization (KEK).
Only when a data file existed in both Chicago and Oslo or KEK did we delete the file from the site.
Thus we always stored two redundant copies of the data.
Moreover, we used a Secure Hash Algorithm (SHA-1) checksum to confirm that the data was not corrupted in transit.
(See also \cite{data_flow_plan,data_management_wiki})

\label{sec:DAQ:files}
The Level-0 data files written by \process{cbicontrol} were not in a format optimized for analysis.
All data, whether receiver, mount, or housekeeping were interleaved because the major organization of the file was one frame per second of data.
Moreover, different channels were grouped together by location (or multiplex address) in the readout electronics rather than by physical proximity.
Conversely, a typical pattern in analysis was to access only a few channels that are in physical proximity for a long time interval.
To make data files more amenable to such use, we automatically converted each Level-0 file to a corresponding Level-1 file.

We designed Level-1 data files  for the typical analysis use case.
The files used the common Flexible Image Transport System (FITS) format.
Different parts of the experiment (receiver, mount, housekeeping, and biasing) were stored in separate Header Data Units (HDUs).
Data from different diodes in the same module were stored together.
Data from each channel were contiguous for time period of the data file (rather than 1\,s as in Level-0).
(See also \cite{level1format})

\section{Database}
We stored metadata about each data file, scan, and observer log in a mySQL (Structured Query Language) database (Appendix~\ref{sec:quiet_data}).
For each data file we recorded its storage locations, begin and end time, and checksum.
This information made it easy to check that no files were lost in the transfer process.
For each scan we recorded the type (CMB, calibration, or other), source, status (successful completion or abort), and begin and end time.
By combining the file and scan information we  dynamically generated the list of files with data for any scan or set of scans.
The daily observer log included the date, observer, and comments.
Indexing the observer log made it search-able; this was useful when we discovered problems long after the observation.

\svnid{$Id: observations.tex 150 2012-07-24 22:04:36Z ibuder $}

\chapter{Observations}
\label{sec:observations}
\epigraph{If I have not seen as far as others, it is because giants were standing on my shoulders.}{Harold Abelson}

We conducted W-band science observations between August 2009 and December 2010.
QUIET deployed to the former CBI site in Chile's Atacama Desert.
During regular observation, we observed low-foreground regions of the sky every day and a weekly schedule of astronomical calibration sources.
At the end of the season (December 2010) we performed special calibration observations with astronomical and artificial sources.
Sometimes, noteworthy events interrupted the regular observation schedule.
We routinely monitored the performance to correct problems as early as possible.

\section{Site}
\label{sec:observations:site}
\begin{figure}
\includegraphics[width=0.75\textwidth]{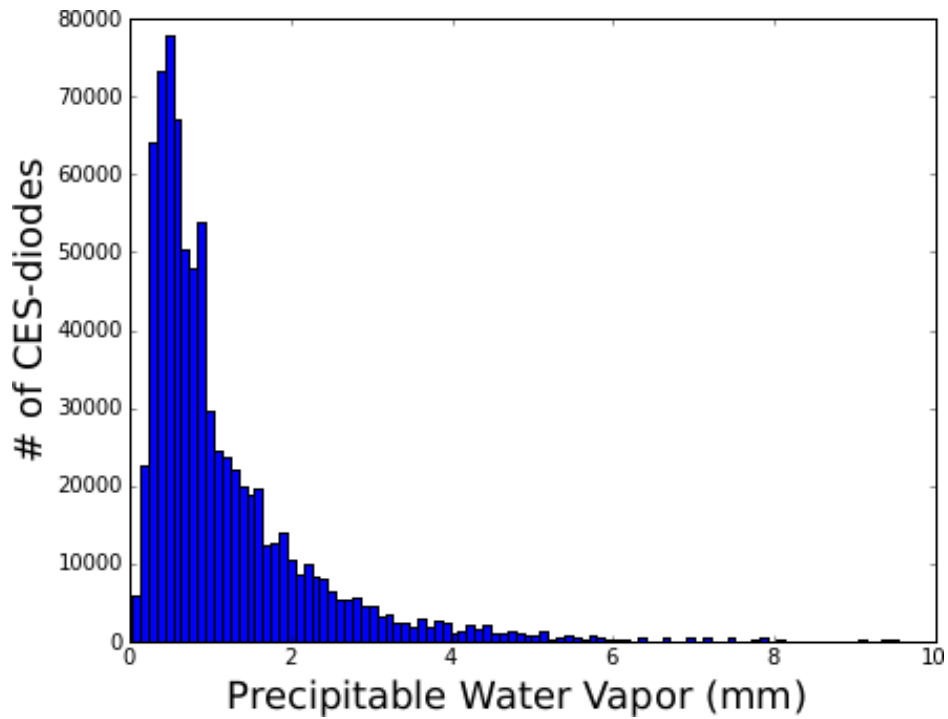}
\caption[Precipitable Water Vapor near the Site]{\label{fig:pwv_wband_season}
The precipitable water vapor (PWV) near the site (CTF) was typically $<1$\,mm.
Low PWV corresponded to good weather.
Our PWV monitoring came from the nearby Atacama Pathfinder EXperiment \citep{apex_instrument}.  (I only show CES-diodes passing data selection.)
}
\end{figure}

The observation site was the Chajnantor Test Facility (CTF, $67\degr45'42''$ W, $23\degr1'41''$ S) on the Chajnantor Plateau in Chile's Atacama Desert.
CBI observed from the same site; we also used it for Q-band observations \citep{quiet_qband_result}.
The low atmospheric water vapor (Figure \ref{fig:pwv_wband_season}) and high elevation (5080\,m, 50\% of sea level pressure)
reduced atmospheric emission and absorption: transmission was 98\% in W band \citep{quiet_instrument, pardo_atm}.
Atmospheric fluctuations (``weather'') tended to be worse in the afternoon and the months December--March.
The air temperature varied between $-15\degr$C (winter at night) and $10\degr$C (summer days).

The site facilities supported observations.
A heated, oxygen-enriched control room contained the Control and Storage PCs and stations for the observers.
Two labs (one heated) provided space for instrument investigation and repair.
Two diesel generators provided power (60\,kW) for the instrument.
(One generator could power the entire site; the other was a backup.)
Water chillers supported the water-cooled compressors needed to keep the cryostat cold.
A fully retractable dome protected the instrument in bad weather (snow, extreme wind).
Atacama Large Millimeter Array (ALMA) generously provided internet access.
There were beds for the observers to sleep at the site; however, typically we returned to the nearby (1\,hour) San Pedro de Atacama every night.

\section{Regular Observations}
\label{sec:regular_observations}

We performed three types of regular observations: CMB, calibration, and Galactic.
We observed the same four sky fields (``patches'') every day for CMB science.
A weekly schedule of calibration observations supported the CMB observations by giving us a regular understanding of the instrument performance.
Galactic observations are outside the scope of this thesis.
We spent 71\%, 15\%, and 14\% of the time on CMB, calibration, and Galactic observations, respectively (Table \ref{tab:observation_summary}).
The observing efficiency (observing time divided by calendar time) was 63\%.
We started regular observations on August 15, 2009 and finished on December 17, 2010.

\begin{deluxetable}{lrr}
\tablecolumns{3}
\tablecaption{Observation Summary  
\label{tab:observation_summary}}
\tablehead{\colhead{Source} & \colhead{Hours Observed} & \colhead{Fraction of Observations (\%)}}
\startdata
CMB-1 (2a) & 1855.0\tablenotemark{a} & 25\\
CMB-2 (4a) & 1443.6\tablenotemark{a} & 19\\
CMB-3 (6a) & 1388.7\tablenotemark{a} & 19\\
CMB-4 (7b) & 649.6\tablenotemark{a} & 9\\
\hline
Total CMB & 5336.9 & 71\\
\hline
Galaxy & 1046.5\tablenotemark{a} & 14\\
\hline
Moon & 125\tablenotemark{b} & 2\\
Tau~A & 324\tablenotemark{b} & 4\\
Jupiter & 168\tablenotemark{b} & 2\\
Venus & 5\tablenotemark{b} & 0.07\\
RCW~38 & 31\tablenotemark{b} & 0.4\\
Sky dip & 250\tablenotemark{b} & 3\\
Offset & 190\tablenotemark{c} & 3\\
\hline
Total Calibration & 1090 & 15\\
\hline
Grand Total & 7470 & 100\\
\enddata

\tablenotetext{a}{SOURCE: \cite{masaya_dataset}}
\tablenotetext{b}{SOURCE: \cite{observation_summary2}}
\tablenotetext{c}{SOURCE: \cite{observation_summary}}
\end{deluxetable}

\subsection{CMB Observations}
\label{sec:obs:cmb}
\begin{figure}
\includegraphics[width=1.0\textwidth]{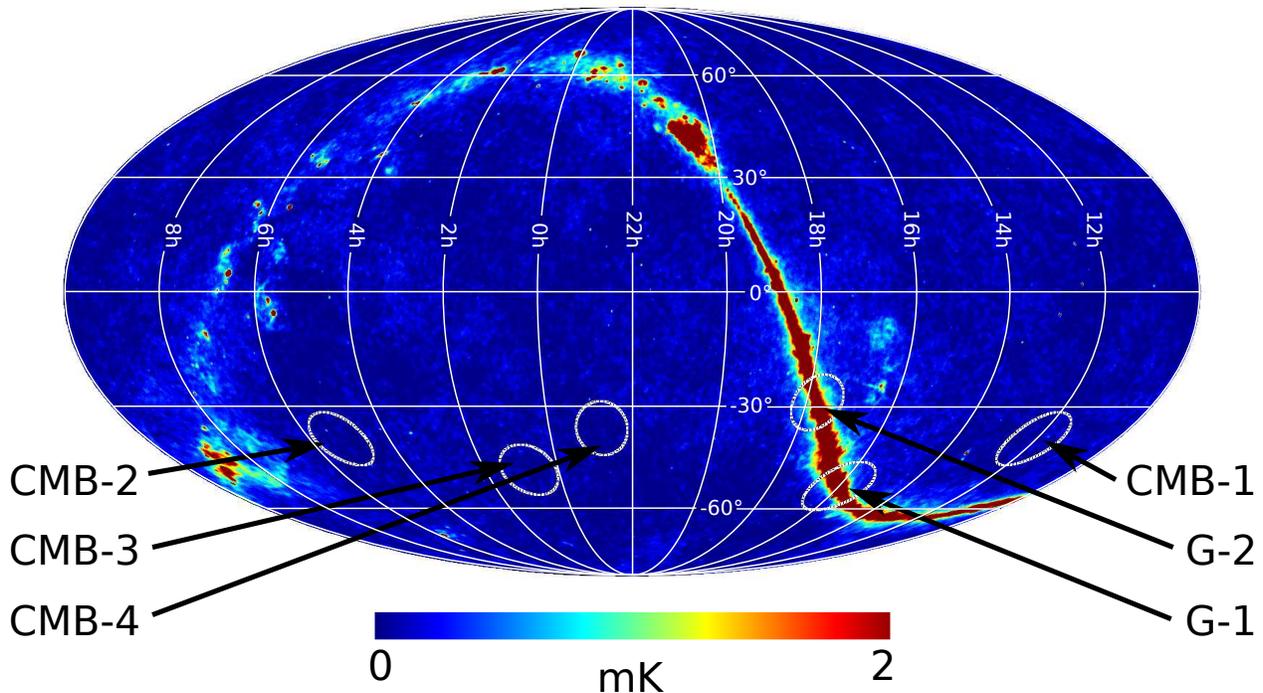}
\caption[Sky Fields]{\label{fig:patches}
We observed four sky fields (``patches,'' CMB-1--CMB-4) for CMB science.
We chose the patches in low-foreground regions of sky.
Two Galactic patches, G-1 and G-2, were for secondary foreground science goals.
Coordinates are Right Ascension and Declination.
The background is the \textit{WMAP} Q-band temperature map.
}
\end{figure}

We regularly observed four patches of sky for CMB science (Figure \ref{fig:patches} and Table \ref{tab:patches}).
Each patch was roughly $15\degr\times15\degr$.
We picked the patches to be in low-foreground regions of the sky, determined from \textit{WMAP} maps and dust templates \citep{dorothea_patch_selection}.
We also required the patches to be evenly distributed in Right Ascension (RA) so that at least one patch would always be above the elevation limit for observing ($43\degr$).
Furthermore, no patch could be within $30\degr$ of the Sun.
We observed patch 2a the most, and patch 7b, the least (Table \ref{tab:observation_summary}).
The total CMB observation time was 5336.9\,hours.
Patches 2a and 4a contained significantly polarized astronomical sources: Centaurus~A and Pictor~A, respectively.
We removed regions near these sources from analysis (\S\ref{sec:spectra_calc}).

\begin{deluxetable}{lrr}
\tablecolumns{3}
\tablecaption{CMB Patch Centers
\label{tab:patches}}
\tablehead{\colhead{Name} & \colhead{Right Ascension} & \colhead{Declination}}
\startdata
CMB-1 (2a) & $12^\textrm{h}04^\textrm{m}$ & $-39\degr00'$\\
CMB-2 (4a) & $05^\textrm{h}12^\textrm{m}$ & $-39\degr00'$\\
CMB-3 (6a) & $00^\textrm{h}48^\textrm{m}$ &$-48\degr00'$ \\
CMB-4 (7b) & $22^\textrm{h}44^\textrm{m}$ & $-36\degr00'$ \\
\enddata
\end{deluxetable}

\begin{figure}
\includegraphics[width=1.0\textwidth]{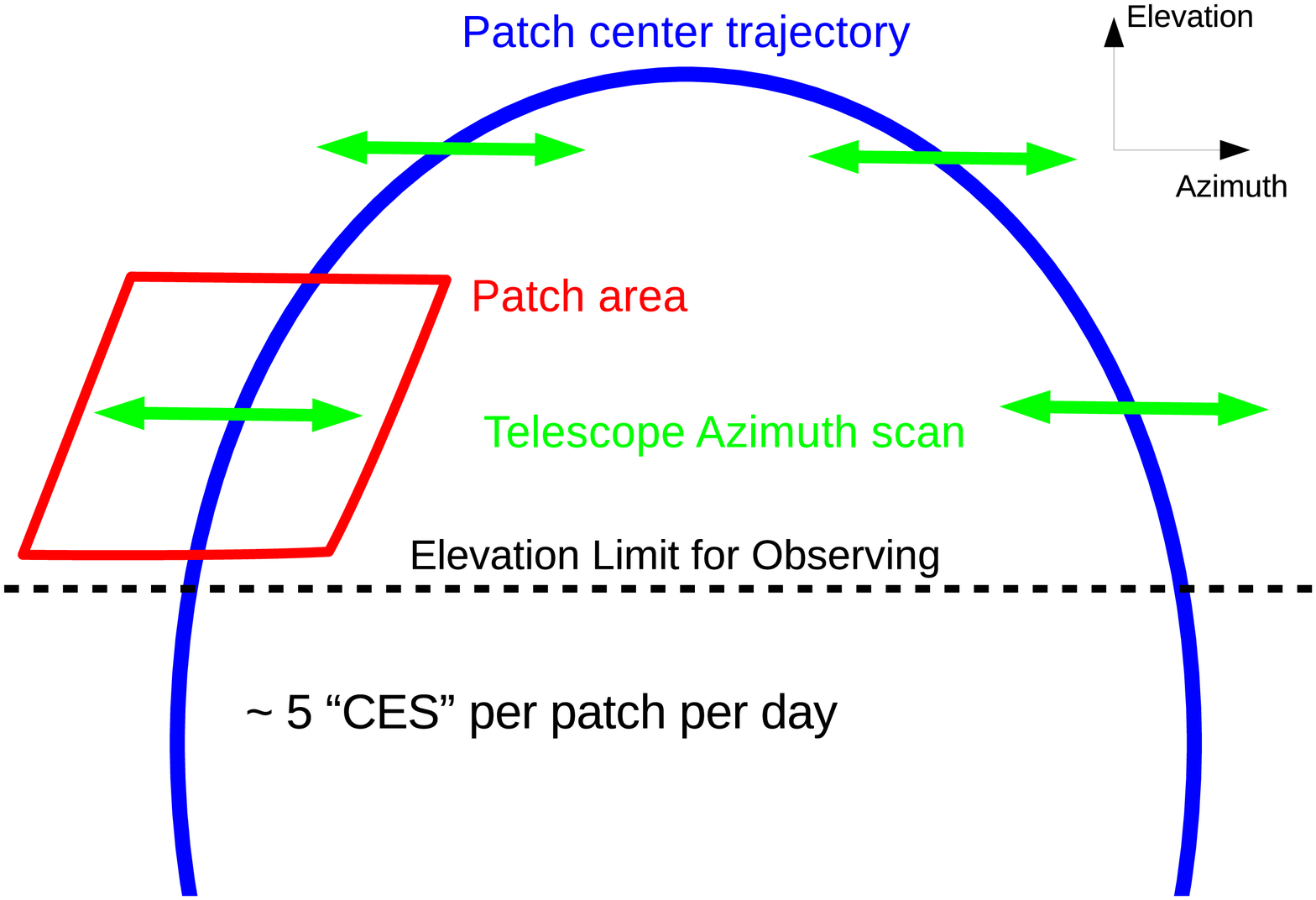}
\caption[Scan Strategy Cartoon]{\label{fig:patch_observing_cartoon}
We observed each patch with a series of constant-elevation azimuth scans.
We pointed the telescope ahead of the patch center in Right Ascension (RA) and fixed the elevation.
The combination of a periodic azimuth scan and sky rotation mapped the patch area in two dimensions.
After $\approx1$\,hour (1 CES), we stopped the scanning, re-pointed the telescope, and repeated the process.
In this way, we accumulated $\approx5$ CESes per patch per day.
}
\end{figure}

\begin{figure}
\includegraphics[width=1.0\textwidth]{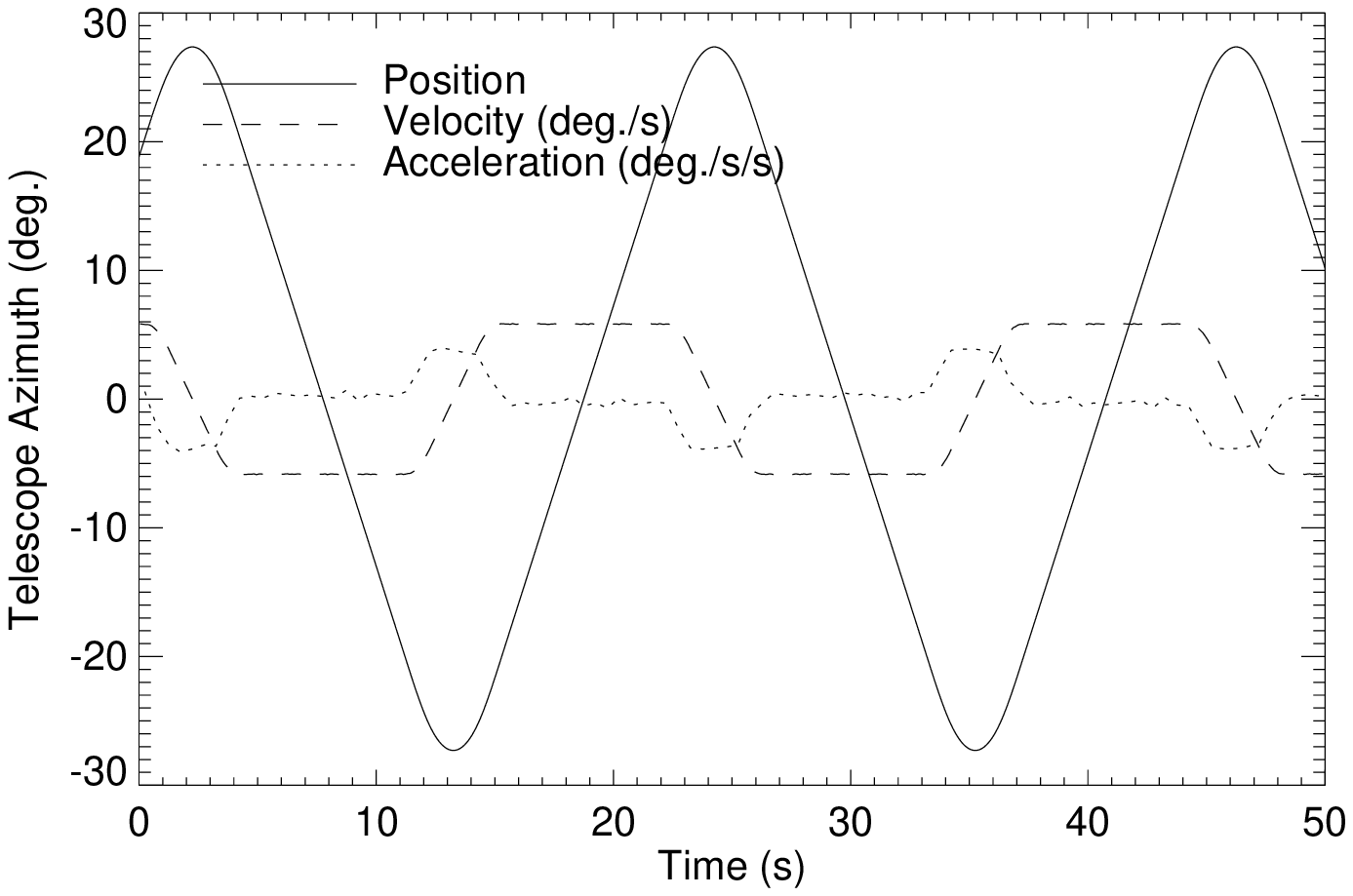}
\caption[Scan Trajectory]{\label{fig:scan_trajectory_example}
The CMB scan combined constant-velocity, constant-acceleration, and constant-jerk segments.
This scan had period 22\,s.
Scans with sufficiently short period did not have constant-velocity segments.
}
\end{figure}

When observing a patch, we conducted a series of constant-elevation azimuth scans (Figure \ref{fig:patch_observing_cartoon}).
First, we pointed the telescope $7.5\degr$ (on the sky) ahead of the patch center in RA.
Then we fixed the elevation and repeated azimuth scans.
The scan amplitude was $15\degr$ (on the sky), and the period was 10--26\,s\footnote{Limitations of the mount control required the scan period to be an integer multiple of 2\,s \citep{mount_motion_control, mount_trajectory}.}.
The scan trajectory combined constant-jerk ($2.25\degr$/s$^3$), constant-acceleration ($\leq4.5\degr$/s$^2$), and constant-velocity ($\leq6.3\degr$/s) segments (Figure \ref{fig:scan_trajectory_example}). 
The typical scan speed was $5\degr$/s in azimuth.
After $\approx1$\,hour, the patch center  drifted by $15\degr$.
At this point we ended the azimuth scans, pointed the telescope ahead of the patch again, and started a new series of azimuth scans.
Each series of azimuth scans was a Constant Elevation Scan (CES).
A uninterrupted (e.g. by the patch setting or a power failure) series of CESes of a single patch was a ``run\footnote{There were a few exceptions.  We defined the final CES boundaries after all observations were complete.
In this process we discovered runs that were incorrectly assigned initially, and some parts of observations had to be discarded.
As a result, we renumbered some runs in violation of the general rule.}.''
Once the patch set below the elevation limit, we finished the run and began a new run of a different patch. 
Every week we rotated the deck angle by $45\degr$ so that observations would be uniformly distributed among deck angles.

\subsection{Calibration Observations}
\label{sec:obs:cal}
We routinely observed astronomical sources to generate calibration data (Table \ref{tab:regular_calibration}).
Many of the sources had limited availability so we interrupted CMB observations when necessary to observe the calibration sources.

\begin{deluxetable}{llrp{2in}}
\tablecolumns{4}
\tablecaption{Regular Calibration Observations
\label{tab:regular_calibration}
}
\tablehead{\colhead{Source} & \colhead{Observation Frequency} & \colhead{Duration (min.)} & \colhead{Calibration Obtained}}
\startdata
Taurus~A & 1/day & 75 & Responsivity, beam shape, polarization angle\\
Large Sky Dip & 1/day & 15 & Responsivity, $I\rightarrow Q/U$ leakage\\
Mini Sky Dip & 1/CES & 3 & Responsivity, $I\rightarrow Q/U$ leakage\\
Moon & 1/day & 30 & Pointing model\\
Moon & 1/week & 60 & Pointing model\\
Jupiter &1/week &60& Pointing model, beam shape, responsivity\\
Jupiter&1/day & 25 & Pointing model, beam shape, responsivity\\
RCW~38 &1/week & 30 & Pointing model, responsivity\\
\enddata
\end{deluxetable}

We observed the supernova remnant Taurus~A (``Tau~A'') every day to calibrate the responsivity and beam shape.
Each day we performed tight ($<1\degr$) raster scans, centering two modules on Tau~A.
Every other day one of these modules was the central module, RQ45.
The other module observations were evenly distributed across the array.
We repeated each raster scan for five deck angles.
Tau~A observations took 75\,minutes each day.
Tau~A was too close to the Sun between May 5 and August 6, creating a gap in observations \citep{taua_run_list}.

Every day we performed a large sky dip.
The large sky dip was a periodic scan of the telescope in elevation between $43\degr$ and $87\degr$ with azimuth and deck held fixed.
The scan repeated six times and used 15\,minutes.
As the elevation changed, so did the effective atmospheric thickness and emission.
We used this changing effective atmospheric temperature as the calibration source.

Before each CES we performed a mini sky dip and offset measurements.
The mini sky dip, like the large sky dip, was a periodic elevation scan, but the amplitude was reduced to $\approx5\degr$. 
During offset measurements, we reverse-bias the phase switches to reduce their transmission.
Then we measured the offset (voltage level with no RF power) of each diode.
The DD offset did not influence the analysis because we only used the modulation of the polarization signal induced by scanning.
However, the TP offset was necessary to relate TP to radiation intensity $I$.

We observed the Moon on both a daily and weekly schedule.
Every day we performed a raster scan with the central row of the array\footnote{RQ45 and whichever modules were aligned with it at the deck angle for CMB observations that week.}.  The raster scan took 30\,minutes.
Every week we performed a full-array scan, with trajectory like that of a CES, that exposed every module to the Moon.
We used the same full-array scan to observe Jupiter every week.

We used Jupiter, Venus, and RCW~38 (an HII region) to calibrate the differential-temperature modules.
These observations contributed to the pointing model.

\section{Season-end Calibrations}
At the end of the season (December 2010) we performed  special calibrations.
We put a rotating sparse wire grid \citep{ltd_wiregrid} in front of the cryostat window.
This calibrator created a modulated, polarized (1\,K) signal in every module.
We used the wire grid with varying cryostat and Electronics Enclosure temperatures to measure the dependence of the responsivity on the temperature.
We used a swept, narrow-band source to measure the band-passes \citep{wband_bandpass}. 
To map the far sidelobes, we performed several large angle scans using the Sun as a source.
We performed large sky dips with additional window material in front of the cryostat to confirm the noise contribution from the window.

\section{Noteworthy Events}
\label{sec:noteworthy_events}
Events throughout the season interrupted or otherwise affected normal observations.
Table~3.3 of \cite{ali_thesis} provides a complete list of such events.
Here I list those that particularly impact this thesis:
\begin{enumerate}
\item On August 15, 2009 we replaced one of the Phase-switch Boards.  This marked the beginning of regular W-band observing.
\item On 16 occasions we changed the CPID regulation parameters because the cryostat temperature was unstable.
\item On 6 occasions the generator failed, forcing a suspension of observations.
\item On 3 occasions we patched holes in the ground screen.
We suspect these were related to residual ground pickup.
\item On 17 occasions the mount motors stalled in the middle of an observation, halting it.
\item On 11 occasions, degradation of the cryostat vacuum made it difficult to maintain the cryogenic temperature, and we suspended observations to pump down the cryostat.
\item On 8 occasions a power supply for one of the Preamp Boards failed.  The result was unacceptable noise increases on a corresponding MAB.  We replaced the power supply as soon as possible.
\item On 6 occasions we changed the Electronics Enclosure regulation parameters to keep the temperature stable.
\item Between September 26 and October 19, 2009, we replaced one of the cryostat cold heads because it exceeded the design lifespan and no longer provided sufficient cooling power.
\item On January 26, 2010 we installed the upper ground screen.
\item On May 30 and June 2, 2010, we found loose bolts in the cryostat window holder.  We suspect these contributed to the vacuum leakage.
\item On December 17, 2010 we removed the upper ground screen for special calibrations.
This marked the end of the W-band season. 
\end{enumerate}

\section{Routine Checks}

We monitored the instrument performance and data flow with routine checks every week.
These checks supplemented the automatic real-time and daily checks (\S\ref{sec:LIVE:QUIET}).
The checks included data transfer and processing status, TP offset level, noise levels, responsivity, and polarization angle.
I particularly checked the Electronics Enclosure temperature regulation and status of the Data Management system.

\subsection{Enclosure Temperature}
\label{sec:routine_checks:enclosure}
\begin{figure}
\includegraphics[width=0.75\textwidth]{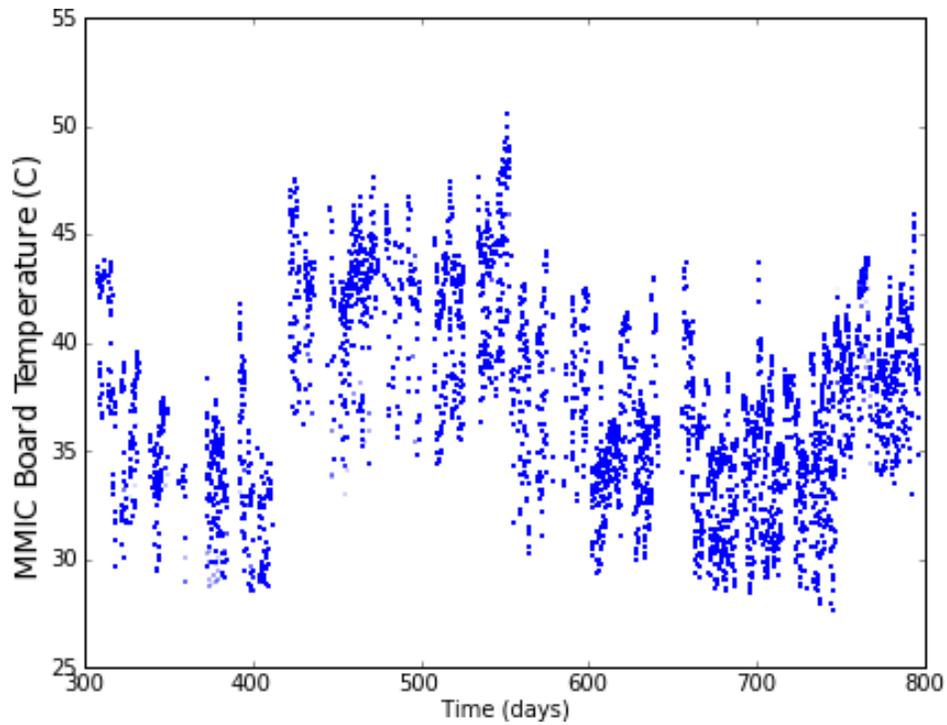}
\caption[Electronics Enclosure Temperature During the Observing Season]{\label{fig:electronics_temp_vs_time}
The Electronics Enclosure temperature changed by $20\degr$\,C during the observing season.
The change was due to two effects.
Diurnal variation caused a $10\degr$C oscillation each day.
The regulation setpoint change throughout the season caused an additional $5\degr$C change on month time scales.
(I only show CESes that passed data selection.)
}
\end{figure}

\begin{figure}
\includegraphics[width=0.75\textwidth]{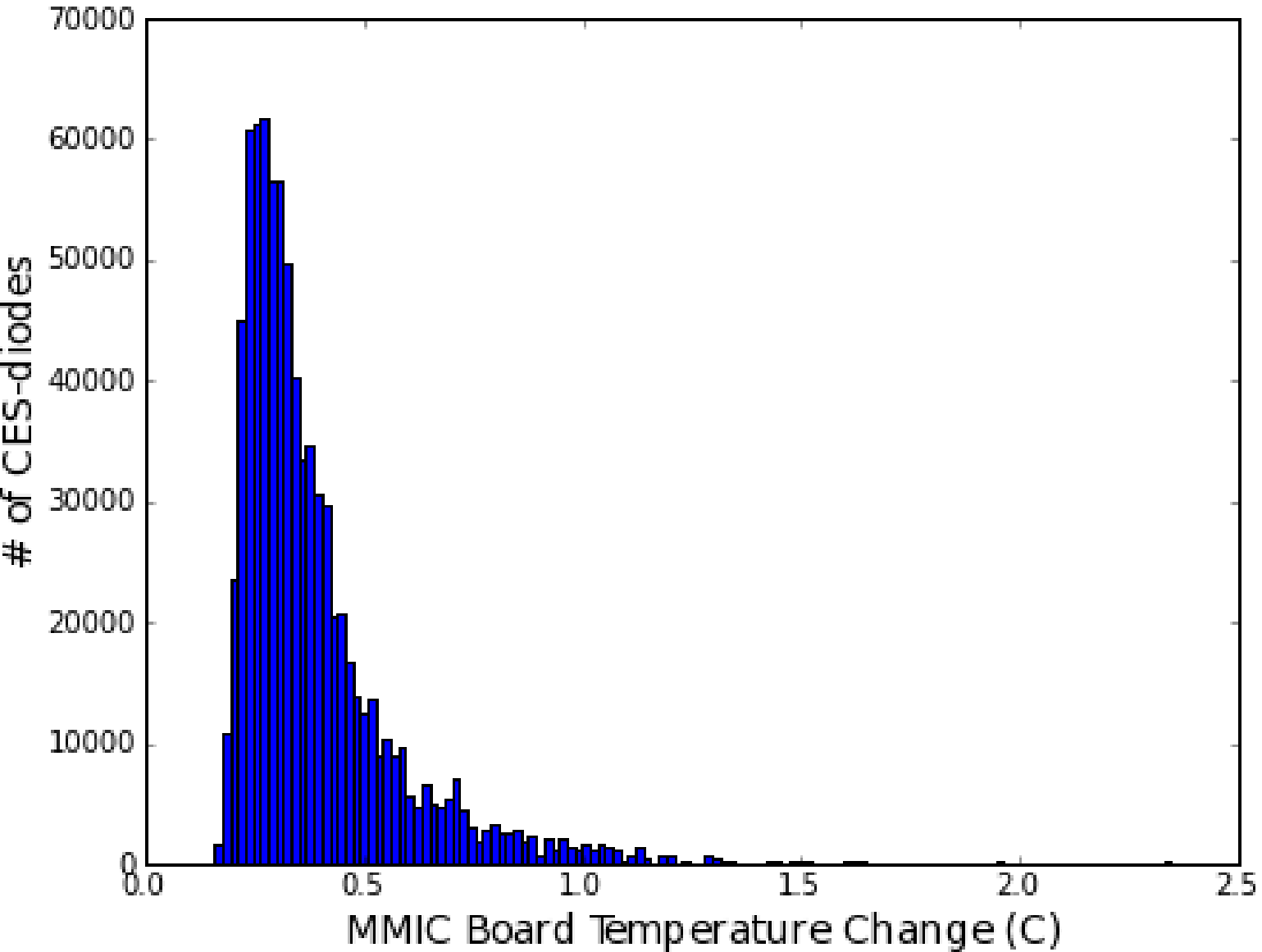}
\caption[Electronics Enclosure Temperature Stability]{\label{fig:electronics_temp_change_hist}
The Electronics Enclosure temperature was stable to $<1\degr$C within a CES.
The change for each CES is the standard deviation of all temperatures recorded within that CES.
I only show CES-diodes that passed data selection.
}
\end{figure}

I monitored the Electronics Enclosure temperature throughout the season.
Typically the temperature at the regulation point (in the Bias-Board Backplane) was stable to $1\degr$C.
However, the temperature was not uniform throughout the Backplane, and some bias boards (particularly those near the walls) experienced large ($10\degr$C) daily variations (Figure \ref{fig:electronics_temp_vs_time}) following the ambient temperature.
In summer, it became too hot during the day to keep the Enclosure regulated---the fan was on 100\% of the time.
At such times I increased the temperature setpoint.
Table \ref{tab:WPID_setpoint} lists the Electronics Enclosure setpoint temperatures during the season.
With these adjustments, the typical temperature change for a bias board during a CES was $<1\degr$C (Figure \ref{fig:electronics_temp_change_hist}).

\begin{deluxetable}{ll}
\tablecolumns{2}
\tablecaption{Electronics Enclosure Temperature Regulation Setpoints
\label{tab:WPID_setpoint}
}
\tablehead{\colhead{Time} & \colhead{Setpoint ($\degr$C)}}
\startdata
August 15--November 29, 2009 & 35\\
November 29, 2009--April 18, 2010 & 40\\
April 18, 2010--October 30, 2010 & 35\\
October 30, 2010--December 17, 2010 & 40\\
\enddata
\end{deluxetable}

\subsection{Data Management Status}
\label{sec:routine_checks:DAQ}

I routinely checked the Data Management system to verify proper recording of the data.
Possible problems included irregular timestamps, missing frames, de-synchronization of the Time-code Reader, Online Software crashes, power supply failures, loss of network connections, and ADC firmware errors.
These problems could manifest incorrect alignment of receiver and pointing data, gaps in the data stream, and corrupt data values.
I determined whether each CES was affected by these problems and whether or not we could use it (see w91\_software in Appendix~\ref{app:db}).
I also reviewed the status flags (Appendix~\ref{sec:DataCompilationStatusFlags}) for non-critical problems that nevertheless warranted action by the observers.
The following critical problems caused CESes to be unusable:
\begin{enumerate}
\item Missing mount encoder data
\item Data length not an integer number of seconds
\item Contains a file with no data (this problem could be fixed by adjusting the CES definition)
\item Missing receiver status data
\item Missing receiver database entry
\item Missing receiver bias data
\item Missing receiver timestamps
\item Missing status flags
\item Missing housekeeping data
\item Missing receiver clock data
\item Uneven demodulation (the number of samples added and subtracted was not the same)
\item Incorrect number of accumulated 800-kHz samples in any 100-Hz sample
\item Frame counter mismatch between ADC Boards
\item Null frames (inserted to fix timing alignment problems, but invalid for properly aligned data)
\item BAD\_TIMING flag (Appendix~\ref{sec:DataCompilationStatusFlags}) 
\item Missing frames (detected by frame counter not incrementing)
\item Timestamps not aligned to beginning of each second
\item Mismatch between radiometer, housekeeping, or mount timestamps
\item Timestamps not incrementing by expected interval
\item Irregularity in 50-Hz (double demodulation) clock
\item Irregularity in firmware 1-Hz clock
\item Incorrect blanking of phase-switch transitions
\item Incorrect housekeeping multiplexing
\end{enumerate}

\svnid{$Id: analysis.tex 150 2012-07-24 22:04:36Z ibuder $}

\chapter{Analysis}
\label{sec:analysis}
\epigraph{I might be blind, but I ain't stupid!}{Ray Charles}

Our analysis method\footnote{QUIET had two parallel analyses.  This thesis describes only pipeline A (PCL).  For discussion of the alternate analysis, see \S\ref{sec:pipeline_comparison} and \cite{sigurd_thesis}.} converted DD (\S\ref{sec:module_principles}) time streams into CMB polarization power spectra.
To organize the analysis, we performed characterization and time-stream analysis separately on each CES, treating it as an independent analysis unit.
The first characterization step was to calibrate the time-stream voltages into physical  units.
Then we modeled the noise in each CES to understand the statistical uncertainty.
We filtered the data in the time domain to remove residual 1/f noise and ground contamination.
We rejected data that had too much contamination or were otherwise unusable.
After this processing in the time domain, we made maps and calculated the power spectra with a cross-correlation technique.
To calibrate the analysis pipeline and understand its uncertainty, we performed the same analysis on hundreds of simulations of the entire experiment.
We validated the analysis  with a blind-analysis strategy.
Following this strategy, we estimated systematic errors before knowing the power-spectrum results.
We tried many analysis configurations (i.e. slightly different methods) before knowing the results.
Unless otherwise noted, the following descriptions apply to the final configuration.
I  mention details of other, preliminary configurations when they had an impact on the final analysis choices.
Appendix~\ref{app:configurations} lists all the configurations we tried.

\label{sec:ces_diode}
The fundamental unit of analysis was the ``CES-diode'': the data from one diode taken during one CES.
For each CES-diode we calculated  statistics for characterization of the data quality (\S\ref{sec:cuts} and Appendix~\ref{app:db}).
These included noise, weather quality, housekeeping, DAQ status, Type-B, and scan information.
We also allowed the calibration parameters (e.g. responsivity) to vary on a CES-diode basis; although, only the pointing model and some Type-B parameters in fact changed between each CES.
We performed filtering and data selection on a CES-diode basis; the smallest unit of data we accepted or discarded was a CES-diode.
Only at the map-making stage did we combine CES-diodes.

\svnid{$Id: calibration.tex 150 2012-07-24 22:04:36Z ibuder $}

\section{Calibration}
\label{sec:calibration}

We needed to calibrate the instrument to understand the mapping between CMB polarization on the sky and time-ordered data (TOD) generated from the detectors.
The relevant information was the beam shape, responsivity, detector angle, and a pointing model.
An additional (compared to other CMB experiments) set of calibration parameters informed the algorithm we used to correct the Type-B differential non-linearity (\S\ref{sec:electronics:readout}).
As much as possible we used astronomical calibration data to measure calibration constants, supplementing them with artificial calibrators when necessary.
In general we considered two or more calibration methods and used the difference between them to calculate the calibration uncertainty.

\subsection{Beam Shape}
\label{sec:cal:beam}
\begin{figure}
\includegraphics[width=1.0\textwidth]{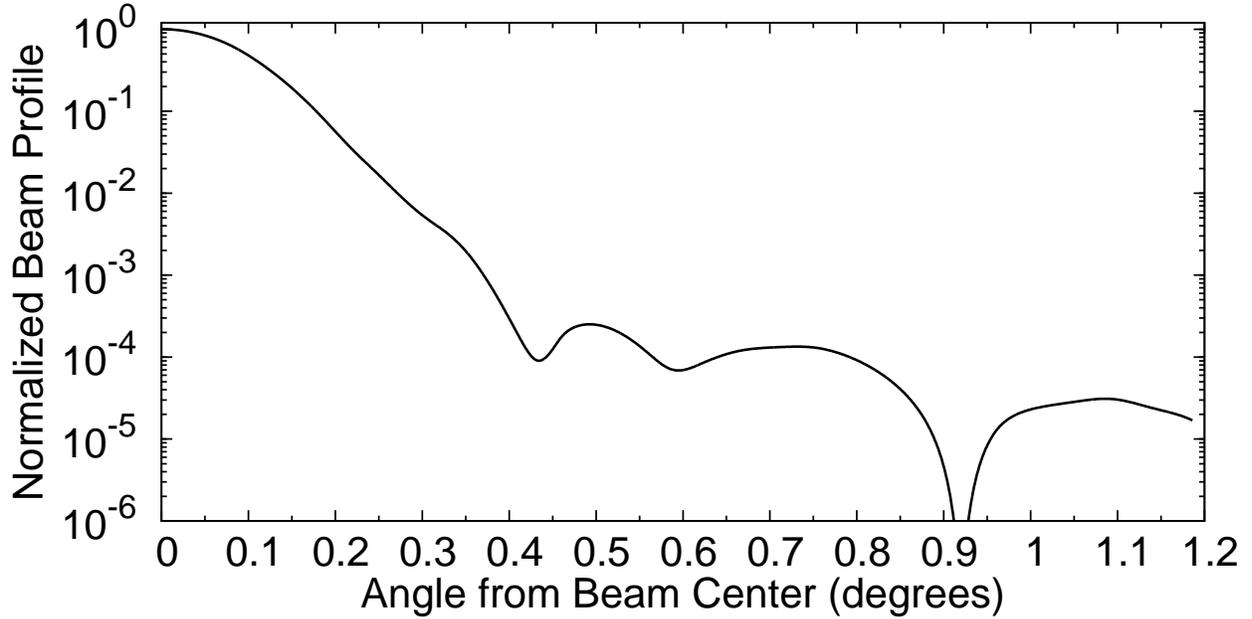}
\caption[Beam Profile]{\label{fig:beam_profile}
The beam shape was approximately Gaussian with a significant non-Gaussian tail.
}
\end{figure}

The beam shape (or point-spread function) effectively smooths the sky signal.
We accounted for this smoothing in analysis (\S\ref{sec:spectra_calc}).
We measured the beam shape from observations of Tau~A and Jupiter as well as pre-deployment antenna-range measurements at Jet Propulsion Laboratory's Mesa Antenna Measurement Facility\footnote{See \cite{raul_thesis} for details on measuring the beam shape.}.
We used the Tau~A measurements to set the beam FWHM and the antenna-range measurements to fix the shape beyond the central Gaussian peak.
After radial averaging, 
the beam shape was approximately Gaussian with a significant non-Gaussian tail (Figure \ref{fig:beam_profile}).
We used a single, average beam shape for all modules; we did not detect any significant differences between the beams of different modules.
We assigned systematic errors from the differences between Tau~A measurements and optics simulations and from the differences between antenna-range measurements and Jupiter measurements with the differential-temperature modules.

The  responsivity calibration sources were much smaller than the angular scales at which we measured the CMB.
We corrected for this difference by scaling the calibration measurements to the angular scales of the CMB using the total beam solid angle, $\Omega$.
We calculated the solid angle by integrating the beam shape above, finding $\Omega= 13.63\pm0.65\,\mu$sr.
We propagated the systematic errors above.
In addition, to estimate the error from the beam tail outside the measurement region, we compared the solid angle when integrating only the central $0\fdg4$ instead of $1\fdg4$ and assigned this difference as an additional (negligible) systematic error for the solid angle.

\subsection{Responsivity}
\label{sec:cal:responsivity}
The responsivity was the coefficient of proportionality between the CMB fluctuations on the sky and the DD voltages.
We modeled it as the product of two factors \citep{gain_model_w}:
the absolute responsivity was the coefficient for a single, well-characterized diode.
The relative responsivity was the ratio between its responsivity and that of all other diodes.

The absolute responsivity factor in the model was the responsivity of the central module diode RQ45Q1.
We measured it from repeated observations of Tau~A.
The maximum polarization signal was $\Gamma P$, where $P=16.52\pm0.84$\,Jy is the polarization flux \citep{taua_input,wmap7_calibrators} and
\begin{equation}
\Gamma = \frac{c^2}{2\nu^2\Omega k_B}
\end{equation}
is the specific intensity to antenna temperature conversion factor.
We calculated the responsivity from the ratio of this expected signal to the observed voltage.
We found this responsivity had significant time dependence due to disruptions of the instrument.
Therefore, we modeled the absolute responsivity as a piecewise-constant function of time.
We fixed the time divisions based on jumps in the responsivity measurements that correlate with known disruptions of the instrument.
Table \ref{tab:gain_model_periods} lists the time divisions.
During the gap in Tau~A observations (\S\ref{sec:obs:cal}), we used large sky dips (\S\ref{sec:obs:cal}) to track the time variation of the responsivity.
We corrected the large-sky-dip data so the responsivities from large sky dips and Tau~A were consistent when both were available \citep{gain_anchor}.
We then applied the antenna-to-thermodynamic-temperature conversion factor (1.25) so subsequent polarization measurements in K had thermodynamic (i.e. observing-frequency-independent) units\footnote{Appendix~E of \cite{yuji_thesis}}.
The absolute responsivity uncertainty was dominated by uncertainty in Tau~A's polarization flux (5.1\%), extrapolation of RQ45Q1 responsivity to other diodes (3.8\%, below), and uncertainty in the beam solid angle (4.8\%).
Uncertainty in the modeling of the time variation was small (0.6\%).
We also considered dependence on the module physical temperature and electronics temperature, but it was negligible.
The total absolute responsivity uncertainty was 8.3\% \citep{gain_systematics}.

\begin{deluxetable}{lll}
\tablecolumns{3}
\tablecaption{Absolute-responsivity--model Time Divisions
\label{tab:gain_model_periods}
}
\tablehead{\colhead{Start Time (UTC)} & \colhead{Responsivity (mV/K, thermodynamic)} & \colhead{Division Reason}}
\startdata
 & 2.43 & Beginning of observations\\
2009-10-19 09:14 & 2.64 & Replaced cold head\\
2009-12-07 19:15 & 2.49 & Fixed CPID oscillation\\
2010-03-02 21:37 & 2.39 & Earthquake\\
2010-05-20 17:11 & 2.15\tablenotemark{a} & Cryostat warmed up\\
2010-05-29 22:39 & 2.23\tablenotemark{a}& Cryostat pumped\\
2010-07-28 22:54 & 2.27\tablenotemark{a}& CPID unstable\\
2010-08-02 12:22 &2.27 &Drain bias instability\\
2010-10-05 20:37& 2.39& Cryostat pumped\\
2010-11-10 19:10& 2.17&Generator failure\\ 
\enddata

\tablenotetext{a}{We used large-sky-dip measurements for the responsivity.}
\end{deluxetable}

We measured the relative responsivity from wire-grid and large-sky-dip observations.
The relative responsivity of diode $X$ to RQ45Q1 at time $t$ is
\begin{equation}
R_X(t) = R_{X,\textrm{WG}}\frac{R_{X,\textrm{LSD}}(t)}{R_{X,\textrm{LSD}}(t_\textrm{WG})},
\end{equation}
where $t_\textrm{WG}$ is the time of the wire-grid measurement,
$R_{X,\textrm{WG}}$ is the responsivity ratio between diode $X$ and RQ45Q1 as measured from the wire grid,
and $R_{X,\textrm{LSD}}(t)$ is the responsivity ratio measured from large sky dips.
We used the median of large-sky-dip measurements in each time division when constructing relative responsivities \citep{relative_gain_w}. 
To estimate the uncertainty in the relative responsivity, we compared our model to the Tau~A measurements for off-center modules\footnote{Except for RQ45Q1, the model does not depend on Tau~A measurements.}.
The mean responsivity shift from this comparison was 1.2\%\footnote{\S3.1 of \cite{gain_systematics}}, and the RMS scatter was 12.5\%.
If we used only Tau~A measurements to model the responsivity, the result would change by 3.8\%.

\subsection{Detector Angle}
\label{sec:cal:angle}
The detector angle was the polarization-sensitive axis of each diode.
Typically\footnote{i.e. if there was no hybrid-coupler imperfection} the Q1 and Q2 diodes had $90\degr$ between their detector angles.
The U1 and U2 diodes were separated by $90\degr$, and each Q diode was offset by $45\degr$ from each U diode.
However, the overall angle remained free\footnote{It was affected by extra phase lengths in the module.}, and imperfections might alter the relationship between the diodes.
Therefore, we measured the axis for each diode.
As with responsivity, we separated the calibration of detector angle into absolute and relative components.

We measured the absolute detector angles for RQ45 from the known polarization direction of Tau~A \citep{aumont, wband_detector_angle}.
Each day observation yielded a precision of $3\degr$.
By combining all measurements, the total measurement uncertainty was $0\fdg3$.
The known polarization direction itself had an uncertainty of $0\fdg2$.
The possible error in extrapolating the angle to non-central modules (see below) was $0\fdg2$.
We also considered possible variation of the detector angle with applied bias, and limited the effect to $0\fdg2$.
Combining these uncertainties, the total uncertainty on the absolute detector angle was $0\fdg5$.

We measured the angles of all non-central detectors relative to RQ45Q1 from the wire-grid season-end calibration.
This calibration simultaneously illuminated all detectors with a modulated polarization signal.
From the relative phases of different diodes, we calculated the relative angles between their axes \citep{wiregrid_analysis}.
We compared the resulting angles to those obtained from the small number of off-center Tau~A measurements.
The resulting mean shift was $0\fdg2$, and we included it in the absolute angle uncertainty above.
The root-mean-square (RMS) scatter was $3\fdg1$.
The uncertainty of the wire-grid measurement was $0\fdg8$.
Combining them, we assigned a total relative uncertainty (i.e. scatter among the diodes) on the polarization angle of $3\fdg2$.

\subsection{Pointing}
\label{sec:cal:pointing}
The pointing model converted the mount encoder positions into angular positions on the sky for each module.
The model had two components: a mount model specifying the position of the central module and an array model specifying the relative positions of the other modules.

We constructed the mount model from observations of Jupiter, RCW~38, and the Galactic center with the differential-temperature modules \citep{wband_mount_model}.
For each observation we had the apparent position (encoder measurements) and true position (ephemeris).
The mount model mapped between them, and we chose the parameters of the model to minimize the difference between the true positions and the positions as reconstructed by the model.
The model itself was a series of rotation matrices,
\begin{equation}
M_\textrm{mount} = M_\textrm{eltilt}M_\textrm{aztilt}M_\textrm{bore}M_\textrm{coll}.
\end{equation}
Using the $zyz$ Euler convention and expressing each matrix as a series of rotation angles $M(\alpha, \beta, \gamma)$, the matrices are
\begin{eqnarray}
M_\textrm{eltilt} &=& M(-(90\degr+\textrm{az}), \theta_E, 90\degr+\textrm{az}) \\
M_\textrm{aztilt} &=& M(-\psi, \theta, \psi)\\
M_\textrm{bore} &=& M(-\textrm{az} - (\Delta_\textrm{az} + \theta_\textrm{ef,az}\cos(-\textrm{az}+\psi_\textrm{ef})), \nonumber\\
&& 90\degr - \textrm{el} - (\Delta_\textrm{el} + \theta_\textrm{ef,el}\cos(-\textrm{az}+\psi_\textrm{ef})), \nonumber\\
&& \textrm{dk} - (\Delta_\textrm{dk} + \theta_\textrm{ef,dk}\cos(-\textrm{az}+\psi_\textrm{ef}))) \\
M_\textrm{coll} &=& M(\psi_c + 180\degr, \theta_c(1 + \theta_\textrm{ec}\cos(\textrm{dk}+\psi_\textrm{ec})),
 -\psi_c-180\degr), 
\end{eqnarray}
where $(\textrm{az, el, dk})$ are the azimuth, elevation, and deck encoder positions.
Table \ref{tab:pointing_param} gives the values and physical meaning of the model parameters.

We performed several tests to evaluate uncertainty in the mount model:
\begin{enumerate}
\item We combined all differential-temperature data to make a map of the source PMN J0538-4405 in CMB patch 4a.
The difference between the true \citep{wmap7_galaxy} and reconstructed positions of this source was $0\farcm4\pm1'$.
The FWHM beam size measured was $14\farcm2\pm1'$.
The increase compared to the beam size found above ($11\farcm66$) was due to scatter in the pointing model.
\item We measured this scatter from the residuals in the pointing model fit.
We fit a two-dimensional (azimuth and elevation) Gaussian to the residuals, and used the difference between the major and minor axes to estimate the uncertainty.
We found that the pointing model had an intrinsic scatter FWHM of $5\farcm1\pm0\farcm25$.
We included this scatter as an additional factor in the beam deconvolution (\S\ref{sec:spectra_calc}).
\item As a check of this additional effective beam, we used the differential-temperature data to calculate the CMB temperature power spectrum, and compared the result to a $\Lambda$CDM model.
This comparison confirmed the additional beam smoothing with an uncertainty of $\pm1\farcm27$.
\item We also used Moon observations to test the pointing model.
The RMS scatter between the position predicted by the model and the actual position was $1\farcm6$ \citep{moon_pointing_check}.
\end{enumerate}

\begin{deluxetable}{llr}
\tablecolumns{3}
\tablecaption{Pointing-model Parameters
\label{tab:pointing_param}
}
\tablehead{\colhead{Name} & \colhead{Meaning\tablenotemark{a}} & \colhead{Value }}
\startdata
$\Delta_\textrm{az}$ & Azimuth encoder offset & $-0\fdg03547$ \\ 
$\Delta_\textrm{el}$ & Elevation encoder offset & $-0\fdg02938$ \\
$\Delta_\textrm{dk}$ & Deck encoder offset & $0\fdg10832$ \\
$\psi$ & Azimuth tilt orientation & $49\fdg62021$ \\
$\theta$ & Azimuth tilt magnitude & $-0\fdg00701$ \\
$\theta_E$ & Elevation tilt & $-0\fdg00371$ \\
$\psi_C$ & Collimation offset direction & $26\fdg67075$ \\
$\theta_C$ & Collimation offset magnitude & $0\fdg29481$ \\
$\psi_\textrm{ec}$ & Collimation ellipticity orientation & $253\fdg82681$ \\
$\theta_\textrm{ec}$ & Collimation ellipticity magnitude& $0.2126$\tablenotemark{b} \\
$\psi_\textrm{ef}$ & Encoder offset flexure direction & $11\fdg35199$ \\
$\theta_\textrm{ef,az}$ & Encoder offset flexure magnitude in azimuth & $-0\fdg02252$\\
$\theta_\textrm{ef,el}$ & Encoder offset flexure magnitude in elevation & $-0\fdg00424 $ \\
$\theta_\textrm{ef,dk}$ & Encoder offset flexure magnitude in deck & $-0\fdg26147 $ \\

\enddata
\tablenotetext{a}{See also \cite{quiet_drive_system, chicago_pointing_convention}.}
\tablenotetext{b}{This parameter is not an angle.}
\end{deluxetable}

We constructed the array model from Moon observations.
In a full-array Moon observation, all modules observed the Moon.
From the time delay between the Moon signal in different modules, we calculated the relative position of each module (Figure \ref{fig:array_layout} and \cite{moon_array}).
The uncertainty as measured by the scatter between different Moon observations was $<0\farcm2$.
We also constrained the offset of the deck angle to $<0\fdg1$\footnote{During part of Q-band observation the deck encoder was loose (\S4.2 of \cite{quiet_qband_result}).  Therefore, we wanted to confirm this problem did not occur in W band.}.

\subsection{Type-B--glitch Correction}
\label{sec:calibration:typeb}

We corrected the data affected by Type-B glitching.
We based the correction algorithm  on a phenomenological understanding of the glitches.
The algorithm had several free parameters to fit or measure.
I address the uncertainty introduced by this correction in \S\ref{sec:type_b_systematics}.

We based the correction on a Type-B glitch model \citep{typeb_formalism}.
Suppose that the voltage at the ADC input was $x^i$.
Then we modeled the ADC operation as 
\begin{equation}
y^i \equiv f(x^i) \equiv x^i + \Delta f(x^i),
\end{equation}
where 
\begin{equation}
\Delta f(x) = \sum_i\Delta f_i(x)
\end{equation}
encodes the effect of glitches, and each glitch is 
\begin{equation}
\Delta f_i(x) = 
\begin{cases} -A_i/2 & \textrm{if } x < x_0^i \\
0 & \textrm{if } x = x_0^i \\
A_i/2 & \textrm{if } x > x_0^i,
\end{cases}
\end{equation}
where $x_0^i$ and $A_i$ are the location and height of the $i$th glitch.
The correct DE and TP data are
\begin{eqnarray}
x_{1(2)} &=& \frac{1}{N}\sum_{i=1}^N x^i_{1(2)}\\
x_\textrm{DE} &=& \frac{x_1-x_2}{2}\\
x_\textrm{TP} &=& \frac{x_1 + x_2}{2},
\end{eqnarray}
where subscripts 1 and 2 indicate samples in the two 4-kHz phase-switch states, and $N=3440$ is the number of 800-kHz samples per 100-Hz sample per phase.
However, the ADC output is 
\begin{eqnarray}
y_{1(2)} &=& \frac{1}{N}\sum_{i=1}^N y^i_{1(2)}\\
y_\textrm{DE} &=& \frac{y_1-y_2}{2}\\
y_\textrm{TP} &=& \frac{y_1 + y_2}{2}.
\end{eqnarray}
Since the 800-kHz noise was Gaussian (\S\ref{sec:800khz_gaussianity}),
we define
\begin{equation}
\label{eq:Fdef}
y = F(x;\sigma)\equiv \int f(s)P_G(s;x,\sigma)ds = x + \int\Delta f(s)P_G(s;x,\sigma)ds,
\end{equation}
where $P_G(s;\mu,\sigma)$ is a Gaussian distribution
\begin{equation}
P_G(s;\mu,\sigma)\equiv\frac{1}{\sqrt{2\pi}\sigma}\exp\left(-\frac{(s-\mu)^2}{2\sigma^2}\right)
\end{equation}
centered at $\mu$ with standard deviation $\sigma$.
The inverse function
\begin{equation}
G(y;\sigma)\equiv F^{-1}(y;\sigma)
\end{equation}
satisfies
\begin{equation}
x = G(F(x;\sigma);\sigma).
\end{equation}

We estimate corrected values
\begin{equation}
x'_{1(2)} = G(y_{1(2)};\sigma).
\end{equation}
In case the correction parameters are accurate and $N\rightarrow\infty$, $x'=x$.
I discuss departures from these idealities later.
Defining $\Delta G(y;\sigma)\equiv G(y;\sigma) - y$, we can calculate the corrected data $x'$ if we know the function $\Delta G$.
From Eq. \ref{eq:Fdef}, 
\begin{eqnarray}
\Delta F(x;\sigma) &\equiv& F(x;\sigma) -x = \nonumber\\ \int\Delta f(s)P_G(s;x,\sigma)ds
&=&  \sum_i A_i\int_0^{x-x_0^i}P_G(s;0,\sigma)ds = \sum_i\frac{A_i}{2}\erf\left(\frac{x-x_0^i}{\sqrt{2}\sigma}\right). 
\end{eqnarray}
Because
\begin{eqnarray}
x &=& G(y;\sigma)\\
&=& y + \Delta G(y;\sigma)\\
&=& F(x;\sigma) + \Delta G(y;\sigma)\\
&=& x + \Delta F(x;\sigma) + \Delta G(y;\sigma),\\
\Delta F(x;\sigma) &=& -\Delta G(y;\sigma).\label{eq:FGrelation}
\end{eqnarray}
Using Eq. \ref{eq:FGrelation} we solve for $\Delta G$ using Newton's method.
If
\begin{equation}
h(s)\equiv s + \Delta F(s+y;\sigma),
\end{equation}
then the value of $s$ that solves $h(s)=0$ is $\Delta G$.
Thus we solved $h(s)=0$ with Newton's method to calculate $\Delta G$ and applied it to $y$ to calculate the corrected data $x'$.
(An additional subtlety is that the measurement of glitch locations $x_0^i$ included glitching.  See Appendix~A.3.1 of \cite{typeb_formalism} for a solution.)

The algorithm above had free parameters $x_0^i$ (glitch locations), $A_i$ (glitch heights), and $\sigma$ (noise level).
We measured the glitch locations and heights by injecting known voltage signals into each ADC channel \citep{typeb_glitch_measurement}.
Although there were several patterns, not all channels had the same glitch locations and heights.
Some channels did not have any glitching.
Because the measurement did not cover the full ADC range, I extrapolated the patterns to cover the range used in observations \citep{typeb_w_v4}.
Moreover, I modified the parameters for a small number (5) of channels for which the ADC measurement was inconsistent with glitching found in CMB data.
Because $\sigma$ is the 800-kHz noise level, the relationship between it and the 100-Hz noise depends on the Preamp Board filtering.
A calculation from the Preamp Board circuit design gave 0.6342 as the ratio between the noise levels.
However, using that to calculate $\sigma$ did not result in good correction.
Instead, I estimated $\sigma$ from CMB data in an iterative procedure.

As an intermediate step in calculating $\sigma$, I defined a ``preamp factor'' for each diode.
This was the ratio between 800-kHz and 100-Hz noise levels that I  estimated empirically.
For each CES-diode I calculated a ``Type-B $\chi^2$'' (\S\ref{sec:cuts}) that quantified the amount of Type-B glitching in it.
First I selected CESes with bad Type-B (ordered by the difference in Type-B $\chi^2$ after correction), but otherwise good data quality.
For each diode I selected the worst 20 such CESes.
For each such CES-diode, I found the preamp factor that minimized the Type-B $\chi^2$.
Then I averaged the preamp factors that made the $\chi^2<3$ and came from CES-diodes with significant Type-B glitching\footnote{The glitching was significant if $\Delta\chi^2>1$.  Some diodes rarely manifested Type-B; for such diodes the worst 20 CESes may contain CESes that did not manifest Type-B.}.
I calculated an uncertainty for the preamp factor from the scatter among CESes.
For a small number (6) of diodes, I could not find any CES from which I could estimate a preamp factor.
For such diodes, I used the average preamp factor of the well-measured diodes.

Once I fixed the preamp factor, I calculated $\sigma$ for each CES-diode in an iterative procedure.
First I made a preliminary guess for the noise level from ``pseudo-quadrature'' 25-Hz data, in which I differenced adjacent DD samples.
These pseudo-quadrature data had the same noise level but no signal.
The noise level guess was the standard deviation of the pseudo-quadrature data.
Then I corrected the CMB data using the guess for $\sigma$.
From the corrected data, I computed a new noise level guess and used it to correct the data again.
I iterated this process until the noise level converged to 0.0002\,bit$\sqrt{\textrm{s}}$.
 (Typical diode noise levels were 0.2\,bit$\sqrt{\textrm{s}}$.)

Because we did not record 800-kHz data all the time, the correction had a statistical uncertainty
\begin{equation}
\delta x_\textrm{DE} = x_\textrm{DE}' - x_\textrm{DE} = \frac{\delta x_1 - \delta x_2}{2},
\end{equation}
where
\begin{equation}
\delta x \equiv G(y;\sigma) - x
\end{equation}
is the miscorrection, in this case due to statistical fluctuation.
\begin{equation}
\sqrt{<\delta x_\textrm{DE}^2>} \lesssim\frac{A}{2\sqrt{N}}.
\end{equation}
For the largest glitches with $A\approx40$, this noise was $0.36\,\mu$V$\sqrt{\textrm{s}}$\footnote{Appendix~A.4.1 of \cite{typeb_formalism}}.
The typical noise level was 3\,$\mu$V$\sqrt{\textrm{s}}$ so this Type-B correction noise was not dominant, and we included it when we measured the noise from the data after correction.
I evaluate the systematic errors due to Type-B in \S\ref{sec:type_b_systematics}.

\section{Noise Modeling}
We modeled the noise in each CES-diode as a combination of white and 1/f noise components.
We fit the model
\begin{equation}
\hat{N} = \sigma^2\left(1+(f_0/f)^\alpha\right)
\end{equation}
to each CES-diode DD data in the Fourier domain.
The model parameters were $\sigma$, the white noise level, $f_0$, the 1/f knee frequency, and $\alpha$, the 1/f slope.
The fitting excluded frequencies between 0.8 and 1.2 times the scan frequency as well as frequencies above 9.6\,Hz (the low-pass filter cutoff, \S\ref{sec:filter}).
We included correlations among the diodes of each module \citep{1f_noise_model}.
To study these correlations, we computed the covariance matrix at each frequency
\begin{equation}
\hat{C}_{ij}(f) = F_i(f)F_j^*(f),
\end{equation}
where $F_i(f)$ is the Fourier transform of diode $i$.
At high frequencies the white noise dominated so
\begin{equation}
\hat{\rho^w_{ij}} = \Re\left[<\frac{\hat{C}_{ij}(f)}{\sqrt{P_i(f)P_j(f)}}>_{\textrm{2--4.5\,Hz}}\right]
\end{equation}
estimates the white noise correlation matrix with minimal bias from residual 1/f noise when we average between 2 and 4.5\,Hz.
$P_i$ is the noise power ($F_iF_i^*$).
We took the real part after checking that the imaginary part of the correlation matrix was negligible.
I also checked that inter-module correlations were negligible.
In two CESes the largest inter-module correlation coefficient was 0.25, and only one pair of modules had such a large correlation.
The expected correlation matrix for an ideal module is
\begin{equation}
\rho^w_{ij} = \begin{pmatrix}
1 & 0.5 & 0.5 & 0\\
0.5 & 1 & 0 & 0.5\\
0.5 & 0 & 1 & 0.5\\
0 & 0.5 & 0.5 & 1
\end{pmatrix},
\end{equation}
where the order of the rows and columns is (Q1, U1, U2, Q2).
However, module imperfections altered the off-diagonal elements \citep{white_noise_correlation}.
The median measured correlation matrix was \citep{wband_sensitivity}
\begin{equation}
\hat{\rho^w_{ij}} = \begin{pmatrix}
1 & 0.495 & 0.502 & 0.058\\
0.495 & 1 & 0.061 & 0.516\\
0.502 & 0.061 & 1 & 0.387\\
0.058 & 0.516 & 0.387 & 1
\end{pmatrix}.
\end{equation}
Note that correlation between Q and U diodes did not cause significant correlation between measurements of $Q$ and $U$ because we effectively differenced Q1 from Q2 in map making.
The difference Q1 $-$ Q2 is uncorrelated with U1 $-$ U2 for any correlation matrix.

\subsection{Correlated 1/f Noise Investigation}
We investigated the 1/f noise correlations in several CESes before deciding on the treatment of them in the noise model.
This section details our investigation and findings;
\S\ref{sec:noise:model} describes the final noise model.
First we defined a 1/f correlation estimator for a single CES.
Then I diagonalized the 1/f correlation matrix for each CES.
I found some significant correlated modes and confirmed the significance with simulations.
Then we showed that for frequencies above the pipeline filter cutoff, the correlated modes do not contribute significantly to the total noise.
We concluded that we could ignore these correlated modes in the noise model, thereby making at most a 1\% error in noise estimation.

Measuring the 1/f noise correlation was more complicated because even at low frequencies the white noise may not be negligible.
Moreover there can be multiple 1/f components with different frequency dependencies.
We used
\begin{equation}
\hat{\rho^{1/f}_{ij}} =  \left(\sum_f\frac{ (C_{ij}(f) - < C_{ij}(f) >_\textrm{2--4.5\,Hz})\sqrt{w_iw_j}}{ \sqrt{\sigma_i^2\sigma_j^2 \left(\frac{f_0}{f}\right)^{\alpha_i+\alpha_j}} }\right)/\sum_f\sqrt{w_iw_j}
\end{equation}
as a 1/f correlation matrix estimator for each CES.
The $w_i$ are weights chosen to optimize the signal to noise
\begin{equation}
w_i(f) = \sigma_i^2\left(\frac{f_0}{f}\right)^{\alpha_i}/\hat{P}_i(f).
\end{equation}
We excluded frequencies between 0.8 and 1.2 times the scan frequency (because there was narrow-band noise from the scan) and above 1.5 times the scan frequency.

\begin{figure}
\includegraphics[width=1.0\textwidth]{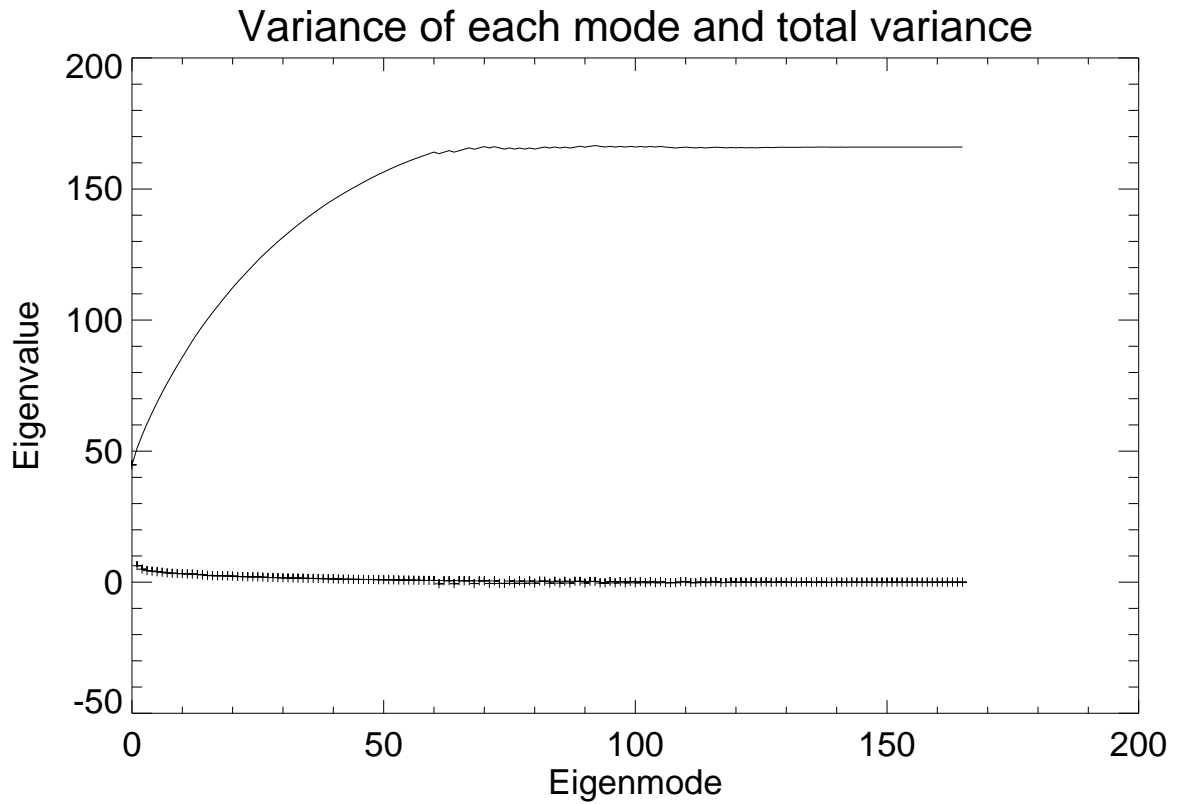}
\caption[1/f Correlation Matrix Eigenvalues]{\label{fig:1f_correlation_matrix_eg}
A small number of modes contained most of the 1/f variance.
The crosses are the eigenvalues of the 1/f correlation matrix for one CES.
I ordered them by absolute value.
The solid line is the sum of the first N eigenvalues.
The negative eigenvalues (crosses slightly below 0 and points where the solid line decreased) were due to fluctuations; with an infinite data length the correlation matrix would be measured perfectly, and there would be no negative eigenvalues.
}
\end{figure}

\begin{figure}
\includegraphics[width=1.0\textwidth]{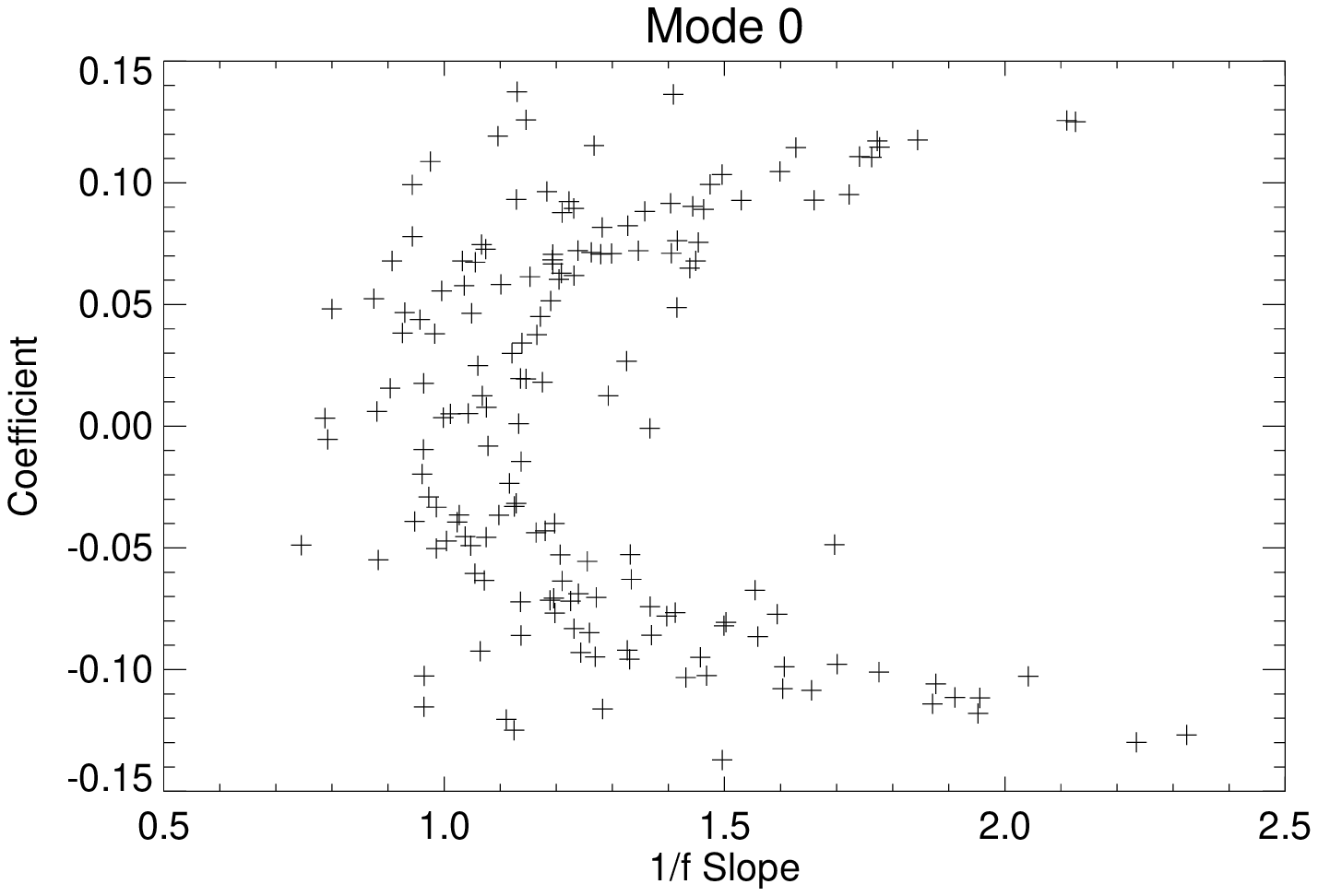}
\caption[1/f Correlation Matrix Eigenvector and Demonstration of two 1/f Noise Components]{\label{fig:1f_coeff_vs_slope}
The coefficients of the first eigenvector of the 1/f correlation matrix were correlated with the 1/f slopes of each diode's noise spectrum.
Each point shows the 1/f slope and eigenvector coefficient for one diode; all points are from the same CES.
There were two 1/f components, and the slope-2 component, which we attributed to weather, was correlated among all diodes.
The slope-2 component generated the first eigenmode.
When that component dominated the 1/f noise of a diode, that diode was strongly correlated with the rest of the array.
Thus the diodes with measured slope $\approx2$ had consistently high coefficients in the first eigenvector.
We attributed the slope-1 component to MMIC-amplifier--gain fluctuations.
}   
\end{figure}

To study the correlation structure, I diagonalized $\hat{\rho^{1/f}_{ij}}$ for several CESes.
Before actually performing the matrix diagonalization, I took several steps to regulate the ``raw'' correlation matrix $\rho^{1/f}_{ij}$.
First, I removed several diodes.  All known non-functional diodes were cut.
Diodes with knee frequency $<10$\,mHz were removed because there was very little 1/f noise so the correlation coefficient had large fluctuations.
I also cut diodes with small ($<0.1$) diagonal elements in the correlation matrix.
Since I used a fit to estimate the variance, it is not guaranteed that the diagonal elements of the raw correlation matrix are 1.  Although I renormalized the matrix (below) so this is the case, extremely small diagonal values here could cause the normalization to become unstable.
The normalized correlation matrix is
\begin{equation}
\rho^\textrm{normalized}_{ij} = \frac{\hat{\rho^{1/f}_{ij}}}{\sqrt{\hat{\rho^{1/f}_{ii}}\hat{\rho^{1/f}_{jj}}}}
\end{equation}
After normalization, I forced off-diagonal elements  to be in the range [-0.99, 0.99] by reducing the absolute value of outliers to 0.99.
The normalization process ensured that the diagonal elements were 1.
After diagonalization, I obtained the eigenvalues and eigenvectors.
Small negative eigenvalues remained even after regulation (Figure \ref{fig:1f_correlation_matrix_eg}).
These were an effect of the difference between the true correlation matrix (which must be positive definite) and the estimated correlation matrix.
For typical CESes, the first mode had a significant coefficient for most diodes.
That is, the mode with the largest 1/f variance was correlated among most diodes in the array.
Figure \ref{fig:1f_coeff_vs_slope} shows the first mode coefficient vs. 1/f slope.
The correlation increased with slope, and all diodes with slope $\gtrsim 1.5$ were correlated.
This was evidence for a two-component model of 1/f noise with the slope-2 component correlated among all diodes.
We attributed the slope-2 component to atmospheric fluctuations and the slope-1 component to MMIC-amplifier--gain fluctuations.

\begin{figure}
\includegraphics[width=1.0\textwidth]{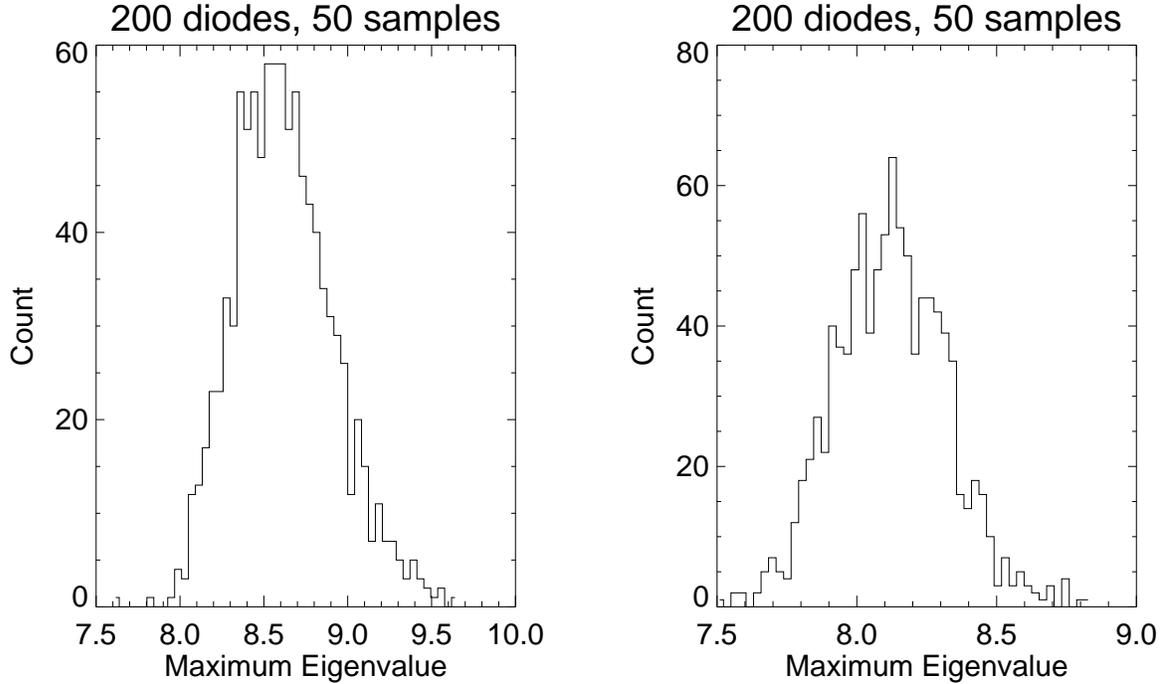}
\caption[Simulated 1/f Correlation Matrix Eigenvalues]{\label{fig:simulated_1f_eigenvalues}
I confirmed that the 1/f correlation measurement was significant.
The left (right) panel shows the first (second) eigenvalues from simulations.
When the 1/f correlation was significant, the data eigenvalues were much larger ($\approx50$).
}
\end{figure}

Since there was statistical noise in the correlation measurement (due to finite data size), I ran a toy simulation of one CES to ensure the measurement was significant.
In the simulation I assumed there were no correlations and that the noise was Gaussian.
A typical CES was 5000\,s so there were 5000 Fourier modes / Hz.
I only considered diodes with a knee frequency above 10\,mHz, and the weight $w$ was small above the knee frequency.
Therefore, I used 10\,mHz as an approximation to the bandwidth used in the correlation measurement.
I simulated 50 independent Gaussian random samples for each diode and computed the resulting correlation matrix.
I diagonalized the simulated correlation matrix and computed the distributions of the first few eigenvalues.
If the eigenvalues from the real data measurement were far outside the simulated distribution, I could conclude that they were not due to statistical fluctuation.
Figure \ref{fig:simulated_1f_eigenvalues} shows the distributions for the first two eigenvalues.
Significant correlated 1/f modes in the real data had much larger eigenvalues than in this simulation; therefore, the measurement was statistically significant.

To determine the importance of correlated 1/f modes, we computed linear combinations of the diode TODs that enhanced the 1/f noise.
We selected the linear combinations by diagonalizing the 1/f noise covariance matrix for a single CES.
This covariance matrix was formed from the full correlation matrix (not a model); we weighted the diodes to have the same white noise level.
Since the 1/f noise was frequency-dependent, we evaluated the covariance matrix at the scan frequency.
The first eigenvector of this matrix corresponded to a linear combination of diode TODs with enhanced 1/f power at the scan frequency.
We calculated 10 TODs that were  linear combinations of diode TODs with coefficients taken from the first 10 eigenvectors of this covariance matrix.
Since our filtering (\S\ref{sec:filter}) eliminated frequencies below twice the scan frequency, we needed to check the modes with power near the cutoff to ensure that our noise model accounted for the 1/f noise remaining after filtering.
For each linear-combination TOD we computed the noise power spectrum and compared it to a noise model.
The noise model included only intra-module correlations.
Even so, the agreement was good above the filter cutoff frequency.
Since only frequencies above the cutoff survived through the pipeline, agreement above the cutoff was sufficient for a noise model.
Since the correlated component had slope 2, it was reasonable for it to be insignificant compared to the slope-1 component at high frequencies.
Thus, we chose to ignore inter-module correlations in the 1/f noise model.

\begin{figure}                                                                  \includegraphics[width=0.8\textwidth]{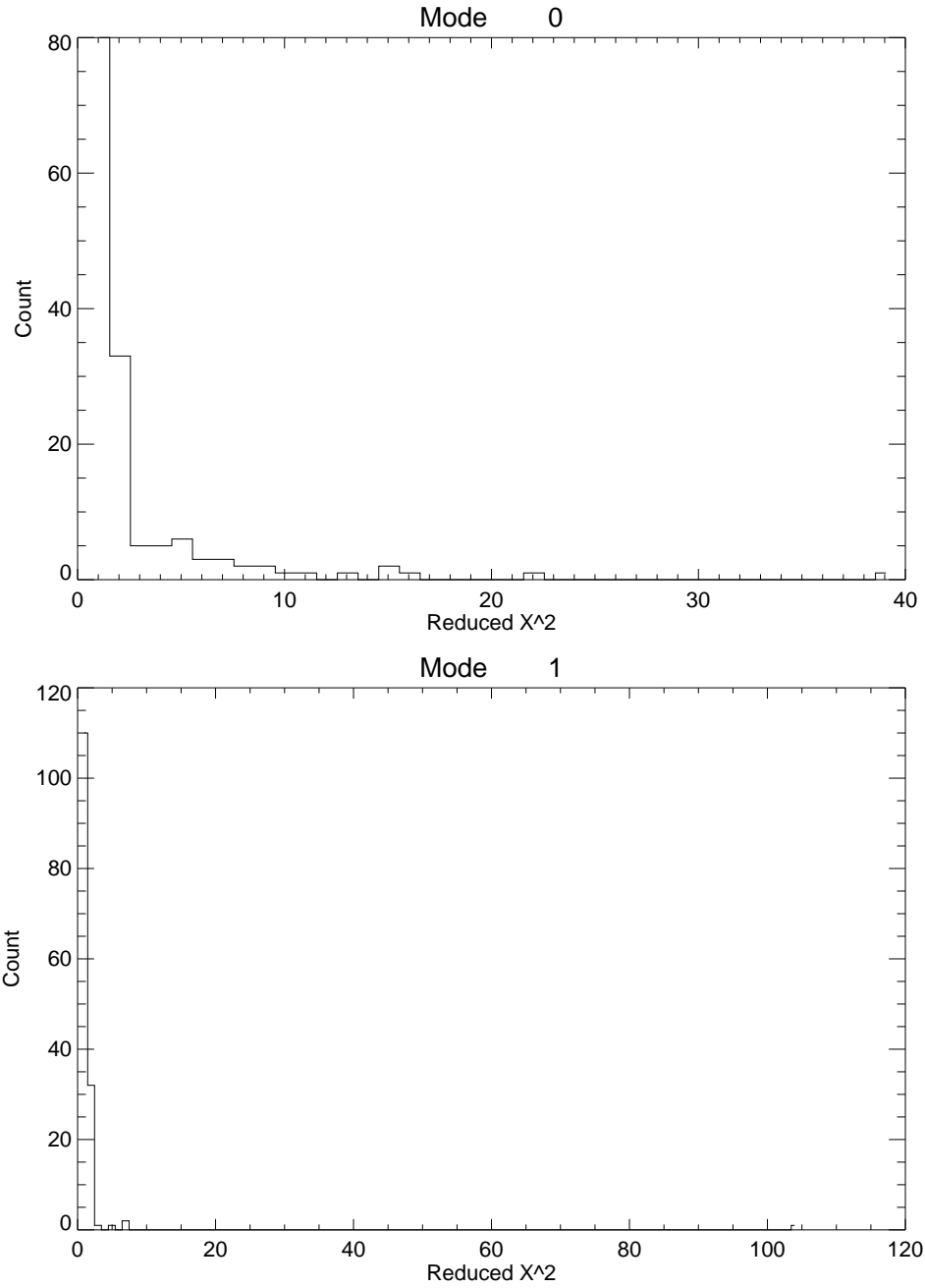}
\caption[1/f Noise Model $\chi^2$]{\label{fig:1f_model_chi2}
Our 1/f model reproduced the 1/f noise above twice the scan frequency.
The two panels show the reduced-$\chi^2$ histograms from the model fit to the two modes with the largest 1/f variance at the scan frequency.
The model did not include inter-module correlations.
The outliers were due to the small number of CESes where the inter-module correlations were important and
 scan-synchronous pickup, which cannot be modeled by a 1/f spectrum.
}
\end{figure}

To check that inter-module correlations can be ignored for all CESes, we repeated the process in the above paragraph for 200 CESes (selected from a randomly chosen 2-week period).  We computed the first 10 eigenmodes of the covariance matrix for each CES.
For each eigenmode we computed the diode--linear-combination TOD.
For each such TOD we computed the noise power spectrum and the $\chi^2$  between that spectrum and the model prediction (including only intra-module correlations) between twice the scan frequency and 1\,Hz.
Figure \ref{fig:1f_model_chi2} shows the distribution of these $\chi^2$ values.
Although most CESes were well-modeled with only intra-module correlations, a few CESes were not, as indicated by the large $\chi^2$ outliers.
Most of the outliers could be eliminated by including in the noise model the first inter-module mode of the correlation matrix.

Since we chose to consider only intra-module correlations in the final noise model, we  estimated the error introduced because of the few CESes that had significant inter-module correlations even above the pipeline filter cutoff.
We picked seven CESes\footnote{CESes 5410.1, 5410.2, 5420.2, 5420.3, 5420.4, 5456.2, and 5475.2} that were outliers of this type.
For these CESes, we prepared two noise models.
The ``array correlated'' model had only the first eigenmode of the 1/f correlation matrix.
In this model, we adjusted the slopes and knee frequencies for this mode in an attempt to extract the slope-2 component.
From the measured slope and knee frequency of each diode, we calculated the effective knee frequencies of slope-1 and 2 components.
Then we used the slope-2--component knee frequency, and the measured slope or 2, whichever was greater.
We made an additional adjustment to the knee frequency
\begin{equation}
f' = f_0\rho_1^{1/\alpha},
\end{equation}
where $f_0$ is the knee frequency before this adjustment and $\rho_1$ is the diagonal element of the correlation matrix constructed from the first mode.
This adjustment accounted  for the fact that the slope-2 component was not perfectly correlated---in the fiducial pipeline noise model we included intra-module correlations.
Therefore in this validation simulation, we did not include the 1/f noise that was not correlated among modules.
The other noise model, ``module correlated,'' used the same knee frequencies and slopes as array correlated, but we set the correlation-matrix elements between different modules to 0.
The difference between ``array correlated'' and ``module correlated'' was the effect of the correlation not included in the fiducial pipeline model.
At low $\ell$, the difference was at most 10\% of the noise bias.
Since at most 10\% of CESes had an important intermodule mode and the noise bias scaled linearly with the number of CESes, the effect was diluted to 1\% when all CESes (including the majority that were not outliers) were included.
We considered 1\% to be an acceptable systematic error in the noise model\footnote{This noise-model error affected only the statistical error estimate, not the power-spectrum central value.  Our cross-correlation method (\S\ref{sec:xcorr}) made the central value insensitive to noise bias.}; 
therefore, we justified using only intra-module correlations in the pipeline.
Note that in this simulation the EE and BB noise-bias differences had opposite signs.
This suggested that due to the limited number of CESes, the inter-module mode happened to couple to E and B differently.
If there was no preferred orientation, we would expect this effect to average down as more CESes were included; although, we did not rely on this reduction in justifying the noise model.

\subsection{Model of Noise Correlations}
\label{sec:noise:model}
\begin{figure}
\includegraphics[width=1.0\textwidth]{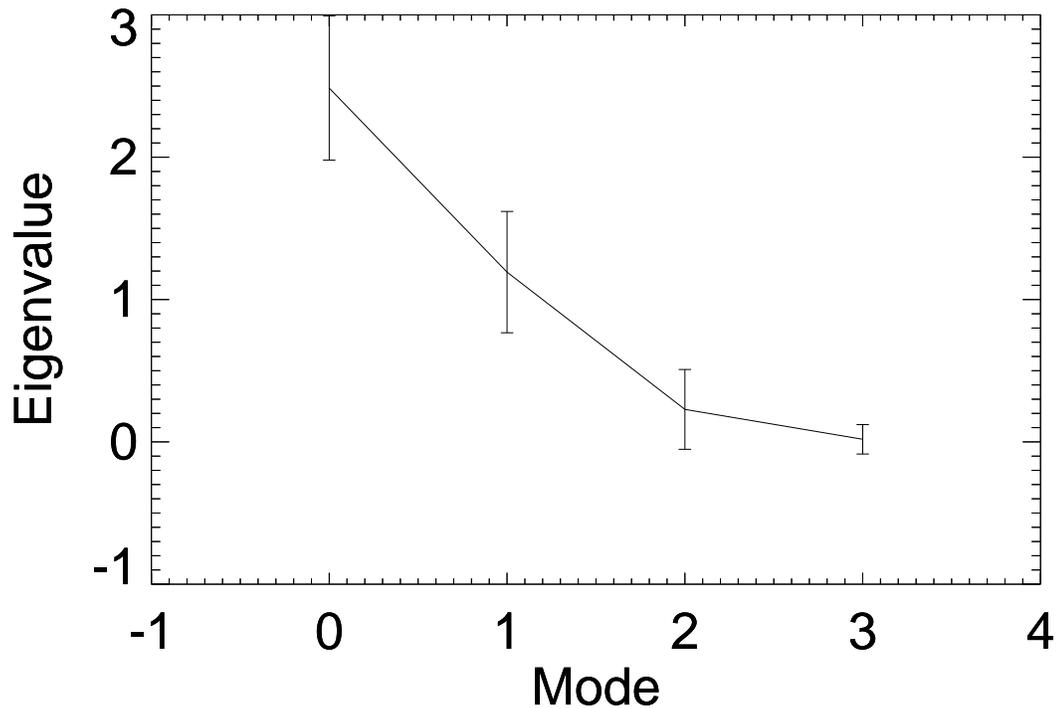}
\caption[1/f Correlation Matrix Eigenvalues in Each Module]{\label{fig:1f_intramodule_eigenvalues}
The first 2--3 1/f modes in each module were significant.
The error bars show the scatter (standard deviation) among modules for one CES.
Note that the last eigenvalue ($0.02\pm0.10$) scatters around 0, indicating that it is dominated by measurement noise.
We attributed the three significant modes to MMIC-amplifier--gain fluctuations (two modes from two amplifier chains) and atmospheric fluctuations (one mode).
}
\end{figure}

\begin{figure}
\includegraphics[width=0.9\textwidth]{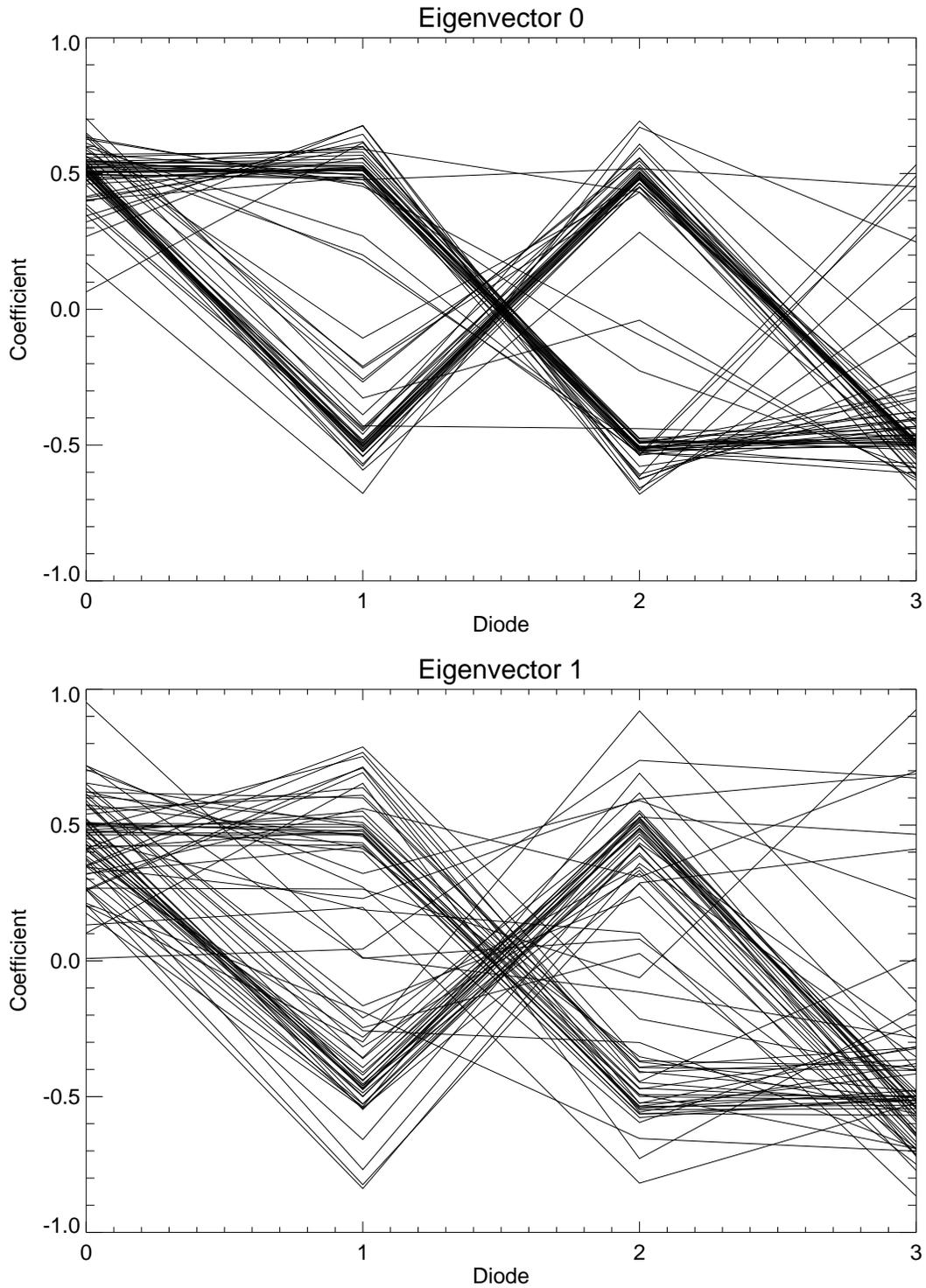}
\caption[1/f Correlation Matrix Eigenvectors]{\label{fig:1f_intramodule_eigenvectors}
We expected opposite signs for Q1 (U1) and Q2 (U2) 1/f noise within each module.
Panels show eigenvectors corresponding to the first two eigenvalues from one CES.
Each line corresponds to one module.
Diodes 0, 1, 2, and 3 are the same as Q1, U1, U2, and Q2, respectively.
}
\end{figure}

Based on the investigation above, we included only intra-module correlations in the noise model.
For white noise, we simply used the noise level for each diode and correlation matrix $\hat{\rho^w_{ij}}$.
For 1/f noise, we used the fitted knee frequency, slope, and correlation matrix $\hat{\rho^{1/f}_{ij}}$.
To generate a noise realization from this model, we needed to diagonalize the correlation matrix.
Because of the small bandwidth in the 1/f noise measurement, the matrix may be ill-conditioned.
In the diagonalization, I applied the same regulation procedure described above except that diodes with knee frequencies in (0, 10]\,mHz were not removed.
Figure \ref{fig:1f_intramodule_eigenvalues} shows the eigenvalues obtained for one CES.
The first 2--3 modes were significant.
Figure \ref{fig:1f_intramodule_eigenvectors} shows the eigenvectors.
The most important modes had negative correlation between the two Q diodes and between the two U diodes.
We expected the $-$ sign because the two diodes had opposite signs.
I included the first three modes of the regulated 1/f correlation matrix in the noise model whenever their eigenvalues were positive.
Negative eigenvalues indicated measurement failures in the correlation matrix because they correspond to unphysical negative variance modes, which cannot be modeled.
The fourth eigenvalue scattered about zero, indicating that it came primarily from measurement noise.
We attributed the three significant modes to MMIC-amplifier--gain fluctuations (two modes from two amplifier chains) and atmospheric fluctuations (one mode).
Except for $\sigma$, the data processing was independent of the noise model.
Therefore, these choices about how to model the noise correlations did not bias the result.
They only affected the estimate of the statistical uncertainty at the percent level.

\section{Time-ordered--data Processing}
After calibration and characterization and before map making, we filtered the TOD.
First, we applied several filters to each CES-diode to remove various noise and contamination effects.
Then I removed CES-diodes from further analysis if contamination remained after filtering.
These steps ensured that the data used in the power-spectrum calculation had well understood noise properties.

\subsection{Filtering}
\label{sec:filter}
We filtered each CES-diode's TOD to remove residual 1/f noise and ground pickup.
The filters were
\begin{enumerate}
\item We subtracted the slope (linear function of time) from each CES-diode.  This prepared the data for Fourier transformation.
\item We removed high-frequency components by multiplying by a filter function in the Fourier domain.  This function was unity below 9.6\,Hz and zero above 9.7\,Hz.
Between those frequencies we smoothed with a cosine-squared function shape.
We chose the filter frequency to remove a variety of high-frequency spikes (including strong contamination at 10\,Hz due to aliasing of the 60-Hz power line).
The signal from the maximum multipole $\ell=1000$ entered the data at $\approx5$\,Hz so the filter did not remove significant signal (see also Figure \ref{fig:transfer_function}).
\item We subtracted the slope (linear function of azimuth) from each scan half-period.
This filter introduced a high-pass cutoff at twice the scan frequency, thereby removing remaining 1/f noise.
\item We subtracted the average (constant during the CES) value from each datum.  For this filter, we binned the data into 40 bins in azimuth.  The subtracted mode was an arbitrary function of azimuth that was constant during each CES.
We used this filter to remove ground pickup caused by  sidelobes hitting the ground.
\end{enumerate}

\subsection{Data Selection}
\label{sec:cuts}
\begin{figure}
\includegraphics[width=0.75\textwidth]{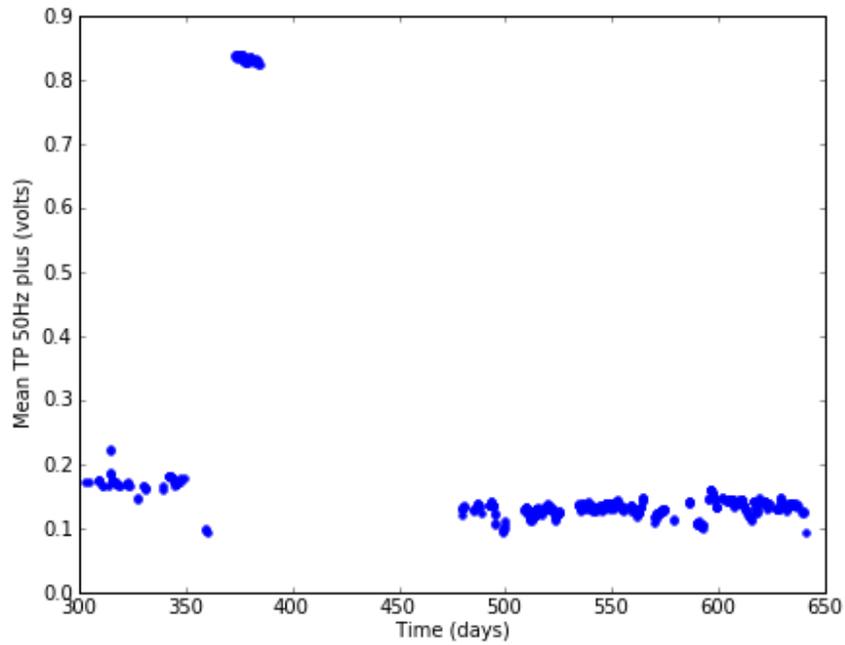}
\caption[RQ57Q1 TP with Two States]{\label{fig:RQ57_vs_time}
RQ57Q1 had two TP states.
I rejected data from it in the high state because the quality was bad.
}
\end{figure}

\begin{figure}
\includegraphics[width=0.75\textwidth]{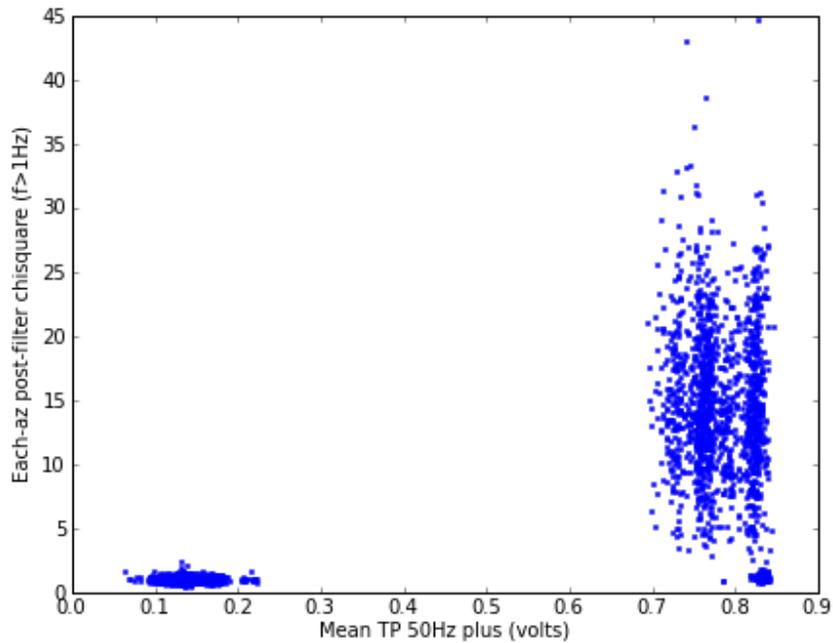}
\caption[RQ57Q1 Noise-model Fit $\chi^2$]{\label{fig:RQ57_quality}
In the high-TP state, RQ57Q1 had very bad noise-model fit $\chi^2$.
}
\end{figure}

\begin{figure}
\includegraphics[width=0.75\textwidth]{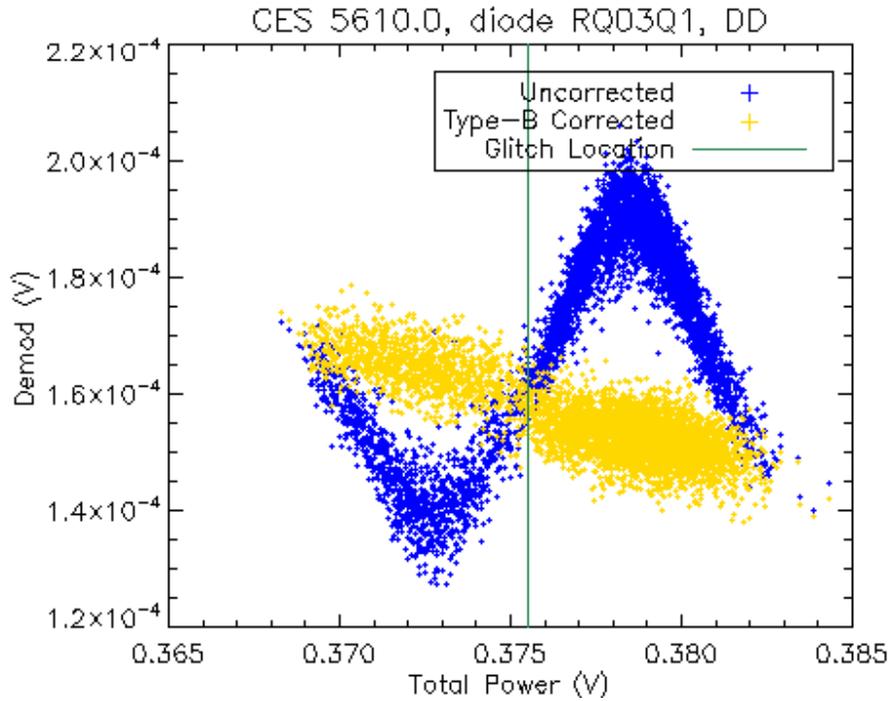}
\caption[DD vs. TP]{\label{fig:typeb_chi2_ex}
I rejected CES-diodes with large $\chi^2$ in a linear fit to the DD vs. TP relation.
Type-B glitching created non-linearity in this relation.
}
\end{figure}

\begin{figure}
\includegraphics[width=0.75\textwidth]{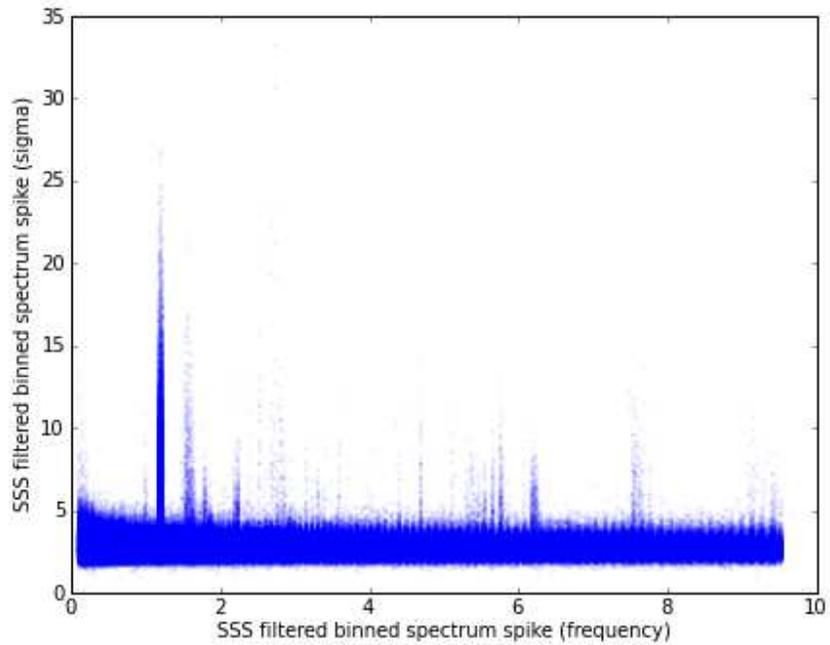}
\caption[Noise-spectrum Spikes]{\label{fig:FFT_spike_distribution}
I rejected CES-diodes with a noise-spectrum--spike statistic $>5\,\sigma$.
The strongest spikes at 1.2\,Hz were due to the refrigerator cycle.
}
\end{figure}

\begin{figure}
\includegraphics[width=0.75\textwidth]{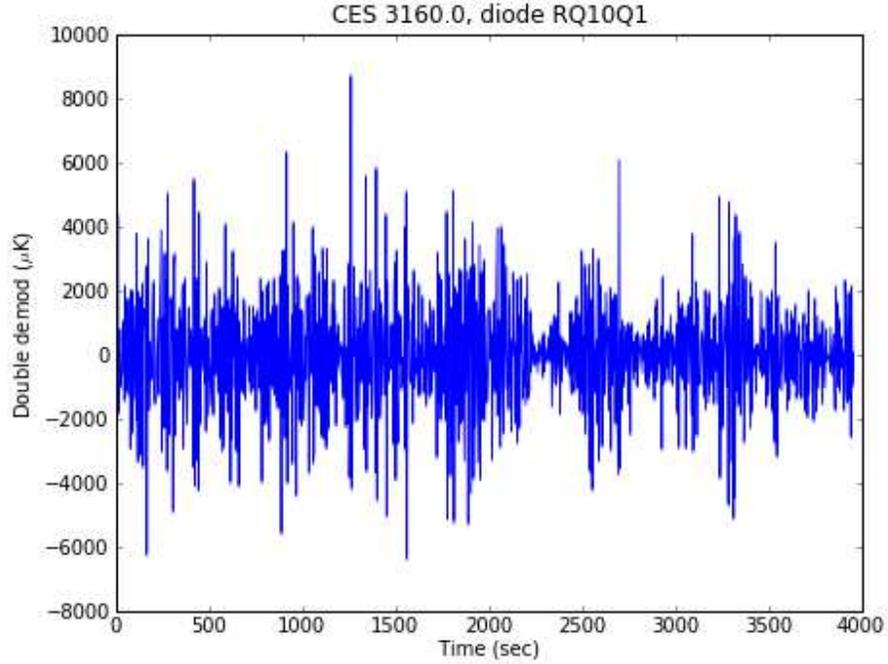}
\caption[CES-diode with Noise Level Changes]{\label{fig:noise_change_eg}
I rejected CES-diodes whose noise level changed within the CES.
}
\end{figure}

The goal of data selection was to remove data that, even with filtering, did not fit our noise model.
I based most of the selection on the statistics calculated for each CES-diode.
For each such statistic, I determined a cut threshold based on its distribution, examination of outliers, and the validation tests
\citep{viewer_study2, viewer_study3, viewer_study1}.
The validation tests served as the final check that data-selection thresholds were adequate.
I removed data in the following steps:
\begin{enumerate}
\item \label{cut:bad_modules} I removed six non-functional modules.  Modules RQ7, 8, 28, and 42 had electrical connection problems, and we could not bias them.
Modules RQ3 and 70 had frequent level shifts: (almost) every CES contained a large (compared to the noise level) jump in the DD level.
\item \label{cut:bad_diodes} I removed four non-functional diodes.  
RQ4U2, RQ38Q1, and RQ40Q1 had connection problems.
RQ35Q1 had a noise spike in the Fourier domain whose frequency was unstable.  This problem was due to a bad ADC channel (the same problem happened in Q band; the ADC channel was the only common element.).
\item \label{cut:RQ57} I removed RQ57Q1 when the TP (\S\ref{sec:module_principles}) was above 0.5\,V.
This diode had two states (Figure \ref{fig:RQ57_vs_time}).
The higher-voltage state had systematically worse quality (Figure \ref{fig:RQ57_quality}).
\item \label{cut:unusable_CES} I removed unusable CESes.  \S\ref{sec:routine_checks:DAQ} describes DAQ problems that caused an entire CES to be unusable.
I also removed CESes during which the mount stalled, deck moved, generator failed, temperatures were unstable, or module bias was unusual.
If a significant fraction of the data could be recovered, I redefined the CES to exclude the bad data.
\item \label{cut:unusable_diodes} I removed CES-diodes for which the Preamp power supply failed \citep{preamp_supply_rejection}.
The symptom of this failure was that the noise levels for one MAB increased greatly.
There were six such failures during the season.
We replaced the failed power supplies within a few days.
\item I removed CES-diodes when one of the sidelobes passed through the Sun \citep{yuji_far_sidelobe}.
This created strong features in the map, and we chose selection criteria based on pointing information to remove CES-diodes that had evidence of such a feature.
\item \label{cut:cryo} I removed CESes with poor cryostat temperature regulation.
I excluded all CESes for which the mean temperature (at RQ9) was $>30$\,K or the RMS fluctuation during the CES was $>0.19$\,K.
\item \label{cut:knee} I removed CES-diodes whose knee frequencies were greater than twice the scan frequency.
Such data had significant 1/f noise above the filter cutoff, often had bad weather, and made the modeling of 1/f noise too important.
\item \label{cut:short} I removed very short CESes with duration $<1000$\,s.
\item I removed data with bad weather.  This cut removed entire CESes based on the array-average fluctuation of TP \citep{wband_weather_stat}.
\item I removed CES-diodes with TOD glitches.  I identified glitches by calculating the maximum excursion (compared to the noise level) on three time scales: 20\,ms (1 sample), 100\,ms, and 1\,s.
If any glitch was $>6$\,$\sigma$, I removed the CES-diode.
\item \label{cut:slopes} I removed CES-diodes for which the dependence on azimuth (slopes removed in filtering) had a large change within the CES.
\item \label{cut:noise_model} I removed CES-diodes with poor fitting of the noise model.  If the reduced $\chi^2$ of the fit was $>3$ below 1\,Hz or $>2$ above 1\,Hz, I removed the CES-diode.
(Because we fit the noise model before filtering but computed this statistic after filtering, we had to compute the effect of the filter on the expected noise \citep{akito_eachaz_noise_effect}.)
\item \label{cut:noise_model_near_scan} I removed CES-diodes whose noise-model fit $\chi^2$ was $>5$ in the 40 modes closest to the scan frequency.  For this statistic, I applied only the ground-subtraction filter.
\item I removed CES-diodes whose Type-B $\chi^2$ was $>10$.
I calculated this statistic as the reduced $\chi^2$ from a linear fit to the unfiltered DD vs. TP relation (Figure \ref{fig:typeb_chi2_ex}).
\item \label{cut:jump} I removed CES-diodes with level shifts during the CES.  We calculated the maximum level shift on three time scales.  If the shift was $>6.5\,\sigma$ in 1\,s, $>9\,\sigma$ in 10\,s, or $>25\,\sigma$ in 100\,s, I discarded the CES-diode.
\item \label{cut:FFT_spike} I removed CES-diodes with narrow spikes in the noise power spectrum.
I averaged the noise spectrum in logarithmically-spaced frequency bins.
If any bin was $>5\,\sigma$ above the noise model, I discarded the CES-diode.
Because I found strong contamination near 1.2\,Hz from the cryogenic refrigerator cycle, I removed CES-diodes with any bin between 1.15 and 1.3\,Hz $>4\,\sigma$.
This cut also removed spikes due to scan frequency harmonics (Figure \ref{fig:FFT_spike_distribution}).

\item \label{cut:noise_stationarity} I found CES-diodes whose noise level changed during the CES (Figure \ref{fig:noise_change_eg}).
I calculated a noise non-stationarity statistic as the standard deviation of the standard deviations of 10-s periods.
If this statistic was $>2.5$ times the expected variation, I removed the CES-diode.
\item \label{cut:sss} I removed CES-diodes with very large scan-synchronous signal.  Although we filtered the scan-synchronous signal with the ground subtraction filter, we found evidence of residual contamination \citep{typeo_modeling}.
Therefore I removed any CES-diode whose noise power at the scan frequency was $>100$ times the expected level. 

\end{enumerate}

Table \ref{tab:selection_efficiency} lists the selection efficiency for each patch.

\begin{deluxetable}{lrrr}
\tablecolumns{4}
\tablecaption{Data-selection Efficiency
\label{tab:selection_efficiency}
}
\tablehead{\colhead{Patch} & \colhead{Baseline CES-diodes\tablenotemark{a}} & \colhead{Selected CES-diodes} & \colhead{Efficiency}}
\startdata
2a & 448607 & 286643 & 64\%\\
4a & 364443 & 247448& 68\%\\
6a & 298943 & 176107& 59\%\\
7b & 173843 & 113038 & 65\%\\
\hline
Total & 1285836 & 823236 & 64\%\\ 
\enddata

\tablenotetext{a}{To compute the baseline I applied only cuts \ref{cut:bad_modules}, \ref{cut:bad_diodes}, \ref{cut:RQ57}, \ref{cut:unusable_CES}, \ref{cut:unusable_diodes}, \ref{cut:cryo}, and \ref{cut:short}.
These corresponded to instrument or observing failures rather than analysis limitations that required a certain data quality.
}
\end{deluxetable}


\label{sec:cuts:ps_spikes}


\section{Power-spectrum Calculation}
\label{sec:spectra_calc}
The power-spectrum calculation was a multi-step process.
First, we made sky maps from the filtered TOD.
Then we computed spectra with a pseudo-$C_\ell$ estimator.
In this process, we divided the maps into cross-correlation units and subtracted the spectra from each unit so the result depended only on cross-correlations.
Then we divided by transfer and window functions to account for the reduction of power from filtering and smoothing by the beam.

\subsection{Map Making}
We made a naive map from each CES-diode.
For each TOD sample, we estimated the polarization
\begin{equation}
\begin{pmatrix}
\hat{Q}\\
\hat{U}
\end{pmatrix} = \left(\mathbf{P}^T\mathbf{N}^{-1}\mathbf{P}\right)^{-1}\mathbf{P}^T\mathbf{N}^{-1}\vec{d},
\end{equation}
where
\begin{equation}
\mathbf{P}\equiv\begin{pmatrix}
\cos2\psi&\sin2\psi\\
\end{pmatrix}
\end{equation}
specifies the polarization angle ($\psi$) of the diode,
\begin{equation}
\mathbf{N}\equiv\begin{pmatrix}
\sigma^2
\end{pmatrix}
\end{equation}
is the variance (including only the white noise  and ignoring correlations), and $\vec{d}$ is the sample.
We assigned each sample to a pixel using the pointing model.
The maps used the HEALPix\footnote{\url{http://healpix.jpl.nasa.gov}} \citep{gorski_healpix} pixelization scheme with the parameter $N_\textrm{side}=512$ corresponding to $0\fdg1$ pixel resolution.
We combined multiple observations of the same pixel with inverse-variance weighting.
This naive combination was not optimal; however, the statistical uncertainty increase was percent-level \citep{akito_map_making}.
We used the same weighting to combine different diodes and CESes.

\subsection{Pseudo-$C_\ell$ Calculation}
\label{sec:pseudocl}
From combined maps of many CES-diodes we estimated the power spectra.
We used the MASTER (Monte Carlo Apodized Spherical Transform Estimator) technique to calculate an unbiased estimator from partial-sky data with filtering \citep{hivon_master,hansen_pseudoCl}.
In this technique, we first calculated ``pseudo power spectra''
\begin{equation}
<\tilde{C}_\ell> = \sum_{\ell'}M_{\ell\ell'}F_{\ell'}B_{\ell'}^2<C_{\ell'}>,
\end{equation}
where $M_{\ell\ell'}$ is a mode-coupling matrix\footnote{This estimator, combined with the limited sky fraction, mixes E and B modes.  The coupling matrix includes this mixing as well; although I suppressed the indices for different polarization modes.} that accounts for the limited sky fraction,
$F_{\ell'}$ is the transfer function that accounts for filtering, and $B_{\ell'}$ is the beam window function that accounts for the instrument's resolution.
The brackets indicate the relation between the true and pseudo spectra holds on average for an ensemble of sky realizations.
We converted the pseudo spectra back into true spectra estimates after binning in $\ell$ to create 19 bandpowers
\begin{equation}
\hat{C_b} = \sum_{b',\ell} F_{b}^{-1}K^{-1}_{bb'}P_{b'\ell}\tilde{C}_\ell,
\end{equation}
where 
$ F_{b}^{-1}$ is the (binned, see \S\ref{sec:anal:sim}) inverse transfer function,
\begin{equation}
K_{bb'} = \sum_{\ell,\ell'} P_{b\ell}M_{\ell\ell'}B^2_{\ell'}Q_{\ell'b'}
\end{equation}
accounts for the mode-mode coupling and beam, and
$P_{b\ell}$ and $Q_{\ell'b'}$ are the binning projection and interpolation functions.
We verified that $\hat{C_b}$ was unbiased with simulations (\S\ref{sec:anal:sim}).

We calculated $B_\ell$ from the calibrated beam shape (\S\ref{sec:cal:beam} and \cite{raul_thesis, wband_beam}).
\begin{equation}
B_\ell = \frac{2\pi}{\Omega}\int b(\theta)P_\ell (\cos\theta)d(\cos\theta),
\end{equation}
where $\Omega$ is the solid angle\footnote{Some conventions differ by not including $\Omega$.}, $b(\theta)$ is the beam shape, and $P_\ell$ are Legendre polynomials.
We multiplied $B_\ell$ calculated from the beam shape by an additional factor corresponding to a Gaussian beam with FWHM = $5\farcm1$ to account for the pointing-model scatter (\S\ref{sec:cal:pointing}).

When we calculated power spectra, we masked known point sources.
In the MASTER technique we weighted pixels based on how often we observed them \citep{Feldman94}.
To avoid contamination from point sources, we reduced the weight to zero for nearby pixels.
We masked pixels within $2\degr$ of Centaurus~A (in patch 2a) and within $1\degr$ of Pictor~A (in patch 4a).
The mask was not apodized i.e. outside those radii the pixel weights were unaffected.

\subsection{Cross-correlation}
\label{sec:xcorr}
We used a cross-correlation technique to eliminate noise bias and reduce systematic errors.
We made separate maps for each pointing division.
Each division included a range of azimuth and deck angles chosen to evenly divide the data.
There were 40 such divisions: 5 in azimuth\footnote{I made the azimuth divisions at $130\fdg157$, $157\fdg52$, $209\fdg935$, and $232\fdg254$ East of North.  I chose these points to equalize the number of CES-diodes in each division.} 
and 8 in deck; although, patches 6a and 7b did not have data in all divisions\footnote{Patch 6a had a lower Declination and, therefore, a smaller azimuth range than the other patches (24 total divisions).  We only observed patch 7b while it was rising, limiting the azimuth range (30 total divisions).}.
We performed the pseudo-$C_\ell$ calculation for each pointing division as well as the cumulative map (sum of all divisions).
Then we subtracted the power-spectra estimates of each division from the cumulative spectra.
Thus, only cross-correlations of different pointing maps contributed to the result.
Because the data in different divisions were taken at different times, the noise between divisions was uncorrelated.
Therefore, there was no noise bias in our estimator\footnote{Power-spectrum estimators that include auto-correlations are subject to ``noise bias''---the power of the noise correlating with itself.  By eliminating noise bias, we removed the possible systematic error from misestimating it.}.
Moreover, systematic contamination that did not correlate between divisions was suppressed.

\section{Simulations}
\label{sec:anal:sim}
A simulation suite is a necessary component of the MASTER technique.
Our suite contained three types of simulations: calibrating, fiducial, and noise.
I based the simulations on the instrument noise model developed above.
I used the simulations to calculate $F_b$ and estimate the statistical uncertainty.
The simulations also verified that the analysis method was unbiased.

The three types of simulations differed in whether I input CMB power and instrumental noise.
In calibrating simulations I input CMB power but no instrumental noise.
For these simulations, I assumed a flat (i.e. $\ell$-independent) input power for both E and B modes.
In fiducial simulations I input both CMB power and instrumental noise.
Usually fiducial simulations had CMB power input drawn from realizations of a $\Lambda$CDM model; however, some simulations had non-zero B-mode power to estimate the sample variance for B modes.
Noise simulations had instrumental noise only.
For the final configuration I performed 200 calibrating, 200 fiducial, and 100 noise simulations.
Each simulation was a complete realization of the experiment observation and TOD.

The simulations included white noise and 1/f noise.
Both noise sources were correlated within each module but not between modules.
For white noise, I added random Gaussian values with correlation matrix $\hat{\rho^w}$.
For 1/f noise, I created random Gaussian values in the Fourier domain with correlation matrix $\hat{\rho^{1/f}}$ after regulation described in \S\ref{sec:noise:model}.
Then I scaled the noise to have the measured 1/f slope and knee frequency.
I generated more samples than the data length of the CES so the 1/f noise at the beginning and end were uncorrelated.

\begin{figure}
\includegraphics[width=0.75\textwidth]{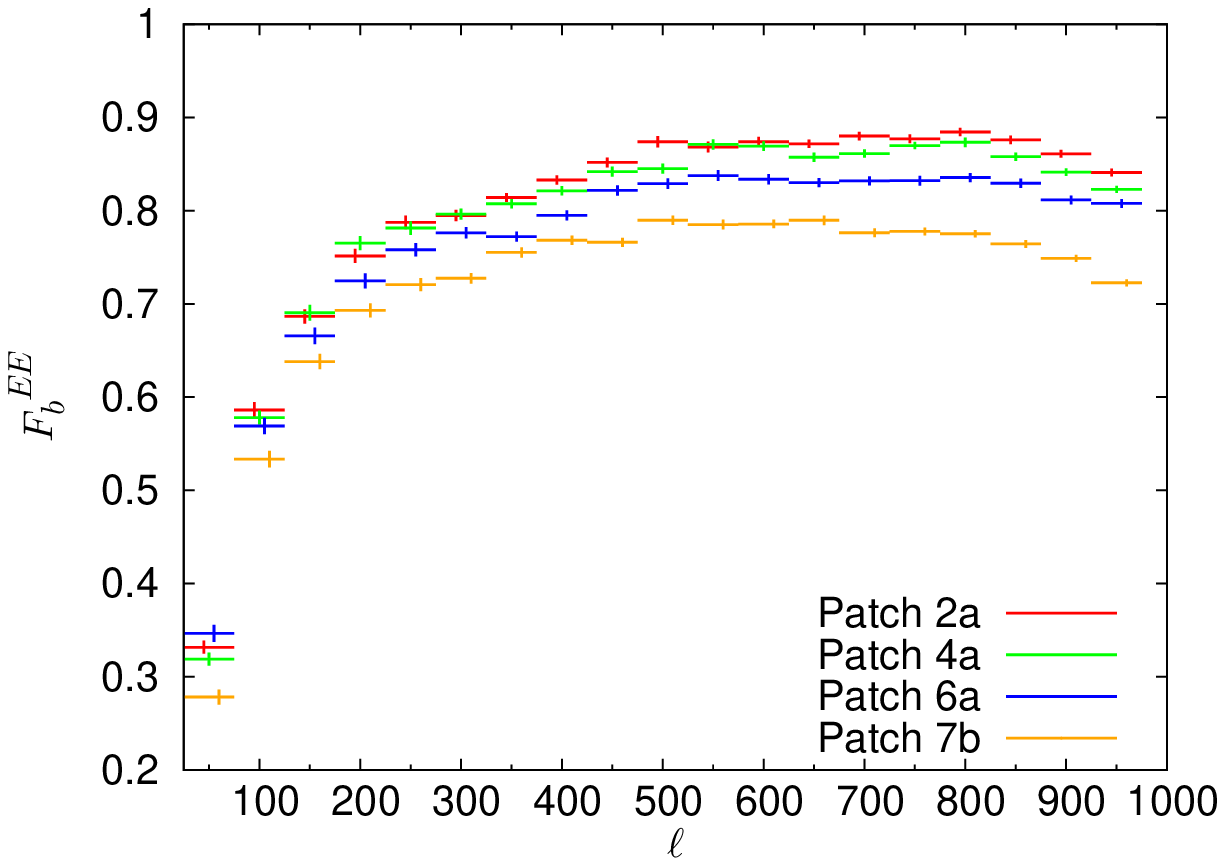}
\includegraphics[width=0.75\textwidth]{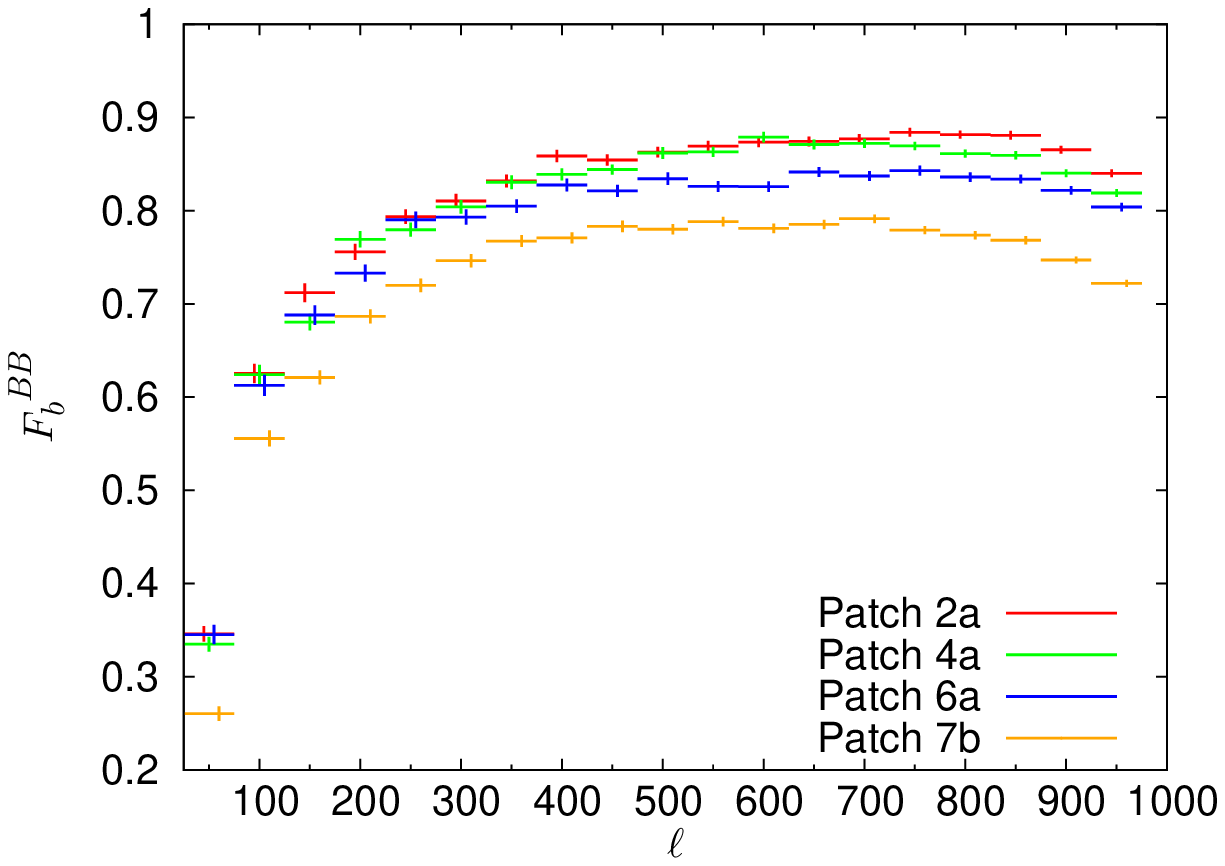}
\caption[Transfer Function]{\label{fig:transfer_function}
The transfer function accounted for filtering in the pipeline.
I computed the transfer function from noiseless simulations; the error bars indicate statistical uncertainty from the 200 simulations of each patch.
Except at low $\ell$, filtering did not affect the CMB power by more than 20\%.
}
\end{figure}

I computed $F_b$ from the calibrating simulations.
The ratio between the simulation input and output spectra is $F_b$ (Figure \ref{fig:transfer_function}).
Since there was no EB power in the simulations, I computed the EB transfer function as the geometric mean of the E and B-mode functions.
The fiducial simulations confirmed that computing the binned transfer function was sufficient; otherwise I would need the transfer function at every $\ell$.
This sufficiency depends on the input spectra.
If I had found a result radically different from $\Lambda$CDM, I would have had to re-evaluate whether the binned transfer function would still provide an unbiased estimator.
We also confirmed that the filtering did not mix different $\ell$ \citep{tod_filter_systematics}; otherwise the transfer function would need to be a matrix including that mixing possibility.
(An additional complication in computing the transfer function is that simulated skies have finite resolution and pixel size, but the true sky has infinite resolution.  Normally the simulation input was pixelized at $N_\textrm{side}=1024$.  To solve this problem I ran additional calibrating simulations with $N_\textrm{side}=2048$.
We also accounted for the finite resolution of TOD samples \citep{akito_high_speed_mc}.)

\begin{figure}
\includegraphics[width=1.0\textwidth]{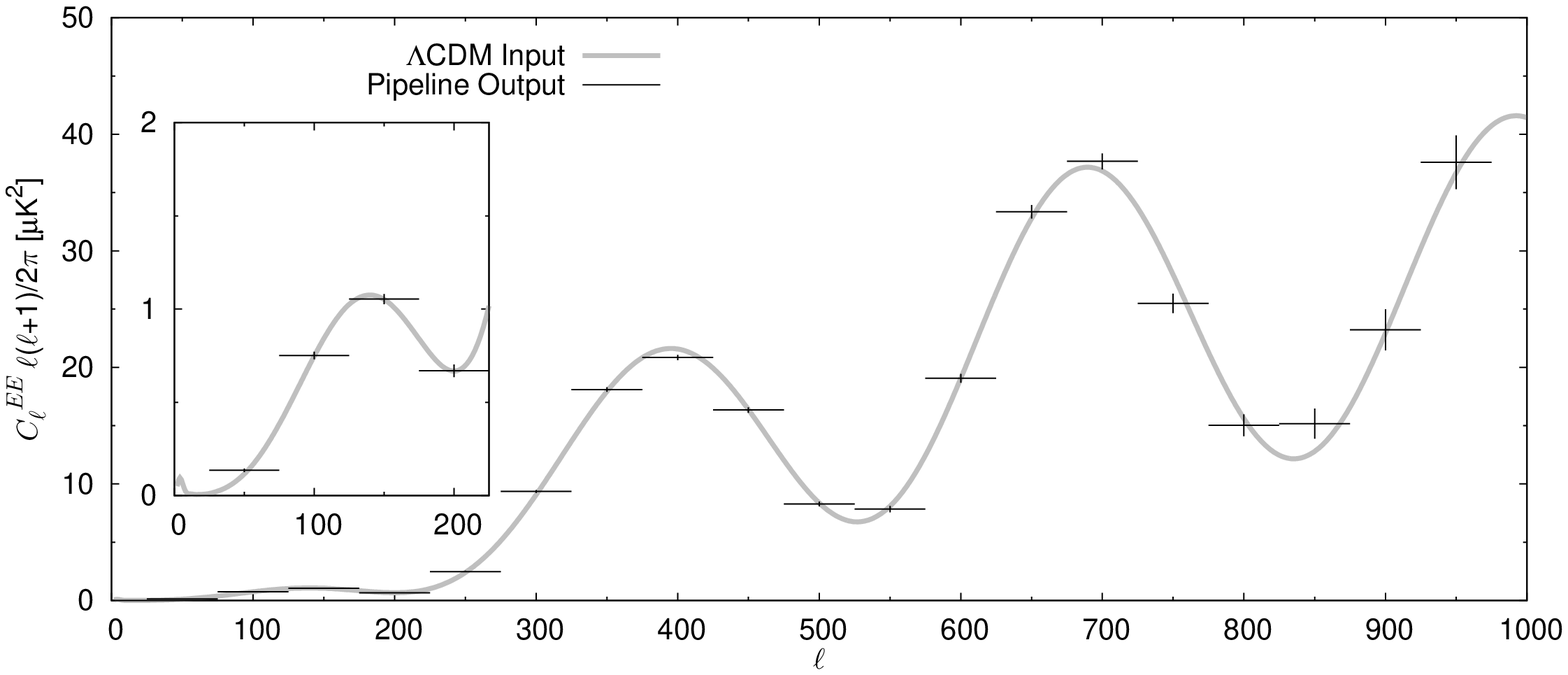}
\caption[Pipeline Fiducial Simulations]{\label{fig:fiducial_ensemble_EE}
Simulations confirmed the analysis pipeline was unbiased.
Error bars indicate statistical uncertainty from the 197 full-season simulations of patch 2a.
}
\end{figure}

The fiducial simulations confirmed the pipeline did not introduce bias.
If I input a $\Lambda$CDM spectrum, I recovered the same spectrum (Figure \ref{fig:fiducial_ensemble_EE}).
The simulations also provided the primary estimate of the statistical uncertainty.
Unless otherwise noted, power-spectra error bars are the standard deviations of the fiducial simulations.

\svnid{$Id: validation.tex 150 2012-07-24 22:04:36Z ibuder $}

\section{Validation Tests}
We performed a set of validation tests to confirm that the analysis was correct.
These tests were part of a blind-analysis strategy wherein we finalized the analysis method without knowing the results.
The most important validation test was the ``null suite,'' a set of tests for consistency between two halves of the data.
As a second validation test, I checked that the spectra computed from different patches were consistent.
Finally, I compared the results of  different analysis configurations.

\subsection{Blind-analysis Strategy}
Blind analysis removes opportunities for observer bias by making the analysis choices independent of our knowledge of the result \citep{blind_analysis}.
There were many choices to make in the analysis method.
If we calculated the results from all possible choices, we could choose the result closest to our a priori beliefs.
These beliefs might include a preference for results close to $\Lambda$CDM or for the lowest B-mode power.
Even if we did not intentionally alter the analysis method to prefer such results, there are more subtle effects such as ``stopping bias'' that can cause results to conform to previous experiments and predictions\footnote{\cite{croft_measurements} provide some evidence that cosmological-parameter measurements have observer bias.}.
Collectively such effects constitute observer bias, a systematic error that is difficult to quantify because it depends on the psychology of physicists.
By performing a blind analysis---one in which we did not see the result until after making all analysis choices---we removed this systematic error.

The method of blind analysis we chose for QUIET was not to know the result (i.e. the CMB polarization power spectra) until we chose the calibration and analysis methods.
We did look at TOD and maps including the CMB signal because the relationship between them and power spectra was not straightforward.
Since we did not let the results inform the analysis method, we used the validation tests to determine whether analysis choices were correct.
For the most important tests (the null suite and patch comparison) we set the criteria to pass the tests in advance \citep{null_test_strategy}.
We also evaluated the systematic errors (\S\ref{sec:systematics}) before seeing the result.

\subsection{Null Suite}
\label{sec:null_suite}
The null suite was the most important validation test.
The suite had 32\footnote{One of the null tests, based on whether the patch was rising or setting during observations, did not work for patch 7b because we always observed it while rising.  There were only 31 null tests for patch 7b.} null tests, each of which divided the data in half to target a known problem or potential systematic error.
We decided in advance to determine the success of the null suite from three criteria for each patch (Table \ref{tab:null_suite_ptes}).
In addition to these main criteria, we examined many other statistics and distributions to look for surprises.
We found two such surprises: a bias in the results including auto-correlations and excess $\chi^2$ at low $\ell$. 
To confirm these effects did not indicate a failure in our noise model, we performed additional, random null tests that validated the noise model with high precision.

\begin{deluxetable}{lrrr}
\tablecolumns{4}
\tablecaption{Null-suite Probabilities to Exceed
\label{tab:null_suite_ptes}}
\tablehead{\colhead{Patch} & \colhead{$\chi$ bias} & \colhead{Largest $(\chi_b^\textrm{null})^2$} & \colhead{Total $(\chi_b^\textrm{null})^2$}}
\startdata
2a & 26 & 14 & 78\\
4a & 46 & 40 & 6\\
6a & 74 & 45 & 50\\
7b & 8 & 18 & 76\\
\enddata
\end{deluxetable}


The null-test calculation followed the normal analysis method including pseudo-$C_\ell$ and cross-correlation.
I divided the CES-diodes\footnote{The two scan null tests divided each CES-diode into left-going/right-going and accelerating/decelerating halves.} into two halves and computed a separate map from each half.
I chose the pixel weighting
\begin{equation}
W(x) = \frac{1}{\sigma_1(x)^2 + \sigma_2(x)^2},
\end{equation}
where $\sigma_1(x)^2$ is the variance for pixel $x$ after accumulating all CES-diodes in half 1.
The choice of $W$ fixed the mode-coupling matrix ($M_{\ell\ell'}$ and $K_{bb'}$, \S\ref{sec:pseudocl}).
Then I defined the null-power estimator
\begin{equation}
\hat{C_b^\textrm{null}} \equiv \frac{K_{bb'}^{-1}P_{b'\ell}\tilde{C}_\ell^{11}}{F_b^{11}} - 2\frac{K_{bb'}^{-1}P_{b'\ell}\tilde{C}_\ell^{12}}{F_b^{12}} + \frac{K_{bb'}^{-1}P_{b'\ell}\tilde{C}_\ell^{22}}{F_b^{22}},
\end{equation}
where
$\tilde{C}_\ell^{11}$ is the pseudo-power of map 1, $\tilde{C}_\ell^{22}$ is the pseudo-power of map 2, $\tilde{C}_\ell^{12}$ is the pseudo-power from the correlation between maps 1 and 2, and $F_b^{ij}$ are the corresponding transfer functions\footnote{I simulated all the null tests and computed the transfer functions from the null-test calibrating simulations.}.
The expected value of $\hat{C_b^\textrm{null}}$ was zero.
Because the two null-test halves had different filtering and scanning, the difference map is not necessarily null\footnote{If and only if all three transfer functions are identical is the difference map a null map.}.
$\hat{C_b^\textrm{null}}$ is the unique null-power estimator \citep{kendrick_nulltests} that
\begin{enumerate}
\item is invariant if the maps are rescaled by constants (during filtering)
\item is invariant if the two halves are interchanged
\item is independent of the (CMB) power spectra 
\end{enumerate}

\begin{figure}
\includegraphics[width=1.0\textwidth]{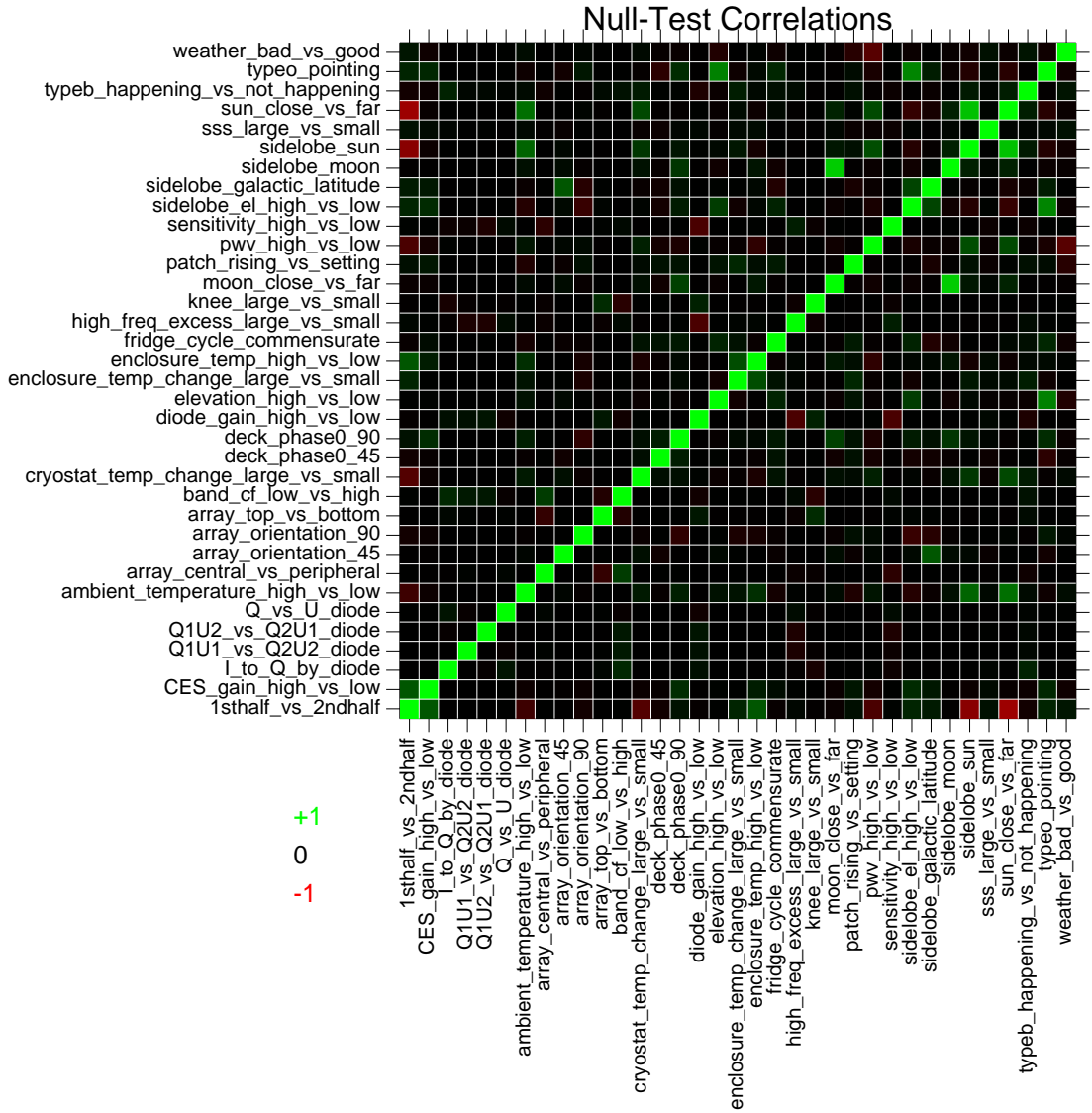}
\caption[Null-test Correlation Matrix]{\label{fig:nt_correlation}
I chose null tests  that were uncorrelated.
Green and red indicate correlated null tests; black indicates no correlation.
Appendix~\ref{app:nt} defines each null test and the correlation calculation.
The correlations shown are for patch 2a.
}
\end{figure}

I chose the null tests to be sensitive to known instrumental problems, systematic-error sources, and noise-model imperfections.
I required that no null test be $>40$\% correlated with any other (Figure \ref{fig:nt_correlation}).
A large number of tests can detect contamination at a level well below the statistical threshold of a single null test. However, performing a very large number of null tests dilutes the well-motivated null tests.
As the number of null tests increases, so does the largest expected statistical fluctuation.
Since any outliers must be compared with the expected fluctuation, a very large number of
null tests implies that any single null test has reduced discriminating ability: the level of
contamination required to cause a statistically significant outlier increases with the number
of null tests. Therefore the choice of the number of null tests represents a compromise between a lower detection threshold for the most well-motivated null tests and a lower detection
threshold for contamination that appears in all null tests\footnote{The random null tests provided a strong, independent test for such contamination.}. Since in practice the number of
null tests was limited by computational resources, I compromised by choosing the $\approx30$ best
motivated null tests.
Appendix~\ref{app:nt} lists the null tests.

\begin{figure}
\includegraphics[width=1.0\textwidth]{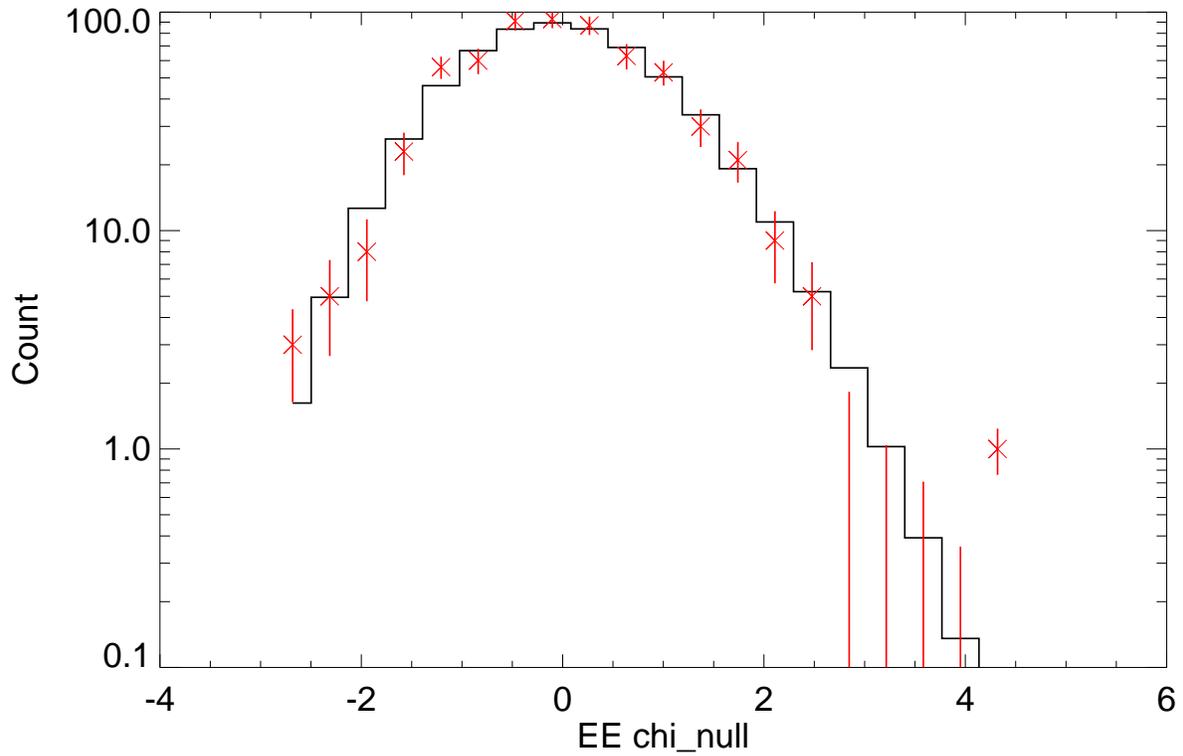}
\caption[Distribution of $\chi^\textrm{null}_b$]{\label{fig:chi_null_distribution}
I compared the distribution of $\chi^\textrm{null}_b$ (red) to simulations (black) to confirm the noise model and limit systematic errors.
The histogram shows all $\chi^\textrm{null}_b$ values for all EE bins $b$ for all null tests of patch 2a.
The error bars are the RMS fluctuations of the histogram from simulations.
}
\end{figure}

I constructed the null-suite passing criteria from
\begin{equation}
\chi^\textrm{null}_b \equiv \hat{C_b^\textrm{null}}/\sigma^\textrm{null}_b,
\end{equation}
where $\sigma^\textrm{null}_b$ is the RMS of $\hat{C_b^\textrm{null}}$ from fiducial simulations.
Based on the experience in Q band, I computed three summary statistics that revealed problems most often:
\begin{enumerate}
\item The average of $\chi^\textrm{null}_b$ for all null tests and $\ell$ bins.  This statistic is called ``$\chi$ bias.''
\item The largest $(\chi^\textrm{null}_b)^2$ for all null tests and $\ell$ bins
\item The total $(\chi^\textrm{null}_b)^2$ for all null tests and $\ell$ bins
\end{enumerate}
Each statistic included only EE and BB null spectra.
Including EB was not an independent test, and EB contamination required correlation contamination in both EE and BB.
(Although I did not include EB spectra in these summary statistics, I still checked that the EB null spectra were zero-consistent.)
To pass, I required each statistic have a probability to exceed (PTE) $>4$\%.
I calculated each PTE by comparing the data statistic to the distribution of the same statistics in fiducial simulations\footnote{I defined the $\chi$ bias PTE to be the fraction of fiducial simulations with difference from zero larger than the data.}.
Each patch had an independent null suite (because the patches did not overlap).
For three independent tests with
4\% probability to fail by statistical fluctuation, the combined failure probability generated
by statistical fluctuations was 12\%. However, the three tests for each patch were not independent.
Therefore, the probability for a patch to fail the null suite by statistical fluctuation was $\lesssim10$\%,
which I considered reasonable.
Even if the null suite passes I am not required to stop the analysis. I can examine other
statistics (e.g. the worst single null test or EE alone), and we should not finish the analysis if we
believe the result will be incorrect. Predefining the null-suite evaluation criteria does not force us
to report an incorrect result if we find a problem outside those criteria.
In the final configuration, all patches passed their null suites (Table \ref{tab:null_suite_ptes}).
Moreover, the full $\chi^\textrm{null}_b$ distribution showed how well the simulations modeled the data (Figure \ref{fig:chi_null_distribution}).

In addition to the main statistics above, I examined many statistics and distributions from the null suite to check for unexpected errors.
These statistics included $\chi^\textrm{null}$ from EE or BB only, the largest total $(\chi^\textrm{null})^2$ for any single null test, the RMS of the  $\chi^\textrm{null}$ distribution, Kolmogorov-Smirnov tests, and $\chi^\textrm{null}_b$ as a function of $b$.
Among all of them I found only two problematic ones: the auto-correlation bias and the low $\ell$ total $\chi^2$.

\begin{figure}
\includegraphics[width=1.0\textwidth]{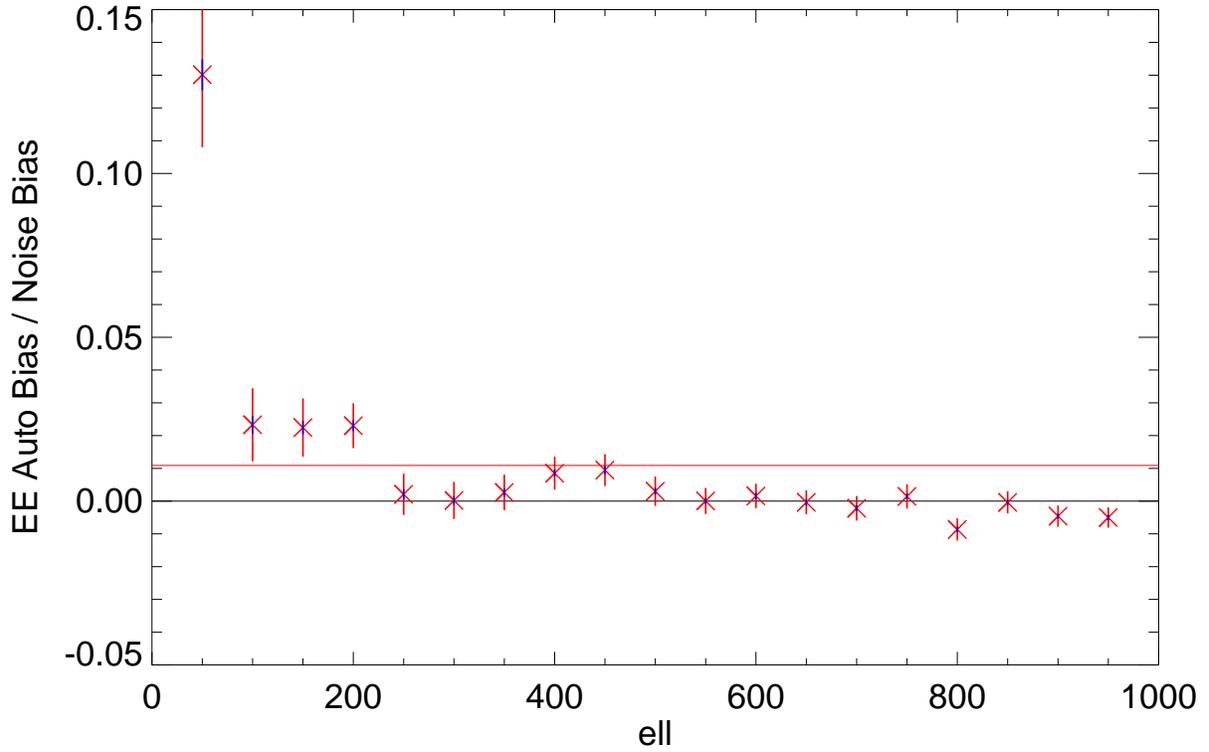}
\caption[Excess Noise in Power Spectra Including Auto-correlations]{\label{fig:low_ell_auto_bias}
The null suite indicated excess noise in power spectra including auto-correlations.
The ``auto bias'' is the difference between null spectra with and without auto-correlations; it is independent of the cross-correlation contribution.
Dividing each bin by the noise bias yields the fractional excess noise as a function of $\ell$.
Our result was free from this excess because it included only cross-correlations.
The plot averages all null tests for patch 2a.
The horizontal red line is the average of all $\ell$.
The red error bars indicate statistical uncertainty; the blue error bars indicate uncertainty from noise modeling due to the finite number of noise simulations.
}
\end{figure}

In addition to the normal analysis using cross-correlation (\S\ref{sec:xcorr}), I also computed the power spectra including the auto-correlations.
Including the auto-correlations makes the result sensitive to noise bias.
I subtracted the noise bias estimated from noise simulations; however, the result remained sensitive to errors in the noise model.
Therefore, the auto-correlation results were an important test of the noise model.
During null-suite evaluation, I computed the difference between the auto-correlation and cross-correlation null tests.
I found that, on average, auto-correlation null tests had higher power, especially at low $\ell$ (Figure \ref{fig:low_ell_auto_bias}).
This excess indicated systematic error in the auto-correlation results or an error in the noise model.
Because the final results used cross-correlation, systematic errors in the auto-correlation results were not concerning.
However, an error in the noise model would cause the statistical error to be misestimated.
Therefore I investigated this excess power in auto-correlation null tests \citep{low_ell_auto_bias}. 
I found that the excess was present in all null tests.
I checked that the excess was not correlated with scan-synchronous signal.
I found that 1/f-noise misestimation was unlikely to be the cause.
Patch 2a (\S\ref{sec:regular_observations}) had the largest excess; if the excess was due to noise misestimation it should be similar for all patches.
I was unable to identify the cause of the excess.

When I totaled $(\chi^\textrm{null}_b)^2$ for each bin $b$, I found an excess in the lowest bin ($\ell=$ 25--75) for EE.
The excess was strongest in patch 4a with total $\chi^2=71$ for 32 degrees of freedom.
Because the distribution of $\hat{C^\textrm{null}}$ was not Gaussian, I calculated the probability of such an excess from the fiducial simulations, finding $<0.5$\%.
However, that was not the proper a priori probability.
We did not decide in advance to focus on the lowest bin, and the probability of finding one outlier among many bins is much higher\footnote{There is no exact probability without deciding which bins to examine, which we did not do.}.
The excess was smaller in patches 6a (22\%) and 7b (14\%), and not present in 2a (53\%).

Because of the surprising statistics above, we used random null tests to confirm the noise model \citep{random_null}.
We created 1000 null tests, each of which divided the CESes randomly into two equal groups.
We compared the resulting null spectra to simulations.
The results confirmed our noise model was accurate to 3.1\% (in $C_\ell$).

\subsection{Patch Comparison}
\begin{figure}
\includegraphics[width=1.0\textwidth]{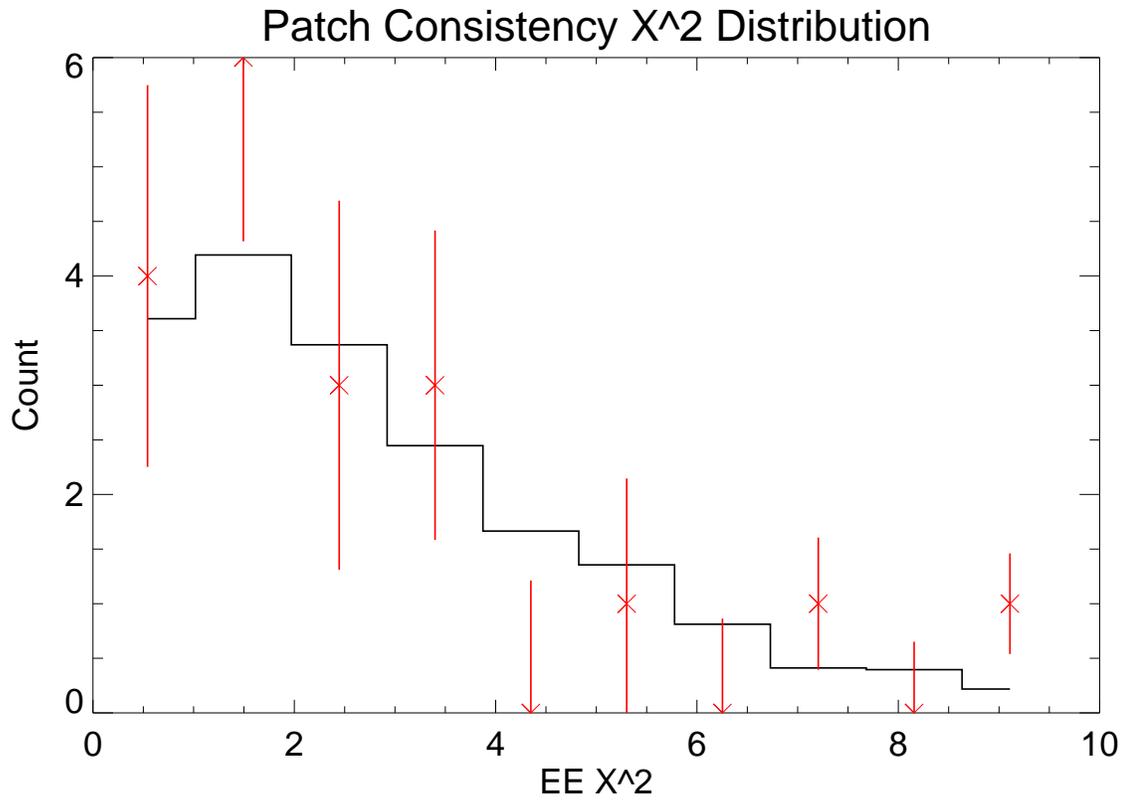}
\caption[Patch Comparison $\chi^2_p$ Distribution]{\label{fig:patch_consistency}
The patch comparison validation test confirmed that the four patches had consistent spectra.
The $\chi^2_p$ distribution from data (red) matched the expectation from simulations (black).
}
\end{figure}

The second most important validation test was the patch comparison.
In this test, I confirmed that the four patches produced consistent power spectra.
I computed the consistency statistic \citep{patch_consistency_def}
\begin{equation}
\chi^2_p\equiv \sum_{i=0}^3\sum_b\left(\frac{\hat{C^i_{b}} - \mu_b}{\sigma_b^i}\right)^2,
\end{equation}
where $\hat{C^i_{b}}$ is the bandpower\footnote{For computational convenience, I used the same transfer function for all patches in this test.  The difference was $<10$\%, and I used simulations to check that there was no resulting bias.}  of patch $i$ in bin $b$,
\begin{equation}
\mu_b\equiv \frac{\sum_i\hat{C^i_{b}}w_b^i}{\sum_iw_b^i}
\end{equation}
is the average of the four patches,
\begin{equation}
w_b^i\equiv1/(\sigma_b^i)^2
\end{equation}
are patch weights, and $\sigma_b^i$ are the fluctuations of each bandpower from fiducial simulations.
I compared $\chi^2_p$ to the expected distribution from fiducial simulations (Figure \ref{fig:patch_consistency}).
Before performing the test, I set passing criteria of 4\% PTE for each of EE, BB, and EB.
The final PTEs were 62\%, 5\%, and 16\% for EE, BB, and EB, respectively.

\subsection{Configuration Comparison}

\begin{figure}
\includegraphics[width=1.0\textwidth]{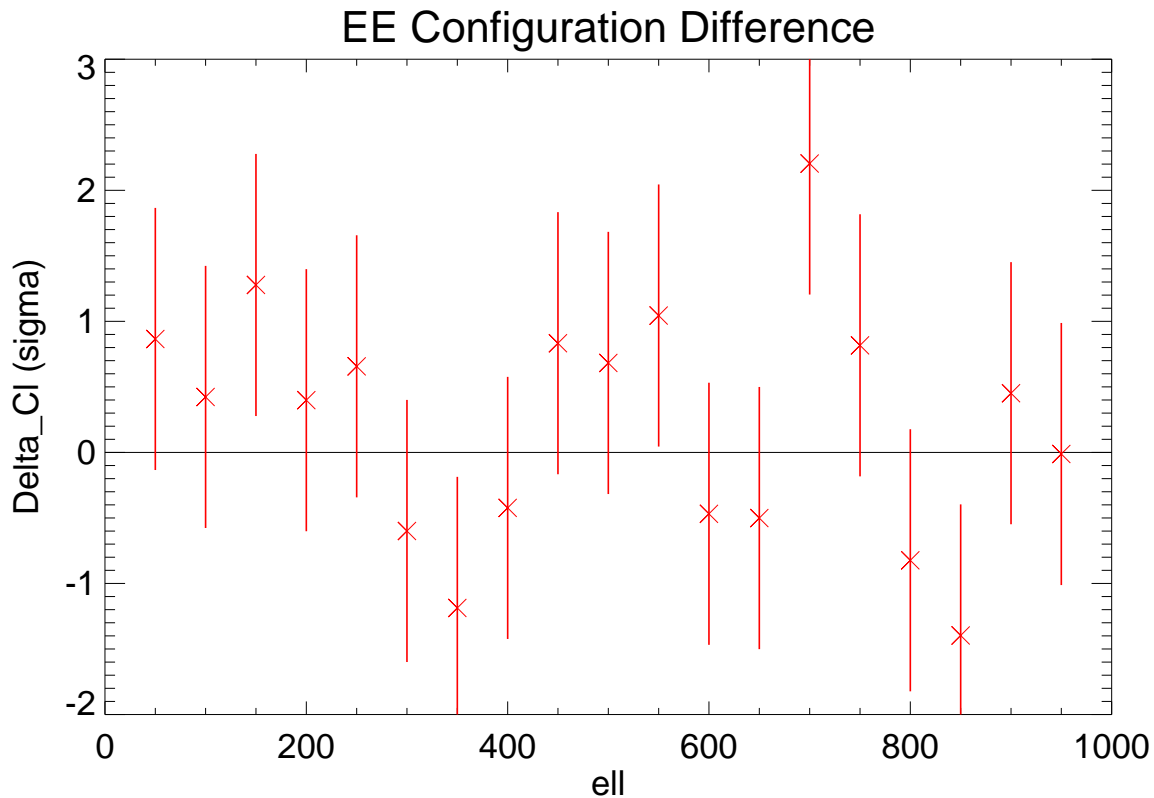}
\caption[Power Spectrum Difference Between Different Data-selection Thresholds]{\label{fig:configuration_comparison_ex}
Comparison of different data-selection thresholds confirmed no contaminated data remained.
The plot shows the difference between two EE spectra from patch 2a, computed with slightly different thresholds for the noise-spectrum--spike  cut (\S\ref{sec:cuts}).
The uncertainty on the difference (error bars) is 17\% of the total statistical uncertainty for one of the configurations.
}
\end{figure}

I tested that the power spectra were independent of the details of the analysis choices.
In this test I computed the difference of spectra from two different configurations of the analysis\footnote{To avoid possible observer bias, I randomized the sign of the difference.}.
Because it was computationally prohibitive to run simulations for this test, I developed a heuristic method to estimate the expected fluctuations (Appendix~\ref{app:error_bar_scaling}).
Usually I compared two data-selection choices with slightly different thresholds.
In such a test the expected fluctuation was well below the statistical error on the final power spectra (Figure \ref{fig:configuration_comparison_ex}).
Moreover, by changing the selection thresholds I directly probed the most contaminated data and confirmed that the result was insensitive to the details of data selection.
I limited such effects to $\approx30$\% of the total statistical error.
I used the same method to test for systematic errors by changing the configuration in ways that would be sensitive to the systematic error sources.
I tested for errors in the pointing model, in ground filtering (\S\ref{sec:systematics:ground}), and in source masking (\S\ref{sec:systematics:point_sources}).

\svnid{$Id: systematics.tex 151 2012-07-25 22:09:40Z ibuder $}

\section{Systematic Errors}
\label{sec:systematics}
\begin{figure}
\includegraphics[width=1.0\textwidth]{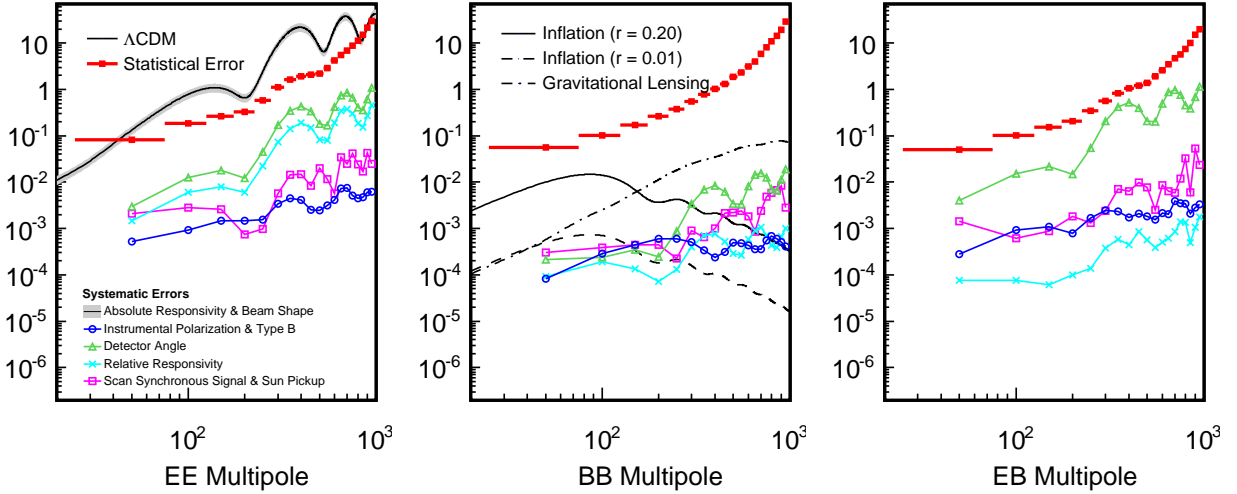}
\caption[Systematic Errors Summary]{\label{fig:syst_error_summary}
We considered many possible systematic errors.
Most were negligible compared to the statistical errors; however, beam-shape and responsivity miscalibration were dominant for $C_\ell^{EE}$.
All systematic errors were below the level of $r=0.01$ for $\ell<125$ B modes.
Units are $C_\ell\ell(\ell+1)/2\pi$ in $\mu$K$^2$.
SOURCE: Osamu Tajima
\fixme{final figure}
}
\end{figure}

We supplemented the validation tests with systematic-error estimates.
We used several techniques to test for systematic errors well below the statistical limits of the validation tests.
We investigated systematic errors from many possible effects:
\begin{enumerate}
\item Several important systematic effects were due to miscalibration---the difference between the actual and assumed calibration parameters.
\item We estimated the effect of spurious polarization created by instrumental imperfections.
\item Errors in Type-B--glitch correction created an additional contribution to this instrumental polarization.
\item We evaluated the impact of Sun contamination through the telescope sidelobes.
\item We discovered and estimated the impact of a type of polarized scan-synchronous signal.
\item Finally, we investigated many other effects and limited the corresponding errors to well below both the statistical uncertainty and the important systematic errors above.
\end{enumerate}
For the effects that were significant, we assigned a corresponding systematic error to the power-spectra results.
Systematic errors were the dominant source of uncertainty in the E-mode spectrum.
We paid special attention to the $\ell\approx100$ region of the B-mode spectrum because that is where we are searching for evidence of inflation.
For the B-mode spectrum, we controlled systematic errors to below the level of $r=0.01$ in the critical $\ell\approx100$ region (Figure \ref{fig:syst_error_summary}).

\subsection{Miscalibration}
\label{sec:systematics:miscal}
The differences between the true calibration parameters and the ones we used in analysis (\S\ref{sec:calibration}) created systematic errors.
Our strategy to address this miscalibration was to compare several alternate calibration methods, and use the difference between them to estimate the possible error in calibration.
The important calibration parameters were the beam shape, responsivity, detector angles, and pointing model.

To estimate the systematic errors from beam-shape miscalibration, we compared the fiducial calibration to optics simulations and pre-deployment measurements.
We calculated the difference in the beam size (FWHM) when we used optics simulations instead of the Tau~A measurements.
We calculated the difference in the beam shape when we used differential-temperature measurements of Jupiter instead of the antenna-range measurements.
We propagated these differences to the window function and assigned them as systematic errors.
Because the window function is a multiplicative factor, these errors are proportional to the power spectra.
We used the same method to assign the error (4.8\%) for the beam solid angle ($13.63 \pm 0.65\,\mu$sr, \S\ref{sec:cal:beam}).

To estimate the systematic errors for responsivity miscalibration,
we propagated the responsivity model uncertainties (\S\ref{sec:cal:responsivity}) to the systematic errors in the power spectra.
The absolute responsivity uncertainty (8.3\%, 17\% in $C_\ell$) created a multiplicative error  independent of $\ell$.
The power spectra are proportional to the responsivity squared so the responsivity uncertainty is double in the spectra.
For the uncertainty on relative responsivity among diodes, we used simulations to compare the effects of different models on the spectra.
The largest effects came from changing the model to use only Tau~A measurements.
We also calculated the effects of fluctuations (which we modeled as random) in the relative responsivity.
Relative responsivity errors are not purely multiplicative and created small, but not negligible, B-mode power \citep{gain_systematics}.

To estimate the systematic errors from detector-angle miscalibration,
we ran simulations with alternate detector-angle--calibration methods (\S\ref{sec:cal:angle} and \cite{angle_systematics}).
For the absolute calibration error (shift in the average angle),
we ran a simulation with all the detector angles shifted by the calibration uncertainty ($0\fdg5$).
We assigned the difference between this simulation and the fiducial angles as a systematic error.
This error was the largest source of spurious B modes.
We used two methods to estimate the error due to miscalibration of the relative angles.
First, we ran a simulation using detector angles from Tau~A instead of the fiducial model.
Second, we randomly varied the detector angles by the relative calibration uncertainty (scatter among the diodes, $3\fdg2$).
These two methods gave consistent estimates for the systematic error.
Finally, we considered possible time variation of the angles by simulating the effects of randomly varying the angles with $2\fdg7$ RMS, based on the scatter of Tau~A measurements.
Detector-angle errors tend to decrease $C_\ell^{EE}$ and generate spurious $C_\ell^{EB}$.
The error for $C_\ell^{EB}$ can have either sign.

To estimate the systematic errors from pointing-model miscalibration,
we compared different measurements of the pointing-model scatter.
Using the same observations of Jupiter, RCW~38, and the Galactic center from which we created the pointing model, the RMS scatter was $2\farcm17$ (\S\ref{sec:cal:pointing}).
If we instead used Moon observations to measure the scatter, the RMS was $1\farcm6$.
We corrected for the pointing miscalibration at first order by including a pointing-smoothing term in the beam window function (\S\ref{sec:pseudocl}).
We assigned the systematic error corresponding to the difference between smoothings of $2\farcm17$ and $1\farcm6$.
We added this uncertainty to the beam-shape systematic errors.
Like the beam-shape miscalibration, the pointing-model systematic error was a multiplicative effect for EE.

\subsection{Instrumental Polarization}
Instrumental imperfections created spurious polarization signals from unpolarized inputs.
We did not correct for this effect in analysis.
Instead, we showed that the resulting systematic error was small enough.
First, we measured the  $I\rightarrow Q/U$ leakage for each diode from the mini sky dips (\S\ref{sec:obs:cal} and \cite{leakage_monopole_syst}).
The median leakage was 0.19\%.
The statistical uncertainty of each measurement was 0.29\%.
The systematic uncertainty was 0.16\%, which was dominated by the dependence on the elevation at which we performed the sky dip (0.14\%).
We confirmed the mini-sky-dip measurements agree with measurements from large sky dips, Jupiter, and Tau~A at the 0.03\% level.
Using the measured leakage, we simulated the corresponding spurious polarization in our analysis assuming CMB spectra from $\Lambda$CDM\footnote{Although \textit{WMAP} provides a high--signal-to-noise map of the CMB temperature, it does not measure the polarization well in our patches.
Therefore, using \textit{WMAP} as the input for this simulation would not provide a good estimate of the correlation between the real and spurious polarization.}.
We conservatively varied the leakage with elevation to account for the systematic error above.
We also fluctuated the leakage based on the statistical measurement uncertainty.
We assigned the spurious polarization in this simulation as a systematic error due to instrumental polarization.
The combination of deck and sky rotation was essential in making the instrumental-polarization systematic error small enough.

The leakage considered above had the simplest beam shape, a monopole.
We also considered leakage that generated spurious polarization signals with a dipole or quadrupole pattern \citep{multipole_leakage}.
We measured each pattern from observations of Jupiter.
The average leakage was 0.47\% for the dipole pattern and 0.32\% for the quadrupole pattern.
As above, we simulated CMB temperature maps and convolved them with these leakage patterns to compute expected spurious polarization maps.
We assigned systematic errors from the power spectra of these maps.
The monopole leakage dominated at low $\ell$ ($\lesssim300$); however, the dipole contribution was comparable at higher $\ell$.
The quadrupole contribution was negligible at all $\ell$.

\subsection{Type-B--glitch Miscorrection}
\label{sec:type_b_systematics}
Miscorrection of Type-B glitching caused additional instrumental polarization.
I used the functional form of the Type-B correction to identify the possible effects.
Then I used simulations to quantify the effects on the power spectra.
I found that Type-B--glitch miscorrection increased the instrumental-polarization systematic error by 8\% \citep{typeb_systematics}.

If I did not use accurate Type-B correction parameters, some Type-B effects remain.
I summarize the Type-B correction (\S\ref{sec:calibration:typeb}) as
\begin{equation}
x'_\textrm{DE} = G( F(x_\textrm{DE}, x_\textrm{TP}; \vec{P}); \vec{P}' ),
\end{equation}
where $x'_\textrm{DE}$ are the demodulated data after correction, $x_\textrm{DE}$ are the demodulated
data that would have been collected without Type-B glitching, $x_\textrm{TP}$ are the TP data without Type-B glitching,
 $\vec{P}$ is the set of glitch parameters, and $\vec{P}'$ is the set of correction parameters.
The collected data with Type-B glitching are ($y_\textrm{DE}$, $y_\textrm{TP}$) $= F(x_\textrm{DE}, x_\textrm{TP}; \vec{P})$.
The function $F$ models the effect of Type-B glitching; $\vec{P}$ specifies the glitch locations and heights.
The function $G$ is the Type-B correction function.
If $\vec{P}= \vec{P}'$  then $x' = x$; however, due to measurement uncertainty, $\vec{P}' \not= \vec{P}$.
Expanding $G$ near $\vec{P}$, the first order miscorrection causes
\begin{eqnarray}
\label{eq:typeb:mis:gain}
\frac{\partial x'_\textrm{DE}}{\partial x_\textrm{DE}} &\not=& 1 \\
\frac{\partial x'_\textrm{DE}}{\partial x_\textrm{TP}} &\not=& 0.
\label{eq:typeb:mis:leakage}
\end{eqnarray}
Eq. \ref{eq:typeb:mis:gain} represents the responsivity change introduced by miscorrection.
Eq. \ref{eq:typeb:mis:leakage} increases the instrumental polarization because $x_\textrm{TP}\propto I$.
Without Type-B, the $I\rightarrow Q/U$ leakage is
\begin{equation}
L_\textrm{normal} = \frac{dx_\textrm{DE}}{dx_\textrm{TP}}.
\end{equation}
With Type-B, the total leakage is $\frac{dx_\textrm{DE}'}{dx_\textrm{TP}}$.
The additional leakage due to Type-B is
\begin{eqnarray}
L_\textrm{Type-B} &=& \frac{dx_\textrm{DE}'}{dx_\textrm{TP}} - \frac{dx_\textrm{DE}}{dx_\textrm{TP}} = \frac{\partial x_\textrm{DE}'}{\partial x_\textrm{DE}}\frac{dx_\textrm{DE}}{dx_\textrm{TP}} + \frac{\partial x_\textrm{DE}'}{\partial x_\textrm{TP}}\frac{dx_\textrm{TP}}{dx_\textrm{TP}} - \frac{dx_\textrm{DE}}{dx_\textrm{TP}}  \nonumber\\ 
&=& \frac{\partial x_\textrm{DE}'}{\partial x_\textrm{TP}} + \left(\frac{\partial x_\textrm{DE}'}{\partial x_\textrm{DE}} - 1\right)\frac{dx_\textrm{DE}}{dx_\textrm{TP}}.
\end{eqnarray}
The first term is the same as Eq. \ref{eq:typeb:mis:leakage}.
The second term is a higher-order correction due to the responsivity change.
Since (as I will show below) the responsivity change is $<1$\% and the normal and Type-B leakages are comparable, I ignored the correction and used Eq. \ref{eq:typeb:mis:leakage} to estimate the Type-B leakage.

\begin{figure}
\includegraphics[width=1.0\textwidth]{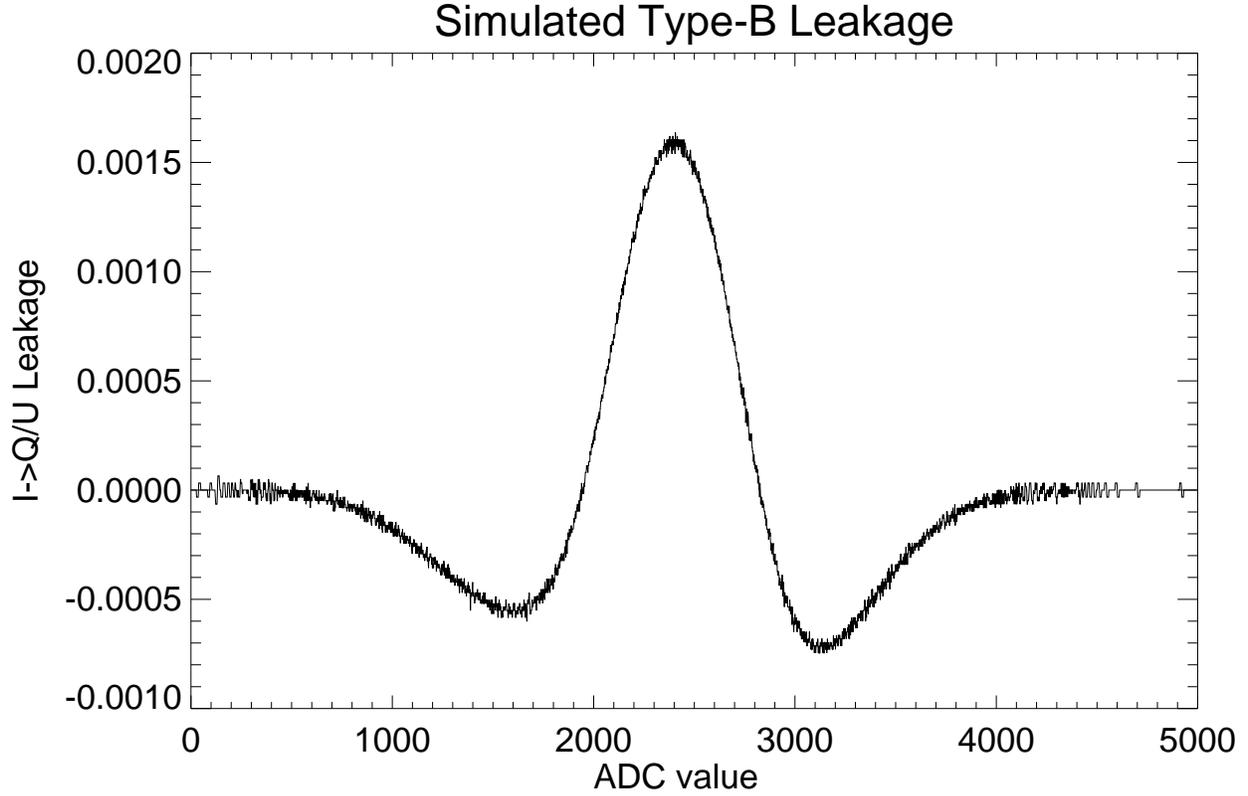}
\caption[Type-B--glitch---miscorrection Instrumental Polarization
]{\label{fig:typeb:mis:leakage_ex}
Type-B--glitch miscorrection created instrumental polarization that varies with the TP offset.
The units of the horizontal axis are arbitrary.
The average polarization is zero; however, the fluctuations create spurious polarization power.
}
\end{figure}

I ran simulations to calculate the Type-B leakage and responsivity change.
I ran two sets of simulations: one with the fiducial correction parameters $\vec{P}$ and another with fluctuated parameters.
The difference between them is the Type-B miscorrection.
I only simulated TOD, not the pipeline map making or power-spectra calculation.
Furthermore, I used only a representative set of CES-diodes with the largest ``Type-B badness'': the absolute value of the difference of the Type-B $\chi^2$ (\S\ref{sec:cuts}) before and after correction.
I tried many sets of fluctuated parameters including glitch locations, heights, and preamp factors differing from the measured ones by more than the measurement uncertainty.
I also checked that the number of iterations (3) in the Newton's method calculation of $\Delta G$ is sufficient.
From the simulated TOD, I calculated the slope $\frac{\partial x'_\textrm{DE}}{\partial x_\textrm{TP}}$ (Figure \ref{fig:typeb:mis:leakage_ex}).
I saved the maximum absolute slope in each simulation as an upper limit on the possible leakage.
I also calculated $\frac{\partial x'_\textrm{DE}}{\partial x_\textrm{DE}}$ and saved the maximum slope change, including its sign\footnote{The absolute responsivity systematic error depends on whether the sign is the same for different CES-diodes.}.
Finally, I calculated the Type-B $\chi^2$ from the simulated TOD.

\begin{figure}
\includegraphics[width=1.0\textwidth]{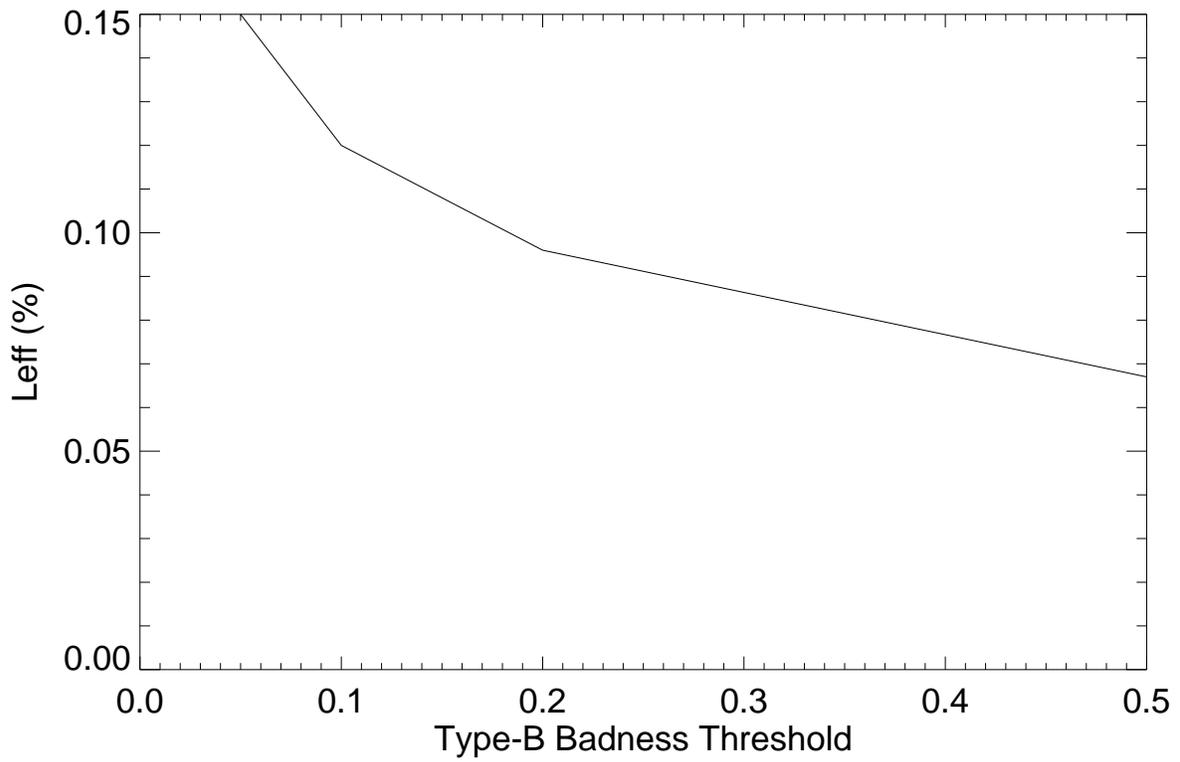}
\caption[Type-B--miscorrection Systematic Error vs. Threshold]{\label{fig:leakage_vs_badnesscut}
The systematic error from Type-B miscorrection did not depend strongly on the threshold I used to exclude CES-diodes without significant Type-B glitching.
}
\end{figure}

I used the relationship between the Type-B $\chi^2$ and simulated leakage to propagate the leakage to the power spectra.
I computed an effective Type-B leakage for each CES-diode from its Type-B $\chi^2$.
Although there are many other reasons for excess $\chi^2$, attributing all of it to miscorrection leads to the most conservative estimate.
I computed the effective leakage due to Type-B miscorrection
\begin{equation}                                                                
L_\textrm{eff}^\textrm{Type-B} = \sum_\textrm{CES-diodes}L_i / N,                 
\end{equation}   
where $N$ is the total number of CES-diodes and $L_i$ is the effective leakage for each CES-diode given by:
\begin{enumerate}
\item If the Type-B badness was less then 0.1, I set the effective leakage to zero.  This excluded CES-diodes that did not have significant Type-B.  
The result did not depend strongly on the threshold (Figure \ref{fig:leakage_vs_badnesscut}).     
\item If the $\chi^2$ of this CES-diode was less then 1, I set the effective leakage to zero.  These CES-diodes were well corrected, and the nature of the $\chi^2$ simulation forced the simulated $\chi^2$ to be greater than 1.
\item If the $\chi^2$ of this CES-diode was greater than the maximum $\chi^2$ in simulation, the maximum simulated leakage was the effective leakage.
Since I simulated extreme parameter fluctuations, if the $\chi^2$ of this CES-diode was larger than any of them, it cannot be the result of miscorrection.
I adopted the maximum leakage as a conservative limit on the miscorrection that may be hidden by whatever other effect caused the bad $\chi^2$.
\item If none of the above apply, I interpolated the simulated leakage-$\chi^2$ relationship to find the effective leakage at this $\chi^2$.
\item I repeated the steps above for each type of parameter fluctuation and used the largest resulting leakage.
This was the conservative choice because the data $\chi^2$ did not indicate which Type-B parameter was inaccurate.
\end{enumerate}

The resulting $L_\textrm{eff}^\textrm{Type-B} = 0.0012$.
In the power spectrum, this leakage effect added in quadrature with the normal leakage.
This is because the average Type-B leakage was zero.
The leakage due to miscorrection is a sum of odd functions\footnote{See Eq. 59 of \cite{typeb_formalism}.} so the average leakage must be zero as long as $x_\textrm{TP}$ was sampled uniformly.
Since the fluctuation of $x_\textrm{TP}$ during the season was larger than the glitch width and there is no reason to prefer one side of the glitch, there was effectively uniform sampling.
The average miscorrection leakage was zero, so there was no correlation with the normal leakage.
The main effect was the variance produced by the correlation of the miscorrection leakage with itself.
I estimated this effect by comparing the effective leakage to the normal leakage (0.19\%).
Since the effect in the power spectra was proportional to the leakage squared, the effect of miscorrection was
$(0.12/0.19)^2\approx40$\% of the normal leakage effect.
Since the two effects added in quadrature, the total systematic error was $\sqrt{1+0.4^2}\approx1.08$ times that from normal leakage alone.

The other effect of Type-B miscorrection was the responsivity change; however, I found that the systematic error was negligible.
I computed an effective responsivity change
\begin{equation}
\Delta R^\textrm{Type-B}_\textrm{eff} = \sum_\textrm{CES-diodes}\Delta R_i/N
\end{equation}
using the same method as for leakage.
$\Delta R$ is the fractional responsivity difference caused by Type-B miscorrection.
I also computed a relative-responsivity fluctuation
\begin{equation}
\sigma R^\textrm{Type-B}_\textrm{eff} = \sqrt{\sum_\textrm{CES-diodes}\Delta R_i^2/N},
\end{equation}
which quantifies the possible relative-responsivity uncertainty introduced in the average CES-diode.
The results were $\Delta R^\textrm{Type-B}_\textrm{eff} = -0.002$ and
$\sigma R^\textrm{Type-B}_\textrm{eff} =  0.005$.
The former created a negligible absolute-responsivity error (0.2\%).
The relative-responsivity error of 0.5\% was small compared to the scatter in the relative-responsivity model (12.5\%).

\begin{figure}
\includegraphics[width=1.0\textwidth]{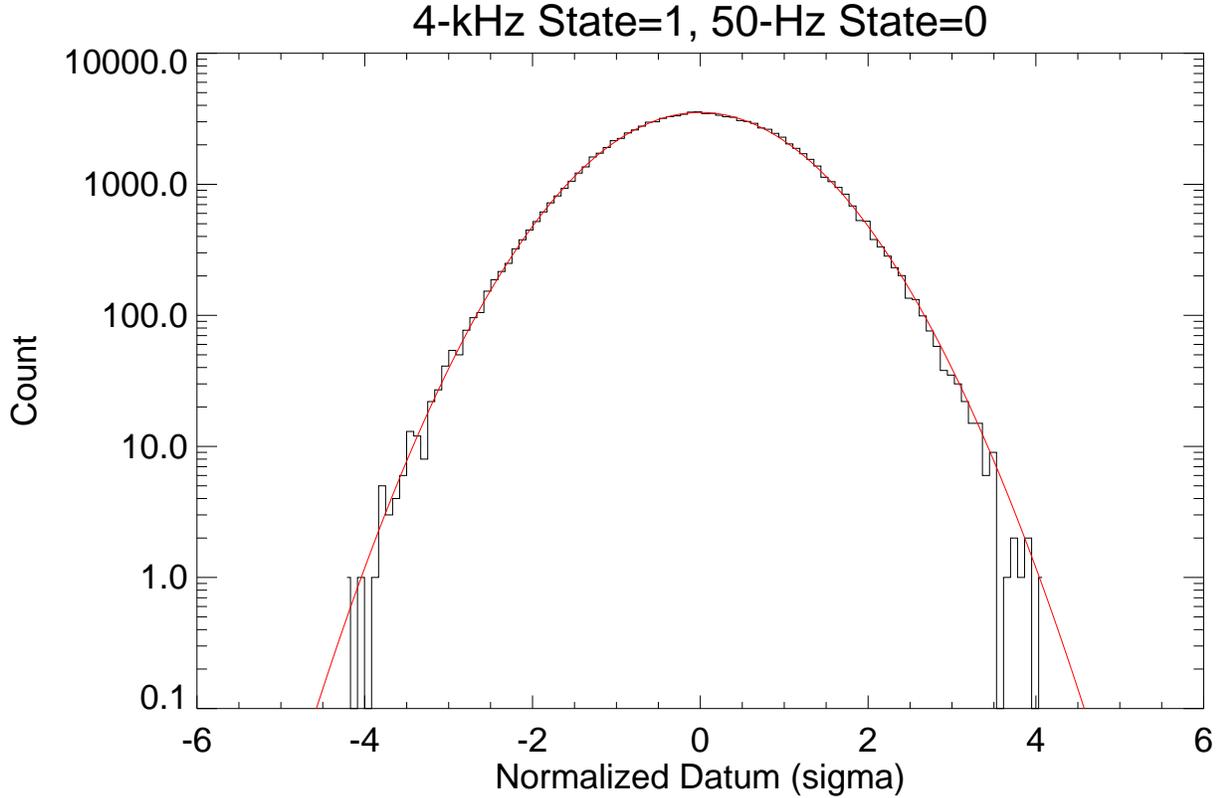}
\caption[800-kHz--data Distribution]{\label{fig:snapshot:gaussian}
The 800-kHz data were nearly Gaussian-distributed.
If the data had significant non-Gaussianity, I would have had to account for it in Type-B--glitch correction.
The red line is a Gaussian distribution with mean 0 and RMS 1.
}
\end{figure}

\label{sec:800khz_gaussianity}
As mentioned in \S\ref{sec:calibration:typeb}, the Type-B correction depended on the noise being Gaussian.
To test this assumption, I stacked 800-kHz snapshot data for one diode \citep{nongaussianity}.
To differentiate between inherent non-Gaussianity and non-Gaussianity introduced by Type-B miscorrection, I chose a diode without Type-B glitching.
I accumulated one month of snapshots (284071 samples).
To normalize the data, I subtracted the mean from each snapshot and divided by the standard deviation.
Although I detected small skewness and excess kurtosis, the normalized data were well-modeled by a Gaussian distribution (Figure \ref{fig:snapshot:gaussian}).

\subsection{Sun Pickup}

Optics sidelobes seeing the Sun could create systematic errors.
We found contamination from the Sun when it passed through known sidelobes before we installed the upper ground screen, which eliminated the contamination.
Although we rejected data with such contamination, the rejection may not be perfect and could have an associated systematic error.
To estimate it, we first made a map of CMB data in Sun-centered coordinates \citep{quiet_instrument, yuji_far_sidelobe}.
This map enhanced the Sun signal compared to the normal analysis and used the Sun as a sidelobe-finding source.
The strongest signal in this map was $<1$\,mK.
After data selection, we did not find any signal in the Sun-centered map.
To estimate the possible systematic error, we performed simulations including Sun contamination drawn from the Sun-centered map.
In these simulations we did not reject the contaminated data, so the error estimate was conservative.
We assigned the power-spectra differences between simulations with and without Sun contamination as the possible systematic errors.
We also considered the effect of the Moon.
However, the Moon was an order of magnitude fainter than the Sun, and the Sun contamination was small.
We concluded that possible Moon-pickup systematic errors were negligible.


\subsection{Type-O Glitching}

We found a type of polarized scan-synchronous signal (``Type-O glitching'') whose residual in analysis created a systematic error.
We first discovered this effect as an elevation-dependent polarization offset in large-sky-dip data \citep{leakage_monopole_syst}.
In CMB data we found a corresponding polarized azimuth structure in some CES-diodes.
The signal depended strongly on the azimuth and deck of the observation \citep{typeo_modeling}.
In filtering, we subtracted an arbitrary function of azimuth from each CES-diode's data (\S\ref{sec:filter}).
This filter removed any scan-synchronous signal whose amplitude and azimuth-dependence were constant for the duration of a CES.
Therefore, only scan-synchronous signals whose amplitude or azimuth-dependence changed during a CES could contaminate the result.
We did not detect any variation of the Type-O signal within a CES; however, because of statistical uncertainty we could only place an upper limit on the possible variation.
We ran simulations with the scan-synchronous signal varying within that limit and assigned the resulting spurious polarization as a systematic error \citep{typeo_syst}.

\subsection{Negligible Effects}
In addition to the assigned systematic errors above, we considered many other effects that we determined were negligible.
Here I discuss some examples.
We considered effects from ground signals that remained after filtering.
We considered the impact of compression, the change of the responsivity as a function of input power\footnote{See Appendix~10.1 of \cite{quiet_instrument}.}.
Finally, we considered whether foregrounds contributed significantly to our power spectra.

\label{sec:systematics:ground}

To limit the possibility of significant ground signal, we removed a slope from each half-scan and an azimuth-dependent signal from each CES-diode's data.
Since we limited the azimuth dependence to 40 bins in azimuth, high--spatial-frequency components of a ground signal might remain after the filter.
Alternatively, a ground signal that varied in time and had more than linear structure would remain after the filter.
I discuss these two possibilities below.

To address the finite binning of the ground filter, I re-ran the analysis with 80 filter bins.
This analysis was consistent with the fiducial one.
The differences were $<2$\% of the statistical uncertainty, on average \citep{ground_nbin_comparison}.

To consider the possible time variation of the ground signal, we ran a simulation based on the CES-diode with the worst ground pickup.
Although we did not detect any time variation in the data,
we created a varying ground signal using that CES-diode as a template and allowing the signal to vary within its statistical uncertainty.
We added that signal to the simulation of one CES and computed the resulting power spectra.
We scaled the simulation to the full season based on the number of CESes and the amount of ground signal in each.
The result was negligible compared to the other systematic errors \citep{ground_vary_syst}.


Emission from diffuse Galactic sources or compact extra-Galactic sources could have contaminated the results with non-CMB polarization power.
We limited the diffuse synchrotron foregrounds in our patches to the level of $r=0.05$ from our previous Q-band measurement \citep{quiet_qband_result}.
Lower-frequency \textit{WMAP} observations supported this limit.
It was well below our statistical uncertainty.
We used the \textit{Planck Sky Model} \citep{psm} to predict the thermal dust emission in our patches.
The prediction was below our statistical uncertainty \citep{psm_x_quiet}.
We also cross-correlated the model with the QUIET maps and did not detect significant power in the cross-correlation.

We masked the two most polarized compact sources in our maps (\S\ref{sec:pseudocl}).
To consider the effect of other known sources, I re-ran the analysis masking other sources known from the \textit{WMAP} \citep{wmap7_galaxy} and \textit{Planck} \citep{planck_ercsc} catalogs.
I masked all pixels within $0\fdg5$ of a \textit{WMAP} source.
I masked all pixels within $0\fdg5$ of a source from the \textit{Planck} 100-GHz catalog; I increased the mask radius (by twice the catalog major axis FWHM) if \textit{Planck} found the source was extended.
I did not find a significant difference between this result and the fiducial analysis.
The largest difference was a $C_\ell^{EB}$ shift corresponding to 17\% of the statistical error in patch 7b, the least observed patch \citep{source_mask_comparison}.
Moreover, the statistical significance of this shift was only 2.5\,$\sigma$.
Even if this effect were real, it would only increase the total error by $<1$\%.
This test confirmed that we did not need to mask the next-brightest class of sources; however,
the sensitivity of \textit{WMAP} and \textit{Planck} limited this test.
The flux density limit of the \textit{Planck} catalog was 344\,mJy, approximately 90\,$\mu$K (antenna temperature) for QUIET.
Assuming a conservative polarization fraction of 10\%, \textit{Planck} would find all sources down to a polarization of 9\,$\mu$K.
We used models of polarized compact sources \citep{battye11, tucci_sources} to estimate the contribution of sources below this detection threshold.
It was below the level of $r=0.01$.
\label{sec:systematics:point_sources}

\svnid{$Id: results.tex 151 2012-07-25 22:09:40Z ibuder $}

\chapter{Results and Discussion}
\label{sec:results}
\epigraph{Scientific discovery and scientific knowledge have been achieved only by those who have gone in pursuit of it without any practical purpose whatsoever in view.}{Max Planck}

After finalizing all analysis choices in the blind stage of analysis, we revealed the CMB power spectra.
We reported these results as frequentist confidence intervals.
We confirmed they were consistent with the results of the alternate QUIET analysis.
We then evaluated the agreement of our results with the standard $\Lambda$CDM cosmological model.
Towards the primary science goal of constraining inflation, we set a limit on the tensor-to-scalar ratio $r$ = \rinterval.
Although this limit was not competitive with other results including temperature anisotropies, it had the lowest systematic error of any limit using CMB polarization information alone.

\section{Confidence Intervals}
Using fiducial simulations of varying CMB power levels, we constructed a likelihood function, $\mathcal{L}$, following the method of \cite{hamimeche_lewis}.
\begin{equation}
\label{eq:likelihood}
-2\ln \mathcal{L}(C_b|\hat{C_b})\equiv\sum_{bb'}[g_b(\hat{x_b})C_{f_b}]
[M_f^{-1}]_{bb'}[C_{f_{b'}}g_{b'}(\hat{x_{b'}})],
\end{equation}
where
\begin{eqnarray}
\hat{x_b} &\equiv& \frac{\hat{C_b}+N_b^\textrm{eff}}{C_b+N_b^\textrm{eff}} \\
g_b(x) &\equiv& \textrm{sign}(x-x_m)\sqrt{2\sigma_b^2\ln\left(\frac{x_mS_b(x_m)}{xS_b(x)}\right)} \\
S_b(x) &\equiv& P^M_{\chi^2}(x-1;\nu_b,\sigma_b) \\
\sigma_b &=& \frac{\sqrt{[M_f]_{bb}}}{C_{f_b}+N_b^\textrm{eff}} \\
 P^M_{\chi^2}(x;\nu,\sigma) &\equiv& \frac{\sqrt{2\nu}}{\sigma} P_{\chi^2}\left(
\nu(x\sqrt{2\nu}/\sigma+1);\nu\right)
\end{eqnarray}
$C_{f_b}$ are the bandpowers from fiducial simulations, 
$M_f$ is the covariance matrix of $\hat{C_b}$ from fiducial simulations,
$x_m$ is the value of $x$ that maximizes $xS(x)$,
$P_{\chi^2}(x;\nu)$ is the probability density function for a $\chi^2$ distribution with $\nu$ degrees of freedom,
$N_b^\textrm{eff}$ is an instrument-noise term from simulations, and
$\nu_b$ is an effective number of degrees of freedom we fit from simulations.
(Eq. \ref{eq:likelihood} is sufficient for EE or BB.  For the EB likelihood we used \cite{hamimeche_lewis} Eq. 47.)
We calibrated $\mathcal{L}$ using simulations without any CMB power, with power equal to $\Lambda$CDM, and with non-zero B-mode power \citep{likelihood_model}.
The limited number of simulations introduced an uncertainty of 10\% on the statistical error.
We combined the four patches by multiplying their likelihood functions.

\begin{figure}
\includegraphics[width=0.5\textwidth]{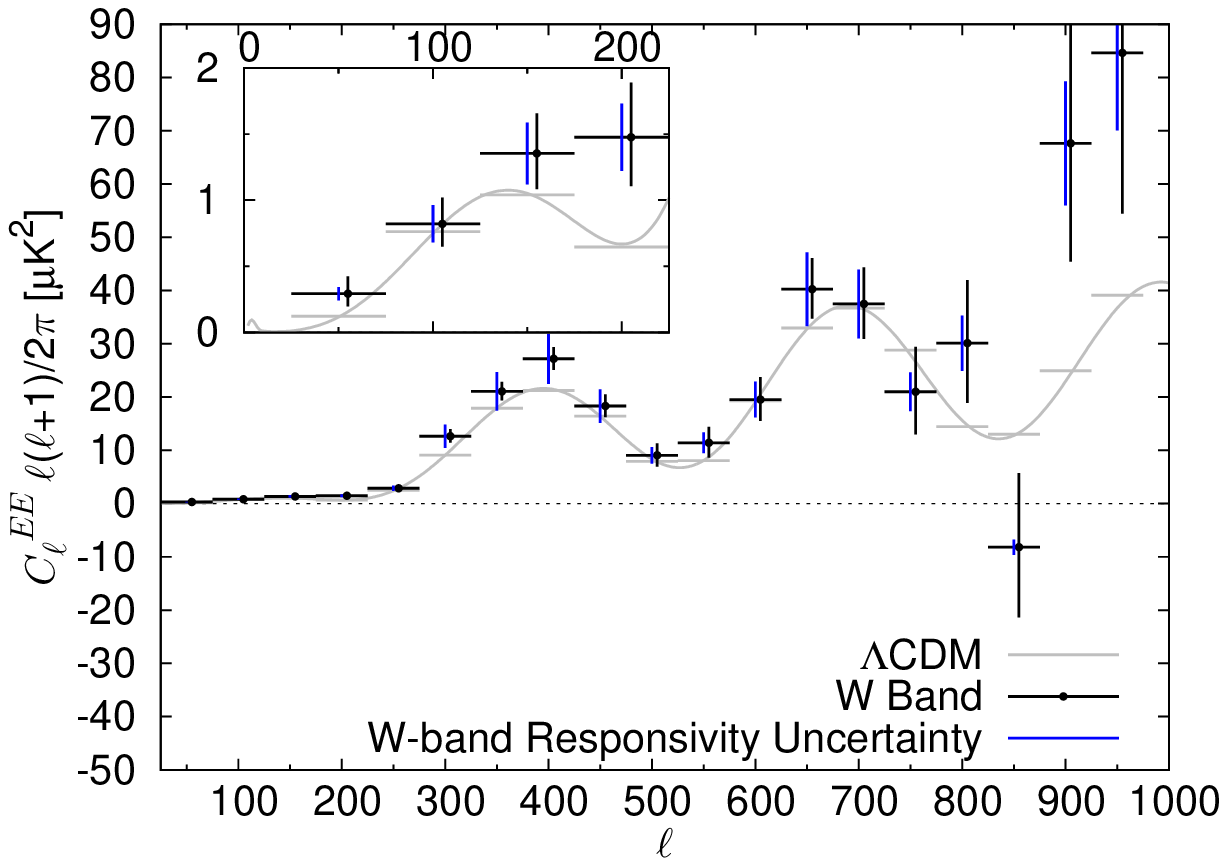}
\includegraphics[width=0.5\textwidth]{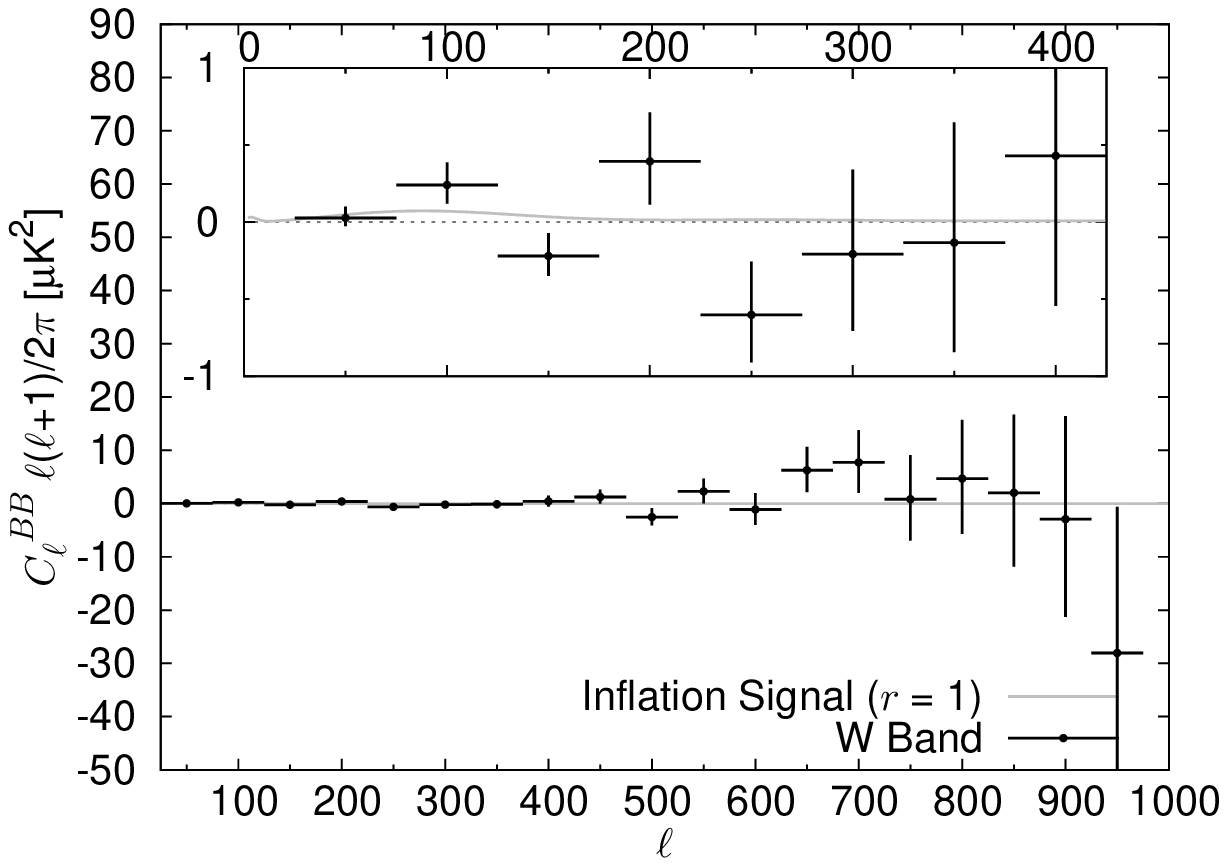}
\includegraphics[width=0.5\textwidth]{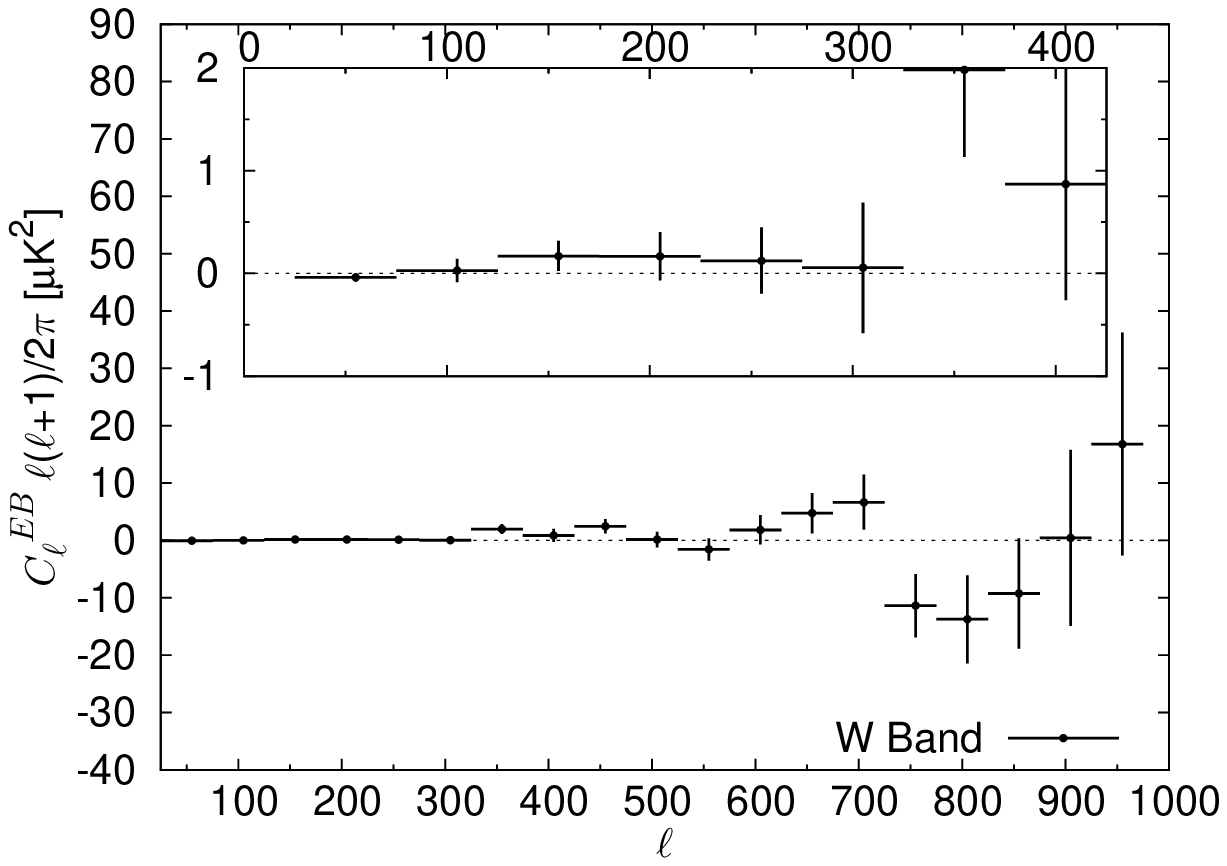}
\caption[Polarization Power Spectra]{\label{fig:cl_results}
We reported the measurements of the CMB power spectra as frequentist two-sided 68\% confidence intervals.
Insets show the low-$\ell$ region.
For EE, the blue error bars show the absolute-responsivity uncertainty, which dominates at some $\ell$.
The horizontal displacement of points is for visualization purposes only.
The horizontal gray lines show the expected $C_b^{EE}$ in the standard model ($\Lambda$CDM).
}
\end{figure}

\begin{figure}
\includegraphics[width=0.5\textwidth]{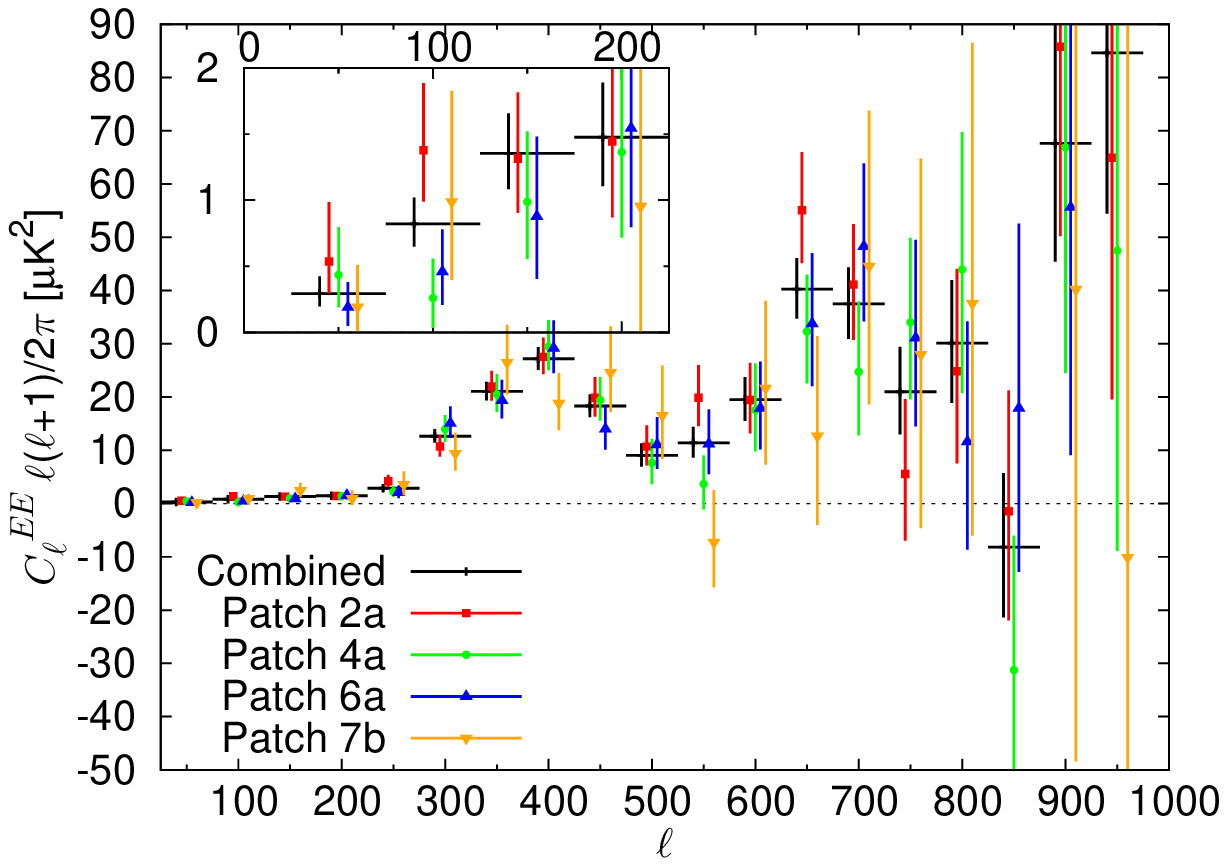}
\includegraphics[width=0.5\textwidth]{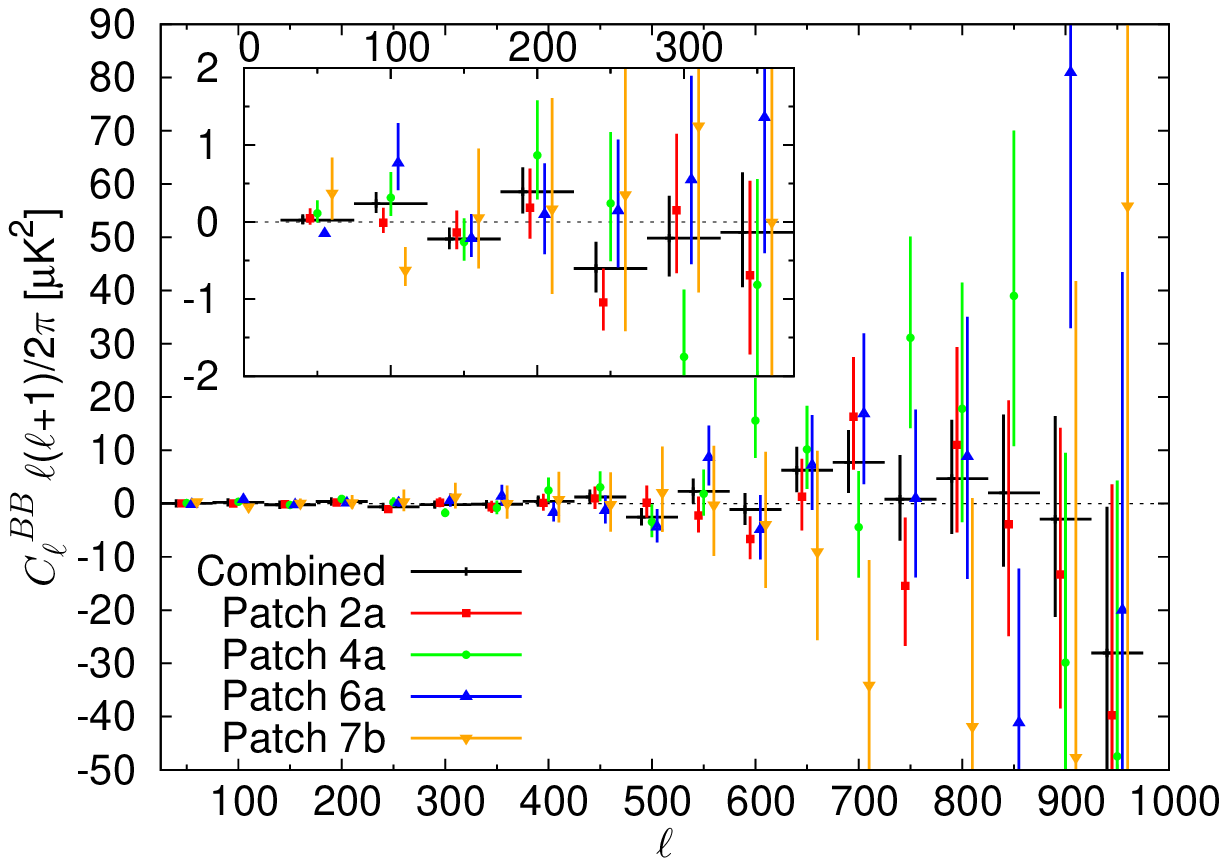}
\includegraphics[width=0.5\textwidth]{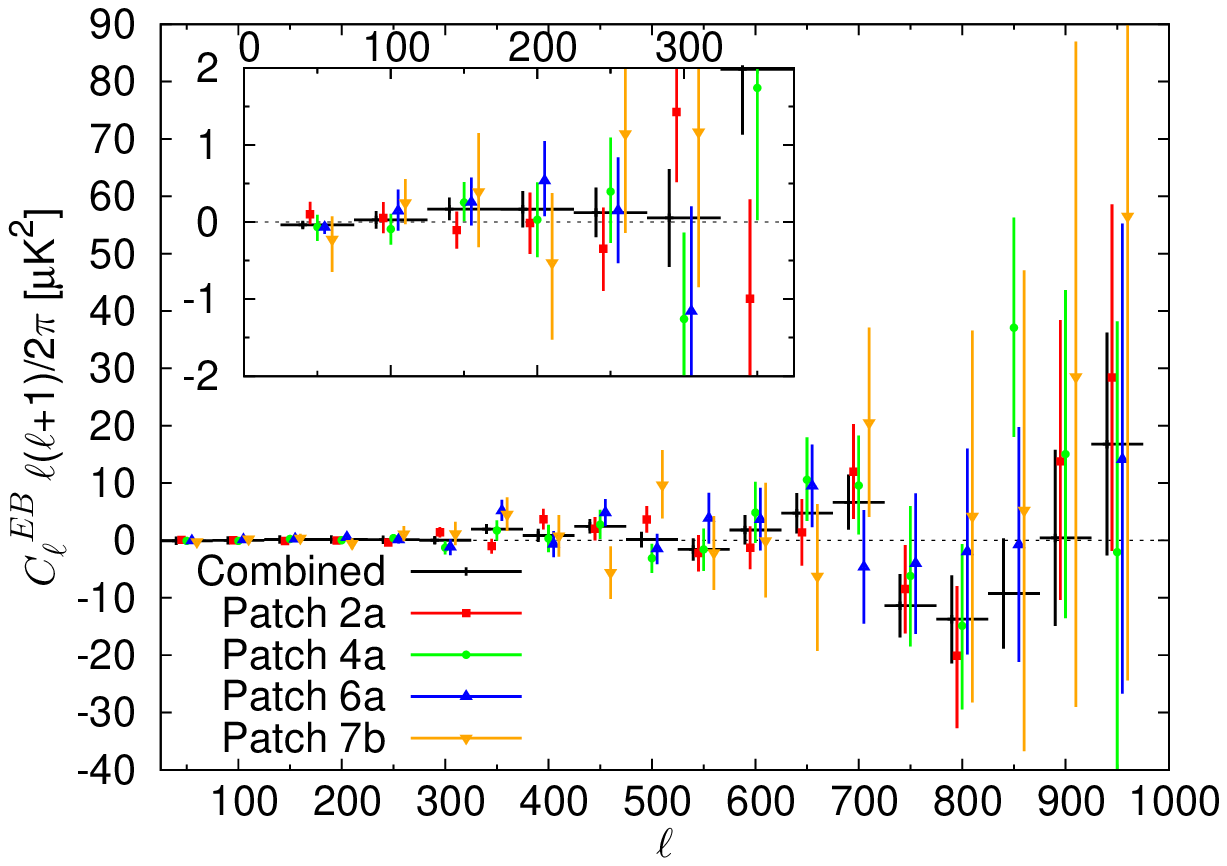}
\caption[Polarization Power Spectra of Each Patch]{\label{fig:cl_each_patch}
The spectra from each patch were consistent with the combined spectra.
The horizontal displacement of points is for visualization purposes only.
}
\end{figure}

\begin{figure}
\includegraphics[width=1.0\textwidth]{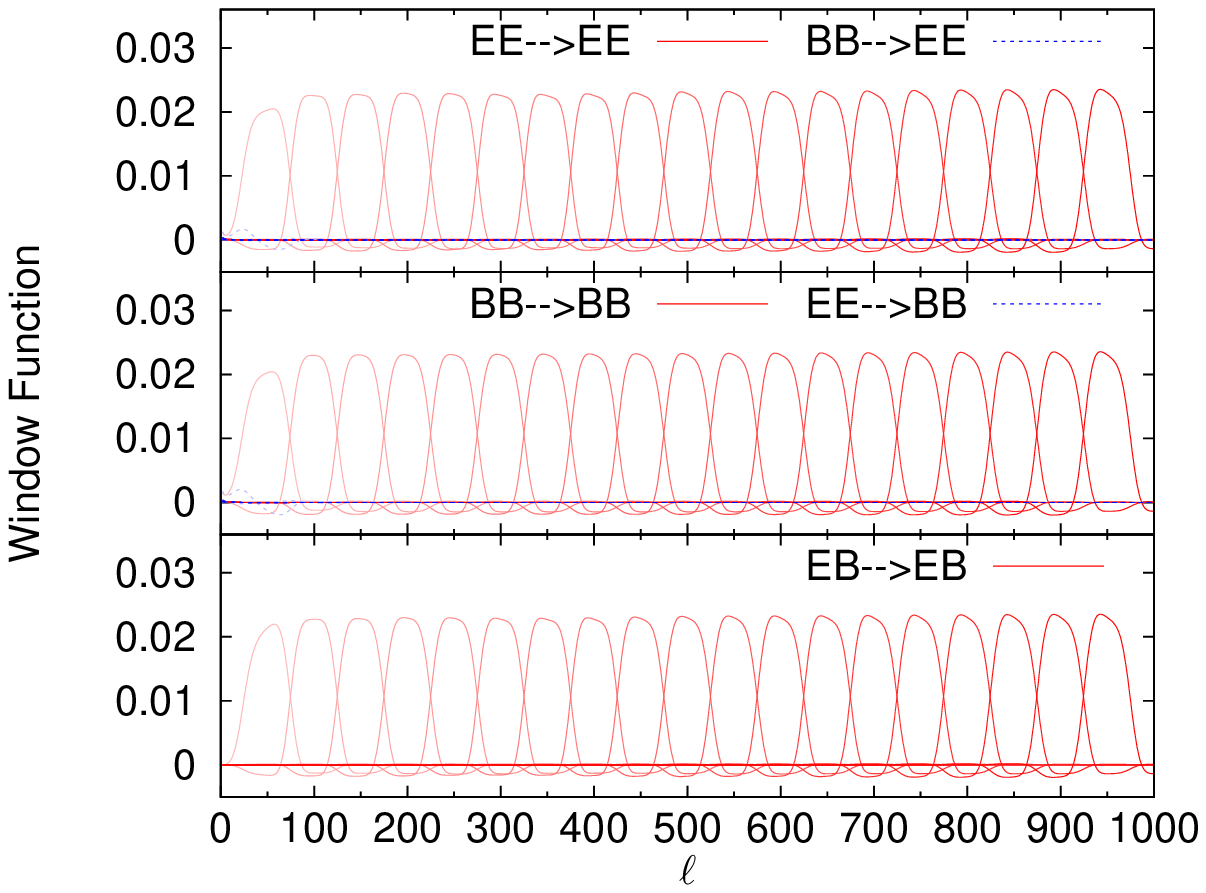}
\caption[Window Functions]{\label{fig:transfer_matrix}
The window functions show the $\ell$ weighting of each bandpower, including the effects of the beam and mode coupling.
Each bandpower had some response in the neighboring band, resulting in 10\% correlation between neighboring $\ell$ bins.
The response of our $C_\ell^{EE}$ estimator to $C_\ell^{BB}$ was small, and vice versa.
I averaged the window functions for all four patches.
}
\end{figure}

From this function we constructed frequentist two-sided 68\% confidence intervals (Figure \ref{fig:cl_results}).
We used the $C_b$ that maximize the likelihood as the central values.
We chose the confidence intervals such that their boundaries are the points where $-2\ln \mathcal{L}$ increases by 1 from its minimum.
When computing the confidence interval for any one $b$, we allowed all $b'\not= b$ to vary to maximize $\mathcal{L}$.
When computing patch-combined intervals for one spectrum (e.g. EE) we did not force the other spectra (e.g. BB and EB) to be the same in all patches.
The goodness of fit for the patch-combined central values was $\chi^2_\textrm{min} \equiv -2\ln \mathcal{L}_\textrm{min} = $ 48.3 (EE), 77.0 (BB), and 64.4 (EB) for $19\times 3 =57$ degrees of freedom.
The corresponding PTEs were 79\%, 4\%, and 23\%, showing that the individual patch results were consistent with each other (Figure \ref{fig:cl_each_patch}).
Because of the limited sky coverage, neighboring $\ell$ bins had $\approx-0.1$ correlation.
Bins that were further apart were effectively uncorrelated.
There were also non-negligible correlations among EE, BB, and EB in the same bin.
The ``window function'' $K^{-1}_{bb'}P_{b'\ell}M_{\ell\ell'}B_{\ell'}^2$ quantifies the dependence on $C_\ell$ of each bandpower (Figure \ref{fig:transfer_matrix}).

\section{Comparison with Alternate Analysis Pipeline}
\label{sec:pipeline_comparison}
\begin{figure}
\includegraphics[width=0.5\textwidth]{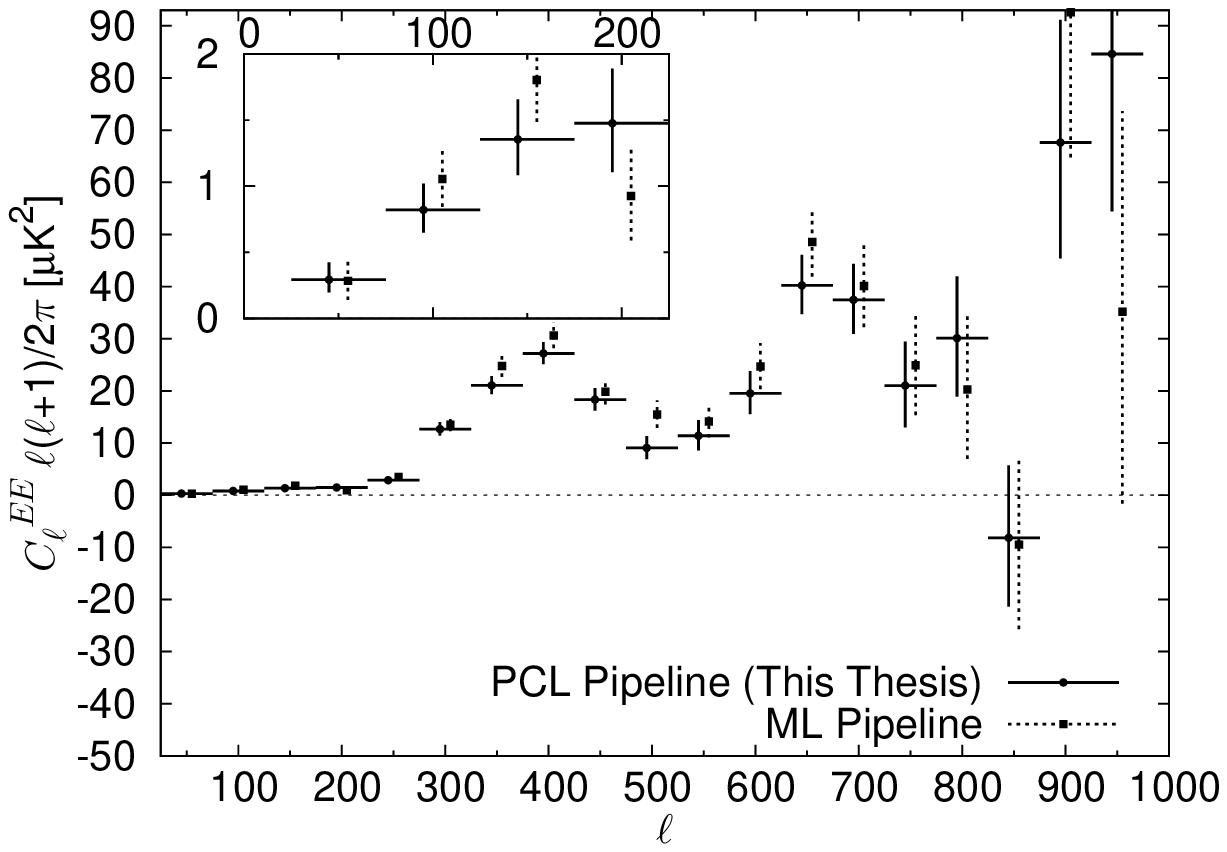}
\includegraphics[width=0.5\textwidth]{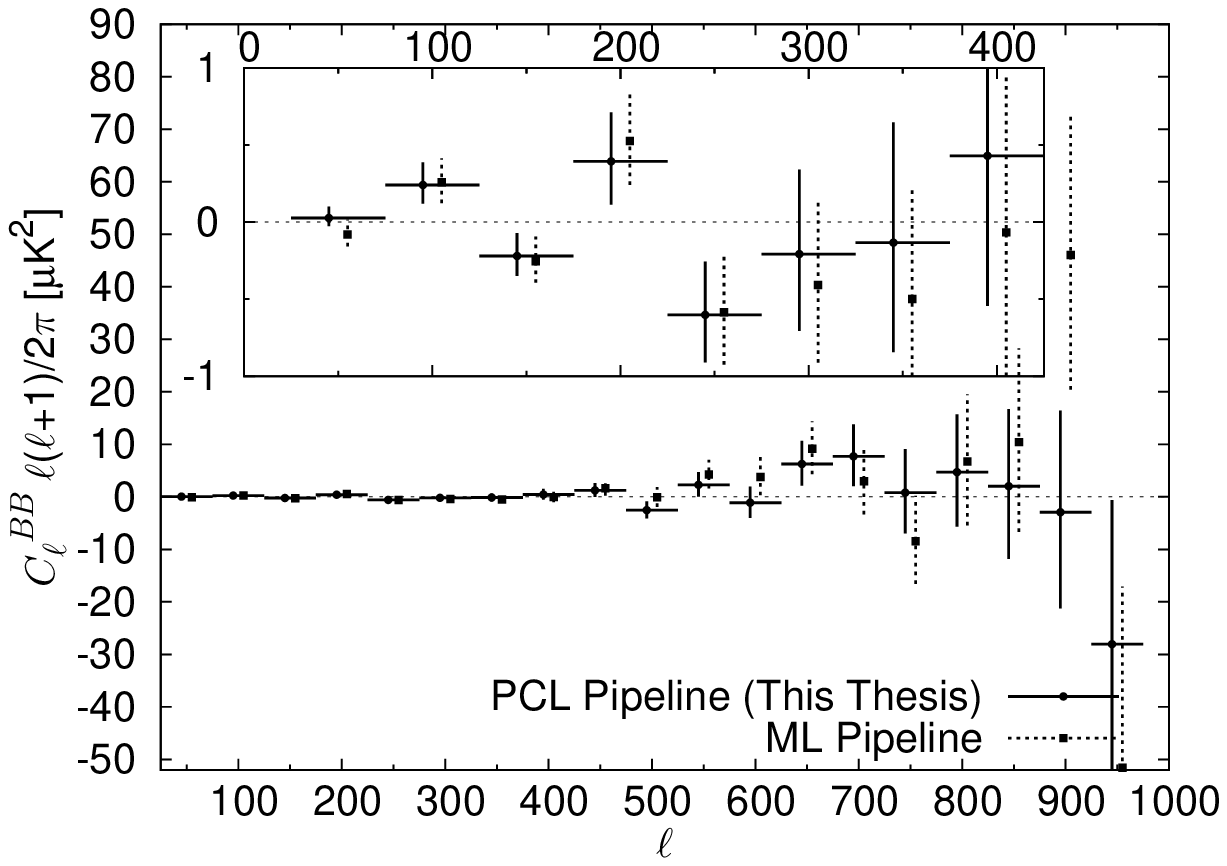}
\includegraphics[width=0.5\textwidth]{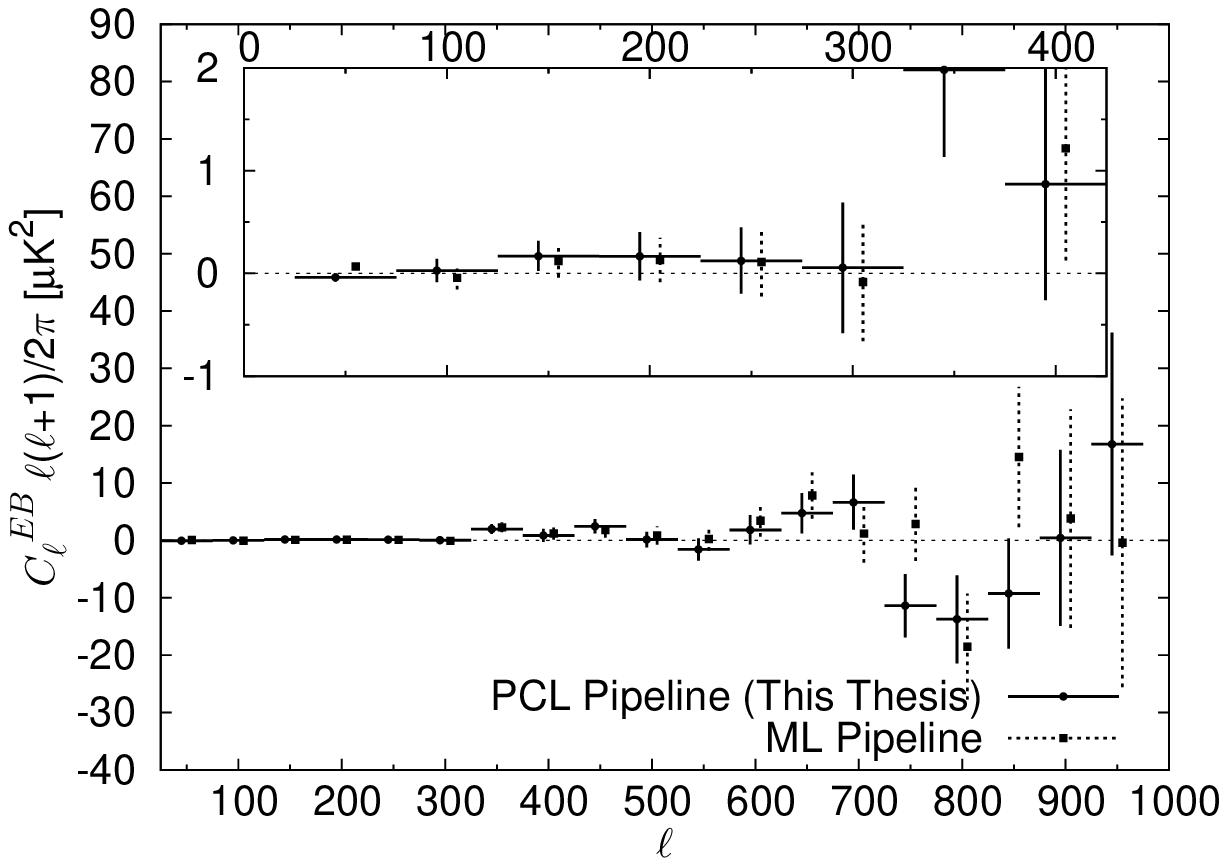}
\caption[Power Spectra from Two Pipelines]{\label{fig:cl_both_pipelines}
We computed the CMB power spectra with two parallel analyses; they found consistent results.
The alternate analysis (ML Pipeline) used a different responsivity model, resulting in 7\% higher absolute responsivity for the power spectra.
The horizontal displacement of points is for visualization purposes only.
}
\end{figure}

As a final check before interpreting the results, I compared the results of the two parallel QUIET analyses
(Figure \ref{fig:cl_both_pipelines}).
I computed the $\chi^2$ for the hypothesis that the two analyses were measurements of the same spectra with independent statistical fluctuations.
The total $\chi^2$ was 8.4, 6.5, and 9.1 for EE, BB, and EB, respectively, each with 19 degrees of freedom.
The corresponding PTEs were 98\%, 100\%, and 97\% indicating that our two analyses of the same data were much more consistent than independent analyses of disjoint data sets.
In this comparison I adjusted the results of the alternate analysis to have the same absolute responsivity\footnote{In the blind stage of analysis, we found that the pipelines' absolute-responsivity calibration differed by 3.5\% (7\% in $C_\ell$).
}.
Without making that adjustment, the PTEs were 89\%, 100\%, and 97\%.
Furthermore, both pipelines found consistent results for all parameter estimates.

\section{Consistency with $\Lambda$CDM}
After confirming that our analysis was internally consistent, we compared our results to the standard cosmological model ($\Lambda$CDM with parameters from \cite{wmap7_cosmology}).
The $\chi^2$ to $\Lambda$CDM was 24.3, 21.5, and 21.5 for EE, BB, and EB, respectively. 
There were 19 degrees of freedom for each spectrum.
The corresponding PTEs were 19\%, 31\%, and 31\%.
Because the responsivity systematic error was significant, we allowed the absolute responsivity to vary when calculating this consistency.
We used our fiducial responsivity as a constraint, assuming it had a Gaussian distribution with RMS equal to the systematic error we assigned.
If we instead allowed the responsivity to be a free parameter, the PTE for EE was 20\%.
Similarly, we included the dominant systematic error due to detector-angle miscalibration when calculating the EB consistency. 
We found no evidence for deviation from $\Lambda$CDM or for non-zero BB or EB power.

\section{Limit on Inflation}
We constrained inflationary models using only the B-mode measurements at $25 \leq\ell\leq 175$ (the first three bandpowers).
We fit a B-mode template spectrum with variable $r$ to these bandpowers.
For simplicity, we fixed $n_t=0$ in the fit; that choice made the template directly proportional to $r$.
We found $r= \rinterval$ as a 68\% confidence interval, corresponding to $r<\rupper$ as a 95\% confidence upper limit.
We calculated the upper limit to be the point where $-2\ln\mathcal{L}$ increased by 2.69 from its minimum.
Statistically, this measurement did not improve on the existing limit $r < 0.2$ \citep{wmap7_cosmology}.
However, our limit used CMB B-mode polarization alone with smaller systematic error than ever before.

\section{Conclusions}

\begin{figure}
\includegraphics[width=0.75\textwidth]{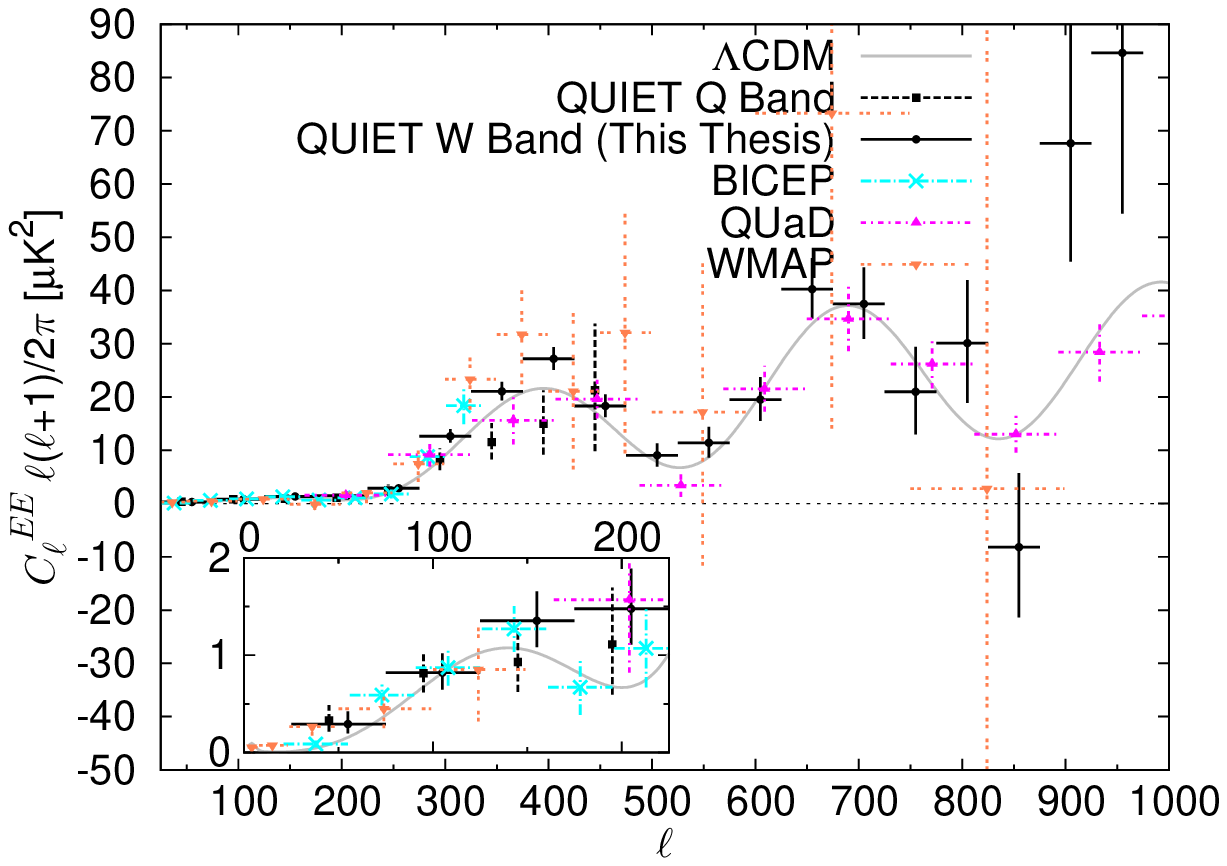}
\includegraphics[width=0.75\textwidth]{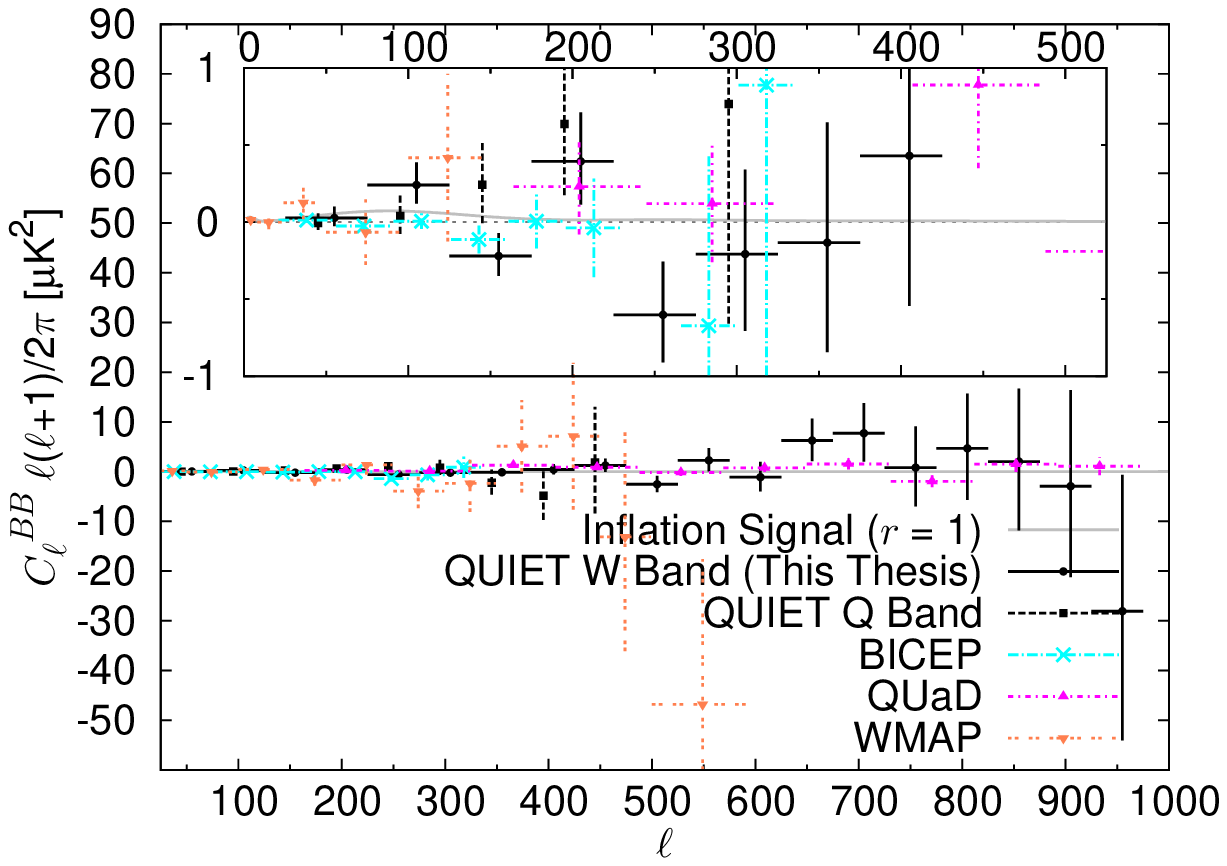}
\caption[Power Spectra Compared to Previous Experiments]{\label{fig:cl_experiment_comparison}
We made the most sensitive measurement of the CMB polarization between $300\lesssim\ell\lesssim500$.
I only show statistical errors.
QUIET Q and W bands share horizontal error bars; the horizontal displacement of points is for visualization purposes only.
}
\end{figure}

We measured the CMB polarization with smaller errors than ever before.
Our $C_\ell^{EE}$ measurement was the most sensitive so far between $300\lesssim\ell\lesssim500$ (Figure \ref{fig:cl_experiment_comparison}).
We placed new limits on $C_\ell^{BB}$ and  $C_\ell^{EB}$.
The most important result from QUIET was a new, low--systematic-error measurement of the tensor-to-scalar ratio $r$ from $C_\ell^{BB}$ alone.
The statistical power of this measurement was insufficient to improve on existing constraints.
However, using a combination of techniques (especially double demodulation, a side-fed Dragonian telescope, an absorbing ground screen, deck rotation, cross-correlation, and blind analysis) we reduced systematic errors by more than an order of magnitude to below the level of $r=0.01$.
Already, new projects are employing these techniques with even more sensitive polarimeter arrays to search for new evidence of inflation in the CMB.

\appendix
\svnid{$Id: status_flags.tex 110 2012-05-09 23:16:27Z ibuder $}

\chapter{\process{DataCompilation} Status Flags}
\label{sec:DataCompilationStatusFlags}

This appendix lists the meaning of the Online Software (\process{DataCompilation}) status flags.
\begin{itemize}
\item DATA\_STATUS\_UNEVEN\_DEMOD means that the number of samples was not the same in the two 4-kHz phase-switch states.  Such data are useless.
\item DATA\_STATUS\_NO\_ADC\_CONN means that \process{DataCompilation} cannot connect to \process{adc\_server}.  Data will not be collected until connectivity is restored.
\item DATA\_STATUS\_NO\_PER\_CONN means that \process{DataCompilation} cannot connect to \process{PeripheralServer}.  Some housekeeping data will be missing until connectivity is restored.
\item DATA\_STATUS\_NO\_BIAS\_CONN means that \process{DataCompilation} cannot connect to \process{BiasServer}.  Biasing data will be missing until connectivity is restored.
\item DATA\_STATUS\_NO\_AGG\_CONN meant that \process{DataCompilation} cannot connect to \process{DataArc}.  Under most circumstances this is of no concern because we only used \process{DataArc} in special circumstances.
\item DATA\_STATUS\_NO\_DB\_CONN means  that \process{DataCompilation} could not read the Receiver Database.  If this flag remains on it may indicate the database is corrupt, missing, etc.
\item DATA\_STATUS\_BUFFER\_FULL means that the \process{DataCompilation} data buffer became full and data were discarded.
\item DATA\_STATUS\_INT\_ERR indicates a critical error in \process{DataCompilation} requiring a restart of the Online Software.
\item DATA\_STATUS\_UNEVEN\_MASK was supposed to indicate that the blanking mask was not the same on all ADC Boards; however, due to firmware limitations there are too many false alarms for this flag to be useful.
\item DATA\_STATUS\_UNEVEN\_CLOCKS was supposed to indicate that clocks were not the same on all ADC Boards; however, due to firmware limitations there
are too many false alarms for this flag to be useful.
\item DATA\_STATUS\_BAD\_SCALE indicates the scale factor (number of 800-kHz samples accumulated in each 100-Hz sample) is impossible.  Usually this means the firmware is not working.
\item DATA\_STATUS\_HK\_PARSE means \process{DataCompilation} could not parse the housekeeping data from \process{adc\_server}.  Data were lost.
\item DATA\_STATUS\_DEMOD\_PARSE means that \process{DataCompilation} could not parse the radiometer data from \process{adc\_server}.  Data were lost.
\item DATA\_STATUS\_SNAP\_PARSE means that \process{DataCompilation} could not parse snapshot data from \process{adc\_server}.  At least 1 snapshot was lost.
\item DATA\_STATUS\_SERVER\_THREAD indicates an unrecoverable error in the \\\process{DataCompilation} \thread{data\_server} thread.  The Online Software should be restarted.
\item DATA\_STATUS\_HK\_THREAD indicates an unrecoverable error in the \\\process{DataCompilation} \thread{hk} thread.  The Online Software should be restarted.
\item DATA\_STATUS\_SERVE\_THREAD indicates an unrecoverable error in the \\\process{DataCompilation} \thread{data\_serve} thread.  The Online Software should be restarted.
\item DATA\_STATUS\_DEMOD\_THREAD indicates an unrecoverable error in the \\\process{DataCompilation} \thread{demod} thread.  The Online Software should be restarted.
\item DATA\_STATUS\_INVALID\_WPID\_VALUE indicates an invalid value received from the WPID.
\item DATA\_STATUS\_INVALID\_CPID\_VALUE indicates an invalid value received from the CPID.
\item DATA\_STATUS\_INVALID\_PRESSURE\_VALUE indicates invalid data from the cryostat pressure sensor.
\item DATA\_STATUS\_INVALID\_TEMP\_VALUE was supposed to indicate invalid data from the thermometers on the ground screen.  However, there were too many false alarms.
\item DATA\_STATUS\_NO\_WPID\_CONN indicated there was no connection to the WPID.  Usually, the Electronics Enclosure temperature regulation also failed.
\item DATA\_STATUS\_NO\_CPID\_CONN indicated there was no connection to the CPID.  Usually, the cryostat temperature regulation also failed.
\item DATA\_STATUS\_NO\_PRESSURE\_CONN indicated there was no connection to the cryostat pressure sensor.  Data were missing.
\item DATA\_STATUS\_NO\_TEMP\_CONN was supposed to indicate connection failure to the ground-screen thermometers.  However, there were too many false alarms.
\item DATA\_STATUS\_CRATE\_PROBLEM indicated a problem with the ADC Backplane.
\item DATA\_STATUS\_WPID\_ALARM\_1 was linked to the WPID ``alarm'' function, which we did not use.
\item DATA\_STATUS\_WPID\_ALARM\_2 was linked to the WPID ``alarm'' function, which we did not use.
\item DATA\_STATUS\_AGG\_THREAD indicated an unrecoverable error in the \\\process{DataCompilation} \thread{archive\_aggregator} thread.  The Online Software should be\\ restarted.
\item ADC\_STATUS\_EXT\_CLOCK\_BAD indicated that the Master ADC could not synchronize with the 10-MHz clock.  Data taken in this situation were not properly synchronized.
\item ADC\_STATUS\_NO\_EXT\_1HZ indicated that the Master ADC was not using the 1-Hz clock for timing synchronization.  Data taken in this situation were not properly synchronized.
\item ADC\_STATUS\_NOT\_TFP indicated that \process{adc\_server} could not provide timestamps from the Time-code Reader.  Such data do not have reliable timing synchronization.
\item ADC\_STATUS\_4HZ\_COUNT\_MATCH indicated that the 4-Hz frame counter was not the same for all ADC Boards.  
Either firmware or \process{adc\_server} problems could cause this flag.
\item TFP\_STATUS\_NOLOCK indicated that the Time-code Reader could not synchronize to the IRIG-B time code.
The data were not properly synchronized.
\item TFP\_STATUS\_BATTFAIL indicated that the Time-code Reader battery failed.
\item TFP\_STATUS\_TIMEOFF indicated that the Time-code Reader reported a time offset.  This created too many false alarms to be useful.
\item TFP\_STATUS\_FREQOFF indicated that the Time-code Reader reported a frequency offset.  This created too many false alarms to be useful.
\item BSERV\_STATUS\_BAD\_COMMAND indicated that \process{BiasServer} could not send a command because it was invalid.
\item BSERV\_STATUS\_NO\_ADC\_CONN indicated that \process{BiasServer} could not connect to \process{adc\_server}.
\item BSERV\_STATUS\_BAD\_ADC\_COMM indicated a communication failure between\\  \process{BiasServer} and \process{adc\_server}.
The most recent bias command failed.
\item BSERV\_STATUS\_NO\_ADC\_REPLY indicated that \process{adc\_server} never indicated success for a `b' request (Appendix~\ref{sec:adc_server_bias_thread_actions}).
The bias command may have failed.
\item BSERV\_STATUS\_ADC\_MINOR\_ERR indicated strange communication with \\\process{adc\_server} that usually does not cause bias commands to fail.
\item BSERV\_STATUS\_NO\_BIAS\_FRAME indicated that \process{BiasServer} did not have memory to store the bias data.
\item BSERV\_STATUS\_MUTEX\_ERROR indicated that \process{BiasServer} needed to be restarted.
\item BSERV\_STATUS\_NO\_RC\_CONN indicated that \process{BiasServer} could not connect to \process{ReceiverControl}.
\item BSERV\_STATUS\_BAD\_RC\_COMM indicated communication failure between \\\process{BiasServer} and \process{ReceiverControl}.
\item BSERV\_STATUS\_BAD\_DC\_COMM indicated communication failure between \\\process{BiasServer} and \process{DataCompilation}.
\item BSERV\_STATUS\_DATA\_THREAD indicated an unrecoverable error in \\\process{BiasServer} requiring restart.
\item BSERV\_STATUS\_COMMAND\_THREAD indicated an unrecoverable error in \\\process{BiasServer} requiring restart.
\item PERSERV\_STATUS\_COMMAND\_THREAD indicated an unrecoverable error in \\\process{PeripheralServer} requiring restart.
\item PERSERV\_STATUS\_BAD\_DC\_COMM indicated nothing.
\item PERSERV\_STATUS\_DATA\_THREAD indicated an unrecoverable error in \\\process{PeripheralServer} requiring restart.
\item PERSERV\_STATUS\_MUTEX\_ERROR indicated an unrecoverable error in \\\process{PeripheralServer} requiring restart.
\item PERSERV\_STATUS\_NO\_PER\_FRAME indicated an unrecoverable error in \\\process{PeripheralServer} requiring restart.
\item DATA\_STATUS\_DARC\_THREAD indicated an unrecoverable error in \\\process{DataCompilation} \thread{data\_arc\_service} thread requiring restart.
\item DATA\_STATUS\_DARC\_BUFFER\_FULL indicated that the buffer of frames to send to \process{DataArc} became full.  During normal  operation this flag could be ignored.
\item DATA\_STATUS\_CBI\_BUFFER\_FULL indicated that the buffer of frames to send to \process{cbicontrol} became full.  Data were lost.
\item DATA\_STATUS\_CBI\_THREAD indicated an unrecoverable error in \process{DataCompilation} \thread{cbi\_service} thread requiring restart.
\item DATA\_STATUS\_SS\_ERROR indicated \process{DataCompilation} encountered an error sending data to \process{cbicontrol}.
\item DATA\_STATUS\_NULL\_FRAME indicated that \\\process{DataCompilation} inserted a null frame for timing alignment.
\item DATA\_STATUS\_BAD\_TIMING indicated that the data were not properly aligned.
\item DATA\_STATUS\_BAD\_QUAD\_SCALE indicated that the scale factor for quadrature data was impossible.
\item DATA\_STATUS\_WPID\_THREAD indicated an error in \\\process{DataCompilation} \thread{wpid\_output} thread.
The WPID output monitoring data were not recorded properly.
\item DATA\_STATUS\_NO\_PDU indicated that \process{DataCompilation} could not receive data from the Electronics Enclosure smart power strip.
\item DATA\_STATUS\_NOISE\_SOURCE\_ON indicated that the (calibration) noise source was on.
\item BSERV\_STATUS\_BAD\_PS\_STATE indicated that \process{adc\_server} sent the wrong phase-switch state information to \process{BiasServer}.
Resetting the phase-switch state usually fixed this problem.

\end{itemize}

\svnid{$Id: online_software_details.tex 138 2012-06-25 21:58:54Z ibuder $}

\chapter{Online Software Details}

\section{\process{adc\_server} Packet Status Bytes}
\label{sec:adc_server_status_bytes}
This section lists the contents of the 16 bytes of status information \process{adc\_server} sent to \process{DataCompilation} in every radiometer and housekeeping packet.
\begin{enumerate}
\item The frame counter, a 4-byte integer.
\item A flag (1 byte) indicating whether the Time-code Reader battery failed.
\item A flag indicating whether the Time-code Reader is locked to the time code.
\item A flag indicating whether the Time-code Reader has a time offset.
\item A flag indicating whether the Time-code Reader has a frequency offset.
\item A flag indicating whether the timestamp came from the Time-code Reader.  In case the Time-code Reader fails, the timestamp will be taken from the SBC system time.  In this case the data are not used.
\item A flag indicating whether all ADC Boards had the same frame counter.
\item A flag indicating whether the Master ADC Board internal clock is working properly.  This flag is not relevant in normal operation because an external clock is used instead.
\item A flag indicating whether the Master ADC Board is receiving an external 10-MHz clock.
\item A flag indicating whether the Master ADC Board is receiving an external 1-Hz clock.
\item Three bytes of padding, which contain no information, to make the total length a multiple of 4 bytes.
\end{enumerate}

\section{\process{adc\_sever} \thread{bias} Thread Actions}
\label{sec:adc_server_bias_thread_actions}
This section details the actions of the \process{adc\_sever} \thread{bias} thread, which depend on the request character received from \process{BiasServer}.

\begin{enumerate}
\item If the request character was `q,' \process{adc\_server} closed.

\item If the request character was `f,' the thread sent the phase switching state information (4 bytes) immediately after reading it from VME address 0x10 on the Master ADC Board.

\item If the request character was `M,' the thread sent the current bias bit values 
(2 bytes each for 16384 bias addresses\footnote{Not all addresses corresponded to a physical bias setting.  
There was no meaning to the values of the non-physical addresses.}) 
immediately after reading them from VME address 0xC00000 on the Master ADC Board.

\item If the request character was `m,' the thread set the demodulation mask to blank the phase-switch transitions.                     
The thread received the mask information (4 bytes) from BiasServer.    
The first two bytes specified the mask offset (typical value $-1$), and the second two specified the mask length (typical value 14).            
From this information, the thread constructed the 200-bit mask to send to the ADC Boards.                                                             
Each bit indicated whether an 800-kHz sample was included in averaging to 4 kHz.   A 1 indicated the corresponding sample was included, and a 0 indicated it was discarded.

The thread also constructed the quadrature mask, which indicated which sign to assign to each 800-kHz sample when constructing quadrature data.            
   The quadrature mask was $90\degr$ out of phase with the demodulation.   
 However, it must take into account the samples lost due to blanking.    
The offset of the quadrature mask (i.e. the location where the quadrature sign changed) was the offset of the demodulation mask plus half the demodulation mask length (because a longer demodulation mask shifted the center of each demodulation phase accounting for blanking) plus $1/4$ of the total cycle length (to make it $90\degr$ out of phase)\footnote{An final offset of $-1$ was applied because the firmware applied a corresponding $+1$ offset.}.                
The length of the quadrature mask (i.e. the length of each quadrature phase) was 100, half the total cycle.
The thread wrote the demodulation mask to VME address 0x20000 on each ADC Board and the quadrature mask to address 0x60000 on each ADC Board.

\item If the request character was `z,' then the thread sent the current demodulation mask offset and length (8 bytes total).
These values were stored in \process{adc\_server} memory and were not read from the ADC Boards.
The \process{adc\_server} set the mask on startup, and there was no other way to change the mask so the ADC Board value was always reflected in \process{adc\_server} memory.

\item \label{command:x} If the request character was `x,' then the thread changed the set of housekeeping addresses to multiplex.
First the thread received the number of addresses (4 bytes).
Then the thread received the address values (2 bytes per address).
Finally the thread wrote the address length to VME address 0xC on the Master ADC and the multiplex addresses to VME address 0x40000.

\item If the request character was `l,' then the thread prepared a bias DAC command.
It received the card, address, and DAC numbers; and the bias value (4 bytes total).
It then wrote them to VME address 0x18 on the Master ADC.
Because these commands often occurred in sequence, the thread paused after executing one in case the \thread{fourhz\_write} thread needed to communicate with the ADC Boards.
The pause ensured that continued bias commands did not interrupt data recording. 

\item If the request character was `b', then the thread instructed the Master ADC Board to send the bias DAC commands (prepared above) to the bias boards.
The thread received a mask (8 bytes total) indicating which cards, addresses, and DACs should be updated.
The first bit was unused
The next 21 bits (corresponding to 21 card locations in the Bias-Board Backplane) indicated whether each card should be updated.
Bits 24--63 indicated which addresses and DACs should be updated\footnote{These masks did not work properly.  I recommend that the masks always be set to update all addresses and DACs.  Failure to do so usually resulted in the bias boards receiving incorrect settings.}.
The thread wrote the mask information to VME addresses 0x1C (first 4 bytes) and 0x14 (second 4 bytes) on the Master ADC Board.
The thread then waited for the Master ADC Board to complete the bias send.
It indicated completion by setting the 8th bit ($2^7$ place) of the word at 0x0.
The thread sent `q' to BiasServer to indicate success.

\item If the request character was `p,' then the thread commanded the Master ADC Board to set the phase-switch state.
It received the setting (4 bytes) from the network and wrote it to VME address 0x10 on the Master ADC.

\item If the request character was `k,' then the thread commanded the housekeeping multiplexing mode.
It received 4 bytes from the network.
If the first bit was 0, then the housekeeping data  multiplexed addresses (see \ref{command:x}).
If the first bit was 1, then the housekeeping data recorded only a single address.
The address was specified by the next 12 bits.
In either case, the command was written to VME address 0x08 on the Master ADC.
\end{enumerate}

\section{\thread{data\_serve} Requests}
\label{sec:data_serve_requests}
This section lists the actions of the \process{DataCompilation} \thread{data\_serve} thread, which depended on the request received.
Table \ref{tab:DataCompilation:DataInterfaceServer:requests} summarizes these requests.

\begin{enumerate}
\item If the request was DATA\_STATUS, the thread sent the \process{DataCompilation} status flags (20 bytes, Appendix~\ref{sec:DataCompilationStatusFlags}).

\item If the request was DATA\_NADC, the thread sent the number of ADC Boards collecting data (\S\ref{sec:adc_server}).

\item If the request was DATA\_DEMOD, the thread sent the latest radiometer data frame.
The frame included the raw DE, TP, and quadrature data (after correcting the sign bit as described in \S\ref{sec:datacompilation}), as well as the metadata: the number of 800-kHz samples accumulated into each 100-Hz sample, the 50-Hz and 1-Hz clock states, and firmware and \process{adc\_server} status flags.

\item If the request was DATA\_SNAP, the thread sent the latest snapshot.

\item If the request was DATA\_HK, the thread sent the latest housekeeping data frame.

\item If the request was DATA\_BIAS, the thread sent the biasing status.
This included the bias board DAC bit values, the phase-switching state, and the demodulation mask.

\item If the request was DATA\_PER, the thread sent the latest monitoring data from \\\process{PeripheralServer}.
These included WPID settings, ADC Backplane monitoring data, CPID settings, cryostat pressure, and warm temperature sensing.

\item If the request was DATA\_NEWSNAP, the thread signaled the \thread{snap} thread with a semaphore to cause it to acquire a new snapshot.
The \thread{data\_serve} thread then waited for the \thread{snap} thread to respond by broadcasting a new snapshot condition \citep{pthread_cond}.
The thread then sent the request result (4 bytes\footnote{Either an acknowledgment that the snapshot was updated or an error code indicating the reason for failure.  See DataInterfaceClient.h in the RCS repository for the possible error codes.}) and the identification number of the new snapshot (4 bytes).

\item If the request was DATA\_ENCLOSURE, the thread sent the Electronics Enclosure smart power strip monitoring data.
These included temperature, humidity, whether power output was enabled, and uptime.

\item If the request was DATA\_PID\_OUTPUT1 (2), the thread sent the status of the first (second) WPID output.
The status was the fraction of the second during which the output was active i.e. driving the heater (fan).
\end{enumerate}

\section{\process{DataCompilation}--\process{DataArc} Communication}
\label{sec:DataCompilationDataArc_communication}
This section details the actions of \process{DataCompilation} when the \thread{data\_arc\_service} thread received a message from \process{DataArc}.
Table \ref{tab:DAQ:AggInterface} summarizes the possible messages.

\begin{enumerate}
\item If the message was AGG\_ACK, then \process{DataCompilation} received the frame number (4 bytes) from \process{DataArc}.
If the frame number matched the current frame the thread sent, then the thread returned to waiting for a frame.
By checking the frame number received by \process{DataArc}, \process{DataCompilation} never discarded data unless they were properly stored by \process{DataArc}.

\item If the message was AGG\_FLUSH\_BUFFER, then the thread removed all frames from the queue without sending them.
This capability allowed \process{DataArc} to start a new data file with only new data\footnote{The \process{DataCompilation} queue could contain arbitrarily old data.}.

\item If the message was AGG\_QUIT, \process{DataCompilation} closed the connection.

\item If the message was AGG\_NACK, then \process{DataCompilation} received the frame number from \process{DataArc}.
If the frame number was the same as the current frame, then the thread sent the
frame.

\item In all other cases (including if there was no message after 50\,ms), \process{DataCompilation} sent AGG\_ACK\_REQ followed by the current frame number.
\end{enumerate}

\section{\process{ReceiverControl} Commands}
Table \ref{tab:ReceiverControl:commands} lists the commands \process{ReceiverControl} could process and the action it took for each.

\svnid{$Id: receiver_control_commands.tex 114 2012-05-11 22:58:23Z ibuder $}

\renewcommand{\thefootnote}{\alph{footnote}}
\begin{center}
\begin{singlespace}

\begin{ThreePartTable}
\begin{TableNotes}
\item[a] Q-band modules require additional arguments \textit{Current3} and \textit{Voltage3}.
\end{TableNotes}

\begin{longtable}{lp{1.5in}p{3in}}
\caption{\process{ReceiverControl} Commands
\label{tab:ReceiverControl:commands}}
\\
\hline \hline \\[-2ex]

\multicolumn{1}{c}{Command} &
\multicolumn{1}{c}{Arguments} &
\multicolumn{1}{c}{Description} \\[0.5ex] \hline
\\[-1.8ex]
\endfirsthead

\multicolumn{3}{c}{{\tablename} \thetable{} -- Continued} \\[0.5ex]
	\hline \hline \\[-2ex]
\multicolumn{1}{c}{Command} &
\multicolumn{1}{c}{Arguments} &
\multicolumn{1}{c}{Description} \\[0.5ex] \hline
\\[-1.8ex]
\endhead

\\[-1.8ex]  \hline
\multicolumn{3}{l}{{Continued on Next Page\ldots}} \\
\endfoot

\\[-1.8ex] \hline
\insertTableNotes
\endlastfoot

PreampSet & \textit{Diode Bias Offset} & Set the Preamp Board bias DAC corresponding to \textit{Diode} to \textit{Bias} bits.  Then adjust the offset DAC with a feedback loop to center the channel output within the ADC range.  \textit{Offset} is an initial guess at the required offset DAC setting.  Multiple commands can be issued in a single call by concatenating \textit{Diode-Bias-Offset} triples with commas.\\
BiasAddr & \it{DAC Bits} & Use the Receiver Database to lookup the card, address, and DAC number of \textit{DAC}.  \textit{Bits} may be optionally specified so that the result can be passed to bias-setting commands.\\
BiasIntAddr & \it{Card DAC Addr} & Convert the arguments to the corresponding bias address.
The address is $DAC + 2^5 Addr + 2^9 Card$ and is the same as the address in Appendix~\ref{sec:adc_server_bias_thread_actions}, bias command `M.'\\
SiteLookep & \it{Site} & Use the Receiver Database to find the module number currently located at electronics \textit{Site}.\\
ModuleLookup & \it{Module} & Use the Receiver Database to find the electronics site where module number \textit{Module} is located.\\
DbArc & & Backup the content of the Receiver Database to disk.\\
BitSet & \it{DAC Bits} & Set bias electronics \textit{DAC} output to \textit{Bits}.  Multiple commands can be issued by concatenating with commas.\\
AverageData & \it{Channel} & Print current average data in Volts.\\
MuxSet & \it{Addresses} & Change housekeeping multiplexing to monitor \textit{Addresses}.  Multiple addresses may be separated by commas, and addresses may contain mySQL wildcards.
ReceiverControl will enlarge the address list with the ground and supply senses related to the input addresses, if applicable.\\
SetMask & \it{Offset Length} & Set the demodulation mask \textit{Offset} and \textit{Length}.\\
PhaseState & \it{States} & Change phase-switching state to \textit{States}.  Four states must be specified (the first two control one module leg; the second two control the other leg).
Table 13 of \cite{QUIET_DAQ_Crate} gives the meaning of the states.\\
AverageHk & \it{Frames} & Acquires \text{Frames} (250\,ms each) of housekeeping data from \process{DataCompilation}.
The number of frames is increased if necessary so that at least 1 full iteration of the housekeeping multiplexing is received.
Print the resulting data, including conversion to physical units.\\
BitZeroNoPreamp & \it{DAC} & Set bias electronics to turn off modules except for preamp biasing (i.e. detector diodes will retain biasing). \textit{DAC} may include mySQL wildcards.  Multiple DACs may be concatenated with commas.\\
BitZeroYesPreamp & \it{DAC} & Same as BitZeroNoPreamp without restriction on preamp biasing.\\
Peripheral & & See \cite{robert_thesis}.\\
DrainVoltSet & \it{Drain Volts} & Adjust the module drain voltage of specified \textit{Drain} to \textit{Volts}.  A feedback loop adjusts the corresponding gate voltages until the commanded drain voltage is reached.  Multiple commands may be concatenated with commas.  This command is only appropriate for Q band and W-band lab testing.  W-band observation used MMIC bias boards with drain voltage control.\\
BiasArc & \it{Site} & Create a new entry in the Receiver Database with the current bias settings for modules matching \textit{Site}, which may contain mySQL wildcards.  Multiple \textit{Sites} may be concatenated with commas. \\
ListBiasArc & \it{Module} & Print the entries in the Receiver Database containing stored bias settings for \textit{Module}.\\
VerifyBiasArcUsed & \it{Id Module Tolerance} & Check that the current bias settings of \textit{Module} match those from the Receiver Database record \textit{Id}.  \textit{Tolerance} controls the allowed difference between the recorded and current housekeeping.  Multiple tuples may be concatenated with commas.\\
UseBiasArcBit & \it{Id Module} & Set the bias settings of \textit{Module} to those recorded in the Receiver Database record \textit{Id}.  Multiple tuples may be concatenated with commas.\\
UseBiasArcHk & \it{Id Module} & Set the bias settings of \textit{Module} to those recorded in the Receiver Database record \textit{Id}.  The settings are adjusted so the current housekeeping matches the recorded values.  Multiple tuples may be concatenated with commas.\\
StartSite & \textit{Site Current1 Current2 Voltage1 Voltage2}\tablenotemark{a} & Turn on bias for modules at \textit{Site}, which may contain mySQL wildcards.  The first and second stage drain currents are set to \textit{Current1} and \textit{Current2} (specified in mA), respectively.  The drain voltages are set to \textit{Voltage1} and \textit{Voltage2} in Volts.  Multiple tuples may be concatenated with commas.\\
PhaseCurrSet & \it{Site A1Current A2Current B1Current B2Current} & Set phase switch currents of module at \textit{Site}, which may contain mySQL wildcards.  Currents are given in mA.  Multiple tuples may be concatenated with commas.\\
PhaseBitSet & \it{Site A1Bits A2Bits B1Bits B2Bits} & Set DACs controlling phase switch currents of module at \textit{Site}, which may contain mySQL wildcards.  Multiple tuples may be concatenated with commas.\\
WhichSites & \it{Site} & Expand \textit{site}, which may contain mySQL wildcards, into a list of unique sites.  Multiple \textit{Sites} may be concatenated with commas.\\
HktoDAC & \it{Address} & Print the identifier of the DAC corresponding to the given housekeeping \textit{Address}.  If there is no corresponding DAC, \textit{Address} is printed.  Multiple \textit{Addresses} may be concatenated with commas.\\
MuxLookup & \it{NumericalAddress} & Print the address (name) of the housekeeping channel with the given \textit{NumericalAddress}.\\
HKidLookup & \it{Address} & Print the numerical address of the housekeeping channel with the given \textit{Address} identifier.\\
HKConvert & \it{Address Raw} & Convert housekeeping voltage \textit{Raw} recorded for \textit{Address} to physical units.  Multiple tuples may be concatenated with commas.\\
HkConvertBatch & \it{Filename} & Convert raw housekeeping voltages from \textit{Filename} to physical units.\\
DDLookup & \it{Diode} & Print ADC channel number recording data from detector \textit{Diode}.\\
AdcChannelLookup & \it{Channel} & Print diode identifier corresponding to ADC \textit{Channel}.\\
GetMask & & Acquire the current demodulation mask from BiasServer and print it.\\
TakeSnapshot & & Command \process{DataCompilation} to take a new snapshot and print the new snapshot number.\\
PerData & \it{Channel} & Print latest peripheral data from \textit{Channel}.  Multiple \textit{Channels} may be concatenated with commas.\\
EnclosureData & & Print latest monitoring data from the Electronics Enclosure smart power strip.\\
WPIDOutput & & Print latest WPID output information.\\

\end{longtable}

\end{ThreePartTable}

\end{singlespace}
\end{center}
\renewcommand{\thefootnote}{\arabic{footnote}}

\svnid{$Id: timing_alignment.tex 72 2012-03-30 21:05:47Z ibuder $}

\chapter{Timing Alignment Algorithm}
\label{app:timing_alignment}

\process{DataCompilation} aligns the radiometer and housekeeping frames:
\begin{enumerate}
\item Alignment of any sort is performed only for housekeeping
and radiometer data.  These are the only data which have
GPS-based timestamps.  For all other data (e.g. biasing), the most recent
data frame available is used.
\item Frame construction waits until the queue of
radiometer data and the queue of housekeeping data each have
at least 4 250-ms frames.
\item Compute TIMING\_FLAG to be true if the Time-code Reader is
synchronized to the IRIG signal and the firmware is synchronized to
the time-code reader and the timestamp source was the Time-code
Reader.
Otherwise, TIMING\_FLAG is false.
\item If the timestamp of the front frame in the radiometer
queue is not 0 (mod 250\,ms) or TIMING\_FLAG is false, the frame is
moved into the first radiometer 250-ms frame of the 1-s frame.
\item Otherwise if the timestamp of the front frame is 0
(mod 250\,ms) but not 250\,ms (mod 1\,s), a null 250-ms frame is
copied into the 1-s frame. The front frame remains in the
queue.
\item The previous two steps are repeated with ``radiometer'' replaced by ``housekeeping.''
\item If the first radiometer 250-ms frame is empty, and the first
housekeeping 250-ms frame is non-empty (or vice versa), the front
radiometer (or housekeeping) 250-ms frame is moved to the 1-s frame.
\item If both first 250-ms frames in the 1-s frame are empty, and the two front
250-ms frames have the same timestamp, both front 250-ms frames are
moved into the 1-s frame.
\item If both first 250-ms frames are empty, and the two front
250-ms frames do not have the same timestamp, the 250-ms frame with
the earlier timestamp is moved, and a null 250-ms frame is copied
into the other 250-ms frame.  The front 250-ms frame with later
timestamp remains in the queue.
\item The preceding 5 steps are repeated with ``first 250-ms frame'' replaced by
 ``second,'' ``third,'' and ``fourth.''
``not 250\,ms'' is replaced by 500\,ms, 750\,ms, and 0\,ms.
\item The latest non-null timestamp\footnote{This algorithm guarantees
at least 1 non-null 250-ms frame.} is copied to frame.utc (the value cbicontrol uses to align receiver and mount data).
\item If TIMING\_FLAG is true and the first housekeeping 250-ms frame's
timestamp is 250\,ms (mod 1\,s) and each housekeeping
250-ms frame has the
same timestamp as the corresponding radiometer 250-ms frame
and the radiometer 250-ms frames' timestamps increment by 250\,ms and frame.utc is 1\,s
greater than the previous value of frame.utc, then increment
the good timing count.  Otherwise, set the good timing count to 0.
\item If the good timing count is at least 10, request (by
setting the time alignment register) that cbicontrol perform
strict alignment using this frame's timestamp.  Otherwise,
request that cbicontrol ignore this frame's timestamp and put
this frame into the earliest available multi-system frame.
\item Move this frame into the buffer of frames to send to
cbicontrol, and return to step 2 to construct the next frame.
\end{enumerate}

\svnid{$Id: receiver_db_tables.tex 120 2012-05-17 03:46:41Z ibuder $}

\chapter{Receiver Database}
\label{app:receiver_db_tables}
This appendix summarizes the information in the Receiver Database and the mapping between modules, MABs, and bias boards.
The Receiver Database was a mySQL database that stored the relationships between the physical configuration of the instrument and the electronics numbering.
The mySQL table  ADC\_MAPPING (Appendix~\ref{sec:ADC_MAPPING}) stored the relationship between modules and ADCs.
The mySQL table BIASES (Appendix~\ref{sec:BIASES}) stored the relationship between a bias voltage or current and the DAC that provided it.
The mySQL table HOUSEKEEPING (Appendix~\ref{sec:HOUSEKEEPING}) stored the relationship between housekeeping addresses and physical monitored items.
The mySQL table mab\_input (Appendix~\ref{sec:mab_input}) stored which modules were in which MABs.
The lists in the following sections give the meanings of the columns in the mySQL tables.

\section{Table ADC\_MAPPING}
\label{sec:ADC_MAPPING}
Each row in table  ADC\_MAPPING corresponded to 1 diode.
\begin{itemize}
\item NAME is the diode name arranged by MAB e.g. M08A6Q2 is the Q2 diode of the module at site 6 of MAB 8.
\item MODULE is the module serial number.
\item TYPE is W1 or W2.  W1 means that the module has two gate controls for the first MMIC.  W2 means that the module has 1 gate control per MMIC.
\item ADC\_CH is the channel number (ADC) recording the diode.
\item PR\_SLOT is the location of the Preamp Board (in the Bias-Board Backplane) for that diode.
\item PR\_SIDE is the side (0 or 1) of the Preamp Board used.
\item ADC\_SLOT is the number (1--13) of the ADC Board recording that diode.
\item PR\_SERIAL is the Preamp Board serial number.
\item ADC\_SERIAL is the ADC Board serial number.
\end{itemize}

\section{Table BIASES}
\label{sec:BIASES}
Each row in table  BIASES corresponded to 1 bias DAC.
\begin{itemize}
\item DESCRIPTION is the identifier (e.g. M12A6Q1B) of a bias setting.
\item SLOT is the location in the Bias-Board Backplane of the Bias Board controlling that setting.
\item DAC is the number of the DAC chip on that Bias Board controlling the bias setting.
\item ADDR is the ``address'' (1--8) of the DAC on the DAC chip.
\item INT\_ADDR is firmware bias memory location storing the bias setting.
\end{itemize}

\section{Table HOUSEKEEPING}
\label{sec:HOUSEKEEPING}
Each row in table  HOUSEKEEPING corresponded to 1 monitored item.

\begin{itemize}
\item NAME is the identifier (e.g. M12A6DA2I) of the voltage, current, temperature, etc. being monitored.
\item DESCRIPTION identifies the locations of thermometers.
\item MUX\_INT is the multiplex address for the identifier.
\item MUX is a string holding the binary digits of MUX\_INT.
\item TYPE identifies the type of measurement:
\begin{enumerate}
\item GN is a ground voltage.
\item G is a gate voltage.
\item DI1 is a first-stage drain current.
\item DV1 is a first-stage drain voltage.
\item DI2 is a second-stage drain current.
\item DV2 is a second-stage drain voltage.
\item S25 is a 2.5-V supply voltage.
\item P is a phase-switch current.
\item CD is a cryogenic thermometer.
\item CD4 is a LakeShore DT-470 cryogenic thermometer.
\item CD6 is a LakeShore DT-670 cryogenic thermometer.
\item TH is a warm (Electronics Enclosure) thermometer.
\end{enumerate}
\item BOARD\_SLOT is the location in the Bias-Board Backplane of the Bias Board controlling the voltage or current being monitored.
\item BOARD\_SIDE is the side of the Bias Board.
\item BOARD\_SN is the serial number of the Bias Board.
\end{itemize}

\section{Table mab\_input}
\label{sec:mab_input}
Table \ref{tab:mab_input} lists the bias boards connected to each MAB.
Table \ref{tab:module_mab_mapping} lists the modules on each MAB and the ADC Board digitizing each.

\begin{deluxetable}{rrrr}
\tablecolumns{4}
\tablecaption{MAB--bias-board Mapping
\label{tab:mab_input}
}
\tablehead{\colhead{MAB} & \colhead{MMIC Board Location/Side} & \colhead{Phase-switch Board} & \colhead{Preamp Board}}
\startdata
0 & 14/1 & 20/1 & 5/1\\
1 & 15/0 & 20/2 & 6/0\\
2 & 14/0 & 20/0 & 5/0\\
3 & 15/1 & 21/0 & 6/1\\
4 & 12/0 & 18/1 & 3/0\\
5 & 16/0 & 21/1 & 7/0\\
6 & 11/1 & 18/0 & 2/1\\
7 & 13/0 & 19/1 & 4/0\\
8 & 16/1 & 21/2 & 7/1\\
9 & 13/1 & 19/2 & 4/1\\
10 & 10/0 & 17/0 & 1/0\\
11 & 11/0 & 17/2 & 2/0\\
12 & 10/1 & 17/1 & 1/1\\
\enddata
\end{deluxetable}

\begin{deluxetable}{rlr}
\tablecolumns{3}
\tablecaption{Modules and ADC Board for Each MAB
\label{tab:module_mab_mapping}
}
\tablehead{\colhead{MAB} & \colhead{Modules (RQ)} & \colhead{ADC Board}}
\startdata
0 & 0, 1, 2, 3, 4, 5, 9 & 9\\
1 & 10, 11, 12, 17, 18, 19, 20 & 10\\
2 & 6, 7, 8, 13, 14, 15, 16 & 8\\
3 & 24, 25, 26, 27, 28, 29, 36 & 11\\
4 & 21, 22, 23, 30, 31, 32, 33 & 5\\
5 & 37, 38, 39, 47, 48, 49, 50 & 12\\
6 & 34, 35, 44, 45, 46, 55, 45 & 4\\
7 & 40, 41, 42, 43, 51, 52, 53 & 6\\
8 & 57, 58, 59, 60, 67, 68, 69 & 13\\
9 & 54, 61, 62, 63, 64, 65, 66 & 7\\
10 & 74, 75, 76, 77, 82, 83, 84 & 1\\
11 & 70, 71, 72, 73, 78, 79, 80 & 3\\
12 & 85, 86, 87, 88, 89, 90 & 2\\
\enddata
\end{deluxetable} 

\svnid{$Id: db.tex 136 2012-06-20 21:17:22Z ibuder $}

\chapter{Database and Data-quality Statistics}
\label{app:db}
This appendix summarizes the QUIET mySQL Database, including the data-quality statistics for each CES-diode.
Database quiet\_cal (Appendix~\ref{sec:quiet_cal}) contained calibration information, particularly the responsivity model.
Database quiet\_data (Appendix~\ref{sec:quiet_data}) contained information about observations and data files.
Database quiet\_quality (Appendix~\ref{sec:quiet_quality}) contained the data-quality statistics we calculated for each CES-diode.
The following sections list the meanings of each column in each mySQL table.

\section{Database quiet\_cal}
\label{sec:quiet_cal}
This database contained calibration information.

\subsection{Table w91\_gain\_model\_each\_runs}
Table w91\_gain\_model\_each\_runs contained the latest responsivity model, currently identical to w91\_gain\_model\_each\_run\_2012\_0404.
Each row contained the responsivity for 1 CES-diode.
\begin{itemize}
\item ces\_id is a unique identification number for each CES, used internally in the database.
\item run\_id is the CES run number.
\item run\_subid is the number of the CES within its run.  Together run\_id and run\_subid uniquely specify the CES.
\item module is the module (RQ) number.
\item diode is Q1, U1, U2, or Q2.
\item version was used to distinguish different models in the same table; however, the final table contains only 1 model (version 0).
\item G is the CES-diode responsivity in mV/K, thermodynamic temperature.
\end{itemize}

\section{Database quiet\_data}
\label{sec:quiet_data}

This database contained observation, data-file, and log information.
Table w91\_ces contained the definition of each CES.
Table w91\_l0\_ces\_intersection listed the data files for each CES.
Table w91\_level0 contained summary information about each Level-0 data file.
Table w91\_level1 contained summary information about each Level-1 data file.
Table w91\_run\_log contained details about each observation run.

\subsection{Table w91\_ces}
Each row contained the definition of 1 CES.
\begin{itemize}
\item ces\_id, run\_id, run\_subid are as above.
\item begin\_time is the start of the CES in UTC.  We included this second.
\item end\_time is the end of the CES in UTC.  We did not include this second.
\end{itemize}

\subsection{Table w91\_l0\_ces\_intersection}
This table contained the relationship between CESes and Level-0 data files.
Each row corresponded to the fraction of 1 Level-0 file containing data for 1 CES.
\begin{itemize}
\item l0\_id The identification number of the Level-0 file.
\item ces\_id is as above.
\item begin\_time is the number of minutes into the Level-0 file when the CES begins.
\item end\_time is the number of minutes into the Level-0 when the CES ends.
\item l0\_length is the length of the Level-0 file in minutes.
\end{itemize}

\subsection{Table w91\_level0}
Each row contained information about 1 Level-0 file.
\begin{itemize}
\item id is the Level-0 file identification number.
\item path is the filename including directories from the data storage root location.
\item type is one of
\begin{enumerate}
\item RCS indicating the old Level-0 format recorded by \process{DataArc}.
\item ARCHIVE indicating normal Level-0 files created by \process{cbicontrol}.
\item ARC\_PRE indicating old Level-0 files created by \process{cbicontrol} but not including receiver data.
\end{enumerate}
\item hash is the SHA-1 checksum of the file.
\item transfer indicates the data transfer status:
\begin{enumerate}
\item DVD means we recorded the file to Blu-ray disk.
\item SHIP means we sent the disk to Chicago.
\item FAIL means the file was corrupt on Blu-ray disk and we need to send it again.
\item COMPL means we transferred the file successfully.
\end{enumerate}
\item available indicates where the file is available
\begin{enumerate}
\item SITE means it is available at the site storage.
\item CHI means it is available at the Chicago data center.
\item KEK means it is available at the KEK data center.
\item OSLO means it is available at the Oslo data center.
\item CHI\_TAPE means it is available on tape archive at Chicago.
\end{enumerate}
\item begin\_time is the start time of the file in UTC.
\item end\_time is the end time of the file in UTC.
\item begin\_db\_ver is the Receiver Database version in use at the beginning of the file.
\item begin\_db\_hash is the checksum of the Receiver Database in use at the beginning of the file.
\item end\_db\_ver is the Receiver Database version in use and the end of the file.
\item end\_db\_hash is the checksum of the Receiver Database in use at the end of the file.
\item comment contains comments about each file.
\end{itemize}

\subsection{Table w91\_level1}
Each row contained information about 1 Level-1 file.
\begin{itemize}
\item id is the identification number of the Level-1 file.
\item path is as above.
\item hash is as above.
\item l0\_id is the identification number of the corresponding Level-0 file.
\item available is as above
\item flag indicates several problems:
\begin{enumerate}
\item CORRUPT\_FILE means the file cannot be read.
\item CORRUPT\_SUBTIME means the receiver timestamp in the file is impossible e.g. before the experiment started.
\item WRONG\_SUBTIME means the receiver timestamp does not match the expected time when the file was recorded.
\end{enumerate}
\item num\_hdu is the number of FITS HDUs in the file.  Valid files have 1876 HDUs.
\end{itemize}

\subsection{Table w91\_run\_log}
Each row contained information about 1 observation run.
\begin{itemize}
\item id is the identification number of the run.
\item begin\_time is the start of the run in UTC.
\item end\_time is the end of the run in UTC.
\item run\_type is:
\begin{enumerate}
\item CMB for CMB patch observations (CESes).
\item Calib for calibration observations.
\item Other for all other observations.
\end{enumerate}
\item object is the astronomical source.
\item state is
\begin{enumerate}
\item Done means the observation finished successfully.
\item Scheduled means we scheduled the observation but have not confirmed it.
\item Canceled means we canceled the observation before it started.
\item Aborted means we stopped the observation before it completed.
\end{enumerate}
\item daily\_log\_scheduled is the identification number of the observer log when when scheduled the observation.
\item daily\_log\_confirmed is the identification number of the observer log when we confirmed the observation completed.
\item comment contains comments for each run.  Usually we put additional details such as the deck angle, modules, or details of special calibrations.
\end{itemize}

\section{Database quiet\_quality}
\label{sec:quiet_quality}
This database contained the data-quality statistics for each CES-diode.
Table w91\_atm contained weather information.
Table w91\_ces\_usable defined whether each CES was usable based on telescope, mount, and software problems.
Table w91\_diode\_usable defined whether each CES-diode was usable.
Table w91\_diodes contained information about each diode.
Table w91\_full\_correlation contained correlation matrices for the noise model.
Table w91\_housekeeping contained average housekeeping for each CES.
Table w91\_scan contained summarized pointing information for each CES.
Table w91\_software contained details about software problems for each CES.
Table w91\_sun\_spike defined the Sun-sidelobe cut.
Table w91\_timestream contained data-quality statistics for each CES-diode.
Table w91\_typeb contained Type-B--glitching statistics for each CES-diode.

\subsection{Table w91\_atm}
Each row contained weather information for 1 CES.
\begin{itemize}
\item ces\_id, run\_id, and run\_subid are as above
\item humidity\_med is the median humidity during the CES in \%.
\item humidity\_min is the minimum humidity during the CES in \%.
\item humidity\_max is the maximum humidity during the CES in \%.
\item humidity\_rms is the RMS fluctuation of the humidity during the CES in \%.
\item air\_temp\_med, air\_temp\_min, air\_temp\_max, air\_temp\_rms are the same for the air temperature in $\degr$C.
\item pressure\_* are the same for the atmospheric pressure in hPa.
\item wind\_speed\_* are the same for the wind speed in m/s.
\item wind\_dir\_* are the same for the wind direction in $\degr$.
\item apex\_humidity\_* are the same for the APEX weather station.
\item apex\_dewpoint\_* are the same for the dew point in $\degr$C from APEX.
\item apex\_pwv\_* are the same for the PWV in mm from APEX.
\end{itemize}

\subsection{Table w91\_ces\_usable}
Each row defines whether 1 CES was usable.
\begin{itemize}
\item run\_id and run\_subid are as above.
\item usable indicates whether the CES is usable.
\item usable\_daq indicates whether the CES is usable based on software problems.
\item usable\_tel indicates whether the CES is usable based on telescope and mount problems.
\end{itemize}

\subsection{Table w91\_diode\_usable}
Each row defined whether 1 CES-diode was usable.
To be usable, both w91\_ces\_usable.usable and w91\_diode\_usable.usable must be true.
\begin{itemize}
\item ces\_id, run\_id, run\_subid, module, diode are as above.
\item usable indicates whether the CES-diode is usable.
\end{itemize}

\subsection{Table w91\_diodes}
Each row contained information on 1 diode.
\begin{itemize}
\item module and diode are as above
\item knee\_median\_Hz is the median knee frequency in Hz of the diode for all CESes passing data selection.
\item knee\_mean\_Hz is the mean knee frequency.
\item knee\_RMS\_Hz is the RMS fluctuation of the knee frequency.
\item uK\_sqrtsec\_* are the same for the sensitivity in $\mu$K$\sqrt{\textrm{s}}$
\item CES\_count is the number of CESes passing data selection.
\item leakage is the $I\rightarrow Q/U$ leakage in \%.
\item MAB is the MAB that diode is on.
\item gain\_mean is the mean responsivity for all CESes passing data selection.
\item band\_center is the bandpass center frequency in GHz.
\item band\_pass is the bandwidth in GHz.
\end{itemize}

\subsection{Table w91\_full\_correlation}
Each row contained 1 element of the white and 1/f-noise correlation matrices.
\begin{itemize}
\item run\_id and run\_subid are as above.
\item module1 is one of the modules whose noise correlation is measured.
\item diode1 is one of the diodes whose noise correlation is measured.
\item module2 and diode2 specify the other diode.
\item correl\_oneoverf is the 1/f-noise correlation coefficient.
\item correl\_white\_noise is the white-noise correlation coefficient.
\end{itemize}

\subsection{Table w91\_housekeeping}
Each row contained housekeeping information for 1 CES.
\begin{itemize}
\item run\_id and run\_subid are as above.
\item preamp1\_temp\_p3t4\_mean is the mean temperature in $\degr$C of a thermometer on a Preamp Board.
\item preamp1\_temp\_p3t4\_rms is the RMS temperature fluctuation during the CES.
\item preamp1\_temp\_p3t4\_range is the range (maximum $-$ minimum) during the CES.
\item mmic3\_temp\_p3t9\_* are the same for a MMIC Board.
\item psw1\_temp\_p3t10\_* are the same for a Phase-Switch Board.
\item psw2\_temp\_p3t11\_* are the same for a second Phase-Switch Board.
\item primary\_mirror\_temp\_0\_* are the temperatures in K for a thermometer on the primary mirror.
\item secondary\_mirror\_temp\_3\_* are the same for the secondary mirror.
\item groundscreen\_temp\_7\_* are the same for the ground screen.
\item external\_air\_temp\_26\_* are the same for a thermometer exposed to the ambient environment.
\item module\_45\_temp\_p2t19\_* are the same for a thermometer on module RQ45.
\item 80K\_temp\_p2t18\_* are the same for the 80-K radiation shield.
\item platelet1\_temp\_p2t16\_* are the same for the platelet array.
\item interface1\_temp\_p2t14\_* are the same for the interface plate.
\item module\_9\_temp\_p2t8\_* are the same for RQ9.
\end{itemize}

\subsection{Table w91\_scan}
Each row contained pointing information for 1 CES.
\begin{itemize}
\item run\_id and run\_subid are as above.
\item azimuth is the average azimuth of the CES in $\degr$ East of North.
\item elevation is the average elevation of the CES.
\item deck\_angle is the deck angle of the CES.
\item scan\_speed is the average azimuthal scan speed in $\degr/$s.
\item scan\_period is the scan period in s.
\item time is the time in days since an arbitrary zero point.
\item object is the patch name.
\item duration is the CES length in s.
\item sun\_ra is the RA of the Sun at the time of the CES.
\item sun\_dec is the Declination of the Sun at the time of the CES.
\item sun\_phi is the azimuthal angle of the Sun in the telescope-centered coordinate system.
\item sun\_theta is the distance to the Sun in $\degr$.
\item moon\_* are the same for the Moon.
\item hour\_angle is the hour angle in $\degr$.
\item patch\_day\_number is a number for dividing CESes into days for day-by-day cross-correlation.
If two CESes scan the same patch, then they have the same\\ patch\_day\_number iff the patch did not set between the two CESes.
\item hour\_of\_day is the hour into the (UTC) day when the observation was performed.
\item sidelobe\_az is the azimuth at which the triple-reflection sidelobe is pointing.
\item sidelobe\_el is the elevation at which the triple-reflection sidelobe is pointing.
\item sidelobe2\_* are the same for the spillover sidelobe.
\item array\_orientation\_equatorial is the orientation axis of the array on the sky in Equatorial coordinates.
\item scan\_direction\_equatorial is the scan axis on the sky in Equatorial coordinates.
\item sidelobe\_lat is the latitude of the triple-reflection sidelobe in Galactic coordinates.
\item sidelobe\_long is the longitude of the triple-reflection sidelobe in Galactic coordinates.
\item sun\_az is the azimuth of the Sun at the time of CES observation.
\item sun\_el is the elevation of the Sun at the time of CES observation.
\end{itemize}

\subsection{Table w91\_software}
Each row contained software status for 1 CES.
\begin{itemize}
\item run\_id and run\_subid are as above.
\item processing\_status indicates possible failures when the CES was being checked. The possible values are
\begin{enumerate}
\item     GOOD indicates no problems occurred
\item    NO\_TELESCOPE indicates the TELESCOPE HDU was missing or could not be read from at least 1 of the Level-1 files constituting the CES.
\item    PARTIAL\_FRAME indicates that the number of telescope rows read was not a multiple of 100 (i.e. not an integral length in seconds).
\item    EXTRA\_FILE indicates a Level-1 file was in the CES intersection but no corresponding rows could be found in it.
\item    CES\_LOOKUP\_FAIL indicates the CES metadata could not be found. This is more likely to be a problem with the checking code than with the CES itself.
\end{enumerate}
\item num\_frame is the number of 1-Hz frames in the CES (i.e. the length in seconds).
\item num\_uneven\_demod is the number of uneven demodulation events detected in the ADC. This must be 0 if the CES can be used.
\item num\_telescope\_time\_jump is the number of telescope timestamps with irregular intervals.
\item num\_uneven\_mask is the number of mask mismatch events between ADC boards. This is contaminated by snapshot bugs.
\item num\_uneven\_clock is the number of fast clock mismatch events between ADC boards. This is contaminated by snapshot bugs.
\item num\_bad\_scale is the number of bad scale factor events. This must be 0.
\item num\_bad\_quad\_scale is the number of bad quadrature scale factor events. This must be 0.
\item num\_4hz\_mismatch is the number of 4-Hz frame counter mismatch events across ADC boards. This must be 0 unless the ADC was started incorrectly.
\item num\_tfp\_time\_offset is the number of ``time offset'' events detected from the Time-code Reader. 
\item num\_tfp\_freq\_offset is the number of ``frequency offset'' events detected from the Time-code Reader.
\item num\_null\_frame is the number of null frames in the data. This must be 0.
\item num\_bad\_timing\_flag is the number of frames flagged as having bad timing by the Online Software. This must be 0.
\item num\_bad\_wpid\_thread is the number of frames when the WPID output was not being monitored.
\item num\_rcs\_frame\_jump is the number of frame gaps due to missing data lost between the Online Software and data being written to disk. This must be 0.
\item num\_rcs\_frame\_offset is the number of frames whose timestamps are not aligned to the start of the second. This must be 0.
\item num\_adc\_timestamp\_differ is the number of frames having different timestamps between radiometer and housekeeping. This must be 0.
\item num\_adc\_timestamp\_jump is the number of irregular radiometer timestamps. This must be 0.
\item num\_hk\_frame\_jump is the number of frame gaps due to missing housekeeping data. This must be 0.
\item num\_adc\_frame\_jump is the number of frame gaps due to missing radiometer data. This must be 0.
\item fiftyhz\_clock\_bad indicates whether a problem was found with the ADC Board 50-Hz clock. This must be 0.
\item onehz\_clock\_bad indicates whether a problem was found with the ADC Board 1-Hz clock. This must be 0.
\item psw\_state* are the phase-switch switching state parameters.
\item num\_bad\_mask is the number of bad blanking mask events. This must be 0.
\item num\_adc\_count\_mismatch is the number of frames with different radiometer and housekeeping counters. This must be 0.
\item num\_bad\_mux\_pattern is the number of housekeeping samples with irregular multiplexing. This must be 0.
\item num\_mux\_addr is the number of different housekeeping addresses sampled. This must be 1774.
\item num\_downsample\_frame is the number of frames when the Online Software was operating in $10\rightarrow1$ downsampling mode. This must be 0.
\item num\_bad\_psw\_flag is the number of events indicating the phase switching state was not known.
\item num\_telescope\_radiometer\_mismatch is the number of samples where the telescope (mount) and radiometer timestamps do not match. This must be 0.
\item wrong\_len indicates whether the different Level-1 HDUs had compatible lengths. This must be 0.
\item max\_telescope\_time\_jump is the maximum irregularity in the telescope timestamps (in ms). This must be $<5$.
\end{itemize}

\subsection{Table w91\_sun\_spike}
Each row defined whether 1 CES-module was rejected by the Sun-sidelobe cut.
\begin{itemize}
\item run\_id, run\_subid, and module are as above.
\item has\_spike is non-zero if the CES-module should be rejected because of Sun-sidelobe contamination.
\end{itemize}

\subsection{Table w91\_timestream}
Each row contained data-quality statistics for 1 CES-diode.
\begin{itemize}
\item run\_id, run\_subid, module, and diode are as above.
\item uK\_sqrtsec is the white-noise level in $\mu$K$\sqrt{\textrm{s}}$.
\item knee\_frequency is the knee frequency in Hz.
\item spectral\_index is the 1/f slope (in noise power).
\item fit\_flag indicates problems in the noise fit:
\begin{itemize}
\item 0 means fit success
\item 1 means the fitter reached the limit for the number of likelihood function evaluations
\item 10 means the 1/f knee frequency was forced to 0 and the slope to 1.2
\end{itemize}
\item scan\_synchronous\_signal2 is the ratio of the noise power at the scan frequency to the power predicted by the noise model.
\item scan\_synchronous\_signal is the statistical significance ($\sigma$) of scan\_synchronous\_signal2.
\item scan\_synchronous\_signal2\_averg is the same as scan\_synchronous\_signal2, but computed from TP instead of DD.
\item scan\_synchronous\_chi2 is the reduced $\chi^2$ to 0 of the signal removed by the ground subtraction filter.
\item excess\_10mHz\_100mHz is the average difference between the noise power and model in $\sigma$s between 10--100\,mHz.
\item excess\_100mHz\_2Hz is the same between 100\,mHz and 2\,Hz.
\item chisquare\_10mHz\_100mHz is the reduced $\chi^2$ between the data and noise model between 10--100\,mHz.
\item chisquare\_excl\_scan\_10mHz\_200mHz is the same as chisquare\_10mHz\_100mHz except we excluded the scan frequency.
\item chisquare\_100mHz\_2Hz is the same as \\chisquare\_10mHz\_100mHz except between 100\,mHz and 2\,Hz.
\item binned\_spectrum\_spike\_sigma is the maximum difference between the data noise spectrum and the model in $\sigma$.
\item binned\_spectrum\_spike\_frequency is the frequency of the maximum.  I limited the search to the range between twice the scan frequency and 9.6\,Hz which the filter accepts.
\item sss\_filtered\_binned\_spectrum\_spike\_sigma is the same as binned\_spectrum\_spike\_sigma except I applied the ground subtraction filter first.
\item sss\_filtered\_binned\_spectrum\_spike\_sigma\_1\_2Hz is the same as \\sss\_filtered\_binned\_spectrum\_spike\_sigma except the search range is 1.15--1.3\,Hz.
\item glitch\_1sample is the largest 1-sample (20-ms) deviation from the mean in the CES-diode in $\sigma$.
\item glitch\_100msec is the largest deviation of 100-ms moving averages.
\item glitch2\_1sample is the same as glitch\_1sample except we used the data after filtering.
\item jump\_demod\_1sec is the largest change in the 1-s averaged DD data.
\item jump\_averg\_1sec is the same as jump\_demod\_1sec, except for TP data.
\item std\_of\_std\_10sec is the fractional variation in the noise level during the CES computed from the standard deviation of the standard deviations of 10-s data chunks.
\item fiducial\_gain is the responsivity in mV/K used in processing for this table.  It can be different from the final responsivity for CMB analysis.
\item filtered\_chisquare\_10mHz\_200mHz is the same as chisquare\_10mHz\_200mHz except we applied a frequency-domain bandpass filter (not the same as the fiducial analysis filter).
\item sss\_filtered\_chisquare\_nearscan is the reduced $\chi^2$ for the noise model fit for the 40 Fourier modes nearest the scan frequency.  We applied the ground subtraction filter first.
\item each\_az\_chisquare\_below\_1Hz is the reduced $\chi^2$ for the noise model fit below 1\,Hz.  We applied the full analysis filter first.
\item each\_az\_dof\_below\_1Hz is the number of degrees of freedom for \\each\_az\_chisquare\_below\_1Hz.
\item each\_az\_chisquare\_slopes is the $\chi^2$ to 0 for the azimuth slopes subtracted in filtering.
\item five\_min\_max\_raw is the maximum subtracted slope ($\mu$K/$\degr$) in a 150-s moving average.
\item five\_min\_max\_normalized is five\_min\_max\_raw times the width of the scan in azimuth i.e. the maximum signal change in $\mu$K attributed to the slope.
\end{itemize}

\subsection{Table w91\_typeb}
Each row contained Type-B statistics for 1 CES-diode.
\begin{itemize}
\item run\_id, run\_subid, module, and diode are as above.
\item chisquare\_demodav\_uncorrected is the Type-B $\chi^2$ before Type-B--glitch correction.
\item chisquare\_demodav\_corrected is the Type-B $\chi^2$ after Type-B--glitch correction.
\item bits\_sqrtsec\_uncorrected is the noise level in ADC bit$\sqrt{\textrm{s}}$ units before Type-B correction.
\item bits\_sqrtsec\_corrected is the noise level found in iterative correction.
\item rms\_1sec\_uncorrected\_volts is the RMS fluctuation in V after creating 1-s samples by averaging.
\item min\_glitch\_distance\_volts is the minimum distance to any of the Type-B glitches.
\item mean\_tp\_plus\_volts is the average TP for one of the 50-Hz phase-switch states.
\item mean\_tp\_minus\_volts is the average TP for the other phase-switch state.
\item mean\_demod\_plus\_corrected\_volts is the mean DE for one of the 50-Hz phase-switch states.
\end{itemize}

\svnid{$Id: array_sensitivity.tex 121 2012-05-17 21:56:25Z ibuder $}

\chapter{Array Sensitivity Definition and Calculation}
\label{app:array_sensitivity}
This appendix defines the ``QUIET array sensitivity'' in a way that takes the white noise correlations within each module into account.
The definition has several steps.
See \cite{qband_sensitivity, wband_sensitivity} for more detail.

\section{Diode Sensitivity}
I defined the sensitivity of each diode as its mean white noise level, averaged over those CESes for which the diode passed data selection.

\section{Diode-pair Sensitivity}
A ``diode pair'' is two Q diodes or two U diodes in the same module.
The white noise in a diode pair tends to have positive correlation, so the sensitivity to $Q$ or $U$ of a diode pair is better than one would expect if one ignored the correlation and treated the diodes as uncorrelated.
Because the detector angles in a diode pair are nearly $90\degr$ apart, we effectively differenced the diodes in map making.
We weighted diodes by their inverse variance.
Therefore, as an approximation to the sensitivity in the map, I defined the diode-pair sensitivity to be the standard deviation of the inverse-variance--weighted diode difference,
\begin{equation}
d = (Q_1w_1 - Q_2w_2)/w,
\end{equation}
where $Q_i$ is the data of one of the diodes,
\begin{equation}
w_i = 1/\sigma_i^2
\end{equation}
are inverse variance weights,
\begin{equation}
w = w_1 + w_2
\end{equation}
is the weight total, and $\sigma_i$ is the sensitivity of diode $i$.
The variance is
\begin{equation}
<d^2> = \frac{<Q_1^2>w_1^2+<Q_2^2>w_2^2 -2w_1w_2<Q_1Q_2>}{w^2}.
\end{equation}
Neglecting any signal, $<Q_i^2> = \sigma_i^2$. Therefore
\begin{equation}
<d^2> = \frac{w_1 + w_2 - 2\rho_{12}\sigma_1^{-1}\sigma_2^{-1}}{w^2},
\end{equation}
where $\rho_{12}$ is the white noise correlation coefficient.
I used the mean correlation coefficient from CESes where the Q1 or U1 diode of a pair passed data selection.
(The difference if I used CESes where Q2 or U2 passed was $\lesssim0.001$.)
The diode-pair sensitivity is $\sigma_{12} = \sqrt{<d^2>}$.
In case one of the diodes in a pair is non-functional, the diode pair sensitivity is the sensitivity of the functional diode.

\section{Module Sensitivity}

Since the diode pairs in a module were nearly uncorrelated \citep{white_noise_correlation}, I treated the module's $Q$ and $U$ sensitivities as independent.
I defined the combined sensitivity to polarization in a module as
\begin{equation}
\sigma = 1/\sqrt{\sigma_Q^{-2} + \sigma_U^{-2}},
\end{equation}
where $\sigma_Q$ is the sensitivity of the Q diode pair in the module.

\section{Array Sensitivity}
The correlations between modules were negligible, so I combined modules in the same way I combined diode pairs in a module.
The array sensitivity (including all modules that passed data selection at least once) was $87\pm7\,\mu$K$\sqrt{\textrm{s}}$ \citep{wband_sensitivity}.
The mean module sensitivity was 899\,$\mu$K$\sqrt{\textrm{s}}$.
I defined an ``effective average module sensitivity,'' 756\,$\mu$K$\sqrt{\textrm{s}}$, to be the array sensitivity times the square root of the number of modules.
An array of modules each having this sensitivity would have the same sensitivity as the real array.

\svnid{$Id: null_tests.tex 149 2012-07-23 20:51:50Z ibuder $}

\chapter{Null-test Definitions and Correlations}
\label{app:nt}

Here I define each of the null-test divisions.  See also \cite{2a_null_report}.
See \S\ref{sec:null_suite} for general discussion of the null tests.

\renewcommand{\thefootnote}{\alph{footnote}}
\begin{center}
\begin{singlespace}

\begin{ThreePartTable}
\begin{TableNotes}
\item[a] I excluded this test from normal analysis because we developed it after the null suite was finalized.  I checked that the null suite passes even if I include these normally excluded tests.
\item[b] I normally excluded these tests because they were designed to test the white noise correlation.  Once I verified that the noise model was accurate, these tests did not target any suspected systematic error.
\item[c] In order to make the divisions even, the thresholds are patch-dependent for some tests.  For these, I give the thresholds for patch 2a.
\item[d] This null test does not work for patch 7b because we only observed it while rising.
\item[e] I excluded these tests from normal analysis because they were highly correlated with the Sun and Moon proximity tests.
\end{TableNotes}

\begin{longtable}{p{1.5in}p{1.5in}p{3in}}
\caption{Null-test Definitions
\label{tab:null_test_defs}}
\\
\hline \hline \\[-2ex]
\multicolumn{1}{c}{Name} &
\multicolumn{1}{c}{Description} &
\multicolumn{1}{c}{Database Query} \\[0.5ex] \hline
\\[-1.8ex]
\endfirsthead

\multicolumn{3}{c}{{\tablename} \thetable{} -- Continued} \\[0.5ex]
        \hline \hline \\[-2ex]
\multicolumn{1}{c}{Name} &
\multicolumn{1}{c}{Description} &
\multicolumn{1}{c}{Database Query} \\[0.5ex] \hline
\\[-1.8ex]
\endhead

\\[-1.8ex]  \hline
\multicolumn{3}{l}{{Continued on Next Page\ldots}} \\
\endfoot

\\[-1.8ex] \hline
\insertTableNotes
\endlastfoot

typeo\_pointing\tnote{a} & CES in Type-O Effect pointing region vs. others & ( ((w91\_scan.deck\_angle BETWEEN -200. AND -100.) OR (w91\_scan.deck\_angle BETWEEN 160. AND 260.)) AND (w91\_scan.elevation BETWEEN 65. AND 75.) )\\
I\_to\_Q\_by\_diode & Diodes with large vs. small $I\rightarrow Q$ leakage & (ABS(w91\_diodes.leakage) $>$ 0.479999989271164)\\
Q\_vs\_U\_diode & Q diodes vs. U diodes & (w91\_timestream.diode = 'Q1') OR (w91\_timestream.diode = 'Q2')\\
Q1U1\_vs\_Q2U2\_diode\tablenotemark{b} & Q1 and U1 diodes vs. Q2 and U2 diodes & 	(w91\_timestream.diode = 'Q1') OR (w91\_timestream.diode = 'U1')\\
Q1U2\_vs\_Q2U1\_diode\tablenotemark{b} & Q1 and U2 diodes vs. Q2 and U1 diodes & 	(w91\_timestream.diode = 'Q1') OR (w91\_timestream.diode = 'U2')\\
band\_cf\_low\_vs\_high & Module band-pass center frequency low vs. high & (w91\_diodes.band\_center $<$= 93.8)\\
array\_top\_vs\_bottom & Modules at the array top vs. bottom & (w91\_timestream.module $<$= 39)\\
array\_central\_vs\_ peripheral & Central vs. peripheral modules & (w91\_timestream.module BETWEEN 15 AND 18 OR w91\_timestream.module BETWEEN 23 AND 27 OR w91\_timestream.module BETWEEN 32 AND 37 OR w91\_timestream.module BETWEEN 42 AND 48 OR w91\_timestream.module BETWEEN 53 AND 58 OR w91\_timestream.module BETWEEN 63 AND 67 OR w91\_timestream.module BETWEEN 72 AND 75)\\
typeb\_happening\_vs\_ not\_happening & Type-B glitching happening vs. not happening & (ABS(w91\_typeb.chisquare\_demodav\_ uncorrected - w91\_typeb.chisquare\_demodav\_corrected) $>$ 0.1)\\
sss\_large\_vs\_small & Scan synchronous signal large vs. small & (w91\_timestream.scan\_synchronous\_signal $>$ 0.602676)\tablenotemark{c}\\
high\_freq\_excess\_ large\_vs\_small & High frequency noise spikes vs. no spikes & (w91\_timestream.chisquare\_2Hz\_15Hz $>$ 1.244210)\tablenotemark{c}\\
elevation\_high\_vs\_low & Elevation high vs. low & 	(w91\_scan.elevation $>$ 65.150700)\tablenotemark{c}\\
sun\_close\_vs\_far & Sun close vs. far to the main beam & (w91\_scan.sun\_theta $<$ 69.490200)\tablenotemark{c}\\
moon\_close\_vs\_far & Moon close vs. far & (w91\_scan.moon\_theta $<$ 97.029000)\tablenotemark{c}\\
enclosure\_temp\_ high\_vs\_low & Electronics enclosure temperature high vs. low & (w91\_housekeeping.preamp1\_temp\_p3t4\_ mean $>$ 45.595500)\tablenotemark{c}\\
enclosure\_temp\_ change\_large\_ vs\_small & Enclosure temperature change within the CES large vs. small & (w91\_housekeeping.preamp1\_temp\_p3t4\_ range $>$ 0.707993)\tablenotemark{c}\\
cryostat\_temp\_ change\_large\_ vs\_small & Cryostat temperature change large vs. small & (w91\_housekeeping.module\_45\_temp\_ p2t19\_range $>$ 0.707127)\tablenotemark{c}\\
weather\_bad\_vs\_good & Bad weather vs. good & (w91\_weather.tp\_rms\_10sec $>$ 0.098875)\tablenotemark{c}\\
pwv\_high\_vs\_low & PWV high vs. low & (w91\_atm.apex\_pwv\_med $<$ 0.800000 )\tablenotemark{c}\\
1sthalf\_vs\_2ndhalf & First half of the season vs. second & 	(w91\_scan.time $<$= 611)\\
knee\_large\_vs\_small & Diodes with 1/f knee frequency large vs. small & ( (SELECT AVG(w91\_diodes\_knee\_temp.knee\_ median\_Hz) FROM w91\_diodes AS w91\_diodes\_knee\_temp WHERE w91\_diodes\_knee\_temp.knee\_median\_Hz $>$ 0 AND w91\_diodes\_knee\_temp.module = w91\_timestream.module) $>$ 0.00862041371874511)\\
sensitivity\_ high\_vs\_low & Sensitivity high vs. low diodes & 	(w91\_diodes.uK\_sqrtsec\_median $>$ 1494.25)\\
deck\_phase0\_90 & Deck angles $90\degr$ apart & (ROUND(w91\_scan.deck\_angle)=30 OR ROUND(w91\_scan.deck\_angle)=75 OR ROUND(w91\_scan.deck\_angle)=210 OR ROUND(w91\_scan.deck\_angle)=-105 OR ROUND(w91\_scan.deck\_angle)=-150)\\
deck\_phase0\_45 & Deck angles $45\degr$ apart & (ROUND(w91\_scan.deck\_angle)=30 OR ROUND(w91\_scan.deck\_angle)=120 OR ROUND(w91\_scan.deck\_angle)=210 OR ROUND(w91\_scan.deck\_angle)=-60 OR ROUND(w91\_scan.deck\_angle)=-150 OR ROUND(w91\_scan.deck\_angle)=-240)\\
patch\_rising\_ vs\_setting\tablenotemark{d} & Patch observed rising vs. setting & 	(w91\_scan.hour\_angle $<$ 0)\\
CES\_gain\_ high\_vs\_low & CES average responsivity high vs. low & ( (SELECT AVG( timestream\_gain\_temp.fiducial\_gain / diodes\_gain\_temp.gain\_mean) FROM (w91\_timestream AS timestream\_gain\_temp) JOIN (w91\_diodes AS diodes\_gain\_temp) USING (module, diode) WHERE w91\_timestream.run\_id = timestream\_gain\_temp.run\_id AND w91\_timestream.run\_subid = timestream\_gain\_temp.run\_subid) $>$ 0.951830)\tablenotemark{c}\\
diode\_gain\_ high\_vs\_low & Diode average responsivity high vs. low & 	(w91\_diodes.gain\_mean $>$ 2.34083721782973)\\
sidelobe\_el\_ high\_vs\_low & Sidelobe elevation high vs. low & ( w91\_scan.sidelobe\_el $>$ 35.475600 )\tablenotemark{c}\\
array\_orientation\_90 & Array orientation on the sky $90\degr$ apart & ( (w91\_scan.array\_orientation\_equatorial $>$= -90. AND w91\_scan.array\_orientation\_equatorial $<$ 0.) OR (w91\_scan.array\_orientation\_equatorial $>$= 90. AND w91\_scan.array\_orientation\_equatorial $<$ 180.) )\\
array\_orientation\_45 & Array orientation on the sky $45\degr$ apart & ( (w91\_scan.array\_orientation\_equatorial $>$= -180. AND w91\_scan.array\_orientation\_equatorial $<$ -135.) OR (w91\_scan.array\_orientation\_equatorial $>$= -90. AND w91\_scan.array\_orientation\_equatorial $<$ -45.) OR (w91\_scan.array\_orientation\_equatorial $>$= 0. AND w91\_scan.array\_orientation\_equatorial $<$ 45.) OR (w91\_scan.array\_orientation\_equatorial $>$= 90. AND w91\_scan.array\_orientation\_equatorial $<$ 135.) )\\
fridge\_cycle\_ commensurate & Scan period an integer multiple of refrigerator cycle period & ( (w91\_scan.scan\_period = 10) OR (w91\_scan.scan\_period = 20) OR (w91\_scan.scan\_period = 30) OR (w91\_scan.scan\_period = 40) )\\
ambient\_ temperature\_ high\_vs\_low & Ambient temperature high vs. low & 	( w91\_atm.air\_temp\_med $>$ -1.359500 )\tablenotemark{c}\\
sidelobe\_sun\tablenotemark{e} & Either sidelobe near sun vs. far & ( IF(w91\_sidelobe\_distance.sun\_triple $<$ w91\_sidelobe\_distance.sun\_spillover, w91\_sidelobe\_distance.sun\_triple, w91\_sidelobe\_distance.sun\_spillover) $<$= 55.827988 )\tablenotemark{c}\\
sidelobe\_moon\tablenotemark{e} &Either sidelobe near moon vs. far & ( IF(w91\_sidelobe\_distance.moon\_triple $<$ w91\_sidelobe\_distance.moon\_spillover, w91\_sidelobe\_distance.moon\_triple, w91\_sidelobe\_distance.moon\_spillover) $<$= 69.572441 )\tablenotemark{c}\\
sidelobe\_galactic\_ latitude & Sidelobes near Galaxy vs. far & ( IF(ABS(w91\_scan.sidelobe\_lat) $<$ ABS(w91\_scan.sidelobe2\_lat), ABS(w91\_scan.sidelobe\_lat), ABS(w91\_scan.sidelobe2\_lat)) $<$= 10.656031 )\tablenotemark{c}\\
scan\_forward\_and\_ backward & Left-going scans vs. right-going & N/A\\
scan\_accel\_and\_decel & Accelerating vs. Decelerating & N/A\\

\end{longtable}

\end{ThreePartTable}

\end{singlespace}

\end{center}  
\renewcommand{\thefootnote}{\arabic{footnote}}

To compute the correlation coefficient between two null tests, I used Eq. C.9 of \cite{colin_thesis}.

\svnid{$Id: configuration_list.tex 124 2012-05-21 22:54:53Z ibuder $}

\chapter{List of Analysis Configurations}
\label{app:configurations}

This appendix lists all analysis configurations I tried before settling on the final configuration\footnote{I put more details in the report for each configuration at \url{https://cmb.uchicago.edu/~ibuder/null_tests/null*/patch??/analysis/report.html}}.
See \S\ref{sec:analysis} for discussion of the final configuration and how I chose it.
\begin{enumerate}
\item This very early configuration tested the software.  I did not analyze the null-test results.
\item This configuration had no results because it used too much memory and processing crashed.
\item \label{config:3} I reduced the number of null tests to 19 to avoid the processing crash.
I found a negative-direction bias in the auto-correlation results indicating noise overestimation.
The bias was strongest in the Q vs. U diode null test indicating a problem in white-noise correlation estimation.
\item This is the same as Configuration \ref{config:3} except I used day-by-day cross-correlation instead of the usual pointing-based cross-correlation.
It was an attempt to solve the bias, but it failed.
\item Due to a bug in previous configurations, I had not used the measured white-noise correlations.  I fixed this bug, and the bias disappeared.
I added the Q1U2 vs. Q2U2 and Q1U2 vs. Q2U1 null tests for future such noise-model problems.
I updated the detector-angle calibration.
\item \label{config:6} I updated the data selection \citep{viewer_study2}.
To speed processing, I randomly used only 30\% of the CESes.
I removed two null tests (weather and cryostat temperature) because they were highly correlated with other existing tests.
I updated the Type-B correction.
I updated the cross-correlation divisions to 5 in azimuth and 8 in deck.
I updated the beam calibration \citep{beam_config6}.
The total $(\chi^\textrm{null})^2$ was too high (3\% PTE).
I found the auto-correlation bias at low $\ell$ in this configuration.
\item \label{config:7} This is the same as \ref{config:6} except I used all CESes.
\item I placed a tight (7\,mHz) cut on knee frequency in attempt to solve the auto bias, but failed.
\item \label{config:9} This is the same as \ref{config:6} except I updated the data selection \citep{viewer_study3}, increased the number of bins for the ground filter from 20 to 40, updated the relative responsivity model \citep{relative_gain_w}, and included 1/f-noise correlations within each module.
There was a bug in the pointing model causing the collimation offset to be applied twice.
\item \label{config:10} This is the same as \ref{config:9} except I used all CESes. 
Patch 6a failed the null suite with 0\% total $(\chi^\textrm{null})^2$ PTE .
\item I reverted the pointing model to \ref{config:7}; however, I did not analyze the results.
\item The same as \ref{config:10} except I reduced the number of ground-filter bins to 20 in a failed attempt to solve the patch 6a failure.
\item I reverted the responsivity model to \ref{config:7}; however, I did not analyze the results.
\item I reverted the data selection to \ref{config:7}; however, it did not solve the patch 6a failure.
\item The same as \ref{config:10} except I reverted both the data selection and pointing model to \ref{config:7}. It solved the patch 6a failure.
\item The same as \ref{config:10} except I reverted both the data selection and responsivity model to \ref{config:7}.  The patch 6a failure remained.
\item The same as \ref{config:10} except I fixed the collimation-offset bug.
This fixed the patch 6a failure.
\item \label{config:18} The same as \ref{config:10} except I updated the array pointing \citep{config18_pointing}.
All patches passed the null suite.
\item \label{config:19} I loosened the cuts on noise model $\chi^2$ (\ref{cut:noise_model}), allowing 3\% more CES-diodes to pass.
\item The same as  \ref{config:18} except I used day-by-day cross-correlation.
The lack of $\chi$ bias (similar to what we found in Q band) supported our decision to use simple Q--W cross-correlation in the combined analysis.
\item The same as \ref{config:19} except I loosened the level-shift cut (\ref{cut:jump}), allowing 1\% more CES-diodes to pass.
\item The same as \ref{config:19} except I loosened the cut on noise-power-spectrum spikes (\ref{cut:FFT_spike}).
\item \label{config:23} I tightened all cuts by an additional 13\% of the CES-diodes.
\item \label{config:24} The same as \ref{config:18} except I used a frequency-domain high-pass filter instead of the slope subtraction filter.
\item The same as \ref{config:24} except I randomly used only 30\% of the CESes.
\item \label{config:26} The same as \ref{config:18} except I tightened the knee frequency (\ref{cut:knee}), slopes (\ref{cut:slopes}), noise model (\ref{cut:noise_model}), and level-shift (\ref{cut:jump}) cuts.
I removed an additional 8\% of the CES-diodes.
\item The same as \ref{config:18} except I tightened the slopes (\ref{cut:slopes}) and noise-model (\ref{cut:noise_model}) cuts, rejecting an additional 4\% of the CES-diodes.
I updated the Type-B database (w91\_typeb, not correction parameters) to reflect new CES definitions.
\item The same as \ref{config:18} except I tightened the noise-model cut (\ref{cut:noise_model}, below 1\,Hz).
\item The same as \ref{config:18} except I tightened the noise-model cut (\ref{cut:noise_model_near_scan}).
\item \label{config:30} The same as \ref{config:18} except I tightened the cut on noise-power-spectrum spikes (\ref{cut:FFT_spike}). 
\item \label{config:31} The same as \ref{config:18} except I tightened the cut on noise-power-spectrum spikes near 1.2\,Hz (\ref{cut:FFT_spike}).
\item The same as \ref{config:18} except I loosened the noise-model cut (\ref{cut:noise_model}, below 1\,Hz).
\item The same as \ref{config:18} except I loosened the noise-model cut (\ref{cut:noise_model_near_scan}).
\item \label{config:34} The same as \ref{config:30} except I tightened the cut further to 4.1\,$\sigma$. 
\item The same as \ref{config:31} except I tightened the cut further to 2.7\,$\sigma$.
\item The same as \ref{config:34} except I tightened the cut further to 3.9\,$\sigma$.
\item The same as \ref{config:18} except I tightened the 1-s level-shift cut (\ref{cut:jump}).
\item The same as \ref{config:18} except I loosened the 1-s level-shift cut (\ref{cut:jump}).
We also updated w91\_timestream.
\item The same as \ref{config:18} except I tightened the 10-s level-shift cut (\ref{cut:jump}).
\item \label{config:40} The same as \ref{config:18} except I tightened the 100-s level-shift cut (\ref{cut:jump}).
\item \label{config:41} The same as \ref{config:18} except I tightened the knee frequency cut (\ref{cut:knee}).
\item The same as \ref{config:18} except I averaged the two 50-Hz phase-switch states instead of differencing them.
For ideal modules the result would have no signal and the same noise level as DD data.
However, because of the phase-switch imbalance, there is worse 1/f noise than the DD data.
The total $(\chi^\textrm{null})^2$ was too large.
\item The same as \ref{config:18} except I tightened the slopes cut (\ref{cut:slopes}).
\item The same as \ref{config:18} except I loosened the 10-s level-shift cut (\ref{cut:jump}).
\item \label{config:45} The same as \ref{config:40} except I tightened the threshold to 18\,$\sigma$.
\item The same as \ref{config:18} except I tightened the non-stationary noise cut (\ref{cut:noise_stationarity}).
\item  The same as \ref{config:18} except I increased the number of bins for the ground filter to 80 to test for residual ground pickup.
\item The same as \ref{config:41} except I tightened the threshold to 0.9 times the scan frequency.
\item The same as \ref{config:18} except I loosened the slopes cut.
\item The same as \ref{config:45} except I tightened the threshold to 17\,$\sigma$.
\item The same as \ref{config:18} except I loosened the non-stationary noise cut (\ref{cut:noise_stationarity}) threshold to 3.5.
\item \label{config:52} The same as \ref{config:18} except I updated the database (quiet\_quality), array pointing, and Type-B correction to the final values.
\item The same as \ref{config:18} except I updated the database (quiet\_quality) to the final values.
\item The same as \ref{config:52} except I tightened the noise-model, level-shift, power-spectrum spike, knee-frequency, and non-stationary noise cuts.
The largest $(\chi^\textrm{null})^2$ PTE for patch 4a was 3\%.
\item \label{config:55} The same as \ref{config:52} except I added the cut on scan-synchronous signal (\ref{cut:sss}).
\item The same as \ref{config:55} except I tightened the knee-frequency threshold to the scan frequency.
\item The same as  \ref{config:52} except I masked sources from the \textit{WMAP} and \textit{Planck} catalogs.
\item This was a test for new pipeline features.  There were no null-test results.
\item \label{config:59} The same as \ref{config:55} except I updated the beam shape, absolute responsivity, and pointing mount model.
I adjusted the null-test division thresholds to evenly divide the CES-diodes.
The patch 4a total $(\chi^\textrm{null})^2$ was too large.
\item The same as \ref{config:59} except I reverted the pointing model to \ref{config:55}.
This fixed the patch 4a failure.
\item The same as \ref{config:59} except I reverted the null-test thresholds to \ref{config:55}.
\item The same as \ref{config:59} except I reverted the data selection to \ref{config:52}. 
\item \label{config:63} The same as \ref{config:59} except I fixed a bug in the pointing model.
The patch 4a total $(\chi^\textrm{null})^2$ for EE only was unexpectedly large.
\item The same as \ref{config:59} except I reverted the data selection to \ref{config:23}.
\item The same as \ref{config:59} except I removed deck flexure from the pointing model.
\item The same as \ref{config:59} except I removed deck flexure and variable encoder offsets from the pointing model. 
\item The same as \ref{config:63} except I increased the number of ground filter bins to 80.
\item The same as \ref{config:63} except I reverted the data selection to \ref{config:23} and included the scan-synchronous signal cut.
\item The same as \ref{config:63} except I reverted the data selection to \ref{config:26} and included the scan-synchronous signal cut. 
\item The same as \ref{config:63} except I reverted the data selection to \ref{config:19} and included the scan-synchronous signal cut. 
\item The same as \ref{config:63} except I doubled the number of azimuth divisions for cross-correlation (to 10).
\item The same as \ref{config:63} except I changed the azimuth divisions for cross-correlation to separate CESes with different scan periods.
\item The same as \ref{config:63} except I reduced the number of deck divisions for cross-correlation (to 4), pairing divisions $180\degr$ apart.
\item The same as \ref{config:63} except I updated the beam shape and absolute responsivity.  This was the final analysis configuration.
\end{enumerate}

\svnid{$Id: error_bar_scaling.tex 119 2012-05-16 22:48:24Z ibuder $}

\chapter{Statistical Uncertainty Scaling Between Configurations}
\label{app:error_bar_scaling}
When comparing two configurations with different data selection, the statistical uncertainty on the difference is smaller than the uncertainty for either configuration.
To reduce the need for additional simulations, I developed a scaling method to estimate the expected statistical uncertainty.
This method has two parts.
First, I derived an approximate uncertainty for the case of instrument noise only \citep{cut_error_scaling}.
Second, I extended the method to account for sample variance \citep{cmb_instrument_noise_scaling}.

\section{Configuration-difference Error Bar}
\label{sec:config_difference_error_bar}
Let the two configurations be denoted 1 and 2.  Let a denote the data common to 1 and 2; b, the data only accepted in 1; and c, the data only accepted in
 2.
Let $x_1$ be the power spectrum in Configuration 1, and $x_a$, the power spectrum computed from data a alone.
Let $\sigma_1$ and $\sigma_a$ be the corresponding statistical errors.\footnote{In this case the instrument-noise--only error from noise simulations should be used.  }
I assumed
\begin{equation}
\label{eq:scaling:weight}
x_1 = \frac{x_a\sigma_b^2 + x_b\sigma_a^2}{\sigma_a^2 + \sigma_b^2}
\end{equation}
as a heuristic approximation to the weighting of data a and b in Configuration 1.
By symmetry,
\begin{equation}
x_2 = \frac{x_a\sigma_c^2 + x_c\sigma_a^2}{\sigma_a^2 + \sigma_c^2}.
\end{equation}
The power spectrum difference
\begin{equation}
x_1 - x_2 = \frac{(x_a\sigma_b^2 + x_b\sigma_a^2) (\sigma_a^2 + \sigma_c^2) - (x_a\sigma_c^2 + x_c\sigma_a^2) (\sigma_a^2 + \sigma_b^2)}{W},
\end{equation}
where
\begin{equation}
W \equiv (\sigma_a^2 + \sigma_b^2) ( \sigma_a^2 + \sigma_c^2).
\end{equation}
Simplifying,
\begin{equation}
x_1 - x_2 = \frac{x_a\sigma_a^2\sigma_b^2 + x_b\sigma_a^4 + x_b\sigma_a^2\sigma_c^2 - x_a\sigma_a^2\sigma_c^2 - x_c\sigma_a^4 - x_c\sigma_a^2\sigma_b^2}{W}.
\end{equation}
Since a, b, and c are disjoint, they have independent fluctuations.  Therefore their error bars on the difference add in quadrature as
\begin{equation}
\sigma_{x_1-x_2} = \sqrt{A^2 + B^2 + C^2},
\end{equation}
where
\begin{eqnarray}
A &\equiv& \frac{\sigma_a^3 ( \sigma_b^2 - \sigma_c^2)}{W}\\
B &\equiv& \frac{\sigma_b ( \sigma_a^4 + \sigma_a^2\sigma_c^2 )}{W}\\
C &\equiv& \frac{\sigma_c (\sigma_a^4 + \sigma_a^2\sigma_b^2 )}{W}.
\end{eqnarray}
$B$ and $C$ can be attributed to the difference in data between 1 and 2.  $A$ is the fluctuation of data a times the change in its weighting between 1 and 2.  Since I usually investigated small changes in data selection,
\begin{equation}
\sigma_b, \sigma_c \gg \sigma_a
\end{equation}
so typically $B, C \gg A$.

Usually I computed $\sigma_1$ and $\sigma_2$ are from noise simulations.
To be consistent with the weighting used in Eq. \ref{eq:scaling:weight} I assumed
\begin{equation}
\label{eq:sigma1}
\frac{1}{\sigma_1^2} = \frac{1}{\sigma_a^2} + \frac{1}{\sigma_b^2}.
\end{equation}
I further assumed that the error bar for different data selections was inversely proportional to the data size
\begin{equation}
\label{eq:sigmaa}
\frac{\sigma_1}{\sigma_a} = \frac{D_a}{D_a + D_b},
\end{equation}
where $D_a$ is the number of CES-diodes in data a.
I used Eq. \ref{eq:sigmaa} to determine $\sigma_a$ when $\sigma_1, D_a, D_b$ are known\footnote{One could instead replace $1\rightarrow 2, b\rightarrow c$ for a different estimate of $\sigma_a$.  The difference between them measures the error introduced by the heuristic assumptions about $\sigma$.  I checked this difference was $\approx$10\% or less.}.
Combining Eqs. \ref{eq:sigma1} and \ref{eq:sigmaa} I solved for
\begin{equation}
\sigma_b = \sigma_1 \frac{D_a + D_b}{\sqrt{2 D_a D_b + D_b^2}}.
\end{equation}
The corresponding equation for
\begin{equation}
\sigma_c = \sigma_2 \frac{D_a + D_c}{\sqrt{2 D_a D_c + D_c^2}}
\end{equation}
is found by replacing $1\rightarrow 2$ and $b\rightarrow c$.

\section{Impact of CMB--instrument-noise Accidental Correlation}
Above I assumed that sample variance (the impact of non-zero CMB power) was negligible.
However, that is not strictly true.
In the configuration comparison tests, I differenced two power spectra computed from the same patch.
Both configurations have the same CMB realization.
However, the instrument noise is different, and there is accidental correlation between the noise and CMB signal.
Although this correlation was 0 on average, the accidental correlation increased the statistical fluctuation of the difference between the spectra from different configurations\footnote{This effect would not have arisen if I had differenced the maps of two configurations and computed the (null) spectra of the difference map.  For historical reasons, the pipeline did not have this facility until very late in the analysis.  For simplicity, I conducted all configuration difference tests by differencing power spectra.  I suggest that future experiments should conduct this type of test by differencing maps.}.
First I calculated this increased fluctuation for a toy model based on a single spatial mode on the sky.
Then I extended the model for the case of multiple modes.
Finally, I provided additional details for the EB spectrum.

\subsection{CMB--instrument-noise Correlation Toy Model}
\label{sec:cmb-instrument:toy}
Suppose there is only a single mode $a$ on the sky.  We measure it with some noise
\begin{equation}
\hat{a} = a^\textrm{CMB} + a^N.
\end{equation}
The measured spectrum is
\begin{equation}
\hat{C} = \hat{a}^2 = (a^\textrm{CMB})^2 + (a^{N})^2 + 2a^\textrm{CMB}a^N.
\end{equation}
The fluctuation of $(a^\textrm{CMB})^2$ is the sample variance, which I ignore because it is the same for two different configurations.
The fluctuation of $(a^{N})^2$ is the noise variance, which I treated in Appendix~\ref{sec:config_difference_error_bar}.
(Our cross-correlation, \S\ref{sec:xcorr}, eliminates the average value of the noise variance---the noise bias---from the result, but we still have the fluctuation.)
The fluctuation of $2a^\textrm{CMB}a^N$ is the new variance I will solve for.
Since we have a fixed sky realization, I treat $a^\textrm{CMB}$ as a constant.
Thus the unknown is the expected variation $\Delta a^N$.
The heuristic error bar gives $\Delta^\textrm{cut}C^N$, the fluctuation of the noise bias between two configurations.
Now I relate it to $\Delta^\textrm{cut}a^N$.
\begin{eqnarray}
C^N &=& (a^N)^2 \\
\Delta C^N &=& 2a^N\Delta a^N \label{eq:DeltaCN} \\
\Delta a^N &=& \frac{\Delta C^N}{2\sqrt{C^N}} \\
\Delta^\textrm{cut} a^N &=& \frac{\Delta^\textrm{cut}C^N}{2\sqrt{C^N}} \label{eq:DeltaaN}
\end{eqnarray}
The fluctuation of $2a^\textrm{CMB}a^N$ is
\begin{eqnarray}
\Delta 2a^\textrm{CMB}a^N = 2a^\textrm{CMB} \Delta a^N
= 2a^\textrm{CMB} \frac{\Delta C^N}{2\sqrt{C^N}}
= \sqrt{\frac{C^\textrm{CMB}}{C^N}}\Delta C^N, \label{eq:newfluct}
\end{eqnarray}
where $C^\textrm{CMB}$ is the (true) CMB power, $C^N$ is the noise bias, and $\Delta C^N$ is the noise error bar due to the configuration difference.
Both $\Delta C^N$ and this new fluctuation are proportional to $\Delta a^N$; however, the constants of proportionality ($a^N$ and $a^\textrm{CMB}$) are uncorrelated.
Therefore, when averaging over many modes, the two fluctuations add in quadrature.
The total fluctuation is $\sqrt{1 + \frac{C^\textrm{CMB}}{C^N}} \Delta C^N$.

\subsection{Multiple Sky Modes}
\label{sec:modes}
Now I consider the more general case of a bandpower $C_b$ which averages $n_b$ modes.
I assume the modes have identical statistical properties.
(In practice they would have a range of $\ell$ and unequal weighting.)
The measured spectrum becomes
\begin{equation}
\hat{C_b} = <\hat{a_{\ell m}}^2> = <(a^\textrm{CMB})^2 + (a^N)^2 + 2a^\textrm{CMB}a^N>,
\end{equation}
where $<>$ indicates the average over $n_b$ modes.
I modified Eq. \ref{eq:DeltaCN} to
\begin{equation}
\Delta C_b^N = 2a_{\ell m}^N\Delta a_{\ell m}^N / \sqrt{n_b}
\end{equation}
because the $n_b$ modes are independent so their fluctuations add in quadrature.
(Equivalently, $C_b^N$ is an average of $n_b$ measurements so the fluctuation of the average $\sim 1/\sqrt{n_b}$.)
Hence Eq. \ref{eq:DeltaaN} becomes
\begin{equation}
\Delta^\textrm{cut} a_{\ell m}^N = \frac{\sqrt{n_b}\Delta^\textrm{cut}C_b^N}{2\sqrt{C_b^N}}.
\end{equation}

The additional fluctuation is now
\begin{eqnarray}
\Delta <2a_{\ell m}^\textrm{CMB}a_{\ell m}^N> &=& \frac{1}{\sqrt{n_b}} \Delta 2a_{\ell m}^\textrm{CMB}a_{\ell m}^N =
\frac{1}{\sqrt{n_b}} 2a_{\ell m}^\textrm{CMB} \Delta a_{\ell m}^N = \nonumber\\
\frac{1}{\sqrt{n_b}} 2a_{\ell m}^\textrm{CMB} \frac{\sqrt{n_b}\Delta^\textrm{cut}C_b^N}{2\sqrt{C_b^N}} &=& 
\sqrt{\frac{C_b^\textrm{CMB}}{C^N_b}}\Delta^\textrm{cut}C_b^N,
\end{eqnarray}
which is equivalent to Eq. \ref{eq:newfluct}.

\subsection{EB Power Spectrum}
\label{sec:config_difference:EB}
The EB spectrum
\begin{equation}
\hat{C_{EB}} = \hat{a_E}\hat{a_B} = (a_E^N+a_E^\textrm{CMB})a_B^N = a_E^Na_B^N + a_E^\textrm{CMB}a_B^N
\end{equation}
is complicated because while there is no signal (I assume the CMB B mode is negligible compared to our noise level), the accidental correlation between the CMB E mode ($a_E^\textrm{CMB}$) and the noise B mode ($a_B^N$) increases the variance.
In the instrument-noise--only case, the error bar is
\begin{equation}
\Delta^\textrm{noise}C_{EB} = \sqrt{ (a_B^N\Delta a_E^N)^2 + (a_E^N\Delta a_B^N)^2 }
\end{equation}
because $a_E^N$ and $a_B^N$ are independent so the two terms add in quadrature.

The error due to the accidental correlation is
\begin{equation}
\Delta a_E^\textrm{CMB}a_B^N = a_E^\textrm{CMB}\Delta a_B^N = \frac{\sqrt{C_{EE}^\textrm{CMB}}\Delta C^N_{BB}}{2\sqrt{C^N_{BB}}},
\end{equation}
where $\Delta C^N_{BB}$ is the BB spectrum error bar (e.g. due to the cut change) and I used Eq. \ref{eq:DeltaaN} to rewrite $\Delta a_B^N$.
This error is uncorrelated with  $\Delta^\textrm{noise}C_{EB}$ for the reason given in
Appendix~\ref{sec:cmb-instrument:toy}.
The total fluctuation is therefore
\begin{equation}
\sqrt{\left(\Delta^\textrm{noise}C_{EB}\right)^2 +
\left( \frac{\sqrt{C_{EE}^\textrm{CMB}}\Delta C^N_{BB}}{2\sqrt{C^N_{BB}}}\right)^2}. \nonumber
\end{equation}

\svnid{$Id: typeb_format.tex 125 2012-05-22 01:51:23Z ibuder $}

\chapter{Type-B--glitch---correction-parameter File Format}
\label{app:typeb_file_format}
This sample Type-B--glitch---correction-parameter file demonstrates the parameter-file format for the Type-B--glitch---correction software.
\begin{verbatim}
__FORMAT_VERSION2__
#Required at the beginning, sets version of the parameter format

#Any line beginning with `#' is a comment

# Sampling rate (Hz) to be used for noise calculation
__DEMOD_SAMPLING_RATE__         100

# Iteration count for Newton's method (solving Delta G)
__N_ITERATIONS__                3

# Factor to be multiplied to the heights below
__HEIGHT_FACTOR__               2.0

# Multi-glitch correction. How many noise sigma from the glitch to correct.
__MULTI_GLITCH_COVERAGE_SIGMA__ 5.0

# Preamp low-pass filter RMS reduction "Preamp factor"
__PREAMP_RMS_FACTOR__
#	Module	Q1	U1	U2	Q2
       0     0.644467     0.643774     0.637945     0.632578
       1     0.576704     0.572848     0.741313     0.607144
#Modules cannot be skipped


# Specify center-height pattern. -1 is non-glitching.
__CENTER_HEIGHT_PATTERN__
#	Module	Q1	U1	U2	Q2
00      0       0       0       0
01      -1       0       0       1
#Modules cannot be skipped

# Detector RMS (bit*sqrt[s]) to be used as default
# Not used in iterative correction
__DETECTOR_RMS__
#       Module  Q1      U1      U2      Q2
00    0.20    0.20    0.20    0.20
01    0.20    0.20    0.20    0.20
#Modules cannot be skipped

# Center-Height data
# Normal pattern
__CENTERS_HEIGHTS__     0
#Location (bits, 0 is the center of the ADC range) 
#                      error(ignored) 
#                                height(bits) 
#                                               error(ignored)
        -65500.300      0.0      11.732002      0.0
        -64455.700      0.0     -6.8432889      0.0
        -63452.300      0.0     11.732002       0.0
        -62407.700      0.0     -0.97366601     0.0
#Location must increase monotonically

# Pattern for other diodes
__CENTERS_HEIGHTS__     1
        -65501.300      0.0      11.732002      0.0
        -64455.700      0.0     -6.8432889      0.0
        -63452.300      0.0     11.732002       0.0
        -62407.700      0.0     -1.97366601     0.0

#More patterns
\end{verbatim}

\newpage

\special{ps:%
  SDict begin [
    /Producer (Revision r\svnrev)
    /Title (Immanuel Buder Thesis)
    /Subject ( )
    /Creator ( )
    /Created (Revision r\svnrev)
    /Author (Immanuel Buder)
    /Keywords ( )
    /DOCINFO pdfmark
  end
}

\phantomsection
\addcontentsline{toc}{chapter}{References}
\begin{singlespace}
\bibliographystyle{yahapj}
\bibliography{thesis}
\end{singlespace}

\end{document}